\documentclass[aps,rmp,floatfix,preprint,nofootinbib]{revtex4}
\pdfoutput=1

\usepackage{amsmath}
\usepackage{xspace}
\usepackage{listings}
\usepackage{epsfig}
\usepackage{multirow}
\usepackage{lineno}
\usepackage[dvipsnames,svgnames]{xcolor} 
\usepackage{soul} 
\usepackage{graphics}
\usepackage{layouts}

\usepackage{aas_macros}
\usepackage{hyperref}
\usepackage{natbib}
\newcommand{\eg}{e.g.\xspace}%\def\eg{e.g.}

\newcommand{\unit}[1]{\ensuremath{\mathrm{\,#1}}\xspace}

\newcommand{\GeV}{\unit{GeV}}

\newcommand{\cm}{\unit{cm}}

\newcommand{\sr}{\unit{sr}}

\newcommand{\secref}[1]{Section~\ref{#1}}

\newcommand{\figref}[1]{Figure~\ref{fig:#1}}

\newcommand{\Fermi}{\textit{Fermi}\xspace}

\newcommand{\pass}[1]{{\tt Pass }{#1}}
\newcommand{\latcls}[1]{{\tt \uppercase{#1}}}

\newcommand{\code}[1]{\lstinline!#1!\xspace}

\newcommand{\sigmav}{\ensuremath{\langle \sigma v \rangle}\xspace}

\newcommand{\bbbar}{\ensuremath{b \bar b}\xspace}

\newcommand{\mumu}{\ensuremath{\mu^{+}\mu^{-}}\xspace}

\providecommand\physrep{\ref@jnl{Phys.~Rep.}}%
\providecommand\apjs{\ref@jnl{ApJS}}%
\providecommand{\jcap}{\ref@jnl{JCAP}}%
\providecommand\epsscale[1]{\gdef\eps@scaling{#1}}%
\providecommand\plotone[1]{%
 \typeout{plotone included the file #1}
 \centering
 \leavevmode
 \includegraphics[width={\eps@scaling\columnwidth}]{#1}%
}%

\newcommand{\cms}{cm$^3$s$^{-1}$}

\newif\ifcolor
\colortrue
\begin{document}

\title{WIMP searches with gamma rays in the Fermi era: challenges, methods and results}

%\author{The Fermi-LAT Collaboration}
%\affiliation{{\it ...}}
%%%%%%%%%%%%%%%%%%%%%%%%%%%%%%%%%%%%%%%%%%
\author{Jan Conrad \footnote{Wallenberg Academy Fellow}}
\affiliation{Oskar Klein Centre, Department of Physics, Stockholm University, Albanova University Center, SE-10691 Stockholm, Sweden}
\author{Johann Cohen-Tanugi}
\affiliation{Laboratoire Univers et Particules de Montpellier, Universit\'e  de Montpellier, CNRS/IN2P3, Montpellier, France}
\author{Louis E. Strigari}
\affiliation{Mitchell Institute for Fundamental Physics and Astronomy, Department of Physics and Astronomy, Texas A \& M University, College Station, TX 77843-4242, USA}

\begin{abstract}
The launch of the gamma-ray telescope \Fermi Large Area Telescope (\Fermi-LAT) started a pivotal period in indirect detection of dark matter. By outperforming expectations, for the first time  a robust and stringent test of the paradigm of weakly interacting massive particles (WIMPs) is within reach. In this paper, we discuss astrophysical targets for WIMP detection and the challenges they present, review the analysis tools which have been employed to tackle these challenges, and summarize the status of constraints on and the claimed detections in the WIMP parameter space.  Methods and results will be discussed in comparison to Imaging Air Cherenkov Telescopes. We also provide an outlook on short term and longer term developments. 
\end{abstract}

\maketitle

\tableofcontents

\section{Introduction}

\par Throughout the course of the past several decades, observational evidence and theoretical models have converged to strongly suggest the presence of a dark component of the matter content of the Universe, most likely consisting of one or several new particles of Nature. Identifying the nature of dark matter is a primary topic of study in modern science. 

\par A variety of experiments around the world are continually improving their sensitivities, ruling out theoretically well-motivated regimes of parameter space for popular extensions to the standard model of particle physics that include Supersymmetry and Universal Extra Dimensions~\citep{Aprile:2012nq,Akerib:2013tjd,Agnese:2014aze}. As dark matter detection experiments continue to improve in sensitivity, new experimental and theoretical challenges will continue to be confronted. 

\par Of particular interest are dark matter particles with weak interactions, labeled as WIMPs (Weakly-Interacting Massive Particles). We will here briefly discuss the reason for the WIMP's popularity, more in-depth discussions can be found for example in~\citet{Jungman:1995df,Bertone:2004pz,Bergstrom:2012fi,Baer:2014eja}.

\par At high temperatures ($T >>  m_{\rm WIMP}$, where $m_{\rm WIMP}$ is the WIMP mass), particles could be thermally created and destroyed, implying a $T^4$ distribution (the Boltzmann law) for their state density. As the temperature decreases, the density is exponentially suppressed ($\propto \exp{-m_{\rm WIMP}/T}$) with temperature. Chemical equilibrium is left when the temperature is no longer high-enough to pair-create WIMPs, at which point their number density decreases. When the WIMP mean free path is comparable to the Hubble distance, the particles also cease to annihilate and leave thermal equilibrium, commonly referred to as ``freeze-out''. At this point, the co-moving density  remains constant. Solving for the Boltzmann equation, one finds that the temperature for which the freeze-out occurs is about 5\% of the WIMP's mass, thus the density becomes constant (so called ``relic density'')  when the particles are already essentially non-relativistic. Consequently,  considering only s-wave annihilation, the relic density depends only on the total annihilation cross-section and the velocity distribution:
\begin{equation}
\Omega_{\rm WIMP}h^2 \sim \frac{3.1 \cdot 10^{-27}}{<\sigma_A |\vec{v}|>}\quad ,
\label{eq:relicdensity}
\end{equation}
\noindent where the average is taken over velocities and angles. The scale for weak interaction strength ($\sim \alpha^2/m_{\rm WIMP}^2$) implies that $<\sigma_A |\vec{v}|> \sim 10^{-25}$ \cms, where the WIMP mass is taken to be 100 GeV. The resulting relic density for such a particle would be within a factor 3 of the  measured relic density of $\Omega_{\rm WIMP}h^2 = 0.1199 \pm 0.0027$ provided by the most recent PLANCK results \cite{Planck:2015xua}. This remarkable coincidence is sometimes referred to as the ``WIMP miracle.'' A more careful calculation for the cross-section of thermal WIMPs is presented in~\citet{Steigman:2012nb}, but does not alter the main argument.

\par There are a variety of WIMP search methods ongoing, and for dark matter particles more generally. They are broadly classified as follows: 

\bigskip

\noindent 
$\bullet$ {\em Indirect searches} measure the annihilation and/or decay products of dark matter from different astronomical environments in the Universe. They specifically limit the rate at which dark matter particles annihilate or decay into standard model particles. Many indirect searches are now underway to detect the remnants of dark matter particles that have annihilated into standard model final states. The most readily accessible standard model particles are photons, in particular high-energy gamma rays. Neutrinos, antiprotons, and positrons may also be detected by modern experiments. The annihilation cross section probed by indirect searches is most closely related to the process that sets the dark matter abundance in the early universe, assuming that the dark matter was once in thermal equilibrium. 

\noindent 
$\bullet$ {\em Direct searches} measure the scattering of dark matter off of nuclei in low background underground detectors, where the collision of dark matter is deduced through energy input into particles in the detector. Though the WIMP-nucleus scattering cross section is not as simply connected to the dark matter relic abundance as is the annihilation cross section, many modern experiments are now probing well-motivated theoretical models, as the generic WIMP cross section is determined by its de Broglie wavelength and the weak interaction scale. 
%In some instances intriguing results have been reported.

\noindent 
$\bullet$ {\em Collider searches} measure the production of dark matter through the collision of high energy standard model particles, such as protons and electrons. New stable dark matter particles produced are identified either by initial state radiation or through the production of quarks and gluons, which eventually decay down into the lightest stable particle in the spectrum. 

\bigskip

\par Now that we are in an era in which several different experiments are able to test theoretically well-motivated particle dark matter models, it is timely to examine the specific contributions that have been made by the most important experiments. This review article will focus exclusively on the recent advances in indirect dark matter searches, and within this subject concentrate exclusively on indirect dark matter searches using gamma rays. Now that we have had an influx of data over the past few years from these experiments, it is important to assimilate this information into the broader context of particle dark matter searches.

\par Of particular interest are the results that have been obtained by the \Fermi Large Area Telescope (\Fermi-LAT), launched in June 2008 into low Earth orbit. It was designed as the successor of the successful EGRET mission, with an order of magnitude better energy and angular resolution than EGRET. \Fermi consists of two experiments, the Large Area Telescope (LAT), which is the primary instrument aboard \Fermi, and the gamma-ray burst monitor. The LAT is sensitive to photons in the range of approximately $20 \, \textrm{MeV} -300 \, \textrm{GeV}$. The LAT is an imaging, wide field-of-view pair conversion telescope that measures electron and position tracks that result from the pair conversion of an incident high-energy gamma-ray in converter foils. The energy resolution over the energy range of interest is approximately 10\%, and the effective area of the LAT is approximately $10^4$ cm$^2$. 

\par In this article we review the constraints on particle dark matter that have been obtained from the first five years of the \Fermi-LAT and compare them with those obtained by Imaging Air Cherenkov Telescopes (IACT). This article has two main focuses. First, to review the experimental methods that are used to analyze  \Fermi-LAT and IACT. Second, to review the impact that uncertainties in astrophysical dark matter distributions have on the interpretation of limits on the dark matter particle mass and cross section. For a recent analysis of particle dark matter models that are probed by indirect detection experiments, see e.g.~\citet{Bergstrom:2012fi}.

\par In our discussion of the experimental methods that are used to analyze the data, we review in detail the statistical techniques that have been developed, how they are applied to different sources in the universe, and we highlight which of these sources provide the best detection prospects. 

\par In our discussion of astrophysical systematics, we review methods to determine dark matter distributions in all of the various sources that have been studied by \Fermi-LAT. We discuss not only how the mass distribution of the Milky Way, dwarf galaxies, and galaxy clusters are determined, but also look forward to how future astronomical surveys and measurements will improve the understanding of dark matter distributions. We discuss the connection that will continue to be strengthened between astroparticle experiments and larger scale astronomical surveys. 

\par This review is organized as follows. In \secref{sec:targets} we discuss the different astrophysical targets that are used for indirect detection, and the pros and cons of each. In~\secref{sec:analysis}, we review different types of gamma-ray instruments, and discuss the analysis challenges that they face. In~\secref{sec:status}, we review the current status of indirect dark matter searches with gamma rays. In~\secref{sec:perspectives}, we discuss what future gamma-ray experiments will bring to the field of indirect dark matter detection, and in~\secref{sec:conclusions} we present conclusions. 

\section{Targets: promises and challenges}
\label{sec:targets}
\par Astrophysical uncertainties have long presented a systematic uncertainty in searches for new physics. Focusing on gamma-ray searches for dark matter, there are two broad types of systematic uncertainties that must be accounted for: uncertainties in modeling of dark matter distributions, and uncertainties in gamma-ray emission from non-dark matter sources. In this section, we primarily discuss the contribution of the former type of uncertainty, and defer to discussion of the latter type of uncertainty when we present the results of the gamma-ray observations in the following sections.  

\par Within the context of the $\Lambda$-Cold Dark Matter ($\Lambda$CDM) paradigm~\cite{Frenk:2012ph}, observed galaxies are formed within dark matter halos. These halos are formed through a sequence of mergers of lower mass halos and a process of smooth accretion onto the halo. As a result, dark matter halos are complex systems that are not totally ``smooth," but rather have features in their phase space in the form of subhalos, or substructure, and tidal debris that reflects their interaction within the larger halo~\cite{Diemand:2008in,Vogelsberger:2008qb}. Dark matter halos are not predicted to be perfectly spherical, but rather retain shapes that reflect their formation and interaction histories~\cite{Allgood:2005eu}. 

\par Though not spherical, it is often convenient to express the density distribution of dark matter halos in a spherically averaged form, for example as 
\begin{equation} 
\rho(r) = \frac{\rho_s}{(r/r_s)^\gamma (1 + (r/r_s)^b)^{(c-\gamma)/b}}. 
\label{eq:gnfw} 
\end{equation}
Here $\rho_s$ and $r_s$ are scale density and scale radius parameters, respectively, $b$ represents the turnover from the asymptotic power-law slope in the inner regime, $\gamma$, and the asymptotic power-law slope in the outer regime, $c$. The set of parameters $(\gamma, b,c) = (1,1,3)$ give the well-known Navarro-Frenk-White profile \citep[NFW,][]{Navarro:1996gj}. 

\par Though an NFW profile is now a robust prediction of dark matter-only simulations of halos, as we discuss below in this section there is debate as to how well this profile describes observations of the dark matter halos over a wide range of mass scales. In spite of this debate, Equation~\ref{eq:gnfw} provides a useful starting point for phenomenological studies of dark matter halos, and will be referred to throughout the course of this section and review. 

\par With a well-motivated model for dark matter halos in place, we can make predictions for gamma-ray fluxes, and the corresponding uncertainties, from different astrophysical targets. We in particular focus on the uncertainties in the dark matter properties of each of these systems, and thus the observational uncertainties on the quantity,  
\begin{equation} 
J(\Delta \Omega) = \frac{1}{2} \int_0^{\Delta \Omega}  \sin \Psi d\Psi \int_{\ell_{-}}^{\ell_{+}} \rho^2 [r(\ell)] d\ell\quad ,
\label{eq:Jvalue}
\end{equation} 
\noindent where $\Psi$ represents the angular separation from the center of the halo, which is typically deduced from the position of the center of the observed galaxy. Here $D$ is the distance to the center of the galaxy, so that $r^2 = \ell^2 + D^2 - 2 \ell D\cos \Psi$. The upper and lower boundaries to the integral are $\ell_\pm = D \cos \Psi \pm \sqrt{r_t^2 - D^2 \sin^2 \Psi}$, where $r_t$ is the tidal radius of the dark matter halo. Equation~\ref{eq:Jvalue} will be referred to here as the ``$J$-factor." This sets the dependence of the annihilation signal on the dark matter distribution in any astrophysical system. As we highlight in this section, the determination of the density profile from observation is not straightforward, and thus the uncertainty in the $J$-factor translates into a systematic uncertainty in the gamma-ray flux determination. 

\par We begin by examining the dark matter distribution in the Milky Way galaxy, including the Galactic center. We then move on to discuss determination of the dark matter distributions in the dwarf spheroidals of the Milky Way, which are probably the simplest astronomical systems from which the dark matter distributions can be derived. We then follow up with a discussion on dark matter substructures, Galaxy clusters, the Galactic and the extragalactic dark matter distributions. 

\subsection{Galactic Center}
\par Most likely, the largest flux of gamma rays from dark matter annihilation comes from the Galactic center because of its high concentration of dark matter and close proximity. However, a drawback of the Galactic center as a target is that there is a substantial population of gamma-ray sources and there is diffuse emission from cosmic rays that must be well-understood in order to extract the dark matter signal. An additional drawback of the Galactic center as a target for gamma-ray and dark matter studies is that it has proven challenging to extract the dark matter distribution in the Galactic center. In the central few parsecs, the stellar mass from the bulge dominates the dynamics~\cite{Schodel2009}. Even including a weak disk component, the shape of the dark matter distribution is unconstrained, so that it is not possible to tell if the dark matter profile rises to a central cusp-like structure or has a constant density within the bulge region~\cite{Iocco:2011jz}. A conservative upper bound on the contribution of the dark matter may be set by the upper limit deduced from the bulge contribution to the potential.

\par From a theoretical perspective, the dark matter distribution near the black hole at the center of the Milky Way is modified by the black hole itself~\cite{Gondolo:1999ef}. Adiabatic growth of the central black hole may steepen the central cusp of dark matter. Assuming an initial dark matter profile in the Galactic center of the form $\rho \propto r^{-\gamma}$, from conservation of mass and angular momentum, the final mass profile scales as $\rho \propto r^{-A}$, where $A = (9-2\gamma)/(4-\gamma)$~\cite{Ullio:2001fb}.~\citet{Gondolo:1999ef} derive a lower bound of $\sim 0.24$ GeV cm$^{-3}$ of dark matter near the Galactic center. This result relies on assumptions that are not well understood, for example the adiabatic nature for the growth of the black hole itself, and scattering of dark matter off of stars in the central nuclear star cluster~\cite{Bertone:2005xv}. It is also possible for dark matter to interact with the central nuclear star cluster and form a profile that is shallower than the NFW profile~\cite{Gnedin:2003rj}. 

\par Numerical simulations have examined the effect of baryonic physics on the structure of Milky Way-mass dark matter halos.  Though the inclusion of baryons leads to many challenges,  numerical simulations have continued to make progress. Simulations of single halos, in combination with analytic models, show that the central densities of dark matter halos become less steep than those found in pure N-body simulations because the baryons induce repeated epochs of feedback due to star formation activity~\cite{Pontzen:2011ty}. 

\par The lack of consensus from both the observational and theoretical side on the nature of the Milky Way dark matter density profile provides a significant systematic uncertainty that must be accounted for in gamma-ray searches for dark matter. In fact it is likely that, from a pure astronomical perspective, this lack of consensus will remain for some time.~\citet{Iocco:2015xga} have recently compiled known data to analyze the dark matter content within the Solar circle, but are insensitive to the dark matter density profile. Their claim that this estimate constitutes the first evidence for dark matter has however triggered some debate\cite{McGaugh:2015tha}. 

Kinematic data from the GAIA satellite is expected to somewhat improve on the systematic uncertainty in the measurement of the local dark matter density~\cite{Bovy:2013raa}, though it will be more difficult to use this data to determine the shape of the profile near the Galactic center. 

\subsection{Milky Way mass profile}
\par The discussion above focused on the central density profile of the Milky Way, i.e. from the Solar circle towards the Galactic center. Going in the other direction, the density profile of the Milky Way as measured from the Solar radius out to the virial radius is important for determinations of diffuse gamma-ray emission. However, because of our position within the Milky Way's dark matter halo, it is more difficult to determine its dark matter density profile than it is for many external galaxies. 

\par The best measurements of the integrated mass of the Milky Way come from spectroscopy of stars in the outer region of the dark matter halo. The Sloan Digital Sky Survey (SDSS) survey has measured the dark matter mass profile of the Milky Way using the kinematics of a large sample of Blue Horizontal Branch (BHB) stars~\cite{Xue:2008se}. BHB stars are important tracers of the Milky Way mass because they are both intrinsically bright and have accurate distance measurements. Due to the average radius of the BHB stars, they provide the best constraint on the Milky Way mass within a radius of about 60 kpc, measuring a mass of $\sim 4 \times 10^{11} M_\odot$. This is consistent with the results of independent analyses~\cite{Deason:2012wm}, and with estimates at larger radii, which find a mass of $\sim 7 \times 10^{11} M_\odot$ enclosed within 80 kpc~\cite{2010ApJ...720L.108G}. Several authors have used samples of bright BHB stars to extrapolate and determine the total mass of the Milky Way. They have reported a total mass of the Milky Way ranging anywhere from $0.5 - 2.5 \times 10^{12} M_\odot$~\cite{2005MNRAS.364..433B,Xue:2008se,Busha:2010sg}, with the variation depending on the exact analysis method and the sample of stars used.  These results can be compared to updated implementation of the timing argument, which implies a total Milky Way mass of $\sim 2 \times 10^{12}$ M$_\odot$~\cite{Li:2007eg}. 

\par Distant satellite galaxies can also be used as tracers of the mass distribution of the Milky Way~\cite{Little:1987}. For these measurements, there are two systematic uncertainty that are particularly important. The first is the due to the uncertain density distribution of the satellites in the Milky Way; this is mainly due to the small number of satellites that are currently known, about a couple dozen. The second, and perhaps probably the most important systematic uncertainty, involves understanding whether the distant satellite Leo I is bound to the Galaxy.  Because Leo I is moving at a high Galactocentric velocity and is at a Galactocentric radius of 260 kpc, it is not yet clear whether this is an outlier that is bound to the halo or if it is unbound and is on its first pass through the Galaxy. Several recent proper motions do in fact seem to indicate that it may be bound to the Galaxy~\cite{BoylanKolchin:2012xy}. The Milky Way mass measurements from both the BHB stars and the satellite galaxies are in good agreement with the recent measurements of the Milky Way escape velocity using a local sample of high velocity stars~\cite{2014A&A...562A..91P} and constraints using the Sagittarius tidal stream \cite{Gibbons:2014ewa}.

\par Similar to the case of the Galactic center, the factor of a few uncertainty that still lingers in the measurement of the mass of the Milky Way's dark matter halo provides a systematic uncertainty in the predicted gamma-ray emission from dark matter annihilation. 

%Independent of this uncertainty in the measured profile, the gamma-ray analysis is made challenging because of the large regions of sky that must be analyzed. This introduces a systematic uncertainty due to spatial variations in the diffuse gamma-ray background.  As a result, any possible dark matter signal that may arise from the diffuse Galactic gamma-ray emission must be corroborated by independent sources. 

\subsection{Dwarf Spheroidal galaxies}

\par There are now approximately two dozen satellite galaxies of the Milky Way that 
are classified as dwarf Spheroidal (dSphs). The dSphs are conveniently classified 
according to the period of discovery. The first nine dSphs that were discovered before 
the turn of the century have come to be classified as ``classical" satellites, while those 
discovered in the era of the SDSS  are largely referred to as
``ultra-faint" satellites. The dSphs range in Galactocentric distance from approximately 
$15-250$ kpc, and their overall distance distribution in the Galactic halo is much more 
extended than the more centrally-concentrated globular cluster population. Nearly all of 
the dSphs are devoid of gas up to the present observational limits. For a more thorough discussion 
of astrophysical aspects of dSphs see~\citet{2012AJ....144....4M,Walker:2012td,Strigari:2013iaa}.

\par Dwarf spheroidals provide excellent targets for
gamma-ray searches for WIMPs for several reasons. First, theoretically
there is expected to be no gamma-ray point sources and no intrinsic
diffuse emission associated with them. Second, their dark matter
distributions are directly derived from the stellar kinematics.
Third, the boost factor from dark matter sub-structure is predicted to be
negligible in these systems, so the interpretation of the limits (or
detections) from them is much more straightforward than it is for other
astrophysical targets (see more detailed discussion below on boost from halo substructure).

\par Focusing specifically on dSphs, it is convenient to start by assuming that their stellar distributions and dark matter distributions
are spherically-symmetric. It has become standard to model the measured line-of-sight velocities of stars in dSphs with the spherical Jeans equation, 
and thus extract a measure of the dark matter content. Though in the Jeans equation the mass profile is the sum of the contribution from the dark
matter and the stars, in nearly all cases of interest for gamma-ray studies the stars contribute
little to the total gravitational potential. The measured line-of-sight velocity of
a star in a dSph is a mixture of the radial and tangential velocity
components, and therefore depends on the intrinsic velocity anisotropy of
the stars. The anisotropy is traditionally defined as the following parameter
\begin{equation}
\beta = 1 - \frac{\sigma_t^2}{\sigma_r^2}, 
\label{eq:beta} 
\end{equation} 
where $\sigma_r^2$ is the radial velocity dispersion and $\sigma_t^2$ is the tangential velocity dispersion. 
At a projected position $R$, the line-of-sight velocity dispersion is 
\begin{equation}
\sigma_{los}^2(R) = \frac{2}{I_s(R)} \int_R^\infty
\left[1-\beta(r)\frac{R^2}{r^2}\right] \frac{\rho_s \sigma_r^2
r}{\sqrt{r^2-R^2}} dr.
\label{eq:jeans_los}
\end{equation} 
The three-dimensional radial velocity dispersion $\sigma_r^2$ is determined
from the Jeans equation, and the projected stellar density profile, $I_s(R)$, is 
determined from fits to the stellar number counts as a function of projected radius. 
It is manifestly clear from Equation~\ref{eq:jeans_los}
that there is a degeneracy between the mass profile and the velocity
anisotropy of the stars.

\par The half-light radii for the classical dSphs are typically of the
order few hundreds of parsecs, and their luminosities spread over a range
of nearly two orders of magnitude, approximately $10^5 - 10^7$ L$_\odot$.
The photometric profiles as derived from star counts are typically
consistent with a cored model in projection, followed by a turnover into
an exponential fall-off in the outer region of the galaxy where the
stellar density blends into the background star counts. For the classical
dSphs there are hundreds, and in some cases thousands, of bright giant
stars that have measured velocities to a precision of a few km/s or less
\cite{Walker:2008ax}. 

\par Constraints on the dark matter distribution of dSphs have been specifically calculated via the following procedure. The kinematic data from a dSph, below defined as ${\cal D}_k$, comprises of $n$ stars each with a measured line-of-sight velocity $v_\imath$ and uncertainty $\sigma_{m,\imath}$ at a projected radius $R_\imath$. Define $\vec a$ as a set of theoretical parameters that are to be extracted from the data. The probability for the data, given the model parameters, is 
\begin{equation}
{\cal P}_{kin}({\cal D}_k | \vec{a}) \propto \prod_{\imath = 1}^n \frac{1}{\sqrt{\sigma_{m,\imath}^2 + \sigma_{los}^2 (R_\imath)}} \exp^{ - 0.5 \frac{(v_\imath - \bar v)^2}{\sigma_{m,\imath}^2 + \sigma_{los}^2 (R_\imath)}},
\label{eq:dsphlike}
\end{equation}
where $\bar v$ is the mean velocity. The projected line-of-sight velocity dispersion, $\sigma_{los}(R)$, which is calculated from the spherical Jeans equation, depends on the parameters of the dark matter density profile. A standard assumption for the dark matter density profile is the generalized double power law model in Eq.~\ref{eq:gnfw}. 

\par With aforementioned improvements in observational data and
theoretical modeling of dSphs around the Milky
Way, it has become timely to search for
the presence of NFW-like dark matter
density profiles predicted by $\Lambda$-CDM. However, in
spite of the high-quality modern data sets, there has been
significant debate as to whether the data are unable to uniquely
specify a model for the dark matter potential. Indeed, the
kinematic data of several dSphs, when modeled as a single
stellar population with spherical Jeans-based models, 
is consistent with both cusped dark matter profiles~\citep{Strigari:2006rd} and cored
profiles~\citep{Gilmore:2007,Richardson:2013lja}.

\par The assumption of the spherical Jeans equation is of course an approximation 
to the true dynamical state of the dSph. Indeed, all of the dSphs are observed to 
be elongated, with an approximate 30\% difference between the length of the major
and minor axis~\cite{McConnachie:2012AJ....144....4M}. Further, the solutions to the 
spherical Jeans equations need not admit fully self-consistent dynamical solutions with 
a positive definite stellar distribution function. Recent work has relaxed the assumption of 
spherical symmetry in the Jeans equations, in most cases finding that the central mass distributions
are consistent with the spherical case~\cite{Hayashi:2012si}. There has also been recent work 
dedicated to establishing self-consistent distribution function models of the dSphs. For example, 
it is possible to find self-consistent solutions in which the orbits of the stars are isotropic, and the 
stars trace the dark matter profile in the central region~\cite{Strigari:2010un}. Orbit-based models
of the dSphs are now being developed; these types of models are ideal in the sense that the 
distribution function is obtained in a model-independent manner~\cite{Jardel:2012am,Breddels:2012cq}. 
All of this modeling indicates that, when considering the observed stars as a single stellar 
population, the mass estimates are in good agreement with those obtained through the Jeans equation, 
and further it is not possible to conclusively establish whether there is a dark matter core or cusp within
these systems. 

\par Some dSphs exhibit evidence for more than a single stellar population, i.e. populations of stars that formed at different epochs. Two well-known examples are Sculptor and Fornax, with Sculptor, which is located at an approximate Galactocentric distance of 80 kpc, receiving a significant amount of attention. Phase space modeling of the kinematics of its stars, when taken as a single stellar population, indicate that it is possible to find NFW-based dark matter profiles with isotropic stellar velocity dispersions~\cite{Strigari:2010un,Breddels:2012cq}. Due to high quality information on both the kinematics and the metallicity of stars in Sculptor, it is possible to break up the stars into distinct populations. In particular,~\citet{Battaglia:2008jz} (B08) have shown that there are two distinct stellar populations; a metal rich population that is centrally-concentrated, and a more extended metal-poor population. Using Jeans-based modeling, B08 showed that in order for both populations to be embedded into a single dark matter halo, both the metal rich and metal poor populations must transition from isotropic in the center to predominantly radially-biased orbits in the outer regions. Following up on the B08 analysis, ~\citet{Agnello:2012uc} apply the projected virial theorem to the Sculptor data, and find that it is not possible to self-consistently embed both populations into a halo with an NFW density profile. Further,~\citet{Amorisco:2011hb} study Michie-King models for the stellar distribution function, and find that even though NFW models provide an acceptable $\chi^2$ fit to the data, in general cored models are preferred. 

\par The analysis of B08 was then soon followed up upon with an independent and larger data set by~\citet{Walker:2011zu} (WP11). With their data set, WP11 also present evidence for two populations, with seemingly similar velocity dispersion and half-light radii to what was obtained by B08. Though there is no observed evidence for two stellar populations in the WP11 data, they extract the dispersion, half-light radii, and metallicity of their populations using a statistical algorithm. WP11 then apply the mass estimator presented in~\citet{Walker:2009zp} and~\citet{Wolf:2009tu} to the two populations, finding that an NFW model is ruled out at the 99\% c.l. While this evidence for a dark matter core is intriguing, and has garnered a substantial amount of attention in the dark matter community, it has not been able to stand up to more rigorous modeling of the two populations in Sculptor. In particular, it is possible to find a self-consistent stellar distribution function model with an NFW dark matter profile that is able to statistically-describe the two populations in Sculptor~\cite{Strigari:2014yea}. 

\par The above discussion underscores the difficulty in determining whether dark matter cores or cusps exist in dSphs. 
While the kinematic data is unable to determine the slope
of the central density of the dark matter in dSphs, it is much
more effective at determining the integrated mass within the
half-light radius, approximately a few hundred
parsecs~\cite{Walker:2009zp,Wolf:2009tu}. This is weakly
dependent on whether there is a central core or cusp in the
dSph, so long as the log-slope is $-d (\log \rho) /d \log r
\lesssim 1.5$. This implies that the constraints on the mass
profile directly translate into constraints on  
the $J$-factor in Equation~\ref{eq:Jvalue} within the same
region~\cite{Strigari:2006rd,Strigari:2007at}. 
For a dSph at a distance of $\sim 50-100$ kpc, the half-light radius corresponds
to less than approximately one degree, which is about the
angular resolution of the \Fermi-LAT over a large energy range
of interest. This is the region within which the integrated
density and the integrated density-squared are the best
constrained from the kinematic data sets. This implies that the
assumption of a core or a cusp for the density profile does not
significantly affect the gamma-ray flux predictions for the
\Fermi-LAT.

\par
\begin{figure}[h]
\centering
\includegraphics[scale=0.3]{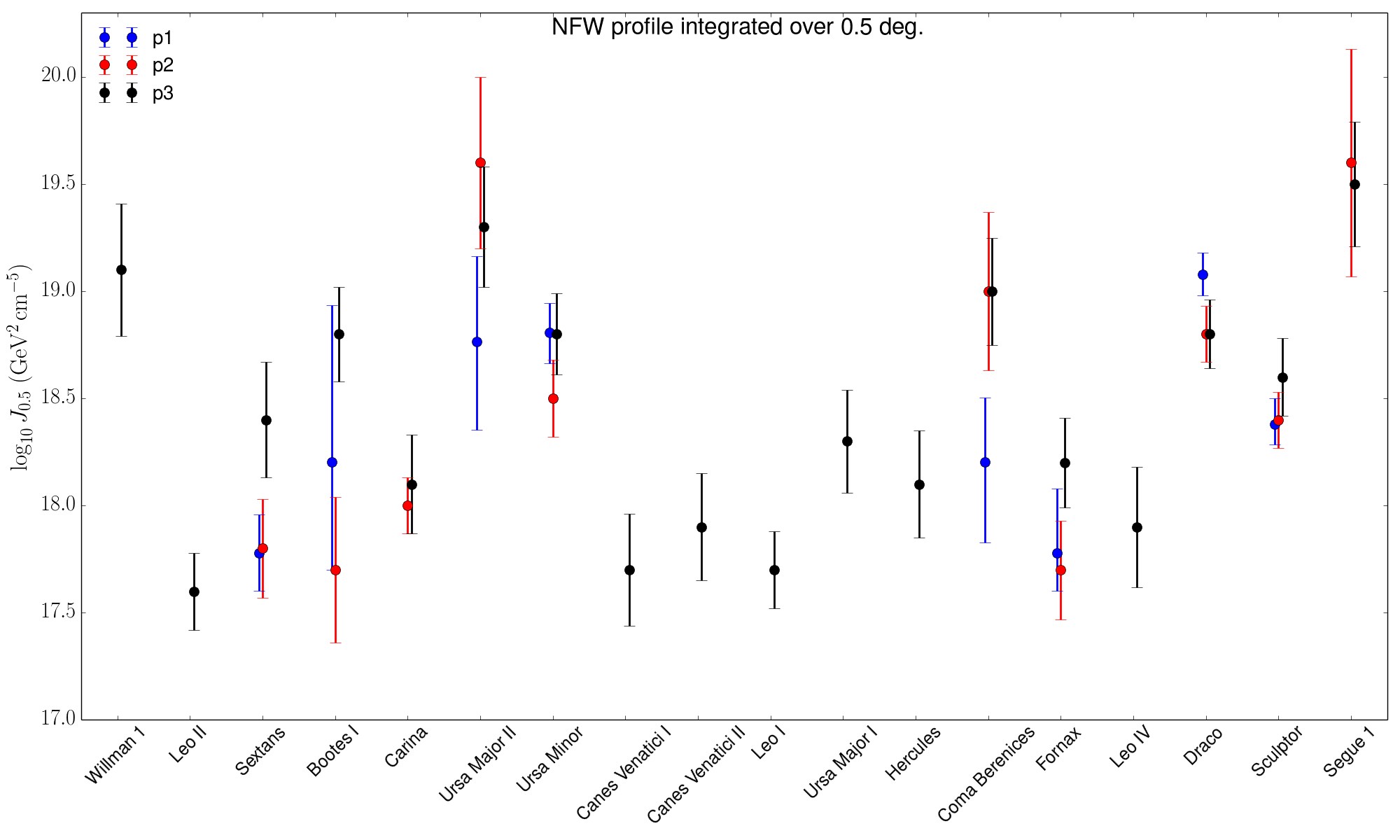}
\caption{$J$-factors within $0.5^\circ$ for dSphs, for three different assumptions for theoretical priors on the dark matter halo parameters. Blue points utilize uniform priors in $\log r_s$ and $\log \rho_s$~\cite{Abdo:2010ex}, red points uses as priors the relation between $\rho_s, r_s$ that is determined in CDM simulations~\cite{Ackermann:2011wa}, and black points uses the hierarchical modeling method that is introduced in Martinez~\cite{Martinez:2013els}. Though these calculations assume an NFW profile, for $J$-factors determined within an angular region of $0.5^\circ$, the results are weakly-dependent on the assumption for the dark matter density profile.   
} 
\label{fig:jvalues}
\end{figure}

 Several authors have now published an analysis of the dark matter distributions
in dSphs using the spherical Jeans method and examined their implications for gamma-ray experiments 
~\cite{Strigari:2006rd,Strigari:2007at,Essig:2009jx,Essig:2010em,Charbonnier:2011ft,Martinez:2013els,Geringer-Sameth:2014yza}. A typical approach is to assume a model for the dark matter profile, such as a generalized double power law model, combined with the likelihood function in Eq.~\ref{eq:dsphlike}. Within a Bayesian framework, the model parameters $\rho_s, r_s,a,b,c$ are then marginalized over assuming priors on these parameters. In the literature there have been several approaches to handle these priors. 
Strigari et al. utilized priors from CDM simulations~\cite{Strigari:2006rd,Strigari:2007at}, 
which effectively weighted Eq.~\ref{eq:dsphlike} 
with a function describing the relation between $\rho_s, r_s$ derived from CDM simulations. 
Several authors have considered ``uninformative'' priors, 
equivalent to flat priors on $\log r_s$ and $\log \rho_s$~\cite{Essig:2009jx,Essig:2010em,Charbonnier:2011ft}. 
Martinez introduced a hierarchical modeling method that uses a relationship between that mass at the half-light radius and the luminosity of a dSph~\cite{Martinez:2013els}. As shown in Figure~\ref{fig:jvalues}, for dSphs with well-measured kinematics, 
the $J$-factors that are derived for each prior are typically consistent with one another, 
though there is a larger spread for dSphs with smaller samples of stars.
As noted above these results are weakly dependent on whether a cored or
cusped central density profile is assumed for the dark matter. 

\par~Figure \ref{fig:jvalues} clearly indicates which dSphs are the most interesting targets for
indirect dark matter detection experiments. The two dSphs with the largest
$J$-factors, Segue 1 and Ursa Major II, are ultra-faint satellites with
sparse samples of stars associated to them (about 60 and 20 stars,
respectively). Though these dSphs have the largest mean flux, they also
have the greatest uncertainty due to the small stellar samples. After
Segue 1 and Ursa Major II, the dSphs with the next largest $J$-factors are
Ursa Minor and Draco, at 66 and 80 kpc, respectively. These $J$-factors are
determined from samples of hundreds of stars so their corresponding
uncertainties are much lower than the uncertainties on the $J$-factors for
Segue 1 and Ursa Major II.

\par In sum, because of the substantial theoretical and observational work that has gone into understanding the kinematics of dSphs and their underlying dark matter distributions, they are a unique target in gamma-ray searches for dark matter. Any possible detection of a signal in other sources must be corroborated by a detection in dSphs. As discussed in the sections below, dSphs are targets for several gamma-ray experiments which hope to probe the WIMP mass range from a few GeV up to tens of TeV over the next decade. 

\subsection{Substructures}
\par N-body simulations of Galactic halos predict that approximately $10-50\%$ of the dark matter mass of the Milky Way is bound up in the form of substructure, or subhalos. According to the $\Lambda$CDM model, some fraction of these subhalos should host the observed dSphs that were discussed above. However, there is not a consensus understanding as to what the mass and radial distribution is for the subhalos that ``light up" with stars-- this issue is strongly intertwined with the classical missing satellites problem~\cite{Kauffmann:1993gv,Moore:1999,Klypin:1999uc} and the more recent too-big-to-fail issue of $\Lambda$CDM~\cite{BoylanKolchin:2011de}. Given that subhalos without any associated stars have yet to be conclusively detected around the Milky Way, a gamma-ray signal from these objects, while intriguing, is still subject to a substantial amount of theoretical assumptions. Nonetheless, it is informative to examine what the modern theory predicts for this population of subhalos around the Milky Way, and beyond. 

\par The numerical simulations of Milky Way-mass galaxies are now complete in their measurement of subhalos down to a mass of approximately $10^6$ $M_\odot$, corresponding to approximately $10^{-6}$ times the total mass of the host halo~\citep{Diemand:2008in,Springel:2008cc}. The mass function of subhalos may in fact extend down to Earth masses or even below, which the simulations are not sensitive to at present~\citep{Green:2003un,Loeb:2005pm,Profumo:2006bv,Martinez:2009jh}. For for subhalos with mass greater than $\sim 10^6$ $M_\odot$, the mass function of subhalos is a power law that scales as $dN/dM \propto M^{-\alpha}$, with $\alpha = 1.9$. Because $1 < \alpha < 2$, subhalos at the low mass end of the mass function dominate the distribution by number, while the subhalos at the high mass end of the mass function dominate the total mass in substructure. 

\par The internal density profiles of the dark subhalos is important, and has been a subject of theoretical debate. More recent simulations find that dark matter halos and subhalos are better described by an appropriate shallower Einasto density profile~\cite{Springel:2008zz}. There are suggestions from numerical simulations that the density profiles of the smallest, $\sim 10^{-6}$ M$_\odot$ halos, are steeper than the NFW form~\cite{Ishiyama:2010es,Anderhalden:2013wd}. The gamma-ray signal from WIMP annihilation from small dark matter halos is also sensitive to the concentration of low mass halos, where here the concentration is defined as the ratio of the scale radius of the halo to its virial radius. There is a substantial amount of uncertainty on the extrapolation of the concentration versus halo mass relation down to low halo mass scales; numerical simulations are at present only able to measure this relationship down to halo mass of about $10^8$ M$_\odot$~\cite{Springel:2008zz,Gao:2011rf}. There is at least an order of magnitude uncertainty in the gamma-ray signal from dark matter annihilation in small halos due to the unknown extrapolation of the concentration versus mass relation to the smallest halos~\cite{Martinez:2009jh,Sanchez-Conde:2013yxa}. 

\par Dark matter substructures are very much ``high risk, high reward" targets for gamma-ray searches. While a confirmed detection of an object that shines only in gamma rays would be a spectacular confirmation of both the $\Lambda$CDM and the WIMP paradigm, for a canonical thermal relic scale WIMP cross section predictions for signal detectability vary by several orders of magnitude~\cite{Koushiappas:2003bn,Pieri:2007ir,Springel:2008zz}. Realistic a priori predictions for a signal from dark matter substructure will only likely be available once simulations are able to resolve substructure near the Earth mass scale, or it becomes observationally feasible to detect substructure at this mass scale via gravitational lensing. Since there are substantial hurdles in reaching these scales in both the simulations and observations, probably the best these types of searches can hope for at the stage is to find a signal in gamma rays from a substructure candidate that can be followed up on and found not to be observed at other wavelengths. 

\subsection{Galaxy Clusters}\label{sec:targets_clusters}
\par The study of dark matter in galaxy clusters has a long history, dating back to the original discovery of Zwicky. The systematic uncertainty in the determination of dark matter mass profiles and $J$-factors in clusters is similar in many ways to the case of dSphs discussed above. The dominant uncertainty in the $J$-factors from clusters arises from two orthogonal aspects of astrophysics. First, there is an uncertainty in the empirically-measured cluster mass profiles, which are derived from a combination of x-ray temperature profiles and gas kinematics. Second, there is significant uncertainty in the predicted gamma-ray luminosity that arises from the contribution of dark matter substructure in the clusters.

%\subsubsection{Dark matter density profiles} 

\par Cluster masses and density profiles are measured through x-ray emission, galaxy dynamics, or gravitational lensing~\cite{Allen:2011zs}. For nearby clusters, masses are generally derived under the assumption of hydrostatic equilibrium, 
\begin{equation}
M(r) = - \frac{kT(r)r}{\mu m_p G} \left[ \frac{d \ln \rho(r)}{d \ln r} + \frac{d \ln T(r)}{d \ln r} \right], 
\label{eq:hydrostatic_eq_clusters}
\end{equation}
where $T(r)$ is the temperature profile, and $\mu$ is the mean mass per particle in units of the proton mass. The assumption of hydrostatic equilibrium is subject to systematic uncertainty depending on the physical state of the cluster. For instance,~\citet{Nagai:2006sz} used numerical simulations to show that systematic uncertainties due to non-thermal pressure introduce a $\sim 10\%$ systematic uncertainty in the mass determination. 

\par Furthermore, there are only of order tens of clusters that have measured temperature and density profiles allowing the use of Equation~\ref{eq:hydrostatic_eq_clusters}, and only a handful of nearby clusters have dark matter mass profiles constrained by redundant estimates. For the majority of known clusters, we are instead left with indirect observational proxies that are calibrated by low-redshift clusters. Such standard observational proxies are the average temperature, the mass of the hot gas, and the product of these two, the total thermal energy~\citep{Vikhlinin:2008cd}. Recent studies have combined all of the aforementioned mass measurement techniques to obtain an estimate of the slope of the dark matter density profile in clusters~\cite{Newman:2012nw}. In these studies, the measured mean slope is $\langle \gamma \rangle \simeq 0.50$, which is less shallow than the central slope of the standard NFW profile. 

\par Of particular importance for gamma-ray analyses are the clusters that are the appropriate combination of the most nearby and the most massive. From their measured mass distributions and known distances, several recent studies have come to the agreement that the Fornax, Coma, and Virgo clusters are the brightest source of gamma rays from dark matter annihilation~\cite{Pinzke:2011ek,Ando:2012vu,Han:2012uw}. 

\par Dark matter-only simulations of clusters of galaxies are only able to resolve dark matter substructure with mass $10^{-5}$ times the mass of the host cluster; this is perhaps 10 orders of magnitude larger than the minimum mass dark matter subhalo that is predicted in $\Lambda$CDM theory. This uncertainty due to the dark matter substructure, along with the fact that only a relatively small number of clusters has been simulated, introduces a significant uncertainty in the contribution of cluster substructure to the gamma-ray luminosity that arises from dark matter annihilation.

\par Nevertheless, a significant increase in the gamma-ray signal from clusters is expected from the presence of such dark matter substructure. Numerical simulations provide the most reliable method for determining the gamma-ray emission from subhalos in clusters. From a sample of nine cluster-mass dark matter halos, which have individual particle masses of $\sim 10^6$ $M_\odot$ and identify subhalos down to a mass scale of $\sim 10^7$ $M_\odot$, \citet{Gao:2011rf} directly determine the overall boost factor and the surface density profile of the substructure component for clusters as a function of the cluster mass. Though it relies on an extrapolation, the substructure likely implies a substantial increase in the gamma-ray luminosity over the smooth component in clusters, perhaps up to several orders of magnitude. 

\par In sum, clusters represent a unique target for dark matter searches. Like the dSphs, they can be localized in space, and their dark matter distributions can be robustly measured from astronomical data sets. However, they are different from dSph targets because there is expected to be a large contribution from dark matter substructure within them. While this is expected to increase the gamma-ray signal, in the case of a null detection it does provide a substantial systematic uncertainty when attempting to set upper limits on the annihilation cross section. 

\par Clusters are also distinct from dSphs because there is expected to be a significant flux of gamma rays due to cosmic ray processes. The gamma-ray luminosity due to cosmic rays in clusters is expected to trace the gas density, so it is more centrally-concentrated than an expected dark matter annihilation signal, since the dark matter signal is more extended because of the emission from subhalos. As we discuss in the sections below, though very plausible theoretical predictions show that gamma rays from both cosmic rays and dark matter could have been detected by gamma-ray observatories, there has been no conclusive signal reported to date. 

%There is expected to be a significant flux of gamma-rays due to cosmic ray processes in clusters \FIXME{CITATION}. However, no gamma-ray emission has been conclusively associated with clusters in the Fermi-LAT one year data, assuming that they are point sources \FIXME{CITATION}. Modeling clusters as extended sources, no gamma-ray emission was yet found in two year data \FIXME{CITATION}. 

%\FIXME{(it lacks a concluding section.... Something summing up the main points : clusters were expected to constitute a gamma-ray emitter class, but have not been detected as such; historically they are the signpost of the dark matter hypothesis to explain the dynamics of luminous matter; predictions especially when accounting for substructure should make clusters a prime candidate for ID searches; but the uncertainties on the smooth profile and small subhalo mass distribution makes prediction very uncertain, and only very few clusters, massive and nearby, currently have their profile sufficiently well constrained. Are there perspectives to improve this state of affairs? The other analysis challenge is that the massive and nearby clusters have a large angular scale on the sky (7° for Virgo?), which makes LAT studies very difficult due to the number of sources to account for and the uncertainty in morphological aspects of the diffuse bacground at such scale.... It seems to me that we thus can put a bit more soul on this section.)}

%\FIXME{Can we have a figure that illustrates the spread in J factor predictions for clusters?}

\subsection{Cosmological mass function of dark matter halos} 

\par The discussion above has focused on measurements of the dark matter distribution within the Milky Way, and nearby identified dwarf galaxies and galaxy clusters. Gamma-ray searches for dark matter are also sensitive to the accumulated emission from dark matter annihilation in all the dark matter halos that have formed in the universe. Though this emission is ``unresolved," it is possible to model given a mass function of dark matter halos. It may be possible to deduce this mass function from the observed luminosity function of galaxies, though this method is hindered by uncertainties in the mapping of galaxy luminosity to dark matter halo mass~\cite{2010MNRAS.404.1111G,Moster:2009fk,Behroozi:2010rx,2014arXiv1401.7329K}. 

\par A more robust estimate of the dark matter halo mass function comes from large scale cosmological simulations. The two largest volume cosmological simulations to date provide a statistically-complete sample of dark matter halos down to a maximum circular velocity of about 50 km/s, or a mass of approximately $10^9$ M$_\odot$~\cite{2005MNRAS.364.1105S,2009MNRAS.398.1150B,2011ApJ...740..102K}. These studies find that the ``halo multiplicity function," which is derived from the halo mass function, appears to have a near-universal form at all redshift. This implies that it is possible to compute the halo multiplicity function at any redshift from well-measured cosmological parameters. 

\par As discussed above in the context of galaxy clusters, dark matter annihilation signals are sensitive to substructure within dark matter halos. This implies that it is important to determine not only the mass function of dark matter halos, but also the mass function of dark matter substructure. Both the shape and the scatter in the subhalo mass function in simulations is well-characterized down to subhalo masses at least three orders of magnitude less than the his halo mass. Semi-analytical models are now able to reproduce the trends observed in the simulations~\cite{2014arXiv1403.6835V}. 

\par Given the uncertainties associated with the distribution of dark matter substructure in halos, in particular in the extrapolation of substructure down to the smallest mass scales below the resolution limit of simulations, theoretical predictions for the diffuse gamma-ray background from dark matter halos is difficult to precisely pinpoint~\cite{Ullio:2002pj,Sefusatti:2014vha}. Searches for diffuse gamma rays from dark matter annihilation must of course also be differentiated from gamma rays that are produced from cosmic rays and other astrophysical sources such as AGN and supernova remnants. Both the uncertainties in the intrinsic emission and the uncertainties in the astrophysical gamma-ray backgrounds provide a substantial challenge for signal detection.

                    \section{\Fermi-LAT and IACT analyses.}
\label{sec:analysis}

\par Over the course of the past decade, the field of indirect dark matter searches with gamma rays has matured substantially. This is in large part due to the performance of the present-generation space and ground instruments, notably the \Fermi\ gamma ray space observatory and several IACT telescopes, such as H.E.S.S., VERITAS, and MAGIC. The results from these experiments are important both when viewed independently and as results complementary to those from collider and direct dark matter searches. In this section, after a brief presentation of these instruments, we discuss the analysis challenges that   they face.

\subsection{Current gamma-ray instruments}\label{sec:inst}
\par As the atmosphere is opaque to gamma rays, space borne telescopes are a priori necessary to observe the sky at high photon energy. Above a few tens of MeV, their interactions are completely dominated by pair production, so that all recent gamma-0ray space instruments rely on a pair converter associated to a tracker to detect the trajectory of the produced electron and positron pair. The tracker is supplemented with an electromagnetic calorimeter to contain the shower and allow for total energy estimate, and an instrumented shield to veto the much more frequent incident charged cosmic rays. The most important of such instruments currently active is the Large Area Telescope \citep[LAT,][]{Atwood:2009ez} onboard the \Fermi\ space satellite, which also includes a gamma-ray burst detector (GBM). Since June 2008, the LAT surveys the whole sky every 90 minutes, in the energy range between 20 MeV to greater than 300 GeV. The upper end of this range is here determined only by sparse statistics for conventional (power-law) astrophysical sources and the availability of reliable instrument response functions, i.e. can be much higher.

The LAT is likely planned for operation until at least 2018.

\par Above a few hundered GeV, the fast decreasing flux of typical astrophysical sources and the necessarily limited effective area of a space instrument combine to significantly degrade the sensitivity. Despite the atmospheric opacity, ground instruments have proven to be able to ``take over'' in this very-high energy range. A similar argument holds for a hypothetical WIMP signal, as the flux from WIMP annihilation is inversely proportional to the WIMP mass squared.
To detect photons at such high energies, Imaging Air Cherenkov Telescopes detect the Cherenkov radiation produced by the charged particle cascade that a high energy gamma ray initiates in the upper layers of the atmosphere. The Cherenkov light is reflected by large ($\sim$ 10 m diameter) mirrors onto cameras consisting of arrays of photomultipliers. As the interaction takes place high in the atmosphere ($\sim$ 10 km), and the shower needs to reach a certain size to be detectable, the analysis threshold is usually close to 100 GeV, though larger telescope mirror size (and the corresponding possibility to detect fainter Cherenkov light) can push the threshold down to about 30 GeV. The Cherenkov telescopes currently in operations are H.E.S.S \citep[5 telescopes, including the recently added 28-meter telescope]{Hinton:2004eu}, VERITAS \citep[4 telescopes]{2006APh....25..391H} and MAGIC \citep[2 telescopes]{Lorenz:2004ah}. 
At slightly higher threshold energy, a different detection technique consists in building an array of water tanks instrumented with photomultipliers, and in recording the passage of the shower particles themselves, throught the Cherenkov light that they emit while traversing the water. This technique is exemplified by MILAGRO \cite{2003ApJ...595..803A}, and more recently HAWC (see~\secref{sec:perspectives}). 

\par The LAT and IACTs have very distinctive operation modes and face quite different challenges. While the LAT surveys the whole sky continually without any need to point in a specific direction (but for the case of transients), IACTs operate in a pointed mode and with a limited duty-cycle (though this is not true for water Cherenkov instruments) and have to rely on the larger effective area (by a factor of $10^4$ to $10^5$) as compared to the LAT. As a result of the field of view, the LAT can a priori provide observations on any of the targets discussed in~\secref{sec:targets}, while in the case of IACTs, these targets are in competition with other astrophysical sources of interest to the community. Furthermore, the LAT has an excellent cosmic-ray rejection power, so that its background in conventional analyses is dominated by gamma-ray events from various sources in the field of view. The complexity of the LAT analysis is due to the multi-dimensional response functions and the need to properly model the sky in the region of interest. Less efficient background rejection implies that IACTs are dominated by an isotropically-distributed charged cosmic ray background. In addition, the fact that the atmosphere is used as a detector volume implies that varying atmospheric conditions as well as the choice of atmospheric conditions and night-sky background  induce important and difficult to handle systematics in acceptance corrections, which eventually (together with the very hard to distinguish component of cosmic ray electrons) provide the fundamental limitation to detecting dark matter.

\subsection{Astrophysical challenges}

\par The astrophysical challenges outlined in~\secref{sec:targets} call for advanced analysis methods that allow robust inference about the particle physics properties of dark matter. \Fermi-LAT analyses are largely based on a maximum likelihood method, where spatial and spectral models of both signal and background components are fit to data after convolution with a parameterized instrument response. For a dedicated review of statistical aspects of these analyses, see~\citet{Conrad:2014nna}.  Modeling the background can in certain cases introduce significant systematics, with likely the most problematic case being modeling the diffuse gamma-ray background towards the Galactic center. Until recently, IACT searches for WIMP dark matter focused on comparing integral flux predictions with the data, instead of using the full spectral information predicted by a single or combination of annihilation or decay channels. Modeling of the background has not been commonly performed but rather the background expectation was obtained from an off-source region of interest. The significance of an excess is then commonly inferred from the maximum likelihood ratio test statistic~\cite{Li:1983fv} applied to the case of OFF estimation of the background, and confidence intervals using either a counting experiment profile likelihood~\cite{Rolke:2004mj} or using a Neyman construction~\cite{Feldman:1997qc}. In the remainder of this section, we will discuss in detail the different analysis approaches for the different astrophysical targets.  

\subsection{Galactic Center}
\par As discussed in~\secref{sec:targets} the Galactic Center (GC) is probably the strongest source of gamma-ray radiation due to dark matter annihilation. However, the GC is crowded with conventional gamma-ray sources: H.E.S.S.and \Fermi-LAT sources at the GC are consistent with each other and known sources~\cite{cohen-tanugi_fermi-lat_2009,cohen-tanugi_gev-band_2009,Hooper:2011ti}. For this reason, the GC is typically not directly studied in search for dark matter annihilation. Modeling the diffuse emission in the inner Galaxy, defined here as the region interior to the Solar circle, also poses a significant theoretical challenge-- for this reason the \Fermi-LAT has not published dark matter constraints from either the GC or inner Galaxy, leaving aside searches for spectral features (see below).

\par At IACT energies the systematics of the diffuse emission are less significant. The analysis presented in~\citet{Abramowski:2011hc} applies an ON-OFF technique, where the background is determined by OFF-source observations, and in this specific case the OFF-source region is defined within the field of view. The disadvantage of defining the OFF-source region within the field of view (i.e. relatively close to the GC) is that a potential signal will be subtracted as part of the background. This means that the sensitivity might be reduced for relatively cored dark matter profiles. An alternative is therefore to define the off region from separate pointings, i.e. pointings that are truly off-source. The likelihood for this approach is: 
\begin{equation}
\mathcal{L}(n_{ON},n_{OFF}|s,b,\alpha) = Pois(n_{ON}|s+b) Pois (n_{OFF}|\alpha\, b)
\label{eq:alpha}
\end{equation}
where $n_{ON}$ and $n_{OFF}$ denote the number of counts in ON and OFF regions, respectively, $\alpha$ is the ratio between the total ON region acceptance and OFF region acceptance and $b$ as usual denotes the background expectation. There are two assumptions going into this approach: (a) the background expectation is the same in the on source and off source region, and the ratio between the on-source and off-source acceptance is known. The latter is a systematic which will become important for future IACTs, such as CTA, where the statistical uncertainties are sub-dominant as compared to systematic. For a potential way of handling these uncertainties see~\citet{Dickinson:2012wp,Spengler:2015dda}. 

\par To circumvent the aforementioned disadvantage for the GC that in standard observation mode the OFF regions are within the few-degree field of view of the camera, a technique which obtains OFF data from a truly separate pointing has to be implemented. However, this introduces new systematic uncertainties that have to be carefully addressed; using this technique competitive limits have nonetheless been presented in~\cite{HESS:2015cda}. 

\subsection{Galactic diffuse emission}\label{sec:galdiff}

\par The Galactic diffuse emission provides a potentially powerful target for dark matter searches, e.g. \cite{Fargion:1998ya}\cite{Zeldovich:1980st}, For these searches, the gamma ray spectrum and spatial distribution can both be used as discriminants. However, astrophysical-induced diffuse emission from the Galaxy is very difficult to model. Parameters to be considered are halo height, diffusion coefficients and indices, Alvfen velocity, power-law indices of the injection spectrum of cosmic rays and spatial distribution of cosmic ray sources as well as maps of the interstellar radiation field and gas distributions. The most detailed model of Galactic diffuse emission and cosmic-ray propagation is provided by the GALPROP code~\cite{strong_cosmic-ray_2007}\footnote{http://galprop.stanford.edu}. The more than 20 parameters entering this modeling (and in principle also versions of the input gas maps and interstellar radiation fields) constitute a set of nuisance parameters that must be handled in order to extract a potential dark matter contribution. The most advanced attempt at dealing with these parameters is presented in~\citet{Ackermann:2012rg}, which uses a profile likelihood. A subset of the full parameter space was considered in this work. For the linear parameters a fit to the data provides a profile estimate, while for the non-linear parameters a grid of likelihood points was considered to map the complete likelihood function. However, the performance of this mapping is still difficult to assess. The complexity of the problem suggests that scanning algorithms designed for Bayesian inference could be more suitable for the problem. This has been applied to cosmic-ray data to provide constraints on diffusion parameters~\cite{Trotta:2010mx}.

\subsection{Dwarf spheroidals}\label{sec:dwarfs_analysis}
\par For reasons highlighted in~\secref{sec:targets}, dSphs are valuable targets for dark matter searches. There is now a substantial body of literature devoted to understanding how to extract a gamma-ray signal from dSphs. For this reason, we will give here detailed account of the methods that are used in these searches. 

\subsubsection{Formalism and illustration}
\par Here we describe in detail the full likelihood analysis method used in dSph searches with the \Fermi-LAT and more recently also with H.E.S.S.~\citep{Abramowski:2014tra}, and MAGIC~\citep{2012JCAP...10..032A} \footnote{for an alternative statistical approach see \citep{Geringer-Sameth:2014qqa}.}. In order to illustrate the advantages of the full likelihood analysis, we start by fixing our notation. We consider a dataset $D_1$ selected in a region of observation ``1'', and a statistical model which contains a set of parameters of interest $\mathbf{p^i}$ and a set of nuisance parameters $\mathbf{p_1^n}$. We further write the logarithm of the likelihood function  ${\cal L}_1(D_1|\mathbf{p^i},\mathbf{p_1^n})$. While the maximum likelihood inference based on the observation $D_i$ results from minimizing ${\cal L}_1$ with respect to the parameters, it can be generalized in a straightforward manner to the case where different, disjoint, observations $D_1$ and $D_2$ are modeled using the same parameters of interest. 

\par The combined log-likelihood function to be used for inference now simply reads 
\begin{equation} 
{\cal L}_\mathrm{comb}(D_1,D_2|\mathbf{p^i},\mathbf{p_1^n},\mathbf{p_2^n}) = {\cal L}_1(D_1|\mathbf{p^i},\mathbf{p_1^n})+{\cal L}_2(D_2|\mathbf{p^i},\mathbf{p_2^n}). 
\end{equation}
Such a property of ``additivity'' is an extremely appealing feature of the likelihood inference process, that we now illustrate with the case of a WIMP gamma ray annihilation analysis in the direction of two dwarf spheroidal galaxies, each in either region ``1'' or ``2''. The parameters of interest are {$\sigmav,m_{\mathrm{WIMP}}$}, and we single out, for the sake of the discussion, the $J$-factors $J_1$ and $J_2$ from the rest of the nuisance parameters.\footnote{In practice, the WIMP mass is kept fixed during the likelihood inference process, so that we omit it in the rest of this discussion.} Taking note of the fact that the likelihood model in such an analysis actually includes only the product of $\sigmav$ and $J_{1,2}$, the combined log-likelihood now reads 
\begin{align}
{\cal L}_\mathrm{comb}(D_1,D_2|\sigmav \times J_1,\sigmav \times J_2, J_1,J_2,\mathbf{p_1^n},\mathbf{p_2^n}) = &&{\cal L}_1(D_1|\sigmav \times J_1,J_1,\mathbf{p_1^n})\\
&+&{\cal L}_2(D_2|\sigmav \times J_2,J_2,\mathbf{p_2^n}).\nonumber
\end{align}

\par In order to infer a best-fit value for $\sigmav$ it is thus necessary to fix $J_{1,2}$ to some values; this bars any propagation of the corresponding uncertainties into the analysis. This can actually be circumvented by considering that any value of a $J$-factor comes from a set of stellar observations $d$ and an inference process based on a posterior distribution or a likelihood function, so that we generically write $P_J(d|J,I)$. Here $I$ stands for any other information needed to construct this function, for instance the choice of a parameterized model for the dark matter profile. The combined likelihood analysis can be extended to read:
\begin{eqnarray} 
%\begin{equation}
 {\cal L}_\mathrm{comb}(D_1,D_2,d_1,d_2 | \sigmav \times J_1,\sigmav \times J_2, J_1,J_2,\mathbf{p_1^n},\mathbf{p_2^n})  &=& 
{\cal L}_1(D_1|\sigmav \times J_1,J_1,\mathbf{p_1^n})  \\
&& +{\cal L}_2(D_2|\sigmav \times J_2,J_2,\mathbf{p_2^n}) \nonumber \\
&& + P_J(d_1|J_1,I)+P_J(d_2|J_2,I). \nonumber 
% \end{equation} 
\end{eqnarray}
As a result the degeneracy between the $J$-factor and $\sigmav$ is removed, and there is an automatic propagation of the statistical uncertainty in the stellar analysis down to the inference about \sigmav. Given the shape of the Bayesian posterior probability densities commonly obtained in $J$-factor derivations from stellar analyses~\cite{Strigari:2007at,Martinez:2009jh},~\citet{Ackermann:2011wa} actually use a Log-Parabola function as an ansatz for the $P_J$ function.
 
\par
\begin{figure}[h]
     \centering
   \includegraphics[scale=0.35]{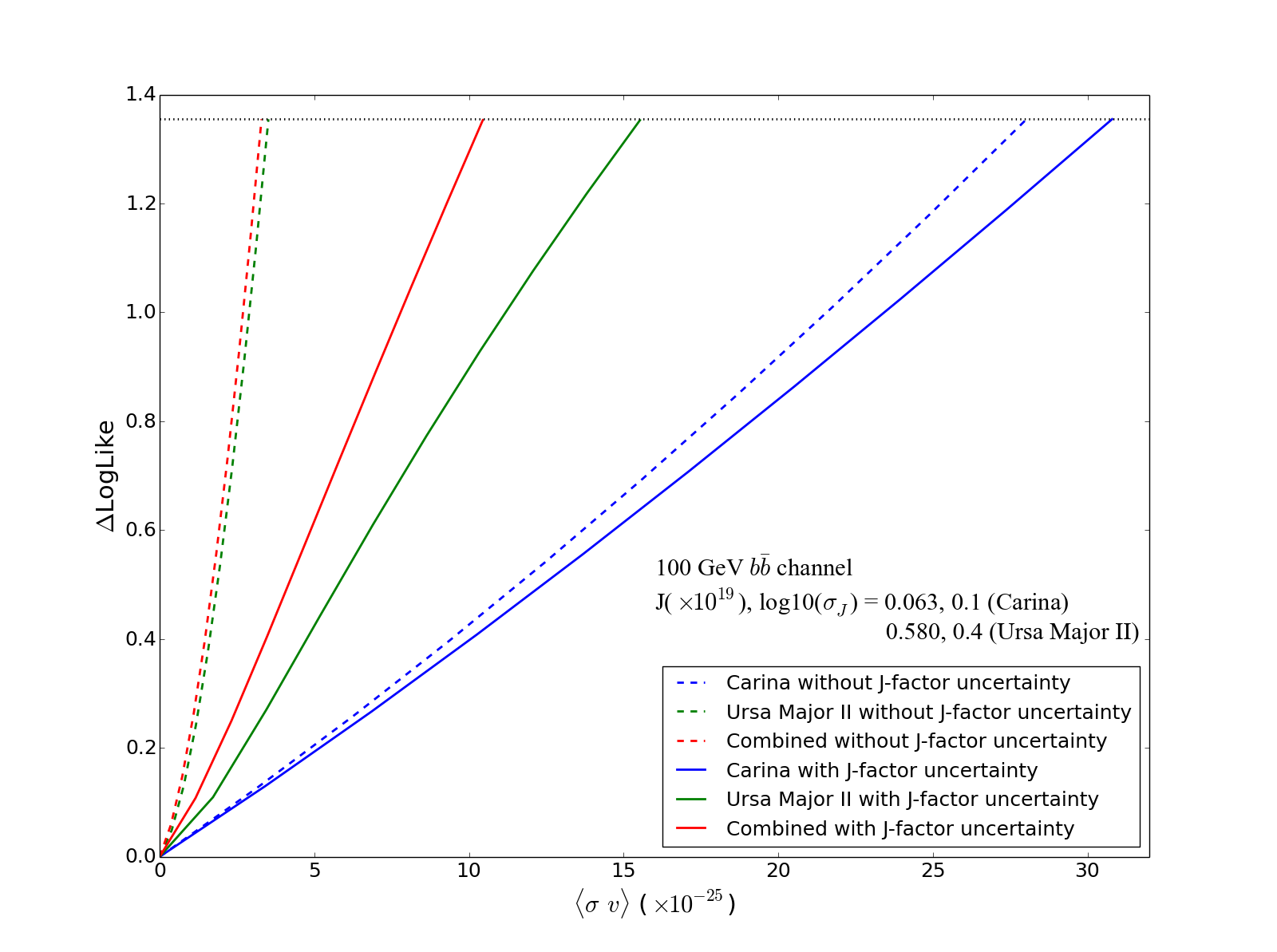}
    % veritas_expected.png: 575x392 pixel, 96dpi, 15.21x10.37 cm, bb=0 0 431 294
   \caption{Relative likelihood function versus velocity averaged annihilation cross-section. The effect of a combined likelihood analysis with inclusion of the $J$-factor uncertainty, in the case of two dSphs,  Carina and Ursa Major II. is illustrated.
   }\label{fig:twodwarfs}
   \end{figure}

 We can now illustrate the power of this combined analysis. For this purpose we consider the two dSphs, Carina and Ursa Major II. For those two dSphs, the central and sigma values for the Log-Parabola function are $(6.3\times 10^{17},10^{0.1})$ and $(5.8\times 10^{17},10^{0.4})$, respectively. Thus Carina has a much lower $J$-factor than Ursa Major II, but its distribution is much narrower, so the uncertainty incurred in choosing the central value as fiducial $J$-factor is much lower.~\figref{twodwarfs} presents the likelihood curves for the single-dSph and the combined analyses, profiled over the nuisance parameters, including the $J$-factors.

\par The mass and channel used for this illustration are 100 GeV and $\bbbar$, respectively. The intersection between the horizontal dotted curve and a likelihood curve localizes in the x-axis the 95\% C.M. value for \sigmav, so that the lower its value, the stronger the upper limit. Thus, as expected when no account is taken of the $J$-factor uncertainties, Ursa Major II (green dashed curve) outperforms Carina (blue dashed curve) by and large. On the other hand, the combined analysis (red dashed curve) performs barely better than Ursa Major II alone, due to the modest statistical power of Carina. When $J$-factor uncertainties are taken into account (see plain curves, with identical color coding), one immediately notices that Carina's limit does not degrade by much, while limits from Ursa Major II are significantly weaker than before. And furthermore, the combined likelihood result is now significantly improved compared to Ursa Major II : when $J$-factor uncertainties are taken into account, the statistical power of Carina is no longer negligible. Comparison of the dashed and plain red curves shows the effect of the $J$-factor uncertainty on the combined upper limit.

\subsubsection{Dwarf spheroidals and the global fit}
\par The likelihood method described above can be extended to attempt statistical inference on a specific supersymmetric model, including taking into account collider experiments~\cite{Scott:2009jn}. The result of this type of analysis is a likelihood or posterior maps in the parameter space of constrained supersymmetry. However, because of the sparsity of the data, inference on the physical model is difficult. Nonetheless, this result serves as a proof of concept for future attempts to combine results of different dark matter probes in a consistent likelihood or Bayesian inference.

\subsection{Dark satellites}
\par Dark satellites are intriguing objects for dark matter searches. In an ideal scenario, they would constitute a gamma ray source without counterpart in any other wavelength. For a preliminary estimate, the list of unassociated \Fermi-LAT sources provides a catalogue of potential dark matter satellites. 

\par In order to be classified as a dark satellite candidate, first and foremost the energy spectrum should be consistent with a dark matter induced spectrum and distinguishable from more common astrophysical-induced spectra, such as a power-law, or a more difficult to deal with pulsar spectrum~\citep[see e.g.][]{Baltz:2006sv}. Additionally, the source may be extended, which would be in particular true if the source were very near to the solar system. From a statistical perspective, a likelihood ratio test statistic provides a mean to address this question of whether the energy spectrum of the source is consistent with dark matter. The problem though is that the hypotheses, in particular the question whether the source exhibits a power-law as compared to a dark matter spectrum, constitutes a non-nested model comparison for which the usual asymptotic theorems~\citep[such as][]{wilks_large-sample_1938} do not hold. This implies that the significance of a potential detection (rejection of the null hypothesis) cannot be robustly calculated. Techniques that are proposed to address this problem are known in literature~\cite{James:2006zz}, but it is unclear if they perform in a satisfactory manner~\citep[see e.g.][and references therein]{Conrad:2014nna}. Thus, the null distribution of the test statistic is derived from Monte Carlo simulations of the experiment or from random region of interests in the sky~\cite{Ackermann:2012nb}. For IACTs, the list of unassociated sources provides potential targets for deep follow-up observations, and the same criteria for claiming a detection would apply.  

\subsection{Galaxy Clusters}
\par The search for dark matter in Galaxy clusters utilizes analysis methods similar to those used for dSphs. The main difference is that galaxy clusters are in all likelihood extended sources of gamma ray emission. Indeed, if substructures and their corresponding flux dominates the dark matter contribution, it is the outer parts of clusters that dominate.  As a specific example, the Virgo cluster may have an extension of approximately 6 degrees. The search for gamma ray emission from clusters therefore has to use extension as source model. The major drawback of this is that the results are in general sensitive to the modeling of the diffuse emission, and uncertainties in this diffuse emission must be accounted for. A full mapping of the likelihood space of this component is very difficult (see also next subsection). As a result, in the analysis presented in~\citet{2013arXiv1308.5654A}
%~\footnote{For the moment does not consider dark matter models, but rather models of cosmic-ray induced gamma-ray emission} 
the systematic effects due to modeling of the diffuse emission are accounted for by considering a set of fiducial diffuse models and recalculating limits separately for each of these fiducial models.

\subsection{Searches for dark matter line signal}
\par Searching for spectral features is in principle a simpler task than searching for a continuum signal. In this case, the main question is whether there is a local (in energy space) deviation from a background, which can be inferred from the data itself. This means that no physics modeling of the background is necessary.  The line signal is described by a delta function convoluted with the energy dispersion introduced by the instrument. 

\par The analysis of line signals typically proceeds by applying a sliding window technique, i.e. the window is centered on the particle mass to be tested, and the background is determined either from the whole window or excluding some signal region. The size of the window is chosen such that the background is described by a simple empirical fit (in best case a power-law).  Statistical inference is often performed using an extended likelihood fit, with application of Wilks theorem~\cite{wilks_large-sample_1938} or Chernoff theorem \citep{Chernoff:1954}  to obtain significance and the profile likelihood to infer confidence intervals. At this stage, most analyses do not attempt to model the spatial distribution of the gamma rays, but instead optimize the region of interest using a signal and background prediction and then treating the problem only with the spectral likelihood. The above approaches have been applied in \citet{Abdo:2010nc,Ackermann:2012qk,Weniger:2012tx,Ackermann:2013uma}.~\citet{Ackermann:2013uma} also takes into account the quality of the event reconstruction by considering not only the likelihood of the parameter of interest but by also considering the distribution of photons in a quality parameter. The analysis in~\citet{Su:2012ft} uses both spectral and spatial distributions.

\par While peak finding of this sort is as old as particle physics, challenges remain. Potential challenges come from choosing the side band window large enough to allow a good determination of the background, while still being small enough that a simple empirical function will describe it. Another important aspect is that very good knowledge of the energy dispersion is necessary as inference is done on a steep spectrum where small biases in our knowledge of the energy dispersion can have large effects. Finally, in searches for dark matter, mass is a parameter that is not defined under the null hypothesis and will thus introduce a trial correction, which in practical applications is not simply calculated by applying the binomial distribution, but correlations have to be taken into account. In these more complicated cases there are two approaches to follow: employing Monte Carlo simulations or resampling from off-source data distributions. The main challenge here is to be able to simulate the null distribution to sufficient accuracy for the high significance needed (usually 5 $\sigma$), i.e. the need for a large ($\sim 10^8$) number of independent experiments. Solutions to this have emerged, the simplest being to fit an empirical function to the obtained test statistic distributions~\citep[e.g. done by][]{Weniger:2012tx}. Another one proposed by~\cite{Gross:2010bma} extrapolates to higher significance from a lower number of simulations~\citep[see also][for a more detailed discussion]{Conrad:2014nna}. It should be noted that the latter is only strictly applicable in the case of continuous trials, i.e. the whole spectrum is fit with the mass as a free parameter. Finally, in case of not very conclusive data, it might be difficult again to distinguish a line feature from any other feature (e.g. a broken power-law). Again, non-nested hypothesis testing would have to be performed for rigorous results.

\section{Status of dark matter searches with gamma rays}
\label{sec:status}

\par In this section we review the current status of indirect dark matter searches with gamma rays. We assimilate the information on astrophysical targets (Section~\ref{sec:targets}) with the information on instrumental sensitivities and statistical methodology (Section~\ref{sec:analysis}). For the two main types of gamma ray experiments, the all-sky \Fermi-LAT and the IACTs, our goal is to review the statistical analysis method that is most appropriate for each target, and review the results that have been obtained. 

\par As was reviewed in Section~\ref{sec:analysis}, the nominal all-sky survey mode of the \Fermi-LAT is ideally suited to explore essentially all potential targets in the gamma-ray sky. From an astrophysical and experimental perspective, though, these targets incur significantly different systematic uncertainties. 

\par  As we detail in this section, in the past few years the \Fermi-LAT collaboration has extended the statistical framework of the official {\tt ScienceTools} to include a joint likelihood formalism that allows for the combination of several source regions of interest into a single analysis. Such a combined analysis was introduced in~\citet{Ackermann:2011wa}, and was reviewed in~\secref{sec:dwarfs_analysis} above.  Though this framework is not yet able to completely solve the complexity of combining LAT analyses of all possible targets, it has for the first time produced sensitivities to the nominal thermal relic value of $\sigmav\sim3\times10^{-26}$ \cms. 

\par At higher energies, IACTs have gained ground. The constraints obtained from observations of the Milky Way halo in the vicinity of the Galactic Center constitute a large step forward for constraining WIMP masses above about a TeV. 

\par In the following sections, we provide a summary of the most relevant gamma ray analyses and results from \Fermi\-LAT and IACTs, including observations of the Galactic center and diffuse Galactic halo, dSphs, dark satellites, galaxy clusters, and the diffuse isotropic background. 

The spectrum of gamma rays depends on the annihilation channel. Unless otherwise noted we will follow the convention to present results for annihilation into b-quark pairs, which can be seen as representative for quark annihilation in general. Annihilation to $\tau$-leptons generically results in a harder spectrum. Majorana WIMPs in general do not preferably annihilate to light leptons (electrons or muons) as the annihilation is helicity suppressed, except if for example virtual internal bremsstrahlung or final state radiation is considered for supersymmetric particles, e.g. \cite{Bringmann:2007nk}.

\subsection{Galactic Center}
\par As a prime target for indirect dark matter searches, the Galactic center has been the subject of intense scrutiny for well over a decade.~\citet{2004APh....21..267C} proposed a dark matter interpretation to the gamma ray source detected with EGRET~\cite{1998A&A...335..161M}. Near the same time IACTs also detected a TeV source \cite{2004A&A...425L..13A,2006ApJ...638L.101A}, which seriously challenged a dark matter interpretation of the EGRET source~\cite{2006PhRvL..97v1102A}. Independent of dark matter, these observations were important because they showed that near the Galactic center, in addition to the diffuse background, there exist strong gamma ray emitters. More recent analyses of the \Fermi-LAT data \citep[e.g.][]{2011PhLB..697..412H,2013PhRvD..87l9902A,Daylan:2014rsa,Abazajian:2014fta,Calore:2014xka,Abazajian:2014hsa} have reported the detection of an extended excess, compatible in shape and spectrum with a $\lesssim\,30$ GeV WIMP annihilating to \bbbar. 

\par Several groups have followed up and confirmed the existence of this inner Galaxy excess in the \Fermi-LAT data.~\citet{Abazajian:2014fta} pointed out the critical dependence on foreground modeling, as well as the possibility that the excess arises from unresolved millisecond pulsars.~\citet{Carlson:2014cwa,Petrovic:2014uda} show that this excess can be explained by  local population of cosmic-ray protons, potentially due to a burst-like injection several thousand years ago. There is also the yet unsettled characterization of the \Fermi bubbles  at low latitude (see~\citet{Calore:2014xka} for a discussion of this point). At the time of writing, there is no clear resolution to this excess emission, and it is quite possible that there won't be a resolution for some time. It is certainly true that a confirmation of this excess in one of the other targets, for example clusters or dwarf galaxies, would be needed to further strengthen the case for dark matter. Taking a best guess nominal $J$-factor of the GC, the most recent \Fermi-LAT combined dSph search could have confirmed the excess, but only reported constraints \citep{2015arXiv150302641F} (see below). Probably the most conservative stance at this point assumes that all of the excess photons are not due to dark matter annihilation; this assumption implies stringent constraints on a contracted NFW scenario at the Galactic Center~\citep{2013JCAP...10..029G}.

\subsection{Galactic Halo}
\par The diffuse Galactic halo has long been advocated as an interesting target for dark matter searches.~\citet{2003MNRAS.345.1313S,2008APh....29..380S} first discussed the idea of searching in an annulus around the center of the Galaxy. This would bypass aforementioned systematic uncertainties that hinder such a Galactic center analysis. The HESS collaboration put such a strategy in practice in~\citet{PhysRevLett.106.161301}, where an annular region from $0.3^\circ$ to $1^\circ$ radius was analyzed, excluding the Galactic plane ($|b|<0.3^\circ)$. Using 112 hours of livetime observations and a generic continuum spectrum from \citet{PhysRevD.66.083006}, they derived, for NFW or Einasto density profiles,  the most stringent limits to date in the 1 TeV WIMP  mass range. For instance \sigmav in excess of $3\times10^{-25}$ \cms\ is excluded assuming an Einasto density profile.

\par As discussed in \secref{sec:analysis}, defining the background region within the field of view is not efficient, especially if the dark matter distribution in the central galaxy is cored. A proof of concept for treating this situation is published in~\cite{HESS:2015cda} providing competitive limits with only 9 hours of data. The main conceptual novelty is that the OFF region is constructed from separate pointings of the telescope array. To make sure that atmospheric conditions are preserved as much as possible an offset in right ascension is chosen to define the OFF-source pointings. However even in this case a novel treatment of the acceptance correction for the background estimate had to be introduced. In this particular case the center of the OFF regions is about 6 degrees away from the center of the signal region, and the $J$-factor is reduced by at least a factor of 3.

\par More recently, \Fermi-LAT has analyzed data on a larger scale within a region of $|l|<15^\circ$ and $5<|b|<80^\circ$~\cite{Ackermann:2012rg}. This analysis used \pass{7}  and masked 1FGL point sources. The analysis proceeds with the \latcls{clean} class and thus the {\tt P7CLEAN\_V6}  instrument response functions (irfs) to infer stringent limits on \sigmav, while discussing in great details the dependence of the analysis on the Galactic foreground modeling, which is the main issue with such a halo analysis when comparing to the dwarf spheroidal constraints.

\bigskip

\begin{figure}[!h]
  \centering
\includegraphics*[scale=0.5]{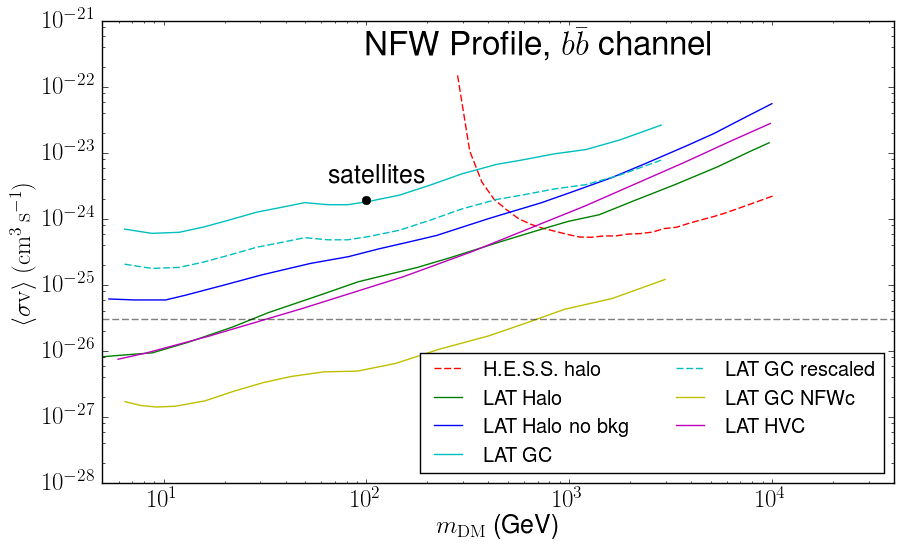}
\caption{ Upper limits (UL) on velocity averaged annihilation cross-section versus WIMP mass from indirect searches in $\gamma$ rays, for analyses focusing on the Milky Way. In dashed red (``HESS halo''), the 95\% CL UL in the Galactic Center halo analysis from \citet{PhysRevLett.106.161301}; in red (``LAT halo no bkg''), the LAT halo analysis \citep{0004-637X-761-2-91} corresponding to 3$\sigma$ CL upper limits, when no diffuse background modeling is performed; in green (``LAT halo''), same but in the case where proper modeling of the diffuse gamma ray background is performed; in cyan (``LAT GC''), the LAT diffuse-model-free 3$\sigma$ ULs at the Galactic Center \citep{2013JCAP...10..029G}; in magenta (``LAT GC rescaled''), same but rescaled to the same local DM density as the LAT halo analysis; in yellow (``LAT GC NFWc), same but in the case of a contracted NFW profile; in magenta (``LAT HVC''), the 95\% C.L. limits obtained with the Smith cloud \citep{0004-637X-790-1-24}. The dot marker with the label ``satellites'' corresponds to the 95\% CL upper limit for a 100 GeV WIMP mass obtained in the unidentified LAT source analysis~\citep{Ackermann:2012nb}.
}\label{fig:galaxy_bb_nfw}
\end{figure}

To summarize the previous discussion and the current observational situation, we gather in Figure~\ref{fig:galaxy_bb_nfw} the most relevant current limits obtained for analyses of regions within the Milky Way. 
This figure clearly illustrates the power that a deep observation with a Cherenkov instrument has to set competitive upper limits for energies about $\sim$1 TeV. It also emphasizes the advantage of a halo analysis with the LAT, with respect to satellite or GC blind searches, though we stress again that background systematics affect the limits much worse in this case. In addition, the halo upper limits are proportional to the squared normalization of the dark matter density distribution of the Galaxy, usually estimated at the position of the Sun. As discussed in Section~\ref{sec:targets} there is still a substantial uncertainty in the estimation of the local dark matter density,  0.2 to 0.9 GeV cm$^{-3}$, so that the resulting uncertainty on a dark matter annihilation flux is in fact larger than the uncertainty just contributed by the modeling of the Galactic diffuse emission~\citep[see][for further details]{Ackermann:2012nb}. 

\subsection{Dwarf spheroidals} 
\par As discussed in~\secref{sec:targets}, with their lack of high-energy emission processes and large dark to luminous mass ratio, dwarf spheroidal galaxies are a prime target for gamma ray instruments. As the expected flux scales with the inverse square of the distance, only the satellites of the Milky Way have been investigated so far, which amounts to about 25 targets. While \Fermi is surveying the whole sky continuously, IACTs need to allocate observation time to point in the direction of any dSph. As of this work, a total of 9 dSphs have been observed by ACTs. The HESS collaboration initially presented constraints from $\sim$11 hours of observation toward Sagittarius dwarf \citep{2008APh....29...55A, 2010APh....33..274A}, which have recently been largely revised with 90 hours accumulated on target and with a major modification to the density profile of this tidally stripped nearby galaxy \cite{2013arXiv1307.4918L}. The HESS collaboration also published results from Canis Major ($\sim$ 10 hours) \cite{2009ApJ...691..175A}, Sculptor ($\sim$ 12 hours) and Carina ($\sim$ 15 hours) \cite{2011APh....34..608H}. As mentioned previously, HESS also has presented an analysis combining previous observations of dSphs with a new long exposure of the Sagittarius dwarf, employing a combined likelihood technique inspired by the \Fermi-LAT analysis~\cite{Abramowski:2014tra}. The MAGIC collaboration presented results from 10 to 20 hours of observations towards Willman I \citep{Aliu:2008ny}, and Draco \citep{Albert:2007xg}, but devoted most of the allocated time to the observation of Segue 1, initially with 43 hours on one telescope \citep{2011JCAP...06..035A}, and recently updated with a 160-hour analysis \citep{2014JCAP...02..008A}, that also makes use of a full likelihood technique, akin to what the \Fermi-LAT collaboration is routinely doing. Finally, the Veritas collaboration presented results for Bootes I, Draco, Ursa Minor and Willman I~\cite{1006.5955v2}, and for Segue 1 (50 hours)~\cite{1202.2144v1}. 

\par These observations are to be compared to the 1000-hour equivalent of 11 months of LAT survey, which resulted in flux limits set in the direction of 20 dSphs, and dark matter constraints derived for a subset of 8 of these, based on robust $J$-factor values \citep{Abdo:2010ex}. Focusing specifically on the constrained minimal super-symmetric model,~\citet{Scott:2009jn} added the analysis of Segue 1 with approximately the same amount of data. Thus, quite generically given the plausible allocation time on dSphs by IACTs, LAT constraints are expected to dominate below WIMP masses of order 1 TeV, at which point the fast degradation of LAT sensitivity quickly limits its performance. As a consequence, while programs are ongoing to increase the total duration of IACT observations in the direction of dSphs~\citep[\eg][]{1307.8367v1}, it is also crucial to maximize the statistical power of the analyses by using the full-likelihood technique, as considered for instance in~\citet{2012JCAP...10..032A,2014JCAP...02..008A} and \citet{Abramowski:2014tra}.

\par Along these lines,~\citet{Ackermann:2011wa} introduced a joint-likelihood formalism that combines the statistical power provided by several dSphs into a single inference (see Section~\ref{sec:dwarfs_analysis}, and also~\citet{PhysRevLett.107.241303} for an alternative combined methodology). Using 24 months of \pass{6} ``diffuse'' data and the corresponding P6V3 irfs, the \Fermi-LAT collaboration derived a single 95\% C.L. exclusion curve using 10 dwarfs, that for the first time reached the ``standard'' thermal-relic value of $\sigmav = 3 \times 10^{-26}$ cm$^{-3}$ s$^{-1}$, for annihilation into quarks. Furthermore, the \Fermi-LAT collaboration introduced for the first time a scheme to directly account for the $J$-factor uncertainty into the upper limits on $\sigmav$ stemming from the statistical nature of the dark matter profile derived from stellar data. The analysis presented in~\citet{Ackermann:2011wa} was later updated to 4 years of \pass{7} reprocessed data~\cite{2014PhRvD..89d2001A}, in which a much more thorough analysis of the data in the direction of 25 dSphs was done. A subsample of 15 dSphs with robust $J$-factors were retained to derive upper limits on the annihilation cross-sections in different channels. In the latter analysis, it was found that the test statistic distribution, naturally defined as the likelihood ratio between the null hypothesis (background-only) best fit and the DM+background fit did not follow a $\chi^2$ distribution, implying a correction reducing the apparent significance of an excess. The most likely cause of this deviation is a population of unresolved point sources, confirmed by the most recent incarnation of the \Fermi-LAT analysis~\cite{2015arXiv150302641F}.

\par For the case of dSph searches it is worth noting that the advantage of the full likelihood formalism developed lies in the fact that it opens up the possibility to eventually share the likelihood functions across collaborations to add as many dSphs as possible, or even combine targets. As a first step in this direction, the \Fermi-LAT collaboration has released the likelihood functions used in~\citet{2014PhRvD..89d2001A}. Finally, to our knowledge  the possibility to add $J$-factor uncertainties to a full-likelihood formalism, as proposed by the LAT collaboration~\cite{2012JCAP...10..032A,2014PhRvD..89d2001A}, has only been considered in the most recent search for dSph emission performed by the HESS collaboration \citep{Abramowski:2014tra}.

\par On the topic of satellite galaxies of the Milky Way, it also should be noted that the Large Magellanic Cloud (LMC) and Small Magellanic Cloud (SMC) also provide targets for dark matter searches. Extraction of gamma-rays from dark matter must however compete with the detected gamma-ray emission that arises from cosmic rays~\cite{Abdo:2010pq,2010A&A...523A..46A}.~\citet{Buckley:2015doa} have recently performed an analysis of the LMC searching for a dark matter signal with \Fermi-LAT data, and report a null detection. 

\subsection{Dark satellites} 
%{\bf Galactic Substructures\quad\ } 
\par The $\Fermi$-LAT has performed a search for a gamma-ray signal from the predicted population of dark satellites of the Milky Way~\cite{2012ApJ...747..121A}. In this analysis the 231 high-latitude ($|b|>20^\circ$) unidentified 1FGL sources were augmented with 154 candidate detections using a dedicated search for potentially extended sources at high latitude, and one year of \pass{6} \latcls{diffuse}-class LAT data (in the energy range 200 MeV to 300 GeV). A spectral and spatial selection was applied to check compatibility with a dark matter signal, resulting in no remaining candidate and, after comparison with simulations, a constraint on \sigmav is derived which is about $1.95\times10^{-24}$ \cms\ for a 100 GeV WIMP annihilating into \bbbar. Starting from the 1FGL point source catalog,~\citet{Buckley:2010vg} also undertake a search for dark matter subhalos, and similarly report no conclusive detection. 

\par Some theories suggest that high-velocity clouds (HVC) may be embedded within dark matter subhalos~\cite{1999ApJ...514..818B}, and if so, they provide a target for gamma ray searches.~\citet{0004-637X-790-1-24} focused on the Smith Cloud, a massive low-metallicity HVC located at a distance of about 12 kpc from the Sun (this is one of the few HVCs with a known distance). Five years of \pass{7} reprocessed~\latcls{clean}-class data in the energy range 500 MeV to 300 GeV yielded no detection, resulting in constraints on \sigmav which are comparable to the dSph combined analysis \cite{2014PhRvD..89d2001A} for an assumed NFW profile. Of course it should be emphasized that the dark matter content of HVCs is very controversial~\cite{2013ApJ...777..119F}, and the constraints from the Smith Cloud strongly depend on the dark matter profile. Indeed the limits degrade by a factor 40 when using a Burkert profile~\cite{1995ApJ...447L..25B} rather than a NFW or Einasto profile~\cite{0004-637X-790-1-24}. 

\subsection{M31}
%\label{sec:localgroup}
%\noindent{\bf M31 \quad} 
In much the same way as the Milky Way, the Andromeda or M31 galaxy, our closest spiral galaxy neighbor, is expected to shine in gamma rays. The primarily emission is due to its gas and cosmic-ray content; these components provide a background from which a potential dark matter signal must be disentangled. In 2010 the $\Fermi$-LAT collaboration announced the detection of M31 using about two years of \pass{6} \latcls{diffuse}-class (P6DIFFUSE\_V3 irfs) data in a $10^\circ\times10^\circ$ squared region centered on M31~\cite{2010A&A...523L...2A}. Using a spatial template derived from the IRIS 100\,$\mu$m far infrared map \citep{2005ApJS..157..302M}, the spectrum derived from a fit to the LAT data is consistent with a rescaled spectrum obtained by a GALPROP run to model the Milky Way gamma ray emissivity~\cite{Vladimirov:2010aq}. Thus the LAT detection of M31 is compatible with star formation rates and gas content, and a conservative 95\% C.L. upper limit on \sigmav for a 100 GeV WIMP in the $b\bar{b}$ channel is derived at about $5\times10^{-25}$ \cms, assuming a smooth Einasto density profile derived from \citet{2002ApJ...573..597K}. 
%The LAT flux results, majored by 2 sigmas, were also used by \citet{2010JCAP...12..015D} to constrain decaying dark matter models, resulting for instance into a decay lifetime in the \bbbar channel of $\sim6\times10^{25}$ and $\sim1.2\times10^{26}$ s, in the case of a point-source and extended model, respectively. 

\par The M31 analysis has since been updated with a 4.5 year \pass{7} dataset~\citep{2013arXiv1312.7609L}. These authors add a dark matter profile to the model that already includes a template for the emission from M31 due to its gas content. This makes the resulting upper limit less conservative but the analysis more sensitive. Under the assumption of a smooth halo, the cross section upper limit is very similar, $\sim5\times10^{-25}\;\mathrm{cm}^3\mathrm{s}^{-1}$.  More recently there has also been the claim of an extendend excess attributed to a cosmic-ray halo \cite{Pshirkov:2015hda}.

\par Finally, in the TeV energy range, early searches have been performed with CELESTE~\citep{2006A&A...450....1L}, and HEGRA~\citep{2003ICRC....3.1685H}, without detection and with uncompetitive upper limits derived.

\bigskip 

\par
\begin{figure}[!h]
  \centering
\includegraphics*[scale=0.5]{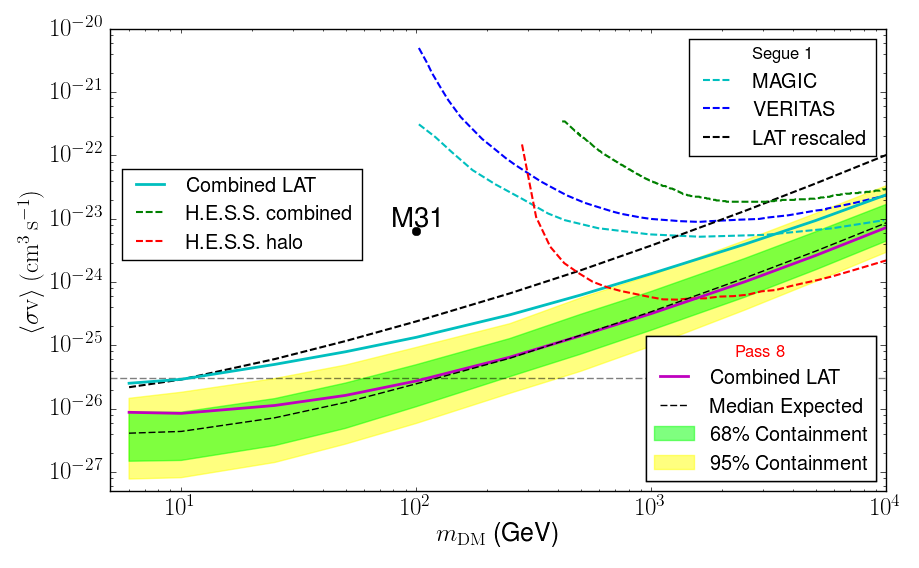}
\caption{ Comparison  of constraints on the velocity averaged annihilation cross section (\bbbar channel) versus WIMP mass derived from the LAT combined analysis of 15 dwarf galaxies (assuming an NFW profile), 160-hour observations of Segue~1 by MAGIC~\cite{2014JCAP...02..008A}, 48-hour observations of Segue~1 by VERITAS (assuming an Einasto profile)~\cite{1202.2144v1}, and 112-hour observations of the Galactic center by HESS, assuming an Einasto profile \citep{PhysRevLett.106.161301}. In the interest of a direct comparison, we also show the LAT constraints derived for Segue~1 alone assuming an Einasto dark matter profile consistent with that used by VERITAS \citep{1202.2144v1}. For this rescaling, the $J$-factor of Segue~1 is calculated over the LAT solid angle of $\Delta\Omega \sim 2.4 \times 10^{-4} \sr$ and yields a rescaled value of $1.7 \times 10^{19} \GeV^2 \cm^{-5} \sr$ (uncertainties on the $J$-factor are neglected for comparison with VERITAS). The \pass{8} limits (in magenta) and related expected-sensitivity bands are from \citet{2015arXiv150302641F}. Finally, the green dashed curve is from the 5-dwarf combined analysis of HESS data by \citet{Abramowski:2014tra}.}
\label{fig:local_group}
\end{figure}

 Figure~\ref{fig:local_group} summarizes the status of the dark matter searches with gamma rays in the Local Group. Similar to the constraints obtained from the Milky Way, this figure clearly indicates the complementarity between \Fermi-LAT and IACT constraints. 

\subsection{Galaxy clusters and isotropic emission}
\label{sec:dmclusters}
\subsubsection{Galaxy clusters}
As discussed in \secref{sec:targets_clusters}, galaxy clusters are anticipated to be gamma ray sources. Using EGRET data, no detection was reported in~\citet{2003ApJ...588..155R} for 58 clusters selected from an X-ray-bright sample, resulting in an average 95\% C.L. flux upper limit of $\sim 6\times10^{-9}$ cm$^{-2}$~s$^{-1}$ above 100 MeV. Null detections were also reported above 400 GeV in the direction of the Perseus and Abell 2029 clusters with the Whipple telescope, using $\sim$14 and $\sim$6 hours on source, respectively~\cite{2006ApJ...644..148P}. More recent null results in searches with ACTs include \citet{2009A&A...502..437A,2009A&A...495...27A,2010ApJ...710..634A,2012ApJ...757..123A,2012A&A...541A..99A}. With an 18-month generic gamma ray analysis toward 33 clusters \citep{2010ApJ...717L..71A} that improved upon previous analyses~\cite{2003ApJ...588..155R}, the LAT collaboration presented early constraints on a dark matter induced gamma ray signal in a subset of six clusters and for \bbbar and \mumu channels \citep{2010JCAP...05..025A}. In the latter case, the cluster analysis confirmed the tension between the LAT $e^+e^-$ spectrum and a generic leptophilic dark matter scenario that would aim to explain the Pamela positron excess~\citep{2009Natur.458..607A}. The \bbbar limits also showed promise, especially when accounting for expected but unknown cluster substructures. Fornax, the best target among the six clusters studied in~\citet{2010JCAP...05..025A}, was also observed by HESS with 14.5 hours and a threshold at 100 GeV~\citep{2012ApJ...750..123A, 2014ApJ...783...63A}. The Veritas collaboration also presented dark matter constraints in~\citet{2012ApJ...757..123A}, based on 18.6 hrs of observations of the Coma cluster, also studied in~\citet{2010JCAP...05..025A}. 

\par
\begin{figure}[h]
  \centering
\includegraphics*[scale=0.5]{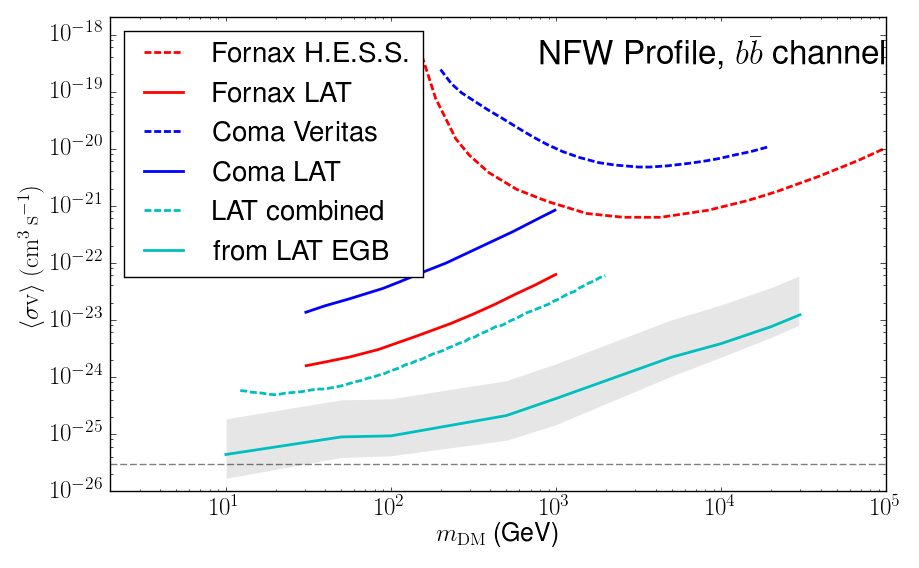}
\caption{Constraints on the velocity averaged annihilation cross-section versus WIMP mass from analyses in the directions of galaxy clusters, for an NFW profile and $\bbbar$ channel. The red and blue curves show the 95\% C.L. upper limits obtained with Fornax and Coma clusters, respectively, and with \Fermi-LAT (thick, \citet{2010JCAP...05..025A}) and IACTs \citep[dashed,][]{2014ApJ...783...63A,2012ApJ...757..123A} instruments. The cyan dashed curve shows the 95\% C.L. upper limits obtained with a combined analysis of 5 clusters \citep{Zimmer2011}. The cyan thick line shows the upper limits derived in \citet{Ajello2015} using the recent LAT EGB observations \citep{EGB2014}. The grey band shows the corresponding uncertainties rising from modeling the expected dark matter signal.}
\label{fig:clusters_bb_nfw}
\end{figure}

 As updated gamma ray constraints have been recently derived by the LAT collaboration in \citet{2013arXiv1308.5654A}, using a combined analysis akin to \citet{2014PhRvD..89d2001A}, it can be expected that a combined dark matter analysis of all or a subset of the clusters considered in this paper will soon be presented. For now we can only rely on the preliminary combined results obtained with 2 years of  \pass{6\_V11} DIFFUSE data that were presented in \citet{Zimmer2011}. In addition to Coma and Fornax, this analysis used M49, Centaurus, and AWM~7. These recent results are illustrated on Figure~\ref{fig:clusters_bb_nfw}. Finally, recent claims of an extended emission toward the Virgo cluster \citep{2012arXiv1201.1003H} have not been confirmed \citep{2012PhRvD..86g6004M}, emphasizing the importance of correctly deriving the background model before drawing conclusions based on the \Fermi-LAT data.

\subsubsection{Isotropic signal}\label{sec:cosmodm} Beneath the conspicuous Galactic diffuse emission and the constellation of resolved gamma ray sources, which are mostly active galactic nuclei (AGN), the gamma ray sky harbors an isotropic signal, the {\it isotropic gamma ray background} (IGRB)
~\cite{Fornasa:2015qua}. This was first detected already by EGRET~\cite{1998ApJ...494..523S}, and recently determined with very good precision by the \Fermi-LAT collaboration from approximately 100 MeV - 800 GeV \citep{2014arXiv1410.3696T}. The IGRB  certainly includes unresolved contributions from standard astrophysical sources, notably AGN. This emission may also contain the so-called ``cosmological'' dark matter signal coming from the summation of the dark-matter annihilation contributions of all dark matter halos across the history of the universe. Approaches to find this signal are based on the spectrum of isotropic component \cite[see][]{Ullio:2002pj} or on a spatial signature exploiting the fact that dark matter-induced anisotropies in the emission should follow the square of the mass density, whereas conventional astrophysical sources should follow the dark matter density linearly, which would reveal itself in differences in the angular power spectrum~\citep[see {\it e.g.}][]{Cuoco:2007sh}. Following the first measurement of the EGB\footnote{Extra-Galactic Background : the sum of the IGRB and resolved LAT extra-galactic sources.} spectrum  with \Fermi LAT data \citep{2010PhRvL.104j1101A}, the LAT collaboration has presented the limits obtained from the spectral shape alone \cite{Abdo:2010dk}. These constraints are very model dependent in two ways. First, the halo and subhalo abundance has to be modeled as function of redshift, and second the contribution of conventional sources has to be modeled. In a situation with a poorly modeled background a conservative way to place model constraints is to assume that the entire detected emission is from the putative signal. With this approach, constraints are about two orders of magnitude above the thermal limits even under moderately optimistic assumptions on substructure properties~\cite{Abdo:2010dk}. A proper background modeling would improve the constraints by maybe one to two orders of magnitude, see e.g.~\citet{2014PhRvD..89b3012B}. Thus, the publication of an updated IGRB spectrum by the LAT collaboration \citep{EGB2014} comes in par with an upcoming reassessment of the contribution of astrophysical sources, notably blazars \citep{Ajello2015}, to the EGB budget. Together with an improved modeling of the dark matter expected signal \citep{2015arXiv150105464T}, this effort leads to potentially competitive limits, as shown on Figure~\ref{fig:clusters_bb_nfw}.

\par Finally, anisotropies in the gamma-ray sky have been investigated in~\citet{2010ApJ...723..277H} and~\citet{2011MNRAS.414.2040C}. Combining the spatial signal and spectral signals may enhance the dark matter sensitivity of these searches~\cite{2013arXiv1312.3945C} and combining isotropic emission measurements with other sources can help to break the degeneracy between substructure boost and annihilation cross-section \citep{Ng:2013xha}.

\par Though employing anisotropy studies in IACTs have potential for very competitive constraints~\cite{Ripken:2012db}, so far these analyses techniques have not been applied to our knowledge. Systematics from the combination of several fields of view might hamper the usefulness of this approach.

\subsection{Searches for spectral features}
\label{lines}

\par
\begin{figure}[h]S
\centering
\includegraphics*[scale=0.5]{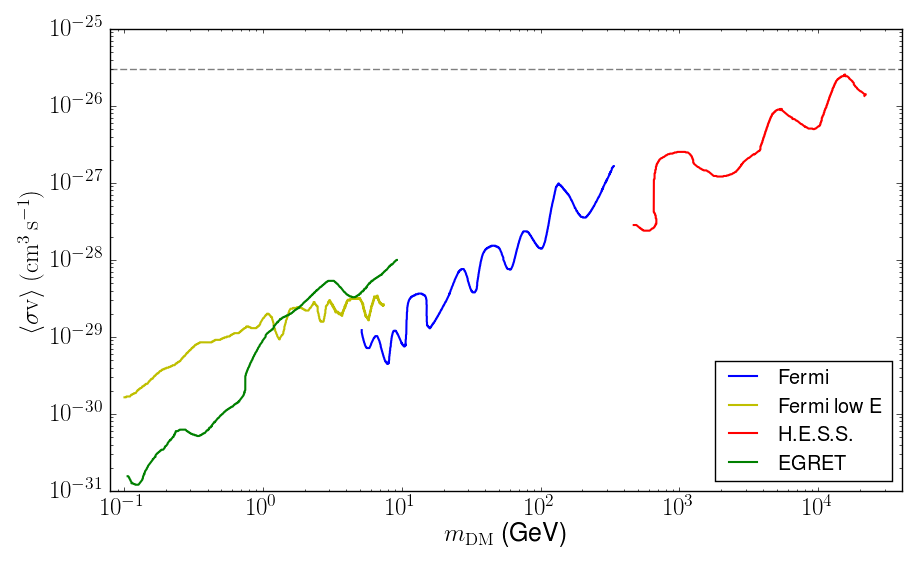}
\caption{Upper limits on the velocity averaged annihilation cross section into gamma-ray pairs versus WIMP mass, derived by several instruments in searches for line features. Blue and yellow lines show the limits obtained with the \Fermi-LAT in a first standard analysis~\cite{Ackermann:2013uma} and in a dedicated low-energy follow-up~\cite{Albert:2014hwa}, respectively.  The red line shows the limits obtained by the HESS collaboration~\cite{Abramowski:2013ax}. For reference, the EGRET limits are also shown~\cite{Pullen:2006sy}.}
\label{fig:glines}
\end{figure}

 Dark matter lines were suggested as a smoking gun signal almost thirty years ago \cite{Bergstrom:1988fp}. Experimentally (as pointed on in~\secref{sec:analysis}), line detection can be established relatively independently from background modeling, and source confusion is unlikely. Therefore, in contrast to continuum searches, the most promising target is the Galactic center or its vicinity.  Constraints on line emission have been presented using EGRET data~\citep{Pullen:2006sy}\citep{Mack:2008wu}, \Fermi LAT~\citep{Abdo:2010nc,Ackermann:2012qk,Albert:2014hwa,GeringerSameth:2012sr} and H.E.S.S. data \citep{Abramowski:2013ax}.  A summary of the most relevant present upper limits on line emission is shown in Figure~\ref{fig:glines}. 

~\citet{Weniger:2012tx} and~\citet{Su:2012ft} claimed detection of a line emission in \Fermi-LAT data. Using an analysis technique similar to~\citet{Abdo:2010nc}, as described in~\secref{sec:analysis}, but with a doubling of the amount of data as well as an optimization of the region of interest for signal over square-root of background,~\citet{Weniger:2012tx} found a (trial corrected) 3.2 $\sigma$ significant excess corresponding to a dark matter mass of  $\sim$ 130 GeV. If interpreted as a signal, this would amount to a cross-section of about $\sigmav \sim 10^{-27}$ \cms. However, this analysis was based on a mis-calibrated data set, and the inferred position of the line feature changes to $\sim$ 133 GeV with a correctly-calibrated sample. The claimed signal is concentrated on the Galactic center with a spatial distribution consistent with an Einasto profile~\cite{Bringmann:2012ez}. This was marginally compatible with the upper limit presented in~\citet{Ackermann:2012qk}. 

In~\citet{Ackermann:2013uma}, about 3.7 years of \Fermi-LAT data were analyzed, including a careful study of systematic uncertainties and a discussion of the alleged line feature. Apart from using correct calibration, this analysis also employed event-by-event reconstruction quality to improve the energy dispersion description. In terms of the line feature this analysis also finds an excess at 133 GeV, but the global significance is estimated to be only 1.5$\sigma$ when including information on the event-by-event energy reconstruction. 

\par Since the first report of a line-like feature, the main challenge to the claim that a line feature originates from dark matter annihilation was the fact that an excess at the same position also appears in gamma rays observed from the Earth's limb. The signal there appears after an appropriate zenith cut is applied that accounts for the fact that the \Fermi-LAT is usually not `head-on' exposed to the Earth limb. Gamma rays in the Earth limb are caused by cosmic-ray interaction in the Earth atmosphere and their spectrum is featureless, thus providing a useful control sample for spectral feature searches. While originally the limb signal was observed at similar significance as the feature in the Galactic center, the latest analysis of the \Fermi-LAT yields somewhat lower significance. As the limb data can be observed for different event selection criteria, it can be used to calibrate the Monte Carlo-based estimate of the effective area. A feature might appear if the effective area is underestimated (or overestimated) in limited energy ranges, and indeed indications are reported that the efficiency is overestimated around the position of the line~\cite{Ackermann:2013uma}. However, the effect seems too small to explain the presence of the feature in the limb data. 

\par As of the writing of this review, the origin of the line feature is not explained. However, its significance at the Galactic center is decreasing, as pointed out by~\citet{Weniger:2013dya}. Should an instrumental origin of the signal be ruled out (or results be inconclusive) independent confirmation will be necessary. The currently operational IACT HESS II may be sensitive enough to soon provide such confirmation~\cite{Bergstrom:2012vd}. The \Fermi-LAT collaboration has also presented a dedicated search for line features at energies between 100 MeV and 10 GeV~\citep{Albert:2014hwa}. 
%The resulting limits are presented in Figure~\ref{fig:glines}. 
The analysis is particularly instructive as the number of expected events in that energy range is large and systematic uncertainties start to dominate.

\section{Perspectives}
\label{sec:perspectives} 
\subsection{Future instrumentation}
\par While many gamma ray features identified by  \Fermi are intriguing, and in addition the recent measurements of the electron and positron spectra by Pamela, ATIC, \Fermi and AMS-02 are sufficiently intriguing to generate a wealth of speculations, they also emphasize the need for improved energy resolution, position resolution, and background rejection up to the TeV range and even beyond. In this section, we discuss what future gamma-ray experiments will bring to the field of indirect dark matter detection, and the improvement that we can expect in our understanding of this field over the course of the next decade, and longer. 

\subsubsection{Space Telescopes}\label{future_space}
%CALET pubs : http://calet.phys.lsu.edu/Public_Documents.php
%HERD vidéo : http://cds.cern.ch/record/1491650?ln=en
%chinese program : www.unoosa.org/pdf/pres/stsc2013/tech-27E.pdf‎
%DAMPE switzerland : http://dpnc.unige.ch/dampe/collaboration.html
% PANGU  IS MISSING : http://arxiv.org/abs/1407.0710
\par Over the next several years, the CALET and DAMPE experiments are expected to begin taking data. The CALorimetric Electron Telescope\footnote{\url{http://calet.phys.lsu.edu/index.php}}, a Japan-led project that involves Italian and American institutes, is planned for launch in 2015 and will be installed on the Japanese Experiment Module on board the International Space Station. DAMPE (DArk Matter Particle Explorer, formerly known as TANSUO)\footnote{\url{http://dpnc.unige.ch/dampe/index.html}}, is one of the five satellite missions selected by Chinese SPRPSS/CAS program, and is scheduled for launch slightly later than CALET, in 2015-2016. Italy and Switzerland take part in the DAMPE collaboration. Both projects feature a deep calorimeter to reach a total of 30 to 33 radiation lengths, in order to provide excellent energy resolution (better than 3\% above 100 GeV) and electron/proton separation ($\sim10^5$ rejection power) in the energy range of interest (1 GeV to 10 TeV). Both instruments will have roughly comparable performance, with a slightly larger electron geometrical factor for DAMPE (0.3 m$^2$sr) than CALET (0.12 m$^2$sr). Further information on the status of each project can be found in recent conference proceedings \citep{2012PhLB..715...35L,CALET_ICRC2013,DAMPE_ICRC2013}.

\par On a longer time scale (2018 and beyond), two other experiments will probe the high-energy electromagnetic sky: GAMMA-400 and HERD. Gamma-400\footnote{\url{http://gamma400.lebedev.ru/indexeng.html}} is a Russian-led satellite observatory, planned for launch in 2018-2019. Building upon the successes of \Fermi\ and AGILE, it features a gamma ray telescope reminiscent of the LAT, supplemented with a Konus-FG gamma ray burst monitor. The baseline design covers the range from 100 MeV to 10 TeV and is optimized for best performance around 100 GeV, where a very deep electromagnetic calorimeter (25 radiation lengths compared to $\sim$ 8.5 for the LAT), associated with a silicon strip tracker, will provide excellent energy and angular resolutions at such energies (a factor ten better angular resolution at 100 GeV than either DAMPE or CALET, and comparable energy resolution). Among the various gamma ray and cosmic ray science topics, dark matter searches, and especially the hunt for gamma ray lines, are a prime focus of the science case \cite{2013arXiv1307.2345M,2013AIPC.1516..288G}. Further information on the GAMMA-400 design and science case can be found in~\citet{2013AIPC.1516..288G,2013AdSpR..51..297G}.

\par The High Energy cosmic Radiation Detection (HERD) is an observatory planned for deployment on board the future China space station~\footnote{\url{http://english.ihep.cas.cn/rs/fs/sm/SM/SM\_aboutherd/}}. Design studies are still at an early stage of development, though the two primary science goals are already defined as the search for a dark matter signal and the origin of Galactic cosmic rays. At this time, this mission is not in competition with any other Chinese project. A recent development moved the baseline concept closer to the GAMMA-400 design, with a massive 3-dimensional calorimeter covered on five sides by tracker silicon planes (more details can be found on the second HERD international collaboration meeting, see \url{http://indico.ihep.ac.cn/conferenceDisplay.py?confId=3808}). 

\par Another experiment worth mentioning is PANGU~\cite{Wu:2014tya}. PANGU (the PAir-productioN Gamma-ray Unit) is a small  mission optimized for spectro-imaging, timing and polarization studies in gamma rays in the still poorly explored energy band from 10 MeV to a few GeV. The present design is a pair conversion telescope with detector resolution about a of factor 2 better than previous instruments and a pointing resolution of a factor 3 to 5 better than \Fermi-LAT. 

\par While all the future space experiments mentioned above involve the high-energy regime, one should not forget that the phase space is still open for a dark matter candidate in the MeV domain~\citep[often coined sterile neutrino or LDM for Light Dark Matter, see {\it e.g.} ][]{2013JHEP...11..193E}. This notoriously difficult energy regime is the focus of current intense research and development efforts geared toward the launch of a general-purpose astrophysical MeV space observatory by 2025 (see for instance the AstroMeV site http://astromev.in2p3.fr/).

\subsubsection{Ground Telescopes}
\par Two future instruments are expected to dominate the landscape of ground instruments: CTA\footnote{\url{https://portal.cta-observatory.org/Pages/Home.aspx}} and HAWC\footnote{\url{http://www.hawc-observatory.org/}}. CTA (Cherenkov Telescope Array) is the next-generation IACT~\citep[see {\it e.g. }][]{Acharya:2013sxa,Consortium:2010bc}. The energy range of this array is envisaged to be from a few tens of GeV to hundreds of TeV and the sensitivity is expected to be improved by one order of magnitude relative to current IACTs. The angular and energy resolution are expected to lie between 0.1 (0.05) deg and 25 (10)\% at low (high) energies, respectively. Technologically this energy range and sensitivity is achieved by combining a large number of single IACTs of different size. Currently, one of the baseline designs foresees four H.E.S.S. II size telescopes ($\sim$ 23 meter diameter), about 30 medium size telescopes ($\sim$ 12 meter diameter) and 30 to 70 small size telescopes ($\sim$ 7 meter). The US part of the consortium envisages to later extend the array with a large ($\sim$ 60) number of medium size telescopes, with particular view on high mass WIMPs (see e.g.~\citet{Wood:2013taa}.)

\par HAWC \citep[High Altitude Water Cherenkov,][]{2013arXiv1310.0074H} is a second generation cosmic ray and gamma ray observatory that builds upon the successful water Cherenkov technique pioneered by Milagro~\cite{2003ApJ...595..803A}. HAWC is designed to continuously ($\sim$ `24/7' duty-cycle) survey during 10 years the 100 GeV to 100 TeV sky with a 1.8 $sr$ instantaneous field-of-view telescope consisting of 300 water tanks, each instrumented with 4 photomultiplier tubes. HAWC is a joint Mexico-U.S.A. project located at 4100 meters on the flanks of the Sierra Negra volcano near Puebla, Mexico. It started science operation in August 2013 with about 100 tanks, and the full array has very recently been completed. While the HAWC astrophysical science case is very strong thanks to its synoptic surveying nature, it may also prove competitive for indirect dark matter searches for WIMP masses larger than about 1 TeV. 

\subsection{Until 2018}
\par Indirect detection of dark matter with gamma rays is entering a pivotal period. Upcoming instruments will reach sensitivities starting to probe into the most relevant parameter space at least for the most generic WIMP models. The \Fermi Large Area Telescope will collect data until potentially beyond 2018 and with an updated event selection, known as ``Pass 8'' \cite{2013ApJ...774...76A,2013arXiv1303.3514A}. The most anticipated results from the updated combined dSph analysis have been recently published \citep{2015arXiv150302641F}, and confirm that the dSph constraints are in mild tension with a dark matter interpretation of the GeV excess.~\figref{future} shows a simplistic forecast of the LAT combined dSph constraints with 10 years of \pass{8} data and 3 more dSphs than currently known.
Such an increase in the number of dSphs is motivated by the anticipation of new ultra-faint discoveries in the southern hemisphere with, for instance, surveys such as DES\footnote{http://www.darkenergysurvey.org/} and LSST\footnote{http://www.lsst.org/lsst/}. Indeed the DES collaboration has completed its first year of observations, and has --at the same time as other groups-- recently announced the discovery of 8 potential new dSphs~\citep{2015arXiv150302584T}\citep{Koposov:2015cua}\footnote{see also the recent discovery by PanSTARRS \citep{2015arXiv150305554L}},  supporting an optimistic view of the final number of dSphs that will eventually be available to a gamma-ray analysis. As a matter of fact, the \Fermi-LAT and DES collaborations released upper limits on these new candidates \citep{2015arXiv150302632T}, showing the promise for gamma-ray indirect searches of future optical surveys in the southern hemisphere. Despite no signifcant excess detected by the \Fermi-LAT analysis with \pass{8} data selection, \citet{Geringer-Sameth:2015lua} claim evidence (based on a 2-3 $\sigma$ excess) using the less sensitive \pass{7} event selection. To our knowledge, the various future experiments reviewed in~\secref{future_space} have not shown sensitivity curves that could be overlaid on~\figref{future}.

\par For IACTs, 2012 showed the first light for H.E.S.S.~II, and early results have been published recently. The addition of the 5th telescope will push the threshold down to about 50 GeV, with corresponding impact on derived dark matter constraints. World-leading limits on WIMP annihilation can be expected from the Galactic center search for masses above about 800 GeV. Such a limit has been estimated by~\citet{2014arXiv1411.1925C}, and is reported in~\figref{future} as well. H.E.S.S.~II will also likely provide new insight into the issue of the possible line emission discussed in~\secref{lines}. Following another path, VERITAS aims to accumulate a total of 1000 hours on Segue 1~\citep{2013arXiv1304.6367S} until 2018, a program that may rival the H.E.S.S.~II constraints, as can also be seen in~\figref{future}. Finally, the HAWC collaboration presented early results for such a search in the direction of Segue 1 with only 30 water tanks and 82.8 days of observation (HAWC-30), and projection with the full array (HAWC-300) and one year of observation \citep{2013arXiv1310.0073H}.
%\begin{figure}[h]
%\centering
%\includegraphics[scale=0.7]{pictures/hawc}
%\caption{ Current limits on \sigmav vs. WIMP mass $M_\chi$ in the annihilation
%channels $\chi\bar{\chi}\rightarrow\tau\bar{\tau}$ and $\chi\bar{\chi}\rightarrow b\bar{b}$ from observations of
%Segue 1 made by Fermi-LAT (24 months) \cite{Ackermann:2011wa} and
%VERITAS (50 hrs) \cite{1202.2144v1}. Preliminary HAWC-30 sensitivity
%for 82.8 day of observation and anticipated sensitivity of
%HAWC-300 for 1 year of observation. Taken from \citet{2013arXiv1310.0073H}.}
%\label{fig:hawc}
%\end{figure}
Using a $J$-factor of $7.7\,10^{18}\,\text{GeV}^{-2}\,\text{cm}^{-5}\,\text{sr}$ they obtain limits, shown on \figref{future}, that could become competitive above 1 TeV, when taking into account that HAWC does not need to allocate dedicated time ``on source',' contrary to imaging Cherenkov telescopes (the latter do have a much better angular and energy resolution, though).~\citet{2014arXiv1405.1730A} recently investigated more thoroughly HAWC sensitivity to a dark matter signal, based on simulations.

\subsection{Beyond 2018: future satellites and CTA}
\par
\begin{figure}[h]
 \centering
\includegraphics*[scale=0.5]{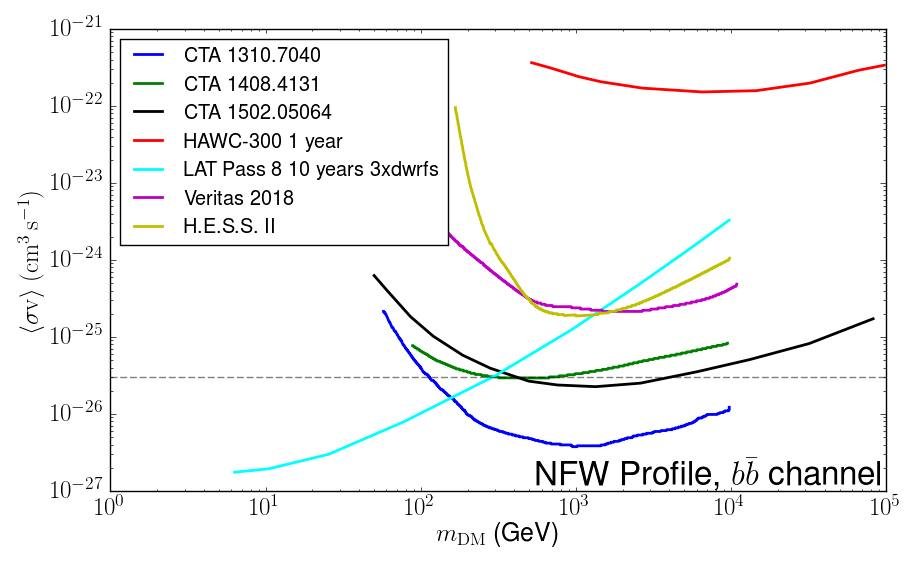}
\caption{Selection of sensitivities to velocity averaged annihilation cross-section versus WIMP mass reachable with current or future experiments. The blue curve is from a Galactic center analysis with CTA, from \citet{2013arXiv1310.7040B}. The green and black curves come from similar analyses, but attempting to account for the degradation of the limits due to uncertainties in the Galactic diffuse emission~\citep{Silverwood:2014yza,2015arXiv150205064L}. The red curve corresponds to one-year of observation of Segue~1 with the full HAWC instrument~\citep{2013arXiv1310.0073H}. The cyan curve shows the~\Fermi-LAT Pass 8 limits by 2018, accounting for more dwarf spheroidals, based on a preliminary analysis of 5 years of data. The magenta curve shows the VERITAS expectations from 1000 hours on Segue~1~\citep{2013arXiv1304.6367S}, and the yellow curve is an estimate by~\citet{2014arXiv1411.1925C} of the limits H.E.S.S.~II could reach on the Galactic center.
\label{fig:future}
}
\end{figure}

 Future satellite missions have been discussed in~\secref{future_space}. GAMMA-400/DAMPE will be mostly contributing to searches for spectral features due to their superior energy resolution. HERD will provide a significant step forward with respect to~\Fermi~-LAT also in terms of acceptance. For a dSph combined analysis, assuming the number of dSphs to be constant (which is a very conservative assumption), it seems that an exclusion of the thermal WIMP cross section should be possible up to WIMP masses of about 1 TeV by 2025. Detailed studies of the potential reach are however not known to us. This is not the case with CTA, for which the sensitivity reach has been studied in several publications, mostly focusing on the Galactic Center~\cite{Doro:2012xx,Wood:2013taa,2013arXiv1310.7040B,Pierre:2014tra,Silverwood:2014yza,Roszkowski:2014iqa}. dSphs and other approaches have been studied in~\citet{Doro:2012xx}.~\figref{future} presents the limits anticipated at the Galactic center from~\citet{2013arXiv1310.7040B}.  The one important caveat in interpreting the constraints is of course the systematics of the diffuse emission. As CTA lowers the threshold to about 10 GeV, constraints obtained from CTA will be prone to the same systematics as the \Fermi-LAT Galactic center analysis. \citet{Silverwood:2014yza} propose a quantitative discussion of this issue, and derive an upper limit curve that attempts to include such uncertainties, which is reproduced in~\figref{future}.

%\begin{figure}
% \includegraphics[height=.35\textheight]{pictures/future_280814}
%  \caption{Same as in figure 1, except that predictions for CTA are presented together with a (pessimistic) expectation for the Fermi-LAT.}
%\label{fig:CTA_forecast}
%\end{figure}

\subsection{Progress on nuisance parameters}
As emphasized throughout this article, determination of particle dark matter limits and extraction of a signal in the future requires an understanding of astrophysical systematics. These systematics include measurements of dark matter distributions in the different targets and measurements of the diffuse gamma-ray backgrounds. In this subsection, we review the progress that is expected to be made with these systematics over the next several years. 

\subsubsection{dSph mass distributions} 
\par Much of this article has emphasized the methods of setting limits on the dark matter annihilation cross section using dSphs. These limits are constrained by the systematic uncertainties in the measurements of the $J$-factors, or the dark matter mass distributions in the dSphs. In this analysis above, we highlighted how this systematic uncertainty is treated as a statistical uncertainty in the gamma-ray analysis. 

\par In the future it is optimistic to expect that this systematic uncertainty in the $J$-factor will be reduced. For the brightest, classical dSphs, several thousands of line-of-sight velocities have now been measured. Order-of-magnitude improvements on these results will require 30-meter class telescopes~\footnote{\href{http://www.gmto.org/}{http://www.gmto.org/},\quad\href{http://www.tmt.org/}{http://www.tmt.org/},\quad\href{http://www.eelt.org.uk/}{http://www.eelt.org.uk/}}, which are expected to come online within the next decade. These larger samples of resolved stars with kinematic information will continue to improve the measurements of their dark matter distributions. A larger sample of line-of-sight velocities will be important to further test the initial results that have been obtained from multiple populations of stars in a small number of dSphs. 

\par A larger sample of line-of-sight velocities is important because forthcoming larger scale IACTs with angular resolution $\lesssim 0.1^\circ$ will be very sensitive to the dark matter distribution in the center of the dSphs. This is the regime in which the shape of the central dark matter profile begins to significantly affect the determinations of $J$-factors~\cite{Charbonnier:2011ft,Strigari:2013iaa}. In addition to the improvement in the kinematics, it will also be important to improve the photometric data in the center of the dSphs. This is because there is a significant degeneracy between the central density of the stellar profile and the central density of the dark matter profile~\cite{Strigari:2010un}. 

\par In addition to the increase in the sample of line-of-sight velocities from known dSphs, new ultra-faint satellites that will be discovered in future surveys will require kinematic follow up. It is intriguing to note that a forthcoming survey, such as the DES or LSST, may detect an object that is as massive and nearby as Segue~1. Beyond measurements of line-of-sight velocities, in the future it will also be possible to obtain transverse velocities of bright stars in several dSphs. This will go a long way towards breaking the degeneracy between the velocity anisotropy and the dark matter mass profiles. Adaptive optics systems on 30-meter class telescopes may be suited for these measurements. 

\subsubsection{Local dark matter density distribution}
\par Gaia, an astrometric mission launched in 2013, will have an astrometric accuracy of approximately 1 mas and a photometric accuracy of 60 mmag for stars brighter than 20th magnitude ~\cite{Perryman:2001sp}. Because of the improvement in the measurements of the phase space distribution of local stars, Gaia will by extension improve upon current measurements of the dark matter distribution in the Milky Way~\cite{2013ApJ...779..115B}, and especially the local dark matter density~\cite{Read:2014qva}. Measurements along these lines will be especially important for interpretation of the annihilation cross section limits from diffuse gamma-ray measurements, in addition to its obvious importance for local dark matter searches. 

\subsubsection{Backgrounds}
\par As alluded to in~\secref{sec:inst}, IACTs and space borne instruments suffer from quite distinctive background. Until recently, the former could rely on a small field of view and steeply decreasing gamma-ray diffuse spectra to model the background below any source analysis as arising dominantly from an isotropic charged cosmic-ray component. With the detection of the Galactic diffuse emission by H.E.S.S.~II~\citep{2014PhRvD..90l2007A}, in the near future this may be no longer true, at least for analyses at low Galactic latitude.

\par On the contrary, cosmic-ray charged background is effectively rejected by an instrument like the LAT, and plays a potential role only at very high latitude, where the isotropic component of a standard LAT sky model may dominate~\footnote{this isotropic component is thus derived by the LAT collaboration for major event classes, and distributed at the FSSC : \href{http://fermi.gsfc.nasa.gov/ssc/data/access/lat/BackgroundModels.html}{http://fermi.gsfc.nasa.gov/ssc/data/access/lat/BackgroundModels.html}}, or for analyses related to the extragalactic gamma-ray background, especially at low energy (see~\secref{sec:cosmodm}). On the other hand, the gamma ray sky as seen by the LAT is extremely complex, and its imperfect angular resolution often demands a model over a much larger fraction of the sky than the immediate vicinity of a source of interest. 

\par While un-modeled sources above threshold can seriously alter the results of a dark matter search~\cite{Han:2012uw} and unresolved sources can bias its significance, the most difficult component to account for remains the Galactic diffuse background. The uncertainties that enter its derivation are mentioned in~\secref{sec:galdiff} and detailed in the review on indirect searches by~\citet{porter_dark_2011}. The effect on LAT all-sky residuals of variations in some of the components that define the model has been studied in~\citet{Ackermann:2012rg}, while a Bayesian investigation of the posterior probability distribution of cosmic-ray-related parameters is the subject of~\citet{Trotta:2010mx}. Whether predicted with {\it ab initio} propagation codes like GALPROP or estimated with template fitting procedures as for the standard diffuse model of the LAT collaboration, Galactic diffuse models always leave residuals that need dedicated treatment. Nevertheless, these residuals are sufficiently low on average to guarantee that a nominal diffuse model is useful for the large majority of LAT analyses. What remains very difficult to achieve is proper propagation of uncertainties, which often result from an ad hoc study, as in the EGB determination~\citep{EGB2014}, the construction of a catalog of supernova remnants~\citep{2013arXiv1307.6570H}, or the extraction of upper limits on a dark matter signal from the EGB~\citep{2015arXiv150105464T,Ajello2015,2015arXiv150105316D}. In the future, new insight may come from next-generation propagation codes, like PICARD \citep{kissmann_picard:_2014},  or new cosmic-ray data, notably from AMS-02\footnote{http://www.ams02.org/}. Also improved constraints are expected on the interstellar radiation field, a critical ingredient to the determination of the large-scale inverse Compton emission of the Galaxy, coupled to further investigations on statistical ways to properly account for these uncertainties.

 \subsection{Beyond 2024}
\par Beyond the next generation satellite experiments (DAMPE, HERD, GAMMA-400, PANGU etc.) and CTA, we are not aware of concrete proposals for next to next generation gamma ray telescopes for dark matter detection. Indeed, the question can be asked if a ``Super-CTA'' or ``Dark Matter Array,'' is possible. This was for example proposed in~\citet{Bergstrom:2010gh}, where CTA's sensitivity was improved by a factor 10 as a ``Gedanken-experiment.'' In many ways, CTA will be the ultimate indirect detection experiment. Progress significantly below the thermal cross-section is hampered by the irreducible electron background and the systematic uncertainties in the knowledge of the acceptance.

\section{Conclusions}
\label{sec:conclusions} 
\par In this article we have reviewed the methods that are utilized to search for gamma rays from dark matter annihilation, discussed the challenges in extracting the signal, and reviewed the results that have been obtained by different experiments. We have in particular focused on the great deal of progress over the past few years that has resulted from the analyses of \Fermi-LAT data. Without question, during the course of its over 6 year mission lifetime, the \Fermi-LAT has been a phenomenal success, breaking new ground in indirect dark matter searches and far exceeding the pre-launch projections for its reach~\cite{Baltz:2008wd}. Analysis of several systems, most notably dwarf spheroidals, are now providing robust upper limits on WIMPs in the mass range $\sim 10-100$ GeV, ruling out velocity independent annihilation cross sections near the cosmologically-motivated thermal relic cross section scale of $\langle \sigma v \rangle \sim 3 \times 10^{-26}$ cm$^3$ s$^{-1}$. 

\par Analysis of \Fermi-LAT data has not, however, been without controversy. After over 6 and a half years of science operations, probably the biggest debate revolves around the analysis of data from the inner Galaxy, in particular the nature of the possible extended emission observed at a few GeV. When interpreting this emission as due to dark matter, at present it is possible to fit this extended excess to a wide range of WIMP masses, cross sections, and annihilation channels~\cite{Calore:2014nla}. Given this wide range of models that seemingly work to fit this emission, as well as the uncertainties in Milky Way dark matter distribution and astrophysical sources of gamma rays in the Galactic center, it is clear that establishing this result as due to dark matter will require a confirmation from other sources. The dwarf spheroidals will provide an especially important confirmation, and indeed the current limits from dwarf spheroidals are able to rule out regions of parameter space that explain the extended emission. 

\par Even after over six years of tremendously successful data taking, there is still much to be gained from future \Fermi-LAT data. WIMPs with mass in the range 10-100 GeV with velocity-independent annihilation cross sections are now strongly ruled out by dwarf spheroidal analysis. WIMPs with mass greater than about 100 GeV have been more difficult to constrain because of the photon counting statistics above about a few GeV. In this regime the improvement in sensitivity will scale linearly with time. Forthcoming \Fermi-LAT data will thus prove vital in extending WIMP limits to this higher mass regime. IACTs main target is the Galactic center and analyses here might be systematics limited at masses below 1 TeV, and thus LAT data with a 10 year exposure will be competitive up to a WIMP mass of about 1 TeV. 
 
\par Over the entire detected energy regime, we have repeatedly seen that potential dark matter signals in the \Fermi-LAT data have been compromised by unresolved point sources that are near the threshold limit for point source detection. Because of the foreground model, there is a flux threshold below which it will not be possible to identify individual point sources, as point sources below this threshold contribute to the diffuse foreground. Future \Fermi-LAT data will be important for better identifying faint point sources, and thereby understanding their contribution to the diffuse Galactic model. 

\par Turning to IACTs, as already mentioned the mass range above a few  TeV will be covered robustly and with unrivaled sensitivity by CTA. This will be especially important if the LHC searches do not find hints of new physics, thereby pushing potential WIMP masses to above 500 GeV to 1 TeV. 

\par Taking an even bigger step back, in order to verify any signal that is obtained from indirect detection experiments, it will be crucial to examine the signal in the context of other dark matter searches, in particular those from direct detection and colliders. While the detailed discussion of results of direct searches~\cite{Cushman:2013zza} and accelerator searches~\cite{Dutta:2014mya} --as well as other indirect probes-- is beyond the scope of this article, it is important to ultimately emphasize the complementarity of the different methods. 

\par Though a comparison between the sensitivities of the different methods is inherently model dependent, we can highlight it in two different theoretical set-ups. For example in minimal supersymmetry, there is little correlation between scattering cross section and annihilation cross-section~\cite{Bergstrom:2012fi}. Direct detection searches are seen to constrain the model space approximately orthogonal to indirect detection by gamma rays. Extending beyond direct and indirect searches, collider detection and subsequent measurements of the sparticle mass spectrum and splittings of a supersymmetric model providing dark matter might be used to calculate annihilation cross-sections and relic density~\cite{Baltz:2006fm}. In the framework of phenomenological minimal supersymmetry, however, it is quite conceivable that only after a combination with indirect detection experiments the solution providing dark matter can be identified \cite{Bertone:2011pq}.

%Supersymmetry~\cite{Bergstrom:2010gh,Bertone:2011pq} 

\hspace{.5cm}

\noindent
{\bf ACKNOWLEDGEMENTS}\\
JC thanks the Knut and Alice Wallenberg Foundation, Swedish Research Council and Swedish National Space Board for support. LSE acknowledges support from NSF grant PHY-1522717.

\bibliography{bib}

%merlin.mbs aipauth4-1.bst 2010-07-25 4.21a (PWD, AO, DPC) hacked
%Control: key (0)
%Control: author (9) reversed initials
%Control: editor formatted (0) differently from author
%Control: production of article title (-1) disabled
%Control: page (0) single
%Control: year (1) truncated
%Control: production of eprint (0) enabled
\begin{thebibliography}{266}%
\makeatletter
\providecommand \@ifxundefined [1]{%
 \@ifx{#1\undefined}
}%
\providecommand \@ifnum [1]{%
 \ifnum #1\expandafter \@firstoftwo
 \else \expandafter \@secondoftwo
 \fi
}%
\providecommand \@ifx [1]{%
 \ifx #1\expandafter \@firstoftwo
 \else \expandafter \@secondoftwo
 \fi
}%
\providecommand \natexlab [1]{#1}%
\providecommand \enquote  [1]{``#1''}%
\providecommand \bibnamefont  [1]{#1}%
\providecommand \bibfnamefont [1]{#1}%
\providecommand \citenamefont [1]{#1}%
\providecommand \href@noop [0]{\@secondoftwo}%
\providecommand \href [0]{\begingroup \@sanitize@url \@href}%
\providecommand \@href[1]{\@@startlink{#1}\@@href}%
\providecommand \@@href[1]{\endgroup#1\@@endlink}%
\providecommand \@sanitize@url [0]{\catcode `\\12\catcode `\$12\catcode
  `\&12\catcode `\#12\catcode `\^12\catcode `\_12\catcode `\%12\relax}%
\providecommand \@@startlink[1]{}%
\providecommand \@@endlink[0]{}%
\providecommand \url  [0]{\begingroup\@sanitize@url \@url }%
\providecommand \@url [1]{\endgroup\@href {#1}{\urlprefix }}%
\providecommand \urlprefix  [0]{URL }%
\providecommand \Eprint [0]{\href }%
\providecommand \doibase [0]{http://dx.doi.org/}%
\providecommand \selectlanguage [0]{\@gobble}%
\providecommand \bibinfo  [0]{\@secondoftwo}%
\providecommand \bibfield  [0]{\@secondoftwo}%
\providecommand \translation [1]{[#1]}%
\providecommand \BibitemOpen [0]{}%
\providecommand \bibitemStop [0]{}%
\providecommand \bibitemNoStop [0]{.\EOS\space}%
\providecommand \EOS [0]{\spacefactor3000\relax}%
\providecommand \BibitemShut  [1]{\csname bibitem#1\endcsname}%
\let\auto@bib@innerbib\@empty
%</preamble>
\bibitem [{\citenamefont {Abazajian}\ \emph
  {et~al.}(2014{\natexlab{a}})\citenamefont {Abazajian}, \citenamefont {Canac},
  \citenamefont {Horiuchi},\ and\ \citenamefont
  {Kaplinghat}}]{Abazajian:2014fta}%
  \BibitemOpen
  \bibfield  {author} {\bibinfo {author} {\bibnamefont {Abazajian},
  \bibfnamefont {K.~N.}}, \bibinfo {author} {\bibnamefont {Canac},
  \bibfnamefont {N.}}, \bibinfo {author} {\bibnamefont {Horiuchi},
  \bibfnamefont {S.}}, \ and\ \bibinfo {author} {\bibnamefont {Kaplinghat},
  \bibfnamefont {M.}},\ }\href {\doibase 10.1103/PhysRevD.90.023526} {\bibfield
   {journal} {\bibinfo  {journal} {Phys.Rev.}\ }\textbf {\bibinfo {volume}
  {D90}},\ \bibinfo {pages} {023526} (\bibinfo {year} {2014}{\natexlab{a}})},\
  \Eprint {http://arxiv.org/abs/1402.4090} {arXiv:1402.4090 [astro-ph.HE]}
  \BibitemShut {NoStop}%
%%CITATION = ARXIV:1402.4090;%%
\bibitem [{\citenamefont {Abazajian}\ \emph
  {et~al.}(2014{\natexlab{b}})\citenamefont {Abazajian}, \citenamefont {Canac},
  \citenamefont {Horiuchi}, \citenamefont {Kaplinghat},\ and\ \citenamefont
  {Kwa}}]{Abazajian:2014hsa}%
  \BibitemOpen
  \bibfield  {author} {\bibinfo {author} {\bibnamefont {Abazajian},
  \bibfnamefont {K.~N.}}, \bibinfo {author} {\bibnamefont {Canac},
  \bibfnamefont {N.}}, \bibinfo {author} {\bibnamefont {Horiuchi},
  \bibfnamefont {S.}}, \bibinfo {author} {\bibnamefont {Kaplinghat},
  \bibfnamefont {M.}}, \ and\ \bibinfo {author} {\bibnamefont {Kwa},
  \bibfnamefont {A.}},\ }\href
  {http://adsabs.harvard.edu/abs/2014arXiv1410.6168A} {\enquote {\bibinfo
  {title} {{Discovery of a New Galactic Center Excess Consistent with
  Upscattered Starlight}},}\ } (\bibinfo {year} {2014}{\natexlab{b}}),\
  \bibinfo {note} {provided by the SAO/NASA Astrophysics Data System},\ \Eprint
  {http://arxiv.org/abs/1410.6168} {arXiv:1410.6168 [astro-ph.HE]} \BibitemShut
  {NoStop}%
\bibitem [{\citenamefont {Abazajian}\ and\ \citenamefont
  {Kaplinghat}(2013)}]{2013PhRvD..87l9902A}%
  \BibitemOpen
  \bibfield  {author} {\bibinfo {author} {\bibnamefont {Abazajian},
  \bibfnamefont {K.~N.}}\ and\ \bibinfo {author} {\bibnamefont {Kaplinghat},
  \bibfnamefont {M.}},\ }\href {\doibase 10.1103/PhysRevD.87.129902} {\bibfield
   {journal} {\bibinfo  {journal} {\prd}\ }\textbf {\bibinfo {volume} {87}},\
  \bibinfo {eid} {129902} (\bibinfo {year} {2013})}\BibitemShut {NoStop}%
\bibitem [{\citenamefont {Abdo}\ \emph
  {et~al.}(2010{\natexlab{a}})\citenamefont {Abdo}, \citenamefont {Ackermann},
  \citenamefont {Ajello}, \citenamefont {Atwood}, \citenamefont {Baldini} \emph
  {et~al.}}]{Abdo:2010nc}%
  \BibitemOpen
  \bibfield  {author} {\bibinfo {author} {\bibnamefont {Abdo}, \bibfnamefont
  {A.}}, \bibinfo {author} {\bibnamefont {Ackermann}, \bibfnamefont {M.}},
  \bibinfo {author} {\bibnamefont {Ajello}, \bibfnamefont {M.}},  \emph
  {et~al.} (\bibinfo {collaboration} {Fermi LAT Collaboration}),\ }\href
  {\doibase 10.1103/PhysRevLett.104.091302} {\bibfield  {journal} {\bibinfo
  {journal} {Phys.Rev.Lett.}\ }\textbf {\bibinfo {volume} {104}},\ \bibinfo
  {pages} {091302} (\bibinfo {year} {2010}{\natexlab{a}})},\ \Eprint
  {http://arxiv.org/abs/1001.4836} {arXiv:1001.4836 [astro-ph.HE]} \BibitemShut
  {NoStop}%
%%CITATION = ARXIV:1001.4836;%%
\bibitem [{\citenamefont {Abdo}\ \emph
  {et~al.}(2010{\natexlab{b}})\citenamefont {Abdo} \emph
  {et~al.}}]{Abdo:2010dk}%
  \BibitemOpen
  \bibfield  {author} {\bibinfo {author} {\bibnamefont {Abdo}, \bibfnamefont
  {A.}} \emph {et~al.} (\bibinfo {collaboration} {Fermi LAT Collaboration}),\
  }\href {\doibase 10.1088/1475-7516/2010/04/014} {\bibfield  {journal}
  {\bibinfo  {journal} {\jcap}\ }\textbf {\bibinfo {volume} {1004}},\ \bibinfo
  {pages} {014} (\bibinfo {year} {2010}{\natexlab{b}})},\ \Eprint
  {http://arxiv.org/abs/1002.4415} {arXiv:1002.4415 [astro-ph.CO]} \BibitemShut
  {NoStop}%
\bibitem [{\citenamefont {Abdo}\ \emph
  {et~al.}(2010{\natexlab{c}})\citenamefont {Abdo} \emph
  {et~al.}}]{Abdo:2010ex}%
  \BibitemOpen
  \bibfield  {author} {\bibinfo {author} {\bibnamefont {Abdo}, \bibfnamefont
  {A.}} \emph {et~al.} (\bibinfo {collaboration} {Fermi LAT Collaboration}),\
  }\href {\doibase 10.1088/0004-637X/712/1/147} {\bibfield  {journal} {\bibinfo
   {journal} {\apj}\ }\textbf {\bibinfo {volume} {712}},\ \bibinfo {pages}
  {147} (\bibinfo {year} {2010}{\natexlab{c}})},\ \Eprint
  {http://arxiv.org/abs/1001.4531} {arXiv:1001.4531 [astro-ph.CO]} \BibitemShut
  {NoStop}%
\bibitem [{\citenamefont {Abdo}\ \emph
  {et~al.}(2010{\natexlab{d}})\citenamefont {Abdo} \emph
  {et~al.}}]{Abdo:2010pq}%
  \BibitemOpen
  \bibfield  {author} {\bibinfo {author} {\bibnamefont {Abdo}, \bibfnamefont
  {A.}} \emph {et~al.} (\bibinfo {collaboration} {Fermi LAT Collaboration}),\
  }\href {\doibase 10.1051/0004-6361/200913474} {\bibfield  {journal} {\bibinfo
   {journal} {Astron.Astrophys.}\ }\textbf {\bibinfo {volume} {512}},\ \bibinfo
  {pages} {A7} (\bibinfo {year} {2010}{\natexlab{d}})},\ \Eprint
  {http://arxiv.org/abs/1001.3298} {arXiv:1001.3298 [astro-ph.HE]} \BibitemShut
  {NoStop}%
%%CITATION = ARXIV:1001.3298;%%
\bibitem [{\citenamefont {{Abdo}}\ \emph
  {et~al.}(2010{\natexlab{a}})\citenamefont {{Abdo}}, \citenamefont
  {{Ackermann}}, \citenamefont {{Ajello}}, \citenamefont {{Allafort}},
  \citenamefont {{Atwood}}, \citenamefont {{Baldini}}, \citenamefont
  {{Ballet}}, \citenamefont {{Barbiellini}}, \citenamefont {{Bastieri}},
  \citenamefont {{Bechtol}}, \citenamefont {{Bellazzini}}, \citenamefont
  {{Berenji}}, \citenamefont {{Blandford}}, \citenamefont {{Bloom}},
  \citenamefont {{Bonamente}}, \citenamefont {{Borgland}}, \citenamefont
  {{Bouvier}}, \citenamefont {{Brandt}}, \citenamefont {{Bregeon}},
  \citenamefont {{Brigida}}, \citenamefont {{Bruel}}, \citenamefont
  {{Buehler}}, \citenamefont {{Burnett}}, \citenamefont {{Buson}},
  \citenamefont {{Caliandro}}, \citenamefont {{Cameron}}, \citenamefont
  {{Cannon}}, \citenamefont {{Caraveo}}, \citenamefont {{Casandjian}},
  \citenamefont {{Cecchi}}, \citenamefont {{{\c C}elik}}, \citenamefont
  {{Charles}}, \citenamefont {{Chekhtman}}, \citenamefont {{Chiang}},
  \citenamefont {{Ciprini}}, \citenamefont {{Claus}}, \citenamefont
  {{Cohen-Tanugi}}, \citenamefont {{Conrad}}, \citenamefont {{Dermer}},
  \citenamefont {{de Angelis}}, \citenamefont {{de Palma}}, \citenamefont
  {{Digel}}, \citenamefont {{Silva}}, \citenamefont {{Drell}}, \citenamefont
  {{Drlica-Wagner}}, \citenamefont {{Dubois}}, \citenamefont {{Favuzzi}},
  \citenamefont {{Fegan}}, \citenamefont {{Fortin}}, \citenamefont {{Frailis}},
  \citenamefont {{Fukazawa}}, \citenamefont {{Funk}}, \citenamefont {{Fusco}},
  \citenamefont {{Gargano}}, \citenamefont {{Germani}}, \citenamefont
  {{Giglietto}}, \citenamefont {{Giordano}}, \citenamefont {{Giroletti}},
  \citenamefont {{Glanzman}}, \citenamefont {{Godfrey}}, \citenamefont
  {{Grenier}}, \citenamefont {{Grondin}}, \citenamefont {{Guiriec}},
  \citenamefont {{Gustafsson}}, \citenamefont {{Hadasch}}, \citenamefont
  {{Harding}}, \citenamefont {{Hayashi}}, \citenamefont {{Hayashida}},
  \citenamefont {{Hays}}, \citenamefont {{Healey}}, \citenamefont {{Jean}},
  \citenamefont {{J{\'o}hannesson}}, \citenamefont {{Johnson}}, \citenamefont
  {{Johnson}}, \citenamefont {{Johnson}}, \citenamefont {{Kamae}},
  \citenamefont {{Katagiri}}, \citenamefont {{Kataoka}}, \citenamefont
  {{Kerr}}, \citenamefont {{Kn{\"o}dlseder}}, \citenamefont {{Kuss}},
  \citenamefont {{Lande}}, \citenamefont {{Latronico}}, \citenamefont {{Lee}},
  \citenamefont {{Lemoine-Goumard}}, \citenamefont {{Longo}}, \citenamefont
  {{Loparco}}, \citenamefont {{Lott}}, \citenamefont {{Lovellette}},
  \citenamefont {{Lubrano}}, \citenamefont {{Madejski}}, \citenamefont
  {{Makeev}}, \citenamefont {{Martin}}, \citenamefont {{Mazziotta}},
  \citenamefont {{Mehault}}, \citenamefont {{Michelson}}, \citenamefont
  {{Mitthumsiri}}, \citenamefont {{Mizuno}}, \citenamefont {{Moiseev}},
  \citenamefont {{Monte}}, \citenamefont {{Monzani}}, \citenamefont
  {{Morselli}}, \citenamefont {{Moskalenko}}, \citenamefont {{Murgia}},
  \citenamefont {{Naumann-Godo}}, \citenamefont {{Nolan}}, \citenamefont
  {{Norris}}, \citenamefont {{Nuss}}, \citenamefont {{Ohsugi}}, \citenamefont
  {{Okumura}}, \citenamefont {{Omodei}}, \citenamefont {{Orlando}},
  \citenamefont {{Ormes}}, \citenamefont {{Ozaki}}, \citenamefont {{Paneque}},
  \citenamefont {{Panetta}}, \citenamefont {{Parent}}, \citenamefont {{Pepe}},
  \citenamefont {{Persic}}, \citenamefont {{Pesce-Rollins}}, \citenamefont
  {{Piron}}, \citenamefont {{Porter}}, \citenamefont {{Rain{\`o}}},
  \citenamefont {{Rando}}, \citenamefont {{Razzano}}, \citenamefont {{Reimer}},
  \citenamefont {{Reimer}}, \citenamefont {{Ritz}}, \citenamefont {{Romani}},
  \citenamefont {{Sadrozinski}}, \citenamefont {{Saz Parkinson}}, \citenamefont
  {{Sgr{\`o}}}, \citenamefont {{Siskind}}, \citenamefont {{Smith}},
  \citenamefont {{Smith}}, \citenamefont {{Spandre}}, \citenamefont
  {{Spinelli}}, \citenamefont {{Strickman}}, \citenamefont {{Strigari}},
  \citenamefont {{Strong}}, \citenamefont {{Suson}}, \citenamefont
  {{Takahashi}}, \citenamefont {{Takahashi}}, \citenamefont {{Tanaka}},
  \citenamefont {{Thayer}}, \citenamefont {{Thompson}}, \citenamefont
  {{Tibaldo}}, \citenamefont {{Torres}}, \citenamefont {{Tosti}}, \citenamefont
  {{Tramacere}}, \citenamefont {{Uchiyama}}, \citenamefont {{Usher}},
  \citenamefont {{Vandenbroucke}}, \citenamefont {{Vianello}}, \citenamefont
  {{Vilchez}}, \citenamefont {{Vitale}}, \citenamefont {{Waite}}, \citenamefont
  {{Wang}}, \citenamefont {{Winer}}, \citenamefont {{Wood}}, \citenamefont
  {{Yang}},\ and\ \citenamefont {{Ziegler}}}]{2010A&A...523L...2A}%
  \BibitemOpen
  \bibfield  {author} {\bibinfo {author} {\bibnamefont {{Abdo}}, \bibfnamefont
  {A.~A.}}, \bibinfo {author} {\bibnamefont {{Ackermann}}, \bibfnamefont {M.}},
  \bibinfo {author} {\bibnamefont {{Ajello}}, \bibfnamefont {M.}},  \emph
  {et~al.},\ }\href {\doibase 10.1051/0004-6361/201015759} {\bibfield
  {journal} {\bibinfo  {journal} {\aap}\ }\textbf {\bibinfo {volume} {523}},\
  \bibinfo {eid} {L2} (\bibinfo {year} {2010}{\natexlab{a}})},\ \Eprint
  {http://arxiv.org/abs/1012.1952} {arXiv:1012.1952 [astro-ph.HE]} \BibitemShut
  {NoStop}%
\bibitem [{\citenamefont {{Abdo}}\ \emph
  {et~al.}(2010{\natexlab{b}})\citenamefont {{Abdo}}, \citenamefont
  {{Ackermann}}, \citenamefont {{Ajello}}, \citenamefont {{Atwood}},
  \citenamefont {{Baldini}}, \citenamefont {{Ballet}}, \citenamefont
  {{Barbiellini}}, \citenamefont {{Bastieri}}, \citenamefont {{Baughman}},
  \citenamefont {{Bechtol}}, \citenamefont {{Bellazzini}}, \citenamefont
  {{Berenji}}, \citenamefont {{Blandford}}, \citenamefont {{Bloom}},
  \citenamefont {{Bonamente}}, \citenamefont {{Borgland}}, \citenamefont
  {{Bregeon}}, \citenamefont {{Brez}}, \citenamefont {{Brigida}}, \citenamefont
  {{Bruel}}, \citenamefont {{Burnett}}, \citenamefont {{Buson}}, \citenamefont
  {{Caliandro}}, \citenamefont {{Cameron}}, \citenamefont {{Caraveo}},
  \citenamefont {{Casandjian}}, \citenamefont {{Cavazzuti}}, \citenamefont
  {{Cecchi}}, \citenamefont {{{\c C}elik}}, \citenamefont {{Charles}},
  \citenamefont {{Chekhtman}}, \citenamefont {{Cheung}}, \citenamefont
  {{Chiang}}, \citenamefont {{Ciprini}}, \citenamefont {{Claus}}, \citenamefont
  {{Cohen-Tanugi}}, \citenamefont {{Cominsky}}, \citenamefont {{Conrad}},
  \citenamefont {{Cutini}}, \citenamefont {{Dermer}}, \citenamefont {{de
  Angelis}}, \citenamefont {{de Palma}}, \citenamefont {{Digel}}, \citenamefont
  {{di Bernardo}}, \citenamefont {{do Couto e Silva}}, \citenamefont {{Drell}},
  \citenamefont {{Drlica-Wagner}}, \citenamefont {{Dubois}}, \citenamefont
  {{Dumora}}, \citenamefont {{Farnier}}, \citenamefont {{Favuzzi}},
  \citenamefont {{Fegan}}, \citenamefont {{Focke}}, \citenamefont {{Fortin}},
  \citenamefont {{Frailis}}, \citenamefont {{Fukazawa}}, \citenamefont
  {{Funk}}, \citenamefont {{Fusco}}, \citenamefont {{Gaggero}}, \citenamefont
  {{Gargano}}, \citenamefont {{Gasparrini}}, \citenamefont {{Gehrels}},
  \citenamefont {{Germani}}, \citenamefont {{Giebels}}, \citenamefont
  {{Giglietto}}, \citenamefont {{Giommi}}, \citenamefont {{Giordano}},
  \citenamefont {{Glanzman}}, \citenamefont {{Godfrey}}, \citenamefont
  {{Grenier}}, \citenamefont {{Grondin}}, \citenamefont {{Grove}},
  \citenamefont {{Guillemot}}, \citenamefont {{Guiriec}}, \citenamefont
  {{Gustafsson}}, \citenamefont {{Hanabata}}, \citenamefont {{Harding}},
  \citenamefont {{Hayashida}}, \citenamefont {{Hughes}}, \citenamefont
  {{Itoh}}, \citenamefont {{Jackson}}, \citenamefont {{J{\'o}hannesson}},
  \citenamefont {{Johnson}}, \citenamefont {{Johnson}}, \citenamefont
  {{Johnson}}, \citenamefont {{Johnson}}, \citenamefont {{Kamae}},
  \citenamefont {{Katagiri}}, \citenamefont {{Kataoka}}, \citenamefont
  {{Kawai}}, \citenamefont {{Kerr}}, \citenamefont {{Kn{\"o}dlseder}},
  \citenamefont {{Kocian}}, \citenamefont {{Kuehn}}, \citenamefont {{Kuss}},
  \citenamefont {{Lande}}, \citenamefont {{Latronico}}, \citenamefont
  {{Lemoine-Goumard}}, \citenamefont {{Longo}}, \citenamefont {{Loparco}},
  \citenamefont {{Lott}}, \citenamefont {{Lovellette}}, \citenamefont
  {{Lubrano}}, \citenamefont {{Madejski}}, \citenamefont {{Makeev}},
  \citenamefont {{Mazziotta}}, \citenamefont {{McConville}}, \citenamefont
  {{McEnery}}, \citenamefont {{Meurer}}, \citenamefont {{Michelson}},
  \citenamefont {{Mitthumsiri}}, \citenamefont {{Mizuno}}, \citenamefont
  {{Moiseev}}, \citenamefont {{Monte}}, \citenamefont {{Monzani}},
  \citenamefont {{Morselli}}, \citenamefont {{Moskalenko}}, \citenamefont
  {{Murgia}}, \citenamefont {{Nolan}}, \citenamefont {{Norris}}, \citenamefont
  {{Nuss}}, \citenamefont {{Ohsugi}}, \citenamefont {{Omodei}}, \citenamefont
  {{Orlando}}, \citenamefont {{Ormes}}, \citenamefont {{Paneque}},
  \citenamefont {{Panetta}}, \citenamefont {{Parent}}, \citenamefont
  {{Pelassa}}, \citenamefont {{Pepe}}, \citenamefont {{Pesce-Rollins}},
  \citenamefont {{Piron}}, \citenamefont {{Porter}}, \citenamefont
  {{Rain{\`o}}}, \citenamefont {{Rando}}, \citenamefont {{Razzano}},
  \citenamefont {{Reimer}}, \citenamefont {{Reimer}}, \citenamefont
  {{Reposeur}}, \citenamefont {{Ritz}}, \citenamefont {{Rochester}},
  \citenamefont {{Rodriguez}}, \citenamefont {{Roth}}, \citenamefont {{Ryde}},
  \citenamefont {{Sadrozinski}}, \citenamefont {{Sanchez}}, \citenamefont
  {{Sander}}, \citenamefont {{Parkinson}}, \citenamefont {{Scargle}},
  \citenamefont {{Sellerholm}}, \citenamefont {{Sgr{\`o}}}, \citenamefont
  {{Shaw}}, \citenamefont {{Siskind}}, \citenamefont {{Smith}}, \citenamefont
  {{Smith}}, \citenamefont {{Spandre}}, \citenamefont {{Spinelli}},
  \citenamefont {{Starck}}, \citenamefont {{Strickman}}, \citenamefont
  {{Strong}}, \citenamefont {{Suson}}, \citenamefont {{Tajima}}, \citenamefont
  {{Takahashi}}, \citenamefont {{Takahashi}}, \citenamefont {{Tanaka}},
  \citenamefont {{Thayer}}, \citenamefont {{Thayer}}, \citenamefont
  {{Thompson}}, \citenamefont {{Tibaldo}}, \citenamefont {{Torres}},
  \citenamefont {{Tosti}}, \citenamefont {{Tramacere}}, \citenamefont
  {{Uchiyama}}, \citenamefont {{Usher}}, \citenamefont {{Vasileiou}},
  \citenamefont {{Vilchez}}, \citenamefont {{Vitale}}, \citenamefont {{Waite}},
  \citenamefont {{Wang}}, \citenamefont {{Winer}}, \citenamefont {{Wood}},
  \citenamefont {{Ylinen}}, \citenamefont {{Ziegler}},\ and\ \citenamefont
  {{Fermi LAT Collaboration}}}]{2010PhRvL.104j1101A}%
  \BibitemOpen
  \bibfield  {author} {\bibinfo {author} {\bibnamefont {{Abdo}}, \bibfnamefont
  {A.~A.}}, \bibinfo {author} {\bibnamefont {{Ackermann}}, \bibfnamefont {M.}},
  \bibinfo {author} {\bibnamefont {{Ajello}}, \bibfnamefont {M.}},  \emph
  {et~al.} (\bibinfo {collaboration} {Fermi LAT Collaboration}),\ }\href
  {\doibase 10.1103/PhysRevLett.104.101101} {\bibfield  {journal} {\bibinfo
  {journal} {Physical Review Letters}\ }\textbf {\bibinfo {volume} {104}},\
  \bibinfo {eid} {101101} (\bibinfo {year} {2010}{\natexlab{b}})},\ \Eprint
  {http://arxiv.org/abs/1002.3603} {arXiv:1002.3603 [astro-ph.HE]} \BibitemShut
  {NoStop}%
\bibitem [{\citenamefont {{Abdo}}\ \emph
  {et~al.}(2010{\natexlab{c}})\citenamefont {{Abdo}}, \citenamefont
  {{Ackermann}}, \citenamefont {{Ajello}}, \citenamefont {{Baldini}},
  \citenamefont {{Ballet}}, \citenamefont {{Barbiellini}}, \citenamefont
  {{Bastieri}}, \citenamefont {{Bechtol}}, \citenamefont {{Bellazzini}},
  \citenamefont {{Berenji}}, \citenamefont {{Blandford}}, \citenamefont
  {{Bloom}}, \citenamefont {{Bonamente}}, \citenamefont {{Borgland}},
  \citenamefont {{Bouvier}}, \citenamefont {{Brandt}}, \citenamefont
  {{Bregeon}}, \citenamefont {{Brez}}, \citenamefont {{Brigida}}, \citenamefont
  {{Bruel}}, \citenamefont {{Buehler}}, \citenamefont {{Buson}}, \citenamefont
  {{Caliandro}}, \citenamefont {{Cameron}}, \citenamefont {{Caraveo}},
  \citenamefont {{Carrigan}}, \citenamefont {{Casandjian}}, \citenamefont
  {{Cecchi}}, \citenamefont {{{\c C}elik}}, \citenamefont {{Charles}},
  \citenamefont {{Chekhtman}}, \citenamefont {{Cheung}}, \citenamefont
  {{Chiang}}, \citenamefont {{Ciprini}}, \citenamefont {{Claus}}, \citenamefont
  {{Cohen-Tanugi}}, \citenamefont {{Conrad}}, \citenamefont {{Dermer}},
  \citenamefont {{de Palma}}, \citenamefont {{Digel}}, \citenamefont {{Silva}},
  \citenamefont {{Drell}}, \citenamefont {{Dubois}}, \citenamefont {{Dumora}},
  \citenamefont {{Favuzzi}}, \citenamefont {{Fegan}}, \citenamefont
  {{Fukazawa}}, \citenamefont {{Funk}}, \citenamefont {{Fusco}}, \citenamefont
  {{Gargano}}, \citenamefont {{Gasparrini}}, \citenamefont {{Gehrels}},
  \citenamefont {{Germani}}, \citenamefont {{Giglietto}}, \citenamefont
  {{Giordano}}, \citenamefont {{Giroletti}}, \citenamefont {{Glanzman}},
  \citenamefont {{Godfrey}}, \citenamefont {{Grenier}}, \citenamefont
  {{Grondin}}, \citenamefont {{Grove}}, \citenamefont {{Guiriec}},
  \citenamefont {{Hadasch}}, \citenamefont {{Harding}}, \citenamefont
  {{Hayashida}}, \citenamefont {{Hays}}, \citenamefont {{Horan}}, \citenamefont
  {{Hughes}}, \citenamefont {{Jean}}, \citenamefont {{J{\'o}hannesson}},
  \citenamefont {{Johnson}}, \citenamefont {{Johnson}}, \citenamefont
  {{Kamae}}, \citenamefont {{Katagiri}}, \citenamefont {{Kataoka}},
  \citenamefont {{Kerr}}, \citenamefont {{Kn{\"o}dlseder}}, \citenamefont
  {{Kuss}}, \citenamefont {{Lande}}, \citenamefont {{Latronico}}, \citenamefont
  {{Lee}}, \citenamefont {{Lemoine-Goumard}}, \citenamefont {{Llena Garde}},
  \citenamefont {{Longo}}, \citenamefont {{Loparco}}, \citenamefont
  {{Lovellette}}, \citenamefont {{Lubrano}}, \citenamefont {{Makeev}},
  \citenamefont {{Martin}}, \citenamefont {{Mazziotta}}, \citenamefont
  {{McEnery}}, \citenamefont {{Michelson}}, \citenamefont {{Mitthumsiri}},
  \citenamefont {{Mizuno}}, \citenamefont {{Monte}}, \citenamefont {{Monzani}},
  \citenamefont {{Morselli}}, \citenamefont {{Moskalenko}}, \citenamefont
  {{Murgia}}, \citenamefont {{Nakamori}}, \citenamefont {{Naumann-Godo}},
  \citenamefont {{Nolan}}, \citenamefont {{Norris}}, \citenamefont {{Nuss}},
  \citenamefont {{Ohsugi}}, \citenamefont {{Okumura}}, \citenamefont
  {{Omodei}}, \citenamefont {{Orlando}}, \citenamefont {{Ormes}}, \citenamefont
  {{Panetta}}, \citenamefont {{Parent}}, \citenamefont {{Pelassa}},
  \citenamefont {{Pepe}}, \citenamefont {{Pesce-Rollins}}, \citenamefont
  {{Piron}}, \citenamefont {{Porter}}, \citenamefont {{Rain{\`o}}},
  \citenamefont {{Rando}}, \citenamefont {{Razzano}}, \citenamefont {{Reimer}},
  \citenamefont {{Reimer}}, \citenamefont {{Reposeur}}, \citenamefont
  {{Ripken}}, \citenamefont {{Ritz}}, \citenamefont {{Romani}}, \citenamefont
  {{Sadrozinski}}, \citenamefont {{Sander}}, \citenamefont {{Saz Parkinson}},
  \citenamefont {{Scargle}}, \citenamefont {{Sgr{\`o}}}, \citenamefont
  {{Siskind}}, \citenamefont {{Smith}}, \citenamefont {{Smith}}, \citenamefont
  {{Spandre}}, \citenamefont {{Spinelli}}, \citenamefont {{Strickman}},
  \citenamefont {{Strong}}, \citenamefont {{Suson}}, \citenamefont
  {{Takahashi}}, \citenamefont {{Takahashi}}, \citenamefont {{Tanaka}},
  \citenamefont {{Thayer}}, \citenamefont {{Thayer}}, \citenamefont
  {{Thompson}}, \citenamefont {{Tibaldo}}, \citenamefont {{Torres}},
  \citenamefont {{Tosti}}, \citenamefont {{Tramacere}}, \citenamefont
  {{Uchiyama}}, \citenamefont {{Usher}}, \citenamefont {{Vandenbroucke}},
  \citenamefont {{Vasileiou}}, \citenamefont {{Vilchez}}, \citenamefont
  {{Vitale}}, \citenamefont {{Waite}}, \citenamefont {{Wang}}, \citenamefont
  {{Winer}}, \citenamefont {{Wood}}, \citenamefont {{Yang}}, \citenamefont
  {{Ylinen}},\ and\ \citenamefont {{Ziegler}}}]{2010A&A...523A..46A}%
  \BibitemOpen
  \bibfield  {author} {\bibinfo {author} {\bibnamefont {{Abdo}}, \bibfnamefont
  {A.~A.}}, \bibinfo {author} {\bibnamefont {{Ackermann}}, \bibfnamefont {M.}},
  \bibinfo {author} {\bibnamefont {{Ajello}}, \bibfnamefont {M.}},  \emph
  {et~al.} (\bibinfo {collaboration} {Fermi LAT Collaboration}),\ }\href
  {\doibase 10.1051/0004-6361/201014855} {\bibfield  {journal} {\bibinfo
  {journal} {\aap}\ }\textbf {\bibinfo {volume} {523}},\ \bibinfo {eid} {A46}
  (\bibinfo {year} {2010}{\natexlab{c}})}\BibitemShut {NoStop}%
\bibitem [{\citenamefont {{Abeysekara}}\ \emph {et~al.}(2014)\citenamefont
  {{Abeysekara}}, \citenamefont {{Alfaro}}, \citenamefont {{Alvarez}},
  \citenamefont {{{\'A}lvarez}}, \citenamefont {{Arceo}}, \citenamefont
  {{Arteaga-Vel{\'a}zquez}}, \citenamefont {{Ayala Solares}}, \citenamefont
  {{Barber}}, \citenamefont {{Baughman}}, \citenamefont {{Bautista-Elivar}},
  \citenamefont {{Becerra Gonzalez}}, \citenamefont {{Belmont}}, \citenamefont
  {{BenZvi}}, \citenamefont {{Berley}}, \citenamefont {{Bonilla Rosales}},
  \citenamefont {{Braun}}, \citenamefont {{Caballero-Lopez}}, \citenamefont
  {{Caballero-Mora}}, \citenamefont {{Carrami{\~n}ana}}, \citenamefont
  {{Castillo}}, \citenamefont {{Cotti}}, \citenamefont {{Cotzomi}},
  \citenamefont {{de la Fuente}}, \citenamefont {{De Le{\'o}n}}, \citenamefont
  {{DeYoung}}, \citenamefont {{Diaz Hernandez}}, \citenamefont {{Diaz-Cruz}},
  \citenamefont {{D$\backslash$'$\backslash$iaz-V{\'e}lez}}, \citenamefont
  {{Dingus}}, \citenamefont {{DuVernois}}, \citenamefont {{Ellsworth}},
  \citenamefont {{S.~F.}}, \citenamefont {{Fiorino}}, \citenamefont {{Fraija}},
  \citenamefont {{Galindo}}, \citenamefont {{Garfias}}, \citenamefont
  {{Gonz{\'a}lez}}, \citenamefont {{Goodman}}, \citenamefont {{Grabski}},
  \citenamefont {{Gussert}}, \citenamefont {{Hampel-Arias}}, \citenamefont
  {{Harding}}, \citenamefont {{Hui}}, \citenamefont {{H{\"u}ntemeyer}},
  \citenamefont {{Imran}}, \citenamefont {{Iriarte}}, \citenamefont {{Karn}},
  \citenamefont {{Kieda}}, \citenamefont {{Kunde}}, \citenamefont {{Lara}},
  \citenamefont {{Lauer}}, \citenamefont {{Lee}}, \citenamefont {{Lennarz}},
  \citenamefont {{Le{\'o}n Vargas}}, \citenamefont {{Linares}}, \citenamefont
  {{Linnemann}}, \citenamefont {{Longo}}, \citenamefont {{Luna-Garcia}},
  \citenamefont {{Marinelli}}, \citenamefont {{Martinez}}, \citenamefont
  {{Martinez}}, \citenamefont {{Mart$\backslash$'$\backslash$inez-Castro}},
  \citenamefont {{Matthews}}, \citenamefont {{McEnery}}, \citenamefont
  {{Mendoza Torres}}, \citenamefont {{Miranda-Romagnoli}}, \citenamefont
  {{Moreno}}, \citenamefont {{Mostaf{\'a}}}, \citenamefont {{Nellen}},
  \citenamefont {{Newbold}}, \citenamefont {{Noriega-Papaqui}}, \citenamefont
  {{Oceguera-Becerra}}, \citenamefont {{Patricelli}}, \citenamefont {{Pelayo}},
  \citenamefont {{P{\'e}rez-P{\'e}rez}}, \citenamefont {{Pretz}}, \citenamefont
  {{Rivi{\`e}re}}, \citenamefont {{Rosa-Gonz{\'a}lez}}, \citenamefont {{Ryan}},
  \citenamefont {{Salazar}}, \citenamefont {{Salesa}}, \citenamefont
  {{Sandoval}}, \citenamefont {{Schneider}}, \citenamefont {{Silich}},
  \citenamefont {{Sinnis}}, \citenamefont {{Smith}}, \citenamefont {{Sparks
  Woodle}}, \citenamefont {{Springer}}, \citenamefont {{Taboada}},
  \citenamefont {{Toale}}, \citenamefont {{Tollefson}}, \citenamefont
  {{Torres}}, \citenamefont {{Ukwatta}}, \citenamefont {{Villase{\~n}or}},
  \citenamefont {{Weisgarber}}, \citenamefont {{Westerhoff}}, \citenamefont
  {{Wisher}}, \citenamefont {{Wood}}, \citenamefont {{Yodh}}, \citenamefont
  {{Younk}}, \citenamefont {{Zaborov}}, \citenamefont {{Zepeda}}, \citenamefont
  {{Zhou}},\ and\ \citenamefont {{Abazajian}}}]{2014arXiv1405.1730A}%
  \BibitemOpen
  \bibfield  {author} {\bibinfo {author} {\bibnamefont {{Abeysekara}},
  \bibfnamefont {A.~U.}}, \bibinfo {author} {\bibnamefont {{Alfaro}},
  \bibfnamefont {R.}}, \bibinfo {author} {\bibnamefont {{Alvarez}},
  \bibfnamefont {C.}},  \emph {et~al.},\ }\href@noop {} {\bibfield  {journal}
  {\bibinfo  {journal} {ArXiv e-prints}\ } (\bibinfo {year} {2014})},\ \Eprint
  {http://arxiv.org/abs/1405.1730} {arXiv:1405.1730 [astro-ph.HE]} \BibitemShut
  {NoStop}%
\bibitem [{\citenamefont {{Abeysekara}}\ \emph
  {et~al.}(2013{\natexlab{a}})\citenamefont {{Abeysekara}}, \citenamefont
  {{Alfaro}}, \citenamefont {{Alvarez}}, \citenamefont {{{\'A}lvarez}},
  \citenamefont {{Arceo}}, \citenamefont {{Arteaga-Vel{\'a}zquez}},
  \citenamefont {{Ayala Solares}}, \citenamefont {{Barber}}, \citenamefont
  {{Baughman}}, \citenamefont {{Bautista-Elivar}}, \citenamefont {{Belmont}},
  \citenamefont {{BenZvi}}, \citenamefont {{Berley}}, \citenamefont {{Bonilla
  Rosales}}, \citenamefont {{Braun}}, \citenamefont {{Caballero-Lopez}},
  \citenamefont {{Caballero-Mora}}, \citenamefont {{Carrami{\~n}ana}},
  \citenamefont {{Castillo}}, \citenamefont {{Cotti}}, \citenamefont
  {{Cotzomi}}, \citenamefont {{de la Fuente}}, \citenamefont {{De Le{\'o}n}},
  \citenamefont {{DeYoung}}, \citenamefont {{Diaz Hernandez}}, \citenamefont
  {{D{\'{\i}}az-V{\'e}lez}}, \citenamefont {{Dingus}}, \citenamefont
  {{DuVernois}}, \citenamefont {{Ellsworth}}, \citenamefont {{Fernandez}},
  \citenamefont {{Fiorino}}, \citenamefont {{Fraija}}, \citenamefont
  {{Galindo}}, \citenamefont {{Garfias}}, \citenamefont {{Gonz{\'a}lez}},
  \citenamefont {{Gonz{\'a}lez}}, \citenamefont {{Goodman}}, \citenamefont
  {{Grabski}}, \citenamefont {{Gussert}}, \citenamefont {{Hampel-Arias}},
  \citenamefont {{Hui}}, \citenamefont {{H{\"u}ntemeyer}}, \citenamefont
  {{Imran}}, \citenamefont {{Iriarte}}, \citenamefont {{Karn}}, \citenamefont
  {{Kieda}}, \citenamefont {{Kunde}}, \citenamefont {{Lara}}, \citenamefont
  {{Lauer}}, \citenamefont {{Lee}}, \citenamefont {{Lennarz}}, \citenamefont
  {{Le{\'o}n Vargas}}, \citenamefont {{Linares}}, \citenamefont {{Linnemann}},
  \citenamefont {{Longo}}, \citenamefont {{Luna-GarcIa}}, \citenamefont
  {{Marinelli}}, \citenamefont {{Martinez}}, \citenamefont {{Martinez}},
  \citenamefont {{Mart{\'{\i}}nez-Castro}}, \citenamefont {{Matthews}},
  \citenamefont {{Miranda-Romagnoli}}, \citenamefont {{Moreno}}, \citenamefont
  {{Mostaf{\'a}}}, \citenamefont {{Nava}}, \citenamefont {{Nellen}},
  \citenamefont {{Newbold}}, \citenamefont {{Noriega-Papaqui}}, \citenamefont
  {{Oceguera-Becerra}}, \citenamefont {{Patricelli}}, \citenamefont {{Pelayo}},
  \citenamefont {{P{\'e}rez-P{\'e}rez}}, \citenamefont {{Pretz}}, \citenamefont
  {{Rivi{\`e}re}}, \citenamefont {{Rosa-Gonz{\'a}lez}}, \citenamefont
  {{Salazar}}, \citenamefont {{Salesa}}, \citenamefont {{Sanchez}},
  \citenamefont {{Sandoval}}, \citenamefont {{Santos}}, \citenamefont
  {{Schneider}}, \citenamefont {{Silich}}, \citenamefont {{Sinnis}},
  \citenamefont {{Smith}}, \citenamefont {{Sparks}}, \citenamefont
  {{Springer}}, \citenamefont {{Taboada}}, \citenamefont {{Toale}},
  \citenamefont {{Tollefson}}, \citenamefont {{Torres}}, \citenamefont
  {{Ukwatta}}, \citenamefont {{Villase{\~n}or}}, \citenamefont {{Weisgarber}},
  \citenamefont {{Westerhoff}}, \citenamefont {{Wisher}}, \citenamefont
  {{Wood}}, \citenamefont {{Yodh}}, \citenamefont {{Younk}}, \citenamefont
  {{Zaborov}}, \citenamefont {{Zepeda}},\ and\ \citenamefont
  {{Zhou}}}]{2013arXiv1310.0074H}%
  \BibitemOpen
  \bibfield  {author} {\bibinfo {author} {\bibnamefont {{Abeysekara}},
  \bibfnamefont {A.~U.}}, \bibinfo {author} {\bibnamefont {{Alfaro}},
  \bibfnamefont {R.}}, \bibinfo {author} {\bibnamefont {{Alvarez}},
  \bibfnamefont {C.}},  \emph {et~al.} (\bibinfo {collaboration} {HAWC
  Collaboration}),\ }in\ \href@noop {} {\emph {\bibinfo {booktitle}
  {Proceedings of the 33rd International Cosmic-Ray Conference}}}\ (\bibinfo
  {year} {2013})\ p.~\bibinfo {pages} {24},\ \Eprint
  {http://arxiv.org/abs/1310.0074} {arXiv:1310.0074 [astro-ph.IM]} \BibitemShut
  {NoStop}%
\bibitem [{\citenamefont {{Abeysekara}}\ \emph
  {et~al.}(2013{\natexlab{b}})\citenamefont {{Abeysekara}}, \citenamefont
  {{Alfaro}}, \citenamefont {{Alvarez}}, \citenamefont {{{\'A}lvarez}},
  \citenamefont {{Arceo}}, \citenamefont {{Arteaga-Vel{\'a}zquez}},
  \citenamefont {{Ayala Solares}}, \citenamefont {{Barber}}, \citenamefont
  {{Baughman}}, \citenamefont {{Bautista-Elivar}}, \citenamefont {{Belmont}},
  \citenamefont {{BenZvi}}, \citenamefont {{Berley}}, \citenamefont {{Bonilla
  Rosales}}, \citenamefont {{Braun}}, \citenamefont {{Caballero-Lopez}},
  \citenamefont {{Caballero-Mora}}, \citenamefont {{Carrami{\~n}ana}},
  \citenamefont {{Castillo}}, \citenamefont {{Cotti}}, \citenamefont
  {{Cotzomi}}, \citenamefont {{de la Fuente}}, \citenamefont {{De Le{\'o}n}},
  \citenamefont {{DeYoung}}, \citenamefont {{Diaz Hernandez}}, \citenamefont
  {{D{\'{\i}}az-V{\'e}lez}}, \citenamefont {{Dingus}}, \citenamefont
  {{DuVernois}}, \citenamefont {{Ellsworth}}, \citenamefont {{Fernandez}},
  \citenamefont {{Fiorino}}, \citenamefont {{Fraija}}, \citenamefont
  {{Galindo}}, \citenamefont {{Garfias}}, \citenamefont {{Gonz{\'a}lez}},
  \citenamefont {{Gonz{\'a}lez}}, \citenamefont {{Goodman}}, \citenamefont
  {{Grabski}}, \citenamefont {{Gussert}}, \citenamefont {{Hampel-Arias}},
  \citenamefont {{Hui}}, \citenamefont {{H{\"u}ntemeyer}}, \citenamefont
  {{Imran}}, \citenamefont {{Iriarte}}, \citenamefont {{Karn}}, \citenamefont
  {{Kieda}}, \citenamefont {{Kunde}}, \citenamefont {{Lara}}, \citenamefont
  {{Lauer}}, \citenamefont {{Lee}}, \citenamefont {{Lennarz}}, \citenamefont
  {{Le{\'o}n Vargas}}, \citenamefont {{Linares}}, \citenamefont {{Linnemann}},
  \citenamefont {{Longo}}, \citenamefont {{Luna-GarcIa}}, \citenamefont
  {{Marinelli}}, \citenamefont {{Martinez}}, \citenamefont {{Martinez}},
  \citenamefont {{Mart{\'{\i}}nez-Castro}}, \citenamefont {{Matthews}},
  \citenamefont {{Miranda-Romagnoli}}, \citenamefont {{Moreno}}, \citenamefont
  {{Mostaf{\'a}}}, \citenamefont {{Nava}}, \citenamefont {{Nellen}},
  \citenamefont {{Newbold}}, \citenamefont {{Noriega-Papaqui}}, \citenamefont
  {{Oceguera-Becerra}}, \citenamefont {{Patricelli}}, \citenamefont {{Pelayo}},
  \citenamefont {{P{\'e}rez-P{\'e}rez}}, \citenamefont {{Pretz}}, \citenamefont
  {{Rivi{\`e}re}}, \citenamefont {{Rosa-Gonz{\'a}lez}}, \citenamefont
  {{Salazar}}, \citenamefont {{Salesa}}, \citenamefont {{Sanchez}},
  \citenamefont {{Sandoval}}, \citenamefont {{Santos}}, \citenamefont
  {{Schneider}}, \citenamefont {{Silich}}, \citenamefont {{Sinnis}},
  \citenamefont {{Smith}}, \citenamefont {{Sparks}}, \citenamefont
  {{Springer}}, \citenamefont {{Taboada}}, \citenamefont {{Toale}},
  \citenamefont {{Tollefson}}, \citenamefont {{Torres}}, \citenamefont
  {{Ukwatta}}, \citenamefont {{Villase{\~n}or}}, \citenamefont {{Weisgarber}},
  \citenamefont {{Westerhoff}}, \citenamefont {{Wisher}}, \citenamefont
  {{Wood}}, \citenamefont {{Yodh}}, \citenamefont {{Younk}}, \citenamefont
  {{Zaborov}}, \citenamefont {{Zepeda}},\ and\ \citenamefont
  {{Zhou}}}]{2013arXiv1310.0073H}%
  \BibitemOpen
  \bibfield  {author} {\bibinfo {author} {\bibnamefont {{Abeysekara}},
  \bibfnamefont {A.~U.}}, \bibinfo {author} {\bibnamefont {{Alfaro}},
  \bibfnamefont {R.}}, \bibinfo {author} {\bibnamefont {{Alvarez}},
  \bibfnamefont {C.}},  \emph {et~al.} (\bibinfo {collaboration} {HAWC
  Collaboration}),\ }in\ \href@noop {} {\emph {\bibinfo {booktitle}
  {Proceedings of the 33rd International Cosmic-Ray Conference}}}\ (\bibinfo
  {address} {Rio de Janeiro, Brazil},\ \bibinfo {year} {2013})\ \Eprint
  {http://arxiv.org/abs/1310.0073} {arXiv:1310.0073 [astro-ph.HE]} \BibitemShut
  {NoStop}%
\bibitem [{\citenamefont {{Abramowski}}\ \emph {et~al.}(2012)\citenamefont
  {{Abramowski}}, \citenamefont {{Acero}}, \citenamefont {{Aharonian}},
  \citenamefont {{Akhperjanian}}, \citenamefont {{Anton}}, \citenamefont
  {{Balzer}}, \citenamefont {{Barnacka}}, \citenamefont {{Barres de Almeida}},
  \citenamefont {{Becherini}}, \citenamefont {{Becker}}, \citenamefont
  {{Behera}}, \citenamefont {{Bernl{\"o}hr}}, \citenamefont {{Birsin}},
  \citenamefont {{Biteau}}, \citenamefont {{Bochow}}, \citenamefont
  {{Boisson}}, \citenamefont {{Bolmont}}, \citenamefont {{Bordas}},
  \citenamefont {{Brucker}}, \citenamefont {{Brun}}, \citenamefont {{Brun}},
  \citenamefont {{Bulik}}, \citenamefont {{B{\"u}sching}}, \citenamefont
  {{Carrigan}}, \citenamefont {{Casanova}}, \citenamefont {{Cerruti}},
  \citenamefont {{Chadwick}}, \citenamefont {{Charbonnier}}, \citenamefont
  {{Chaves}}, \citenamefont {{Cheesebrough}}, \citenamefont {{Clapson}},
  \citenamefont {{Coignet}}, \citenamefont {{Cologna}}, \citenamefont
  {{Conrad}}, \citenamefont {{Dalton}}, \citenamefont {{Daniel}}, \citenamefont
  {{Davids}}, \citenamefont {{Degrange}}, \citenamefont {{Deil}}, \citenamefont
  {{Dickinson}}, \citenamefont {{Djannati-Ata{\"i}}}, \citenamefont
  {{Domainko}}, \citenamefont {{Drury}}, \citenamefont {{Dubus}}, \citenamefont
  {{Dutson}}, \citenamefont {{Dyks}}, \citenamefont {{Dyrda}}, \citenamefont
  {{Egberts}}, \citenamefont {{Eger}}, \citenamefont {{Espigat}}, \citenamefont
  {{Fallon}}, \citenamefont {{Farnier}}, \citenamefont {{Fegan}}, \citenamefont
  {{Feinstein}}, \citenamefont {{Fernandes}}, \citenamefont {{Fiasson}},
  \citenamefont {{Fontaine}}, \citenamefont {{F{\"o}rster}}, \citenamefont
  {{F{\"u}{\ss}ling}}, \citenamefont {{Gallant}}, \citenamefont {{Gast}},
  \citenamefont {{G{\'e}rard}}, \citenamefont {{Gerbig}}, \citenamefont
  {{Giebels}}, \citenamefont {{Glicenstein}}, \citenamefont {{Gl{\"u}ck}},
  \citenamefont {{Goret}}, \citenamefont {{G{\"o}ring}}, \citenamefont
  {{H{\"a}ffner}}, \citenamefont {{Hague}}, \citenamefont {{Hampf}},
  \citenamefont {{Hauser}}, \citenamefont {{Heinz}}, \citenamefont
  {{Heinzelmann}}, \citenamefont {{Henri}}, \citenamefont {{Hermann}},
  \citenamefont {{Hinton}}, \citenamefont {{Hoffmann}}, \citenamefont
  {{Hofmann}}, \citenamefont {{Hofverberg}}, \citenamefont {{Holler}},
  \citenamefont {{Horns}}, \citenamefont {{Jacholkowska}}, \citenamefont {{de
  Jager}}, \citenamefont {{Jahn}}, \citenamefont {{Jamrozy}}, \citenamefont
  {{Jung}}, \citenamefont {{Kastendieck}}, \citenamefont {{Katarzy{\'n}ski}},
  \citenamefont {{Katz}}, \citenamefont {{Kaufmann}}, \citenamefont {{Keogh}},
  \citenamefont {{Khangulyan}}, \citenamefont {{Kh{\'e}lifi}}, \citenamefont
  {{Klochkov}}, \citenamefont {{Klu{\'z}niak}}, \citenamefont {{Kneiske}},
  \citenamefont {{Komin}}, \citenamefont {{Kosack}}, \citenamefont
  {{Kossakowski}}, \citenamefont {{Laffon}}, \citenamefont {{Lamanna}},
  \citenamefont {{Lennarz}}, \citenamefont {{Lohse}}, \citenamefont
  {{Lopatin}}, \citenamefont {{Lu}}, \citenamefont {{Marandon}}, \citenamefont
  {{Marcowith}}, \citenamefont {{Masbou}}, \citenamefont {{Maurin}},
  \citenamefont {{Maxted}}, \citenamefont {{Mayer}}, \citenamefont {{McComb}},
  \citenamefont {{Medina}}, \citenamefont {{M{\'e}hault}}, \citenamefont
  {{Moderski}}, \citenamefont {{Moulin}}, \citenamefont {{Naumann}},
  \citenamefont {{Naumann-Godo}}, \citenamefont {{de Naurois}}, \citenamefont
  {{Nedbal}}, \citenamefont {{Nekrassov}}, \citenamefont {{Nguyen}},
  \citenamefont {{Nicholas}}, \citenamefont {{Niemiec}}, \citenamefont
  {{Nolan}}, \citenamefont {{Ohm}}, \citenamefont {{de O{\~n}a Wilhelmi}},
  \citenamefont {{Opitz}}, \citenamefont {{Ostrowski}}, \citenamefont {{Oya}},
  \citenamefont {{Panter}}, \citenamefont {{Paz Arribas}}, \citenamefont
  {{Pedaletti}}, \citenamefont {{Pelletier}}, \citenamefont {{Petrucci}},
  \citenamefont {{Pita}}, \citenamefont {{P{\"u}hlhofer}}, \citenamefont
  {{Punch}}, \citenamefont {{Quirrenbach}}, \citenamefont {{Raue}},
  \citenamefont {{Rayner}}, \citenamefont {{Reimer}}, \citenamefont {{Reimer}},
  \citenamefont {{Renaud}}, \citenamefont {{de los Reyes}}, \citenamefont
  {{Rieger}}, \citenamefont {{Ripken}}, \citenamefont {{Rob}}, \citenamefont
  {{Rosier-Lees}}, \citenamefont {{Rowell}}, \citenamefont {{Rudak}},
  \citenamefont {{Rulten}}, \citenamefont {{Ruppel}}, \citenamefont
  {{Sahakian}}, \citenamefont {{Sanchez}}, \citenamefont {{Santangelo}},
  \citenamefont {{Schlickeiser}}, \citenamefont {{Sch{\"o}ck}}, \citenamefont
  {{Schulz}}, \citenamefont {{Schwanke}}, \citenamefont {{Schwarzburg}},
  \citenamefont {{Schwemmer}}, \citenamefont {{Sheidaei}}, \citenamefont
  {{Skilton}}, \citenamefont {{Sol}}, \citenamefont {{Spengler}}, \citenamefont
  {{Stawarz}}, \citenamefont {{Steenkamp}}, \citenamefont {{Stegmann}},
  \citenamefont {{Stinzing}}, \citenamefont {{Stycz}}, \citenamefont
  {{Sushch}}, \citenamefont {{Szostek}}, \citenamefont {{Tavernet}},
  \citenamefont {{Terrier}}, \citenamefont {{Tluczykont}}, \citenamefont
  {{Valerius}}, \citenamefont {{van Eldik}}, \citenamefont {{Vasileiadis}},
  \citenamefont {{Venter}}, \citenamefont {{Vialle}}, \citenamefont {{Viana}},
  \citenamefont {{Vincent}}, \citenamefont {{V{\"o}lk}}, \citenamefont
  {{Volpe}}, \citenamefont {{Vorobiov}}, \citenamefont {{Vorster}},
  \citenamefont {{Wagner}}, \citenamefont {{Ward}}, \citenamefont {{White}},
  \citenamefont {{Wierzcholska}}, \citenamefont {{Zacharias}}, \citenamefont
  {{Zajczyk}}, \citenamefont {{Zdziarski}}, \citenamefont {{Zech}},
  \citenamefont {{Zechlin}},\ and\ \citenamefont
  {{H.~E.~S.~S.~Collaboration}}}]{2012ApJ...750..123A}%
  \BibitemOpen
  \bibfield  {author} {\bibinfo {author} {\bibnamefont {{Abramowski}},
  \bibfnamefont {A.}}, \bibinfo {author} {\bibnamefont {{Acero}}, \bibfnamefont
  {F.}}, \bibinfo {author} {\bibnamefont {{Aharonian}}, \bibfnamefont {F.}},
  \emph {et~al.},\ }\href {\doibase 10.1088/0004-637X/750/2/123} {\bibfield
  {journal} {\bibinfo  {journal} {\apj}\ }\textbf {\bibinfo {volume} {750}},\
  \bibinfo {eid} {123} (\bibinfo {year} {2012})},\ \Eprint
  {http://arxiv.org/abs/1202.5494} {arXiv:1202.5494 [astro-ph.HE]} \BibitemShut
  {NoStop}%
\bibitem [{\citenamefont {{Abramowski}}\ \emph
  {et~al.}(2014{\natexlab{a}})\citenamefont {{Abramowski}}, \citenamefont
  {{Acero}}, \citenamefont {{Aharonian}}, \citenamefont {{Akhperjanian}},
  \citenamefont {{Anton}}, \citenamefont {{Balzer}}, \citenamefont
  {{Barnacka}}, \citenamefont {{Barres de Almeida}}, \citenamefont
  {{Becherini}}, \citenamefont {{Becker}}, \citenamefont {{Behera}},
  \citenamefont {{Bernl{\"o}hr}}, \citenamefont {{Birsin}}, \citenamefont
  {{Biteau}}, \citenamefont {{Bochow}}, \citenamefont {{Boisson}},
  \citenamefont {{Bolmont}}, \citenamefont {{Bordas}}, \citenamefont
  {{Brucker}}, \citenamefont {{Brun}}, \citenamefont {{Brun}}, \citenamefont
  {{Bulik}}, \citenamefont {{B{\"u}sching}}, \citenamefont {{Carrigan}},
  \citenamefont {{Casanova}}, \citenamefont {{Cerruti}}, \citenamefont
  {{Chadwick}}, \citenamefont {{Charbonnier}}, \citenamefont {{Chaves}},
  \citenamefont {{Cheesebrough}}, \citenamefont {{Clapson}}, \citenamefont
  {{Coignet}}, \citenamefont {{Cologna}}, \citenamefont {{Conrad}},
  \citenamefont {{Dalton}}, \citenamefont {{Daniel}}, \citenamefont {{Davids}},
  \citenamefont {{Degrange}}, \citenamefont {{Deil}}, \citenamefont
  {{Dickinson}}, \citenamefont {{Djannati-Ata{\"i}}}, \citenamefont
  {{Domainko}}, \citenamefont {{O'C.~Drury}}, \citenamefont {{Dubus}},
  \citenamefont {{Dutson}}, \citenamefont {{Dyks}}, \citenamefont {{Dyrda}},
  \citenamefont {{Egberts}}, \citenamefont {{Eger}}, \citenamefont {{Espigat}},
  \citenamefont {{Fallon}}, \citenamefont {{Farnier}}, \citenamefont {{Fegan}},
  \citenamefont {{Feinstein}}, \citenamefont {{Fernandes}}, \citenamefont
  {{Fiasson}}, \citenamefont {{Fontaine}}, \citenamefont {{F{\"o}rster}},
  \citenamefont {{F{\"u}{\ss}ling}}, \citenamefont {{Gallant}}, \citenamefont
  {{Gast}}, \citenamefont {{G{\'e}rard}}, \citenamefont {{Gerbig}},
  \citenamefont {{Giebels}}, \citenamefont {{Glicenstein}}, \citenamefont
  {{Gl{\"u}ck}}, \citenamefont {{Goret}}, \citenamefont {{G{\"o}ring}},
  \citenamefont {{H{\"a}ffner}}, \citenamefont {{Hague}}, \citenamefont
  {{Hampf}}, \citenamefont {{Hauser}}, \citenamefont {{Heinz}}, \citenamefont
  {{Heinzelmann}}, \citenamefont {{Henri}}, \citenamefont {{Hermann}},
  \citenamefont {{Hinton}}, \citenamefont {{Hoffmann}}, \citenamefont
  {{Hofmann}}, \citenamefont {{Hofverberg}}, \citenamefont {{Holler}},
  \citenamefont {{Horns}}, \citenamefont {{Jacholkowska}}, \citenamefont {{de
  Jager}}, \citenamefont {{Jahn}}, \citenamefont {{Jamrozy}}, \citenamefont
  {{Jung}}, \citenamefont {{Kastendieck}}, \citenamefont {{Katarzys{\'n}ki}},
  \citenamefont {{Katz}}, \citenamefont {{Kaufmann}}, \citenamefont {{Keogh}},
  \citenamefont {{Khangulyan}}, \citenamefont {{Kh{\'e}lifi}}, \citenamefont
  {{Klochkov}}, \citenamefont {{Klu{\'z}niak}}, \citenamefont {{Kneiske}},
  \citenamefont {{Komin}}, \citenamefont {{Kosack}}, \citenamefont
  {{Kossakowski}}, \citenamefont {{Laffon}}, \citenamefont {{Lamanna}},
  \citenamefont {{Lennarz}}, \citenamefont {{Lohse}}, \citenamefont
  {{Lopatin}}, \citenamefont {{Lu}}, \citenamefont {{Marandon}}, \citenamefont
  {{Marcowith}}, \citenamefont {{Masbou}}, \citenamefont {{Maurin}},
  \citenamefont {{Maxted}}, \citenamefont {{Mayer}}, \citenamefont {{McComb}},
  \citenamefont {{Medina}}, \citenamefont {{M{\'e}hault}}, \citenamefont
  {{Moderski}}, \citenamefont {{Moulin}}, \citenamefont {{Naumann}},
  \citenamefont {{Naumann-Godo}}, \citenamefont {{de Naurois}}, \citenamefont
  {{Nedbal}}, \citenamefont {{Nekrassov}}, \citenamefont {{Nguyen}},
  \citenamefont {{Nicholas}}, \citenamefont {{Niemiec}}, \citenamefont
  {{Nolan}}, \citenamefont {{Ohm}}, \citenamefont {{de O{\~n}a Wilhelmi}},
  \citenamefont {{Opitz}}, \citenamefont {{Ostrowski}}, \citenamefont {{Oya}},
  \citenamefont {{Panter}}, \citenamefont {{Paz Arribas}}, \citenamefont
  {{Pedaletti}}, \citenamefont {{Pelletier}}, \citenamefont {{Petrucci}},
  \citenamefont {{Pita}}, \citenamefont {{P{\"u}hlhofer}}, \citenamefont
  {{Punch}}, \citenamefont {{Quirrenbach}}, \citenamefont {{Raue}},
  \citenamefont {{Rayner}}, \citenamefont {{Reimer}}, \citenamefont {{Reimer}},
  \citenamefont {{Renaud}}, \citenamefont {{de los Reyes}}, \citenamefont
  {{Rieger}}, \citenamefont {{Ripken}}, \citenamefont {{Rob}}, \citenamefont
  {{Rosier-Lees}}, \citenamefont {{Rowell}}, \citenamefont {{Rudak}},
  \citenamefont {{Rulten}}, \citenamefont {{Ruppel}}, \citenamefont
  {{Sahakian}}, \citenamefont {{Sanchez}}, \citenamefont {{Santangelo}},
  \citenamefont {{Schlickeiser}}, \citenamefont {{Sch{\"o}ck}}, \citenamefont
  {{Schulz}}, \citenamefont {{Schwanke}}, \citenamefont {{Schwarzburg}},
  \citenamefont {{Schwemmer}}, \citenamefont {{Sheidaei}}, \citenamefont
  {{Skilton}}, \citenamefont {{Sol}}, \citenamefont {{Spengler}}, \citenamefont
  {{Stawarz}}, \citenamefont {{Steenkamp}}, \citenamefont {{Stegmann}},
  \citenamefont {{Stinzing}}, \citenamefont {{Stycz}}, \citenamefont
  {{Sushch}}, \citenamefont {{Szostek}}, \citenamefont {{Tavernet}},
  \citenamefont {{Terrier}}, \citenamefont {{Tluczykont}}, \citenamefont
  {{Valerius}}, \citenamefont {{van Eldik}}, \citenamefont {{Vasileiadis}},
  \citenamefont {{Venter}}, \citenamefont {{Vialle}}, \citenamefont {{Viana}},
  \citenamefont {{Vincent}}, \citenamefont {{V{\"o}lk}}, \citenamefont
  {{Volpe}}, \citenamefont {{Vorobiov}}, \citenamefont {{Vorster}},
  \citenamefont {{Wagner}}, \citenamefont {{Ward}}, \citenamefont {{White}},
  \citenamefont {{Wierzcholska}}, \citenamefont {{Zacharias}}, \citenamefont
  {{Zajczyk}}, \citenamefont {{Zdziarski}}, \citenamefont {{Zech}},
  \citenamefont {{Zechlin}},\ and\ \citenamefont
  {{H.~E.~S.~S.~Collaboration}}}]{2014ApJ...783...63A}%
  \BibitemOpen
  \bibfield  {author} {\bibinfo {author} {\bibnamefont {{Abramowski}},
  \bibfnamefont {A.}}, \bibinfo {author} {\bibnamefont {{Acero}}, \bibfnamefont
  {F.}}, \bibinfo {author} {\bibnamefont {{Aharonian}}, \bibfnamefont {F.}},
  \emph {et~al.},\ }\href {\doibase 10.1088/0004-637X/783/1/63} {\bibfield
  {journal} {\bibinfo  {journal} {\apj}\ }\textbf {\bibinfo {volume} {783}},\
  \bibinfo {eid} {63} (\bibinfo {year} {2014}{\natexlab{a}})}\BibitemShut
  {NoStop}%
\bibitem [{\citenamefont {Abramowski}\ \emph {et~al.}(2011)\citenamefont
  {Abramowski}, \citenamefont {Acero}, \citenamefont {Aharonian}, \citenamefont
  {Akhperjanian}, \citenamefont {Anton}, \citenamefont {Barnacka},
  \citenamefont {Barres~de Almeida}, \citenamefont {Bazer-Bachi}, \citenamefont
  {Becherini}, \citenamefont {Becker}, \citenamefont {Behera}, \citenamefont
  {Bernl\"ohr}, \citenamefont {Bochow}, \citenamefont {Boisson}, \citenamefont
  {Bolmont}, \citenamefont {Bordas}, \citenamefont {Borrel}, \citenamefont
  {Brucker}, \citenamefont {Brun}, \citenamefont {Brun}, \citenamefont {Bulik},
  \citenamefont {B\"usching}, \citenamefont {Carrigan}, \citenamefont
  {Casanova}, \citenamefont {Cerruti}, \citenamefont {Chadwick}, \citenamefont
  {Charbonnier}, \citenamefont {Chaves}, \citenamefont {Cheesebrough},
  \citenamefont {Chounet}, \citenamefont {Clapson}, \citenamefont {Coignet},
  \citenamefont {Conrad}, \citenamefont {Dalton}, \citenamefont {Daniel},
  \citenamefont {Davids}, \citenamefont {Degrange}, \citenamefont {Deil},
  \citenamefont {Dickinson}, \citenamefont {Djannati-Ata\"\i}, \citenamefont
  {Domainko}, \citenamefont {Drury}, \citenamefont {Dubois}, \citenamefont
  {Dubus}, \citenamefont {Dyks}, \citenamefont {Dyrda}, \citenamefont
  {Egberts}, \citenamefont {Eger}, \citenamefont {Espigat}, \citenamefont
  {Fallon}, \citenamefont {Farnier}, \citenamefont {Fegan}, \citenamefont
  {Feinstein}, \citenamefont {Fernandes}, \citenamefont {Fiasson},
  \citenamefont {Fontaine}, \citenamefont {F\"orster}, \citenamefont
  {F\"u\ss{}ling}, \citenamefont {Gallant}, \citenamefont {Gast}, \citenamefont
  {G\'erard}, \citenamefont {Gerbig}, \citenamefont {Giebels}, \citenamefont
  {Glicenstein}, \citenamefont {Gl\"uck}, \citenamefont {Goret}, \citenamefont
  {G\"oring}, \citenamefont {Hague}, \citenamefont {Hampf}, \citenamefont
  {Hauser}, \citenamefont {Heinz}, \citenamefont {Heinzelmann}, \citenamefont
  {Henri}, \citenamefont {Hermann}, \citenamefont {Hinton}, \citenamefont
  {Hoffmann}, \citenamefont {Hofmann}, \citenamefont {Hofverberg},
  \citenamefont {Horns}, \citenamefont {Jacholkowska}, \citenamefont
  {de~Jager}, \citenamefont {Jahn}, \citenamefont {Jamrozy}, \citenamefont
  {Jung}, \citenamefont {Kastendieck}, \citenamefont
  {Katarzy\ifmmode~\acute{n}\else \'{n}\fi{}ski}, \citenamefont {Katz},
  \citenamefont {Kaufmann}, \citenamefont {Keogh}, \citenamefont {Kerschhaggl},
  \citenamefont {Khangulyan}, \citenamefont {Kh\'elifi}, \citenamefont
  {Klochkov}, \citenamefont {Klu\ifmmode~\acute{z}\else \'{z}\fi{}niak},
  \citenamefont {Kneiske}, \citenamefont {Komin}, \citenamefont {Kosack},
  \citenamefont {Kossakowski}, \citenamefont {Laffon}, \citenamefont {Lamanna},
  \citenamefont {Lennarz}, \citenamefont {Lohse}, \citenamefont {Lopatin},
  \citenamefont {Lu}, \citenamefont {Marandon}, \citenamefont {Marcowith},
  \citenamefont {Masbou}, \citenamefont {Maurin}, \citenamefont {Maxted},
  \citenamefont {McComb}, \citenamefont {Medina}, \citenamefont {M\'ehault},
  \citenamefont {Moderski}, \citenamefont {Moulin}, \citenamefont {Naumann},
  \citenamefont {Naumann-Godo}, \citenamefont {de~Naurois}, \citenamefont
  {Nedbal}, \citenamefont {Nekrassov}, \citenamefont {Nguyen}, \citenamefont
  {Nicholas}, \citenamefont {Niemiec}, \citenamefont {Nolan}, \citenamefont
  {Ohm}, \citenamefont {Olive}, \citenamefont {de~O\~na Wilhelmi},
  \citenamefont {Opitz}, \citenamefont {Ostrowski}, \citenamefont {Panter},
  \citenamefont {Paz~Arribas}, \citenamefont {Pedaletti}, \citenamefont
  {Pelletier}, \citenamefont {Petrucci}, \citenamefont {Pita}, \citenamefont
  {P\"uhlhofer}, \citenamefont {Punch}, \citenamefont {Quirrenbach},
  \citenamefont {Raue}, \citenamefont {Rayner}, \citenamefont {Reimer},
  \citenamefont {Reimer}, \citenamefont {Renaud}, \citenamefont {de~los Reyes},
  \citenamefont {Rieger}, \citenamefont {Ripken}, \citenamefont {Rob},
  \citenamefont {Rosier-Lees}, \citenamefont {Rowell}, \citenamefont {Rudak},
  \citenamefont {Rulten}, \citenamefont {Ruppel}, \citenamefont {Ryde},
  \citenamefont {Sahakian}, \citenamefont {Santangelo}, \citenamefont
  {Schlickeiser}, \citenamefont {Sch\"ock}, \citenamefont {Sch\"onwald},
  \citenamefont {Schwanke}, \citenamefont {Schwarzburg}, \citenamefont
  {Schwemmer}, \citenamefont {Shalchi}, \citenamefont {Sikora}, \citenamefont
  {Skilton}, \citenamefont {Sol}, \citenamefont {Spengler}, \citenamefont
  {Stawarz}, \citenamefont {Steenkamp}, \citenamefont {Stegmann}, \citenamefont
  {Stinzing}, \citenamefont {Sushch}, \citenamefont {Szostek}, \citenamefont
  {Tavernet}, \citenamefont {Terrier}, \citenamefont {Tibolla}, \citenamefont
  {Tluczykont}, \citenamefont {Valerius}, \citenamefont {van Eldik},
  \citenamefont {Vasileiadis}, \citenamefont {Venter}, \citenamefont {Vialle},
  \citenamefont {Viana}, \citenamefont {Vincent}, \citenamefont {Vivier},
  \citenamefont {V\"olk}, \citenamefont {Volpe}, \citenamefont {Vorobiov},
  \citenamefont {Vorster}, \citenamefont {Wagner}, \citenamefont {Ward},
  \citenamefont {Wierzcholska}, \citenamefont {Zajczyk}, \citenamefont
  {Zdziarski}, \citenamefont {Zech},\ and\ \citenamefont
  {Zechlin}}]{PhysRevLett.106.161301}%
  \BibitemOpen
  \bibfield  {author} {\bibinfo {author} {\bibnamefont {Abramowski},
  \bibfnamefont {A.}}, \bibinfo {author} {\bibnamefont {Acero}, \bibfnamefont
  {F.}}, \bibinfo {author} {\bibnamefont {Aharonian}, \bibfnamefont {F.}},
  \emph {et~al.} (\bibinfo {collaboration} {H.E.S.S. Collaboration}),\ }\href
  {\doibase 10.1103/PhysRevLett.106.161301} {\bibfield  {journal} {\bibinfo
  {journal} {Phys. Rev. Lett.}\ }\textbf {\bibinfo {volume} {106}},\ \bibinfo
  {pages} {161301} (\bibinfo {year} {2011})}\BibitemShut {NoStop}%
\bibitem [{\citenamefont {{Abramowski}}\ \emph {et~al.}(2011)\citenamefont
  {{Abramowski}}, \citenamefont {{Acero}}, \citenamefont {{Aharonian}},
  \citenamefont {{Akhperjanian}}, \citenamefont {{Anton}}, \citenamefont
  {{Barnacka}}, \citenamefont {{Barres de Almeida}}, \citenamefont
  {{Bazer-Bachi}}, \citenamefont {{Becherini}}, \citenamefont {{Becker}},
  \citenamefont {{Behera}}, \citenamefont {{Bernl{\"o}hr}}, \citenamefont
  {{Bochow}}, \citenamefont {{Boisson}}, \citenamefont {{Bolmont}},
  \citenamefont {{Bordas}}, \citenamefont {{Borrel}}, \citenamefont
  {{Brucker}}, \citenamefont {{Brun}}, \citenamefont {{Brun}}, \citenamefont
  {{Bulik}}, \citenamefont {{B{\"u}sching}}, \citenamefont {{Carrigan}},
  \citenamefont {{Casanova}}, \citenamefont {{Cerruti}}, \citenamefont
  {{Chadwick}}, \citenamefont {{Charbonnier}}, \citenamefont {{Chaves}},
  \citenamefont {{Cheesebrough}}, \citenamefont {{Chounet}}, \citenamefont
  {{Clapson}}, \citenamefont {{Coignet}}, \citenamefont {{Conrad}},
  \citenamefont {{Dalton}}, \citenamefont {{Daniel}}, \citenamefont {{Davids}},
  \citenamefont {{Degrange}}, \citenamefont {{Deil}}, \citenamefont
  {{Dickinson}}, \citenamefont {{Djannati-Ata{\"i}}}, \citenamefont
  {{Domainko}}, \citenamefont {{Drury}}, \citenamefont {{Dubois}},
  \citenamefont {{Dubus}}, \citenamefont {{Dyks}}, \citenamefont {{Dyrda}},
  \citenamefont {{Egberts}}, \citenamefont {{Eger}}, \citenamefont {{Espigat}},
  \citenamefont {{Fallon}}, \citenamefont {{Farnier}}, \citenamefont {{Fegan}},
  \citenamefont {{Feinstein}}, \citenamefont {{Fernandes}}, \citenamefont
  {{Fiasson}}, \citenamefont {{Fontaine}}, \citenamefont {{F{\"o}rster}},
  \citenamefont {{F{\"u}{\ss}ling}}, \citenamefont {{Gallant}}, \citenamefont
  {{Gast}}, \citenamefont {{G{\'e}rard}}, \citenamefont {{Gerbig}},
  \citenamefont {{Giebels}}, \citenamefont {{Glicenstein}}, \citenamefont
  {{Gl{\"u}ck}}, \citenamefont {{Goret}}, \citenamefont {{G{\"o}ring}},
  \citenamefont {{Hague}}, \citenamefont {{Hampf}}, \citenamefont {{Hauser}},
  \citenamefont {{Heinz}}, \citenamefont {{Heinzelmann}}, \citenamefont
  {{Henri}}, \citenamefont {{Hermann}}, \citenamefont {{Hinton}}, \citenamefont
  {{Hoffmann}}, \citenamefont {{Hofmann}}, \citenamefont {{Hofverberg}},
  \citenamefont {{Horns}}, \citenamefont {{Jacholkowska}}, \citenamefont {{de
  Jager}}, \citenamefont {{Jahn}}, \citenamefont {{Jamrozy}}, \citenamefont
  {{Jung}}, \citenamefont {{Kastendieck}}, \citenamefont {{Katarzy{\'n}ski}},
  \citenamefont {{Katz}}, \citenamefont {{Kaufmann}}, \citenamefont {{Keogh}},
  \citenamefont {{Kerschhaggl}}, \citenamefont {{Khangulyan}}, \citenamefont
  {{Kh{\'e}lifi}}, \citenamefont {{Klochkov}}, \citenamefont {{Klu{\'z}niak}},
  \citenamefont {{Kneiske}}, \citenamefont {{Komin}}, \citenamefont {{Kosack}},
  \citenamefont {{Kossakowski}}, \citenamefont {{Laffon}}, \citenamefont
  {{Lamanna}}, \citenamefont {{Lennarz}}, \citenamefont {{Lohse}},
  \citenamefont {{Lopatin}}, \citenamefont {{Lu}}, \citenamefont {{Marandon}},
  \citenamefont {{Marcowith}}, \citenamefont {{Masbou}}, \citenamefont
  {{Maurin}}, \citenamefont {{Maxted}}, \citenamefont {{McComb}}, \citenamefont
  {{Medina}}, \citenamefont {{M{\'e}hault}}, \citenamefont {{Moderski}},
  \citenamefont {{Moulin}}, \citenamefont {{Naumann}}, \citenamefont
  {{Naumann-Godo}}, \citenamefont {{de Naurois}}, \citenamefont {{Nedbal}},
  \citenamefont {{Nekrassov}}, \citenamefont {{Nguyen}}, \citenamefont
  {{Nicholas}}, \citenamefont {{Niemiec}}, \citenamefont {{Nolan}},
  \citenamefont {{Ohm}}, \citenamefont {{Olive}}, \citenamefont {{de O{\~n}a
  Wilhelmi}}, \citenamefont {{Opitz}}, \citenamefont {{Ostrowski}},
  \citenamefont {{Panter}}, \citenamefont {{Paz Arribas}}, \citenamefont
  {{Pedaletti}}, \citenamefont {{Pelletier}}, \citenamefont {{Petrucci}},
  \citenamefont {{Pita}}, \citenamefont {{P{\"u}hlhofer}}, \citenamefont
  {{Punch}}, \citenamefont {{Quirrenbach}}, \citenamefont {{Raue}},
  \citenamefont {{Rayner}}, \citenamefont {{Reimer}}, \citenamefont {{Reimer}},
  \citenamefont {{Renaud}}, \citenamefont {{de Los Reyes}}, \citenamefont
  {{Rieger}}, \citenamefont {{Ripken}}, \citenamefont {{Rob}}, \citenamefont
  {{Rosier-Lees}}, \citenamefont {{Rowell}}, \citenamefont {{Rudak}},
  \citenamefont {{Rulten}}, \citenamefont {{Ruppel}}, \citenamefont {{Ryde}},
  \citenamefont {{Sahakian}}, \citenamefont {{Santangelo}}, \citenamefont
  {{Schlickeiser}}, \citenamefont {{Sch{\"o}ck}}, \citenamefont
  {{Sch{\"o}nwald}}, \citenamefont {{Schwanke}}, \citenamefont {{Schwarzburg}},
  \citenamefont {{Schwemmer}}, \citenamefont {{Shalchi}}, \citenamefont
  {{Sikora}}, \citenamefont {{Skilton}}, \citenamefont {{Sol}}, \citenamefont
  {{Spengler}}, \citenamefont {{Stawarz}}, \citenamefont {{Steenkamp}},
  \citenamefont {{Stegmann}}, \citenamefont {{Stinzing}}, \citenamefont
  {{Sushch}}, \citenamefont {{Szostek}}, \citenamefont {{Tavernet}},
  \citenamefont {{Terrier}}, \citenamefont {{Tibolla}}, \citenamefont
  {{Tluczykont}}, \citenamefont {{Valerius}}, \citenamefont {{van Eldik}},
  \citenamefont {{Vasileiadis}}, \citenamefont {{Venter}}, \citenamefont
  {{Vialle}}, \citenamefont {{Viana}}, \citenamefont {{Vincent}}, \citenamefont
  {{Vivier}}, \citenamefont {{V{\"o}lk}}, \citenamefont {{Volpe}},
  \citenamefont {{Vorobiov}}, \citenamefont {{Vorster}}, \citenamefont
  {{Wagner}}, \citenamefont {{Ward}}, \citenamefont {{Wierzcholska}},
  \citenamefont {{Zajczyk}}, \citenamefont {{Zdziarski}}, \citenamefont
  {{Zech}}, \citenamefont {{Zechlin}},\ and\ \citenamefont
  {{H.E.S.S.~Collaboration}}}]{2011APh....34..608H}%
  \BibitemOpen
  \bibfield  {author} {\bibinfo {author} {\bibnamefont {{Abramowski}},
  \bibfnamefont {A.}}, \bibinfo {author} {\bibnamefont {{Acero}}, \bibfnamefont
  {F.}}, \bibinfo {author} {\bibnamefont {{Aharonian}}, \bibfnamefont {F.}},
  \emph {et~al.} (\bibinfo {collaboration} {H.E.S.S. Collaboration}),\ }\href
  {\doibase 10.1016/j.astropartphys.2010.12.006} {\bibfield  {journal}
  {\bibinfo  {journal} {Astroparticle Physics}\ }\textbf {\bibinfo {volume}
  {34}},\ \bibinfo {pages} {608} (\bibinfo {year} {2011})},\ \Eprint
  {http://arxiv.org/abs/1012.5602} {arXiv:1012.5602 [astro-ph.HE]} \BibitemShut
  {NoStop}%
\bibitem [{\citenamefont {{Abramowski}}\ \emph
  {et~al.}(2014{\natexlab{b}})\citenamefont {{Abramowski}}, \citenamefont
  {{Aharonian}}, \citenamefont {{Ait Benkhali}}, \citenamefont
  {{Akhperjanian}}, \citenamefont {{Ang{\"u}ner}}, \citenamefont {{Backes}},
  \citenamefont {{Balenderan}}, \citenamefont {{Balzer}}, \citenamefont
  {{Barnacka}}, \citenamefont {{Becherini}},\ and\ \citenamefont
  {et~al.}}]{Abramowski:2014tra}%
  \BibitemOpen
  \bibfield  {author} {\bibinfo {author} {\bibnamefont {{Abramowski}},
  \bibfnamefont {A.}}, \bibinfo {author} {\bibnamefont {{Aharonian}},
  \bibfnamefont {F.}}, \bibinfo {author} {\bibnamefont {{Ait Benkhali}},
  \bibfnamefont {F.}},  \emph {et~al.},\ }\href {\doibase
  10.1103/PhysRevD.90.112012} {\bibfield  {journal} {\bibinfo  {journal}
  {\prd}\ }\textbf {\bibinfo {volume} {90}},\ \bibinfo {eid} {112012} (\bibinfo
  {year} {2014}{\natexlab{b}})},\ \Eprint {http://arxiv.org/abs/1410.2589}
  {arXiv:1410.2589 [astro-ph.HE]} \BibitemShut {NoStop}%
\bibitem [{\citenamefont {{Abramowski}}\ \emph
  {et~al.}(2014{\natexlab{c}})\citenamefont {{Abramowski}}, \citenamefont
  {{Aharonian}}, \citenamefont {{Ait Benkhali}}, \citenamefont
  {{Akhperjanian}}, \citenamefont {{Ang{\"u}ner}}, \citenamefont {{Backes}},
  \citenamefont {{Balenderan}}, \citenamefont {{Balzer}}, \citenamefont
  {{Barnacka}}, \citenamefont {{Becherini}},\ and\ \citenamefont
  {et~al.}}]{2014PhRvD..90l2007A}%
  \BibitemOpen
  \bibfield  {author} {\bibinfo {author} {\bibnamefont {{Abramowski}},
  \bibfnamefont {A.}}, \bibinfo {author} {\bibnamefont {{Aharonian}},
  \bibfnamefont {F.}}, \bibinfo {author} {\bibnamefont {{Ait Benkhali}},
  \bibfnamefont {F.}},  \emph {et~al.},\ }\href {\doibase
  10.1103/PhysRevD.90.122007} {\bibfield  {journal} {\bibinfo  {journal}
  {\prd}\ }\textbf {\bibinfo {volume} {90}},\ \bibinfo {eid} {122007} (\bibinfo
  {year} {2014}{\natexlab{c}})}\BibitemShut {NoStop}%
\bibitem [{\citenamefont {Abramowski}\ \emph {et~al.}(2011)\citenamefont
  {Abramowski} \emph {et~al.}}]{Abramowski:2011hc}%
  \BibitemOpen
  \bibfield  {author} {\bibinfo {author} {\bibnamefont {Abramowski},
  \bibfnamefont {A.}} \emph {et~al.} (\bibinfo {collaboration} {H.E.S.S.
  Collaboration}),\ }\href {\doibase 10.1103/PhysRevLett.106.161301} {\bibfield
   {journal} {\bibinfo  {journal} {Phys.Rev.Lett.}\ }\textbf {\bibinfo {volume}
  {106}},\ \bibinfo {pages} {161301} (\bibinfo {year} {2011})},\ \Eprint
  {http://arxiv.org/abs/1103.3266} {arXiv:1103.3266 [astro-ph.HE]} \BibitemShut
  {NoStop}%
%%CITATION = ARXIV:1103.3266;%%
\bibitem [{\citenamefont {Abramowski}\ \emph {et~al.}(2013)\citenamefont
  {Abramowski} \emph {et~al.}}]{Abramowski:2013ax}%
  \BibitemOpen
  \bibfield  {author} {\bibinfo {author} {\bibnamefont {Abramowski},
  \bibfnamefont {A.}} \emph {et~al.} (\bibinfo {collaboration} {H.E.S.S.
  Collaboration}),\ }\href {\doibase 10.1103/PhysRevLett.110.041301} {\bibfield
   {journal} {\bibinfo  {journal} {Phys.Rev.Lett.}\ }\textbf {\bibinfo {volume}
  {110}},\ \bibinfo {pages} {041301} (\bibinfo {year} {2013})},\ \Eprint
  {http://arxiv.org/abs/1301.1173} {arXiv:1301.1173 [astro-ph.HE]} \BibitemShut
  {NoStop}%
%%CITATION = ARXIV:1301.1173;%%
\bibitem [{\citenamefont {Abramowski}\ \emph {et~al.}(2015)\citenamefont
  {Abramowski} \emph {et~al.}}]{HESS:2015cda}%
  \BibitemOpen
  \bibfield  {author} {\bibinfo {author} {\bibnamefont {Abramowski},
  \bibfnamefont {A.}} \emph {et~al.} (\bibinfo {collaboration} {H.E.S.S.
  Collaboration}),\ }\href {\doibase 10.1103/PhysRevLett.114.081301} {\bibfield
   {journal} {\bibinfo  {journal} {Phys.Rev.Lett.}\ }\textbf {\bibinfo {volume}
  {114}},\ \bibinfo {pages} {081301} (\bibinfo {year} {2015})},\ \Eprint
  {http://arxiv.org/abs/1502.03244} {arXiv:1502.03244 [astro-ph.HE]}
  \BibitemShut {NoStop}%
%%CITATION = ARXIV:1502.03244;%%
\bibitem [{\citenamefont {Acciari}\ \emph {et~al.}(2010)\citenamefont
  {Acciari}, \citenamefont {Arlen}, \citenamefont {Aune}, \citenamefont
  {Beilicke}, \citenamefont {Benbow}, \citenamefont {Boltuch}, \citenamefont
  {Bradbury}, \citenamefont {Buckley}, \citenamefont {Bugaev}, \citenamefont
  {Byrum}, \citenamefont {Cannon}, \citenamefont {Cesarini}, \citenamefont
  {Christiansen}, \citenamefont {Ciupik}, \citenamefont {Cui}, \citenamefont
  {Dickherber}, \citenamefont {Duke}, \citenamefont {Finley}, \citenamefont
  {Finnegan}, \citenamefont {Furniss}, \citenamefont {Galante}, \citenamefont
  {Godambe}, \citenamefont {Grube}, \citenamefont {Guenette}, \citenamefont
  {Gyuk}, \citenamefont {Hanna}, \citenamefont {Holder}, \citenamefont {Hui},
  \citenamefont {Humensky}, \citenamefont {Imran}, \citenamefont {Kaaret},
  \citenamefont {Karlsson}, \citenamefont {Kertzman}, \citenamefont {Kieda},
  \citenamefont {Konopelko}, \citenamefont {Krawczynski}, \citenamefont
  {Krennrich}, \citenamefont {Maier}, \citenamefont {McArthur}, \citenamefont
  {McCann}, \citenamefont {McCutcheon}, \citenamefont {Moriarty}, \citenamefont
  {Ong}, \citenamefont {Otte}, \citenamefont {Pandel}, \citenamefont {Perkins},
  \citenamefont {Pohl}, \citenamefont {Quinn}, \citenamefont {Ragan},
  \citenamefont {Reyes}, \citenamefont {Reynolds}, \citenamefont {Roache},
  \citenamefont {Rose}, \citenamefont {Schroedter}, \citenamefont {Sembroski},
  \citenamefont {Senturk}, \citenamefont {Smith}, \citenamefont {Steele},
  \citenamefont {Swordy}, \citenamefont {Tešić}, \citenamefont {Theiling},
  \citenamefont {Thibadeau}, \citenamefont {Varlotta}, \citenamefont
  {Vassiliev}, \citenamefont {Vincent}, \citenamefont {Wagner}, \citenamefont
  {Wakely}, \citenamefont {Ward}, \citenamefont {Weekes}, \citenamefont
  {Weinstein}, \citenamefont {Weisgarber}, \citenamefont {Williams},
  \citenamefont {Wissel}, \citenamefont {Wood},\ and\ \citenamefont
  {Zitzer}}]{1006.5955v2}%
  \BibitemOpen
  \bibfield  {author} {\bibinfo {author} {\bibnamefont {Acciari}, \bibfnamefont
  {V.~A.}}, \bibinfo {author} {\bibnamefont {Arlen}, \bibfnamefont {T.}},
  \bibinfo {author} {\bibnamefont {Aune}, \bibfnamefont {T.}},  \emph
  {et~al.},\ }\href {http://arxiv.org/abs/1006.5955v2} {\bibfield  {journal}
  {\bibinfo  {journal} {Astrophysical Journal 720 (2010) 1174-1180}\ }
  (\bibinfo {year} {2010})},\ \Eprint {http://arxiv.org/abs/1006.5955v2}
  {arXiv:1006.5955v2 [astro-ph.CO]} \BibitemShut {NoStop}%
\bibitem [{\citenamefont {Acharya}\ \emph {et~al.}(2013)\citenamefont
  {Acharya}, \citenamefont {Actis}, \citenamefont {Aghajani}, \citenamefont
  {Agnetta}, \citenamefont {Aguilar} \emph {et~al.}}]{Acharya:2013sxa}%
  \BibitemOpen
  \bibfield  {author} {\bibinfo {author} {\bibnamefont {Acharya}, \bibfnamefont
  {B.}}, \bibinfo {author} {\bibnamefont {Actis}, \bibfnamefont {M.}}, \bibinfo
  {author} {\bibnamefont {Aghajani}, \bibfnamefont {T.}},  \emph {et~al.},\
  }\href {\doibase 10.1016/j.astropartphys.2013.01.007} {\bibfield  {journal}
  {\bibinfo  {journal} {Astropart.Phys.}\ }\textbf {\bibinfo {volume} {43}},\
  \bibinfo {pages} {3} (\bibinfo {year} {2013})}\BibitemShut {NoStop}%
%%CITATION = APHYE,43,3;%%
\bibitem [{\citenamefont {{Ackermann}}\ \emph
  {et~al.}(2014{\natexlab{a}})\citenamefont {{Ackermann}}, \citenamefont
  {{Ajello}}, \citenamefont {{Albert}}, \citenamefont {{Allafort}},
  \citenamefont {{Atwood}}, \citenamefont {{Baldini}}, \citenamefont
  {{Ballet}}, \citenamefont {{Barbiellini}}, \citenamefont {{Bastieri}},
  \citenamefont {{Bechtol}}, \citenamefont {{Bellazzini}}, \citenamefont
  {{Bloom}}, \citenamefont {{Bonamente}}, \citenamefont {{Bottacini}},
  \citenamefont {{Brandt}}, \citenamefont {{Bregeon}}, \citenamefont
  {{Brigida}}, \citenamefont {{Bruel}}, \citenamefont {{Buehler}},
  \citenamefont {{Buson}}, \citenamefont {{Caliandro}}, \citenamefont
  {{Cameron}}, \citenamefont {{Caraveo}}, \citenamefont {{Cavazzuti}},
  \citenamefont {{Chaves}}, \citenamefont {{Chiang}}, \citenamefont {{Chiaro}},
  \citenamefont {{Ciprini}}, \citenamefont {{Claus}}, \citenamefont
  {{Cohen-Tanugi}}, \citenamefont {{Conrad}}, \citenamefont {{D'Ammando}},
  \citenamefont {{de Angelis}}, \citenamefont {{de Palma}}, \citenamefont
  {{Dermer}}, \citenamefont {{Digel}}, \citenamefont {{Drell}}, \citenamefont
  {{Drlica-Wagner}}, \citenamefont {{Favuzzi}}, \citenamefont {{Franckowiak}},
  \citenamefont {{Funk}}, \citenamefont {{Fusco}}, \citenamefont {{Gargano}},
  \citenamefont {{Gasparrini}}, \citenamefont {{Germani}}, \citenamefont
  {{Giglietto}}, \citenamefont {{Giordano}}, \citenamefont {{Giroletti}},
  \citenamefont {{Godfrey}}, \citenamefont {{Gomez-Vargas}}, \citenamefont
  {{Grenier}}, \citenamefont {{Guiriec}}, \citenamefont {{Gustafsson}},
  \citenamefont {{Hadasch}}, \citenamefont {{Hayashida}}, \citenamefont
  {{Hewitt}}, \citenamefont {{Hughes}}, \citenamefont {{Jeltema}},
  \citenamefont {{J{\'o}hannesson}}, \citenamefont {{Johnson}}, \citenamefont
  {{Kamae}}, \citenamefont {{Kataoka}}, \citenamefont {{Kn{\"o}dlseder}},
  \citenamefont {{Kuss}}, \citenamefont {{Lande}}, \citenamefont {{Larsson}},
  \citenamefont {{Latronico}}, \citenamefont {{Llena Garde}}, \citenamefont
  {{Longo}}, \citenamefont {{Loparco}}, \citenamefont {{Lovellette}},
  \citenamefont {{Lubrano}}, \citenamefont {{Mayer}}, \citenamefont
  {{Mazziotta}}, \citenamefont {{McEnery}}, \citenamefont {{Michelson}},
  \citenamefont {{Mitthumsiri}}, \citenamefont {{Mizuno}}, \citenamefont
  {{Monzani}}, \citenamefont {{Morselli}}, \citenamefont {{Moskalenko}},
  \citenamefont {{Murgia}}, \citenamefont {{Nemmen}}, \citenamefont {{Nuss}},
  \citenamefont {{Ohsugi}}, \citenamefont {{Orienti}}, \citenamefont
  {{Orlando}}, \citenamefont {{Ormes}}, \citenamefont {{Perkins}},
  \citenamefont {{Pesce-Rollins}}, \citenamefont {{Piron}}, \citenamefont
  {{Pivato}}, \citenamefont {{Rain{\`o}}}, \citenamefont {{Rando}},
  \citenamefont {{Razzano}}, \citenamefont {{Razzaque}}, \citenamefont
  {{Reimer}}, \citenamefont {{Reimer}}, \citenamefont {{Ruan}}, \citenamefont
  {{S{\'a}nchez-Conde}}, \citenamefont {{Schulz}}, \citenamefont {{Sgr{\`o}}},
  \citenamefont {{Siskind}}, \citenamefont {{Spandre}}, \citenamefont
  {{Spinelli}}, \citenamefont {{Storm}}, \citenamefont {{Strong}},
  \citenamefont {{Suson}}, \citenamefont {{Takahashi}}, \citenamefont
  {{Thayer}}, \citenamefont {{Thayer}}, \citenamefont {{Thompson}},
  \citenamefont {{Tibaldo}}, \citenamefont {{Tinivella}}, \citenamefont
  {{Torres}}, \citenamefont {{Troja}}, \citenamefont {{Uchiyama}},
  \citenamefont {{Usher}}, \citenamefont {{Vandenbroucke}}, \citenamefont
  {{Vianello}}, \citenamefont {{Vitale}}, \citenamefont {{Winer}},
  \citenamefont {{Wood}}, \citenamefont {{Zimmer}}, \citenamefont {{Fermi-LAT
  Collaboration}}, \citenamefont {{Pinzke}},\ and\ \citenamefont
  {{Pfrommer}}}]{2013arXiv1308.5654A}%
  \BibitemOpen
  \bibfield  {author} {\bibinfo {author} {\bibnamefont {{Ackermann}},
  \bibfnamefont {M.}}, \bibinfo {author} {\bibnamefont {{Ajello}},
  \bibfnamefont {M.}}, \bibinfo {author} {\bibnamefont {{Albert}},
  \bibfnamefont {A.}},  \emph {et~al.} (\bibinfo {collaboration} {Fermi LAT
  Collaboration}),\ }\href {\doibase 10.1088/0004-637X/787/1/18} {\bibfield
  {journal} {\bibinfo  {journal} {\apj}\ }\textbf {\bibinfo {volume} {787}},\
  \bibinfo {eid} {18} (\bibinfo {year} {2014}{\natexlab{a}})},\ \Eprint
  {http://arxiv.org/abs/1308.5654} {arXiv:1308.5654 [astro-ph.HE]} \BibitemShut
  {NoStop}%
\bibitem [{\citenamefont {{Ackermann}}\ \emph
  {et~al.}(2015{\natexlab{a}})\citenamefont {{Ackermann}}, \citenamefont
  {{Ajello}}, \citenamefont {{Albert}}, \citenamefont {{Atwood}}, \citenamefont
  {{Baldini}}, \citenamefont {{Ballet}}, \citenamefont {{Barbiellini}},
  \citenamefont {{Bastieri}}, \citenamefont {{Bechtol}}, \citenamefont
  {{Bellazzini}}, \citenamefont {{Bissaldi}}, \citenamefont {{Blandford}},
  \citenamefont {{Bloom}}, \citenamefont {{Bottacini}}, \citenamefont
  {{Brandt}}, \citenamefont {{Bregeon}}, \citenamefont {{Bruel}}, \citenamefont
  {{Buehler}}, \citenamefont {{Buson}}, \citenamefont {{Caliandro}},
  \citenamefont {{Cameron}}, \citenamefont {{Caragiulo}}, \citenamefont
  {{Caraveo}}, \citenamefont {{Cavazzuti}}, \citenamefont {{Cecchi}},
  \citenamefont {{Charles}}, \citenamefont {{Chekhtman}}, \citenamefont
  {{Chiang}}, \citenamefont {{Chiaro}}, \citenamefont {{Ciprini}},
  \citenamefont {{Claus}}, \citenamefont {{Cohen-Tanugi}}, \citenamefont
  {{Conrad}}, \citenamefont {{Cuoco}}, \citenamefont {{Cutini}}, \citenamefont
  {{D'Ammando}}, \citenamefont {{de Angelis}}, \citenamefont {{de Palma}},
  \citenamefont {{Dermer}}, \citenamefont {{Digel}}, \citenamefont {{Silva}},
  \citenamefont {{Drell}}, \citenamefont {{Favuzzi}}, \citenamefont
  {{Ferrara}}, \citenamefont {{Focke}}, \citenamefont {{Franckowiak}},
  \citenamefont {{Fukazawa}}, \citenamefont {{Funk}}, \citenamefont {{Fusco}},
  \citenamefont {{Gargano}}, \citenamefont {{Gasparrini}}, \citenamefont
  {{Germani}}, \citenamefont {{Giglietto}}, \citenamefont {{Giommi}},
  \citenamefont {{Giordano}}, \citenamefont {{Giroletti}}, \citenamefont
  {{Godfrey}}, \citenamefont {{Gomez-Vargas}}, \citenamefont {{Grenier}},
  \citenamefont {{Guiriec}}, \citenamefont {{Gustafsson}}, \citenamefont
  {{Hadasch}}, \citenamefont {{Hayashi}}, \citenamefont {{Hays}}, \citenamefont
  {{Hewitt}}, \citenamefont {{Ippoliti}}, \citenamefont {{Jogler}},
  \citenamefont {{J{\'o}hannesson}}, \citenamefont {{Johnson}}, \citenamefont
  {{Johnson}}, \citenamefont {{Kamae}}, \citenamefont {{Kataoka}},
  \citenamefont {{Kn{\"o}dlseder}}, \citenamefont {{Kuss}}, \citenamefont
  {{Larsson}}, \citenamefont {{Latronico}}, \citenamefont {{Li}}, \citenamefont
  {{Li}}, \citenamefont {{Longo}}, \citenamefont {{Loparco}}, \citenamefont
  {{Lott}}, \citenamefont {{Lovellette}}, \citenamefont {{Lubrano}},
  \citenamefont {{Madejski}}, \citenamefont {{Manfreda}}, \citenamefont
  {{Massaro}}, \citenamefont {{Mayer}}, \citenamefont {{Mazziotta}},
  \citenamefont {{McEnery}}, \citenamefont {{Michelson}}, \citenamefont
  {{Mitthumsiri}}, \citenamefont {{Mizuno}}, \citenamefont {{Moiseev}},
  \citenamefont {{Monzani}}, \citenamefont {{Morselli}}, \citenamefont
  {{Moskalenko}}, \citenamefont {{Murgia}}, \citenamefont {{Nemmen}},
  \citenamefont {{Nuss}}, \citenamefont {{Ohsugi}}, \citenamefont {{Omodei}},
  \citenamefont {{Orlando}}, \citenamefont {{Ormes}}, \citenamefont
  {{Paneque}}, \citenamefont {{Panetta}}, \citenamefont {{Perkins}},
  \citenamefont {{Pesce-Rollins}}, \citenamefont {{Piron}}, \citenamefont
  {{Pivato}}, \citenamefont {{Porter}}, \citenamefont {{Rain{\`o}}},
  \citenamefont {{Rando}}, \citenamefont {{Razzano}}, \citenamefont
  {{Razzaque}}, \citenamefont {{Reimer}}, \citenamefont {{Reimer}},
  \citenamefont {{Reposeur}}, \citenamefont {{Ritz}}, \citenamefont {{Romani}},
  \citenamefont {{S{\'a}nchez-Conde}}, \citenamefont {{Schaal}}, \citenamefont
  {{Schulz}}, \citenamefont {{Sgr{\`o}}}, \citenamefont {{Siskind}},
  \citenamefont {{Spandre}}, \citenamefont {{Spinelli}}, \citenamefont
  {{Strong}}, \citenamefont {{Suson}}, \citenamefont {{Takahashi}},
  \citenamefont {{Thayer}}, \citenamefont {{Thayer}}, \citenamefont
  {{Tibaldo}}, \citenamefont {{Tinivella}}, \citenamefont {{Torres}},
  \citenamefont {{Tosti}}, \citenamefont {{Troja}}, \citenamefont {{Uchiyama}},
  \citenamefont {{Vianello}}, \citenamefont {{Werner}}, \citenamefont
  {{Winer}}, \citenamefont {{Wood}}, \citenamefont {{Wood}}, \citenamefont
  {{Zaharijas}},\ and\ \citenamefont {{Zimmer}}}]{EGB2014}%
  \BibitemOpen
  \bibfield  {author} {\bibinfo {author} {\bibnamefont {{Ackermann}},
  \bibfnamefont {M.}}, \bibinfo {author} {\bibnamefont {{Ajello}},
  \bibfnamefont {M.}}, \bibinfo {author} {\bibnamefont {{Albert}},
  \bibfnamefont {A.}},  \emph {et~al.} (\bibinfo {collaboration} {Fermi LAT
  Collaboration}),\ }\href {\doibase 10.1088/0004-637X/799/1/86} {\bibfield
  {journal} {\bibinfo  {journal} {\apj}\ }\textbf {\bibinfo {volume} {799}},\
  \bibinfo {eid} {86} (\bibinfo {year} {2015}{\natexlab{a}})}\BibitemShut
  {NoStop}%
\bibitem [{\citenamefont {{Ackermann}}\ \emph
  {et~al.}(2015{\natexlab{b}})\citenamefont {{Ackermann}}, \citenamefont
  {{Ajello}}, \citenamefont {{Albert}}, \citenamefont {{Atwood}}, \citenamefont
  {{Baldini}}, \citenamefont {{Ballet}}, \citenamefont {{Barbiellini}},
  \citenamefont {{Bastieri}}, \citenamefont {{Bechtol}}, \citenamefont
  {{Bellazzini}}, \citenamefont {{Bissaldi}}, \citenamefont {{Blandford}},
  \citenamefont {{Bloom}}, \citenamefont {{Bottacini}}, \citenamefont
  {{Brandt}}, \citenamefont {{Bregeon}}, \citenamefont {{Bruel}}, \citenamefont
  {{Buehler}}, \citenamefont {{Buson}}, \citenamefont {{Caliandro}},
  \citenamefont {{Cameron}}, \citenamefont {{Caragiulo}}, \citenamefont
  {{Caraveo}}, \citenamefont {{Cavazzuti}}, \citenamefont {{Cecchi}},
  \citenamefont {{Charles}}, \citenamefont {{Chekhtman}}, \citenamefont
  {{Chiang}}, \citenamefont {{Chiaro}}, \citenamefont {{Ciprini}},
  \citenamefont {{Claus}}, \citenamefont {{Cohen-Tanugi}}, \citenamefont
  {{Conrad}}, \citenamefont {{Cuoco}}, \citenamefont {{Cutini}}, \citenamefont
  {{D'Ammando}}, \citenamefont {{de Angelis}}, \citenamefont {{de Palma}},
  \citenamefont {{Dermer}}, \citenamefont {{Digel}}, \citenamefont {{Silva}},
  \citenamefont {{Drell}}, \citenamefont {{Favuzzi}}, \citenamefont
  {{Ferrara}}, \citenamefont {{Focke}}, \citenamefont {{Franckowiak}},
  \citenamefont {{Fukazawa}}, \citenamefont {{Funk}}, \citenamefont {{Fusco}},
  \citenamefont {{Gargano}}, \citenamefont {{Gasparrini}}, \citenamefont
  {{Germani}}, \citenamefont {{Giglietto}}, \citenamefont {{Giommi}},
  \citenamefont {{Giordano}}, \citenamefont {{Giroletti}}, \citenamefont
  {{Godfrey}}, \citenamefont {{Gomez-Vargas}}, \citenamefont {{Grenier}},
  \citenamefont {{Guiriec}}, \citenamefont {{Gustafsson}}, \citenamefont
  {{Hadasch}}, \citenamefont {{Hayashi}}, \citenamefont {{Hays}}, \citenamefont
  {{Hewitt}}, \citenamefont {{Ippoliti}}, \citenamefont {{Jogler}},
  \citenamefont {{J{\'o}hannesson}}, \citenamefont {{Johnson}}, \citenamefont
  {{Johnson}}, \citenamefont {{Kamae}}, \citenamefont {{Kataoka}},
  \citenamefont {{Kn{\"o}dlseder}}, \citenamefont {{Kuss}}, \citenamefont
  {{Larsson}}, \citenamefont {{Latronico}}, \citenamefont {{Li}}, \citenamefont
  {{Li}}, \citenamefont {{Longo}}, \citenamefont {{Loparco}}, \citenamefont
  {{Lott}}, \citenamefont {{Lovellette}}, \citenamefont {{Lubrano}},
  \citenamefont {{Madejski}}, \citenamefont {{Manfreda}}, \citenamefont
  {{Massaro}}, \citenamefont {{Mayer}}, \citenamefont {{Mazziotta}},
  \citenamefont {{McEnery}}, \citenamefont {{Michelson}}, \citenamefont
  {{Mitthumsiri}}, \citenamefont {{Mizuno}}, \citenamefont {{Moiseev}},
  \citenamefont {{Monzani}}, \citenamefont {{Morselli}}, \citenamefont
  {{Moskalenko}}, \citenamefont {{Murgia}}, \citenamefont {{Nemmen}},
  \citenamefont {{Nuss}}, \citenamefont {{Ohsugi}}, \citenamefont {{Omodei}},
  \citenamefont {{Orlando}}, \citenamefont {{Ormes}}, \citenamefont
  {{Paneque}}, \citenamefont {{Panetta}}, \citenamefont {{Perkins}},
  \citenamefont {{Pesce-Rollins}}, \citenamefont {{Piron}}, \citenamefont
  {{Pivato}}, \citenamefont {{Porter}}, \citenamefont {{Rain{\`o}}},
  \citenamefont {{Rando}}, \citenamefont {{Razzano}}, \citenamefont
  {{Razzaque}}, \citenamefont {{Reimer}}, \citenamefont {{Reimer}},
  \citenamefont {{Reposeur}}, \citenamefont {{Ritz}}, \citenamefont {{Romani}},
  \citenamefont {{S{\'a}nchez-Conde}}, \citenamefont {{Schaal}}, \citenamefont
  {{Schulz}}, \citenamefont {{Sgr{\`o}}}, \citenamefont {{Siskind}},
  \citenamefont {{Spandre}}, \citenamefont {{Spinelli}}, \citenamefont
  {{Strong}}, \citenamefont {{Suson}}, \citenamefont {{Takahashi}},
  \citenamefont {{Thayer}}, \citenamefont {{Thayer}}, \citenamefont
  {{Tibaldo}}, \citenamefont {{Tinivella}}, \citenamefont {{Torres}},
  \citenamefont {{Tosti}}, \citenamefont {{Troja}}, \citenamefont {{Uchiyama}},
  \citenamefont {{Vianello}}, \citenamefont {{Werner}}, \citenamefont
  {{Winer}}, \citenamefont {{Wood}}, \citenamefont {{Wood}}, \citenamefont
  {{Zaharijas}},\ and\ \citenamefont {{Zimmer}}}]{2014arXiv1410.3696T}%
  \BibitemOpen
  \bibfield  {author} {\bibinfo {author} {\bibnamefont {{Ackermann}},
  \bibfnamefont {M.}}, \bibinfo {author} {\bibnamefont {{Ajello}},
  \bibfnamefont {M.}}, \bibinfo {author} {\bibnamefont {{Albert}},
  \bibfnamefont {A.}},  \emph {et~al.} (\bibinfo {collaboration} {Fermi LAT
  Collaboration}),\ }\href {http://stacks.iop.org/0004-637X/799/i=1/a=86}
  {\bibfield  {journal} {\bibinfo  {journal} {\apj}\ }\textbf {\bibinfo
  {volume} {799}},\ \bibinfo {pages} {86} (\bibinfo {year}
  {2015}{\natexlab{b}})},\ \Eprint {http://arxiv.org/abs/1410.3696}
  {arXiv:1410.3696 [astro-ph.HE]} \BibitemShut {NoStop}%
\bibitem [{\citenamefont {{Ackermann}}\ \emph
  {et~al.}(2015{\natexlab{c}})\citenamefont {{Ackermann}}, \citenamefont
  {{Ajello}}, \citenamefont {{Albert}}, \citenamefont {{Baldini}},
  \citenamefont {{Barbiellini}}, \citenamefont {{Bastieri}}, \citenamefont
  {{Bechtol}}, \citenamefont {{Bellazzini}}, \citenamefont {{Bissaldi}},
  \citenamefont {{Bloom}}, \citenamefont {{Bonino}}, \citenamefont {{Bregeon}},
  \citenamefont {{Bruel}}, \citenamefont {{Buehler}}, \citenamefont {{Buson}},
  \citenamefont {{Caliandro}}, \citenamefont {{Cameron}}, \citenamefont
  {{Caragiulo}}, \citenamefont {{Caraveo}}, \citenamefont {{Cecchi}},
  \citenamefont {{Charles}}, \citenamefont {{Chekhtman}}, \citenamefont
  {{Chiang}}, \citenamefont {{Chiaro}}, \citenamefont {{Ciprini}},
  \citenamefont {{Claus}}, \citenamefont {{Cohen-Tanugi}}, \citenamefont
  {{Conrad}}, \citenamefont {{Cuoco}}, \citenamefont {{Cutini}}, \citenamefont
  {{D'Ammando}}, \citenamefont {{de Angelis}}, \citenamefont {{de Palma}},
  \citenamefont {{Dermer}}, \citenamefont {{Digel}}, \citenamefont {{Drell}},
  \citenamefont {{Drlica-Wagner}}, \citenamefont {{Favuzzi}}, \citenamefont
  {{Ferrara}}, \citenamefont {{Franckowiak}}, \citenamefont {{Fukazawa}},
  \citenamefont {{Funk}}, \citenamefont {{Fusco}}, \citenamefont {{Gargano}},
  \citenamefont {{Gasparrini}}, \citenamefont {{Giglietto}}, \citenamefont
  {{Giordano}}, \citenamefont {{Giroletti}}, \citenamefont {{Godfrey}},
  \citenamefont {{Guiriec}}, \citenamefont {{Gustafsson}}, \citenamefont
  {{Hewitt}}, \citenamefont {{Hou}}, \citenamefont {{Kamae}}, \citenamefont
  {{Kuss}}, \citenamefont {{Larsson}}, \citenamefont {{Latronico}},
  \citenamefont {{Longo}}, \citenamefont {{Loparco}}, \citenamefont
  {{Lovellette}}, \citenamefont {{Lubrano}}, \citenamefont {{Malyshev}},
  \citenamefont {{Massaro}}, \citenamefont {{Mayer}}, \citenamefont
  {{Mazziotta}}, \citenamefont {{Michelson}}, \citenamefont {{Mitthumsiri}},
  \citenamefont {{Mizuno}}, \citenamefont {{Monzani}}, \citenamefont
  {{Morselli}}, \citenamefont {{Moskalenko}}, \citenamefont {{Murgia}},
  \citenamefont {{Negro}}, \citenamefont {{Nemmen}}, \citenamefont {{Nuss}},
  \citenamefont {{Ohsugi}}, \citenamefont {{Orienti}}, \citenamefont
  {{Orlando}}, \citenamefont {{Ormes}}, \citenamefont {{Paneque}},
  \citenamefont {{Perkins}}, \citenamefont {{Pesce-Rollins}}, \citenamefont
  {{Piron}}, \citenamefont {{Pivato}}, \citenamefont {{Raino}}, \citenamefont
  {{Rando}}, \citenamefont {{Razzano}}, \citenamefont {{Reimer}}, \citenamefont
  {{Reimer}}, \citenamefont {{Sanchez-Conde}}, \citenamefont {{Schulz}},
  \citenamefont {{Sgro}}, \citenamefont {{Siskind}}, \citenamefont {{Spandre}},
  \citenamefont {{Spinelli}}, \citenamefont {{Strong}}, \citenamefont
  {{Suson}}, \citenamefont {{Tajima}}, \citenamefont {{Takahashi}},
  \citenamefont {{Thayer}}, \citenamefont {{Thayer}}, \citenamefont
  {{Tibaldo}}, \citenamefont {{Tinivella}}, \citenamefont {{Torres}},
  \citenamefont {{Troja}}, \citenamefont {{Uchiyama}}, \citenamefont
  {{Vianello}}, \citenamefont {{Werner}}, \citenamefont {{Winer}},
  \citenamefont {{Wood}}, \citenamefont {{Wood}},\ and\ \citenamefont
  {{Zaharijas}}}]{2015arXiv150105464T}%
  \BibitemOpen
  \bibfield  {author} {\bibinfo {author} {\bibnamefont {{Ackermann}},
  \bibfnamefont {M.}}, \bibinfo {author} {\bibnamefont {{Ajello}},
  \bibfnamefont {M.}}, \bibinfo {author} {\bibnamefont {{Albert}},
  \bibfnamefont {A.}},  \emph {et~al.} (\bibinfo {collaboration} {Fermi LAT
  Collaboration}),\ }\href {http://adsabs.harvard.edu/abs/2015arXiv150105464T}
  {\bibfield  {journal} {\bibinfo  {journal} {ArXiv e-prints}\ } (\bibinfo
  {year} {2015}{\natexlab{c}})},\ \bibinfo {note} {submitted to \jcap},\
  \Eprint {http://arxiv.org/abs/1501.05464} {arXiv:1501.05464} \BibitemShut
  {NoStop}%
\bibitem [{\citenamefont {{Ackermann}}\ \emph
  {et~al.}(2010{\natexlab{a}})\citenamefont {{Ackermann}}, \citenamefont
  {{Ajello}}, \citenamefont {{Allafort}}, \citenamefont {{Baldini}},
  \citenamefont {{Ballet}}, \citenamefont {{Barbiellini}}, \citenamefont
  {{Bastieri}}, \citenamefont {{Bechtol}}, \citenamefont {{Bellazzini}},
  \citenamefont {{Blandford}}, \citenamefont {{Blasi}}, \citenamefont
  {{Bloom}}, \citenamefont {{Bonamente}}, \citenamefont {{Borgland}},
  \citenamefont {{Bouvier}}, \citenamefont {{Brandt}}, \citenamefont
  {{Bregeon}}, \citenamefont {{Brigida}}, \citenamefont {{Bruel}},
  \citenamefont {{Buehler}}, \citenamefont {{Buson}}, \citenamefont
  {{Caliandro}}, \citenamefont {{Cameron}}, \citenamefont {{Caraveo}},
  \citenamefont {{Carrigan}}, \citenamefont {{Casandjian}}, \citenamefont
  {{Cavazzuti}}, \citenamefont {{Cecchi}}, \citenamefont {{{\c C}elik}},
  \citenamefont {{Charles}}, \citenamefont {{Chekhtman}}, \citenamefont
  {{Cheung}}, \citenamefont {{Chiang}}, \citenamefont {{Ciprini}},
  \citenamefont {{Claus}}, \citenamefont {{Cohen-Tanugi}}, \citenamefont
  {{Colafrancesco}}, \citenamefont {{Cominsky}}, \citenamefont {{Conrad}},
  \citenamefont {{Dermer}}, \citenamefont {{de Palma}}, \citenamefont
  {{Silva}}, \citenamefont {{Drell}}, \citenamefont {{Dubois}}, \citenamefont
  {{Dumora}}, \citenamefont {{Edmonds}}, \citenamefont {{Farnier}},
  \citenamefont {{Favuzzi}}, \citenamefont {{Frailis}}, \citenamefont
  {{Fukazawa}}, \citenamefont {{Funk}}, \citenamefont {{Fusco}}, \citenamefont
  {{Gargano}}, \citenamefont {{Gasparrini}}, \citenamefont {{Gehrels}},
  \citenamefont {{Germani}}, \citenamefont {{Giglietto}}, \citenamefont
  {{Giordano}}, \citenamefont {{Giroletti}}, \citenamefont {{Glanzman}},
  \citenamefont {{Godfrey}}, \citenamefont {{Grenier}}, \citenamefont
  {{Grondin}}, \citenamefont {{Guiriec}}, \citenamefont {{Hadasch}},
  \citenamefont {{Harding}}, \citenamefont {{Hayashida}}, \citenamefont
  {{Hays}}, \citenamefont {{Horan}}, \citenamefont {{Hughes}}, \citenamefont
  {{Jeltema}}, \citenamefont {{J{\'o}hannesson}}, \citenamefont {{Johnson}},
  \citenamefont {{Johnson}}, \citenamefont {{Johnson}}, \citenamefont
  {{Kamae}}, \citenamefont {{Katagiri}}, \citenamefont {{Kataoka}},
  \citenamefont {{Kerr}}, \citenamefont {{Kn{\"o}dlseder}}, \citenamefont
  {{Kuss}}, \citenamefont {{Lande}}, \citenamefont {{Latronico}}, \citenamefont
  {{Lee}}, \citenamefont {{Lemoine-Goumard}}, \citenamefont {{Longo}},
  \citenamefont {{Loparco}}, \citenamefont {{Lott}}, \citenamefont
  {{Lovellette}}, \citenamefont {{Lubrano}}, \citenamefont {{Madejski}},
  \citenamefont {{Makeev}}, \citenamefont {{Mazziotta}}, \citenamefont
  {{Michelson}}, \citenamefont {{Mitthumsiri}}, \citenamefont {{Mizuno}},
  \citenamefont {{Moiseev}}, \citenamefont {{Monte}}, \citenamefont
  {{Monzani}}, \citenamefont {{Morselli}}, \citenamefont {{Moskalenko}},
  \citenamefont {{Murgia}}, \citenamefont {{Naumann-Godo}}, \citenamefont
  {{Nolan}}, \citenamefont {{Norris}}, \citenamefont {{Nuss}}, \citenamefont
  {{Ohsugi}}, \citenamefont {{Omodei}}, \citenamefont {{Orlando}},
  \citenamefont {{Ormes}}, \citenamefont {{Ozaki}}, \citenamefont {{Paneque}},
  \citenamefont {{Panetta}}, \citenamefont {{Pepe}}, \citenamefont
  {{Pesce-Rollins}}, \citenamefont {{Petrosian}}, \citenamefont {{Pfrommer}},
  \citenamefont {{Piron}}, \citenamefont {{Porter}}, \citenamefont {{Profumo}},
  \citenamefont {{Rain{\`o}}}, \citenamefont {{Rando}}, \citenamefont
  {{Razzano}}, \citenamefont {{Reimer}}, \citenamefont {{Reimer}},
  \citenamefont {{Reposeur}}, \citenamefont {{Ripken}}, \citenamefont {{Ritz}},
  \citenamefont {{Rodriguez}}, \citenamefont {{Romani}}, \citenamefont
  {{Roth}}, \citenamefont {{Sadrozinski}}, \citenamefont {{Sander}},
  \citenamefont {{Saz Parkinson}}, \citenamefont {{Scargle}}, \citenamefont
  {{Sgr{\`o}}}, \citenamefont {{Siskind}}, \citenamefont {{Smith}},
  \citenamefont {{Spandre}}, \citenamefont {{Spinelli}}, \citenamefont
  {{Starck}}, \citenamefont {{Stawarz}}, \citenamefont {{Strickman}},
  \citenamefont {{Strong}}, \citenamefont {{Suson}}, \citenamefont {{Tajima}},
  \citenamefont {{Takahashi}}, \citenamefont {{Takahashi}}, \citenamefont
  {{Tanaka}}, \citenamefont {{Thayer}}, \citenamefont {{Thayer}}, \citenamefont
  {{Tibaldo}}, \citenamefont {{Tibolla}}, \citenamefont {{Torres}},
  \citenamefont {{Tosti}}, \citenamefont {{Tramacere}}, \citenamefont
  {{Uchiyama}}, \citenamefont {{Usher}}, \citenamefont {{Vandenbroucke}},
  \citenamefont {{Vasileiou}}, \citenamefont {{Vilchez}}, \citenamefont
  {{Vitale}}, \citenamefont {{Waite}}, \citenamefont {{Wang}}, \citenamefont
  {{Winer}}, \citenamefont {{Wood}}, \citenamefont {{Yang}}, \citenamefont
  {{Ylinen}},\ and\ \citenamefont {{Ziegler}}}]{2010ApJ...717L..71A}%
  \BibitemOpen
  \bibfield  {author} {\bibinfo {author} {\bibnamefont {{Ackermann}},
  \bibfnamefont {M.}}, \bibinfo {author} {\bibnamefont {{Ajello}},
  \bibfnamefont {M.}}, \bibinfo {author} {\bibnamefont {{Allafort}},
  \bibfnamefont {A.}},  \emph {et~al.} (\bibinfo {collaboration} {Fermi LAT
  Collaboration}),\ }\href {\doibase 10.1088/2041-8205/717/1/L71} {\bibfield
  {journal} {\bibinfo  {journal} {\apjl}\ }\textbf {\bibinfo {volume} {717}},\
  \bibinfo {pages} {L71} (\bibinfo {year} {2010}{\natexlab{a}})},\ \Eprint
  {http://arxiv.org/abs/1006.0748} {arXiv:1006.0748 [astro-ph.HE]} \BibitemShut
  {NoStop}%
\bibitem [{\citenamefont {{Ackermann}}\ \emph
  {et~al.}(2010{\natexlab{b}})\citenamefont {{Ackermann}}, \citenamefont
  {{Ajello}}, \citenamefont {{Allafort}}, \citenamefont {{Baldini}},
  \citenamefont {{Ballet}}, \citenamefont {{Barbiellini}}, \citenamefont
  {{Bastieri}}, \citenamefont {{Bechtol}}, \citenamefont {{Bellazzini}},
  \citenamefont {{Blandford}}, \citenamefont {{Bloom}}, \citenamefont
  {{Bonamente}}, \citenamefont {{Borgland}}, \citenamefont {{Bouvier}},
  \citenamefont {{Brandt}}, \citenamefont {{Bregeon}}, \citenamefont
  {{Brigida}}, \citenamefont {{Bruel}}, \citenamefont {{Buehler}},
  \citenamefont {{Buson}}, \citenamefont {{Caliandro}}, \citenamefont
  {{Cameron}}, \citenamefont {{Caraveo}}, \citenamefont {{Carrigan}},
  \citenamefont {{Casandjian}}, \citenamefont {{Cecchi}}, \citenamefont
  {{Charles}}, \citenamefont {{Chekhtman}}, \citenamefont {{Cheung}},
  \citenamefont {{Chiang}}, \citenamefont {{Ciprini}}, \citenamefont {{Claus}},
  \citenamefont {{Cohen-Tanugi}}, \citenamefont {{Cominsky}}, \citenamefont
  {{Conrad}}, \citenamefont {{de Angelis}}, \citenamefont {{de Palma}},
  \citenamefont {{Silva}}, \citenamefont {{Drell}}, \citenamefont
  {{Drlica-Wagner}}, \citenamefont {{Dubois}}, \citenamefont {{Dumora}},
  \citenamefont {{Edmonds}}, \citenamefont {{Farnier}}, \citenamefont
  {{Favuzzi}}, \citenamefont {{Fegan}}, \citenamefont {{Frailis}},
  \citenamefont {{Fukazawa}}, \citenamefont {{Fusco}}, \citenamefont
  {{Gargano}}, \citenamefont {{Gasparrini}}, \citenamefont {{Gehrels}},
  \citenamefont {{Germani}}, \citenamefont {{Giglietto}}, \citenamefont
  {{Giordano}}, \citenamefont {{Glanzman}}, \citenamefont {{Godfrey}},
  \citenamefont {{Grenier}}, \citenamefont {{Guiriec}}, \citenamefont
  {{Gustafsson}}, \citenamefont {{Harding}}, \citenamefont {{Hayashida}},
  \citenamefont {{Horan}}, \citenamefont {{Hughes}}, \citenamefont {{Jeltema}},
  \citenamefont {{J{\'o}hannesson}}, \citenamefont {{Johnson}}, \citenamefont
  {{Johnson}}, \citenamefont {{Kamae}}, \citenamefont {{Katagiri}},
  \citenamefont {{Kataoka}}, \citenamefont {{Kn{\"o}dlseder}}, \citenamefont
  {{Kuss}}, \citenamefont {{Lande}}, \citenamefont {{Latronico}}, \citenamefont
  {{Lee}}, \citenamefont {{Llena Garde}}, \citenamefont {{Longo}},
  \citenamefont {{Loparco}}, \citenamefont {{Lovellette}}, \citenamefont
  {{Lubrano}}, \citenamefont {{Madejski}}, \citenamefont {{Makeev}},
  \citenamefont {{Mazziotta}}, \citenamefont {{Michelson}}, \citenamefont
  {{Mitthumsiri}}, \citenamefont {{Mizuno}}, \citenamefont {{Moiseev}},
  \citenamefont {{Monte}}, \citenamefont {{Monzani}}, \citenamefont
  {{Morselli}}, \citenamefont {{Moskalenko}}, \citenamefont {{Murgia}},
  \citenamefont {{Nolan}}, \citenamefont {{Norris}}, \citenamefont {{Nuss}},
  \citenamefont {{Ohno}}, \citenamefont {{Ohsugi}}, \citenamefont {{Omodei}},
  \citenamefont {{Orlando}}, \citenamefont {{Ormes}}, \citenamefont
  {{Panetta}}, \citenamefont {{Pepe}}, \citenamefont {{Pesce-Rollins}},
  \citenamefont {{Piron}}, \citenamefont {{Porter}}, \citenamefont {{Profumo}},
  \citenamefont {{Rain{\`o}}}, \citenamefont {{Razzano}}, \citenamefont
  {{Reposeur}}, \citenamefont {{Ritz}}, \citenamefont {{Rodriguez}},
  \citenamefont {{Roth}}, \citenamefont {{Sadrozinski}}, \citenamefont
  {{Sander}}, \citenamefont {{Scargle}}, \citenamefont {{Sgr{\`o}}},
  \citenamefont {{Siskind}}, \citenamefont {{Smith}}, \citenamefont
  {{Spandre}}, \citenamefont {{Spinelli}}, \citenamefont {{Starck}},
  \citenamefont {{Strickman}}, \citenamefont {{Suson}}, \citenamefont
  {{Takahashi}}, \citenamefont {{Tanaka}}, \citenamefont {{Thayer}},
  \citenamefont {{Thayer}}, \citenamefont {{Tibaldo}}, \citenamefont
  {{Torres}}, \citenamefont {{Tosti}}, \citenamefont {{Usher}}, \citenamefont
  {{Vasileiou}}, \citenamefont {{Vitale}}, \citenamefont {{Waite}},
  \citenamefont {{Wang}}, \citenamefont {{Winer}}, \citenamefont {{Wood}},
  \citenamefont {{Yang}}, \citenamefont {{Ylinen}}, \citenamefont {{Ziegler}},\
  and\ \citenamefont {{Fermi LAT Collaboration}}}]{2010JCAP...05..025A}%
  \BibitemOpen
  \bibfield  {author} {\bibinfo {author} {\bibnamefont {{Ackermann}},
  \bibfnamefont {M.}}, \bibinfo {author} {\bibnamefont {{Ajello}},
  \bibfnamefont {M.}}, \bibinfo {author} {\bibnamefont {{Allafort}},
  \bibfnamefont {A.}},  \emph {et~al.} (\bibinfo {collaboration} {Fermi LAT
  Collaboration}),\ }\href {\doibase 10.1088/1475-7516/2010/05/025} {\bibfield
  {journal} {\bibinfo  {journal} {\jcap}\ }\textbf {\bibinfo {volume} {5}},\
  \bibinfo {eid} {025} (\bibinfo {year} {2010}{\natexlab{b}})},\ \Eprint
  {http://arxiv.org/abs/1002.2239} {arXiv:1002.2239 [astro-ph.CO]} \BibitemShut
  {NoStop}%
\bibitem [{\citenamefont {Ackermann}\ \emph {et~al.}(2012)\citenamefont
  {Ackermann}, \citenamefont {Ajello}, \citenamefont {Atwood}, \citenamefont
  {Baldini}, \citenamefont {Barbiellini}, \citenamefont {Bastieri},
  \citenamefont {Bechtol}, \citenamefont {Bellazzini}, \citenamefont
  {Blandford}, \citenamefont {Bloom}, \citenamefont {Bonamente}, \citenamefont
  {Borgland}, \citenamefont {Bottacini}, \citenamefont {Brandt}, \citenamefont
  {Bregeon}, \citenamefont {Brigida}, \citenamefont {Bruel}, \citenamefont
  {Buehler}, \citenamefont {Buson}, \citenamefont {Caliandro}, \citenamefont
  {Cameron}, \citenamefont {Caraveo}, \citenamefont {Casandjian}, \citenamefont
  {Cecchi}, \citenamefont {Charles}, \citenamefont {Chekhtman}, \citenamefont
  {Chiang}, \citenamefont {Ciprini}, \citenamefont {Claus}, \citenamefont
  {Cohen-Tanugi}, \citenamefont {Conrad}, \citenamefont {Cuoco}, \citenamefont
  {Cutini}, \citenamefont {D?Ammando}, \citenamefont {de~Angelis},
  \citenamefont {de~Palma}, \citenamefont {Dermer}, \citenamefont {do~Couto~e
  Silva}, \citenamefont {Drell}, \citenamefont {Drlica-Wagner}, \citenamefont
  {Falletti}, \citenamefont {Favuzzi}, \citenamefont {Fegan}, \citenamefont
  {Focke}, \citenamefont {Fukazawa}, \citenamefont {Funk}, \citenamefont
  {Fusco}, \citenamefont {Gargano}, \citenamefont {Gasparrini}, \citenamefont
  {Germani}, \citenamefont {Giglietto}, \citenamefont {Giordano}, \citenamefont
  {Giroletti}, \citenamefont {Glanzman}, \citenamefont {Godfrey}, \citenamefont
  {Grenier}, \citenamefont {Guiriec}, \citenamefont {Gustafsson}, \citenamefont
  {Hadasch}, \citenamefont {Hayashida}, \citenamefont {Horan}, \citenamefont
  {Hughes}, \citenamefont {Jackson}, \citenamefont {Jogler}, \citenamefont
  {Jóhannesson}, \citenamefont {Johnson}, \citenamefont {Kamae}, \citenamefont
  {Knödlseder}, \citenamefont {Kuss}, \citenamefont {Lande}, \citenamefont
  {Latronico}, \citenamefont {Lionetto}, \citenamefont {Garde}, \citenamefont
  {Longo}, \citenamefont {Loparco}, \citenamefont {Lott}, \citenamefont
  {Lovellette}, \citenamefont {Lubrano}, \citenamefont {Mazziotta},
  \citenamefont {McEnery}, \citenamefont {Mehault}, \citenamefont {Michelson},
  \citenamefont {Mitthumsiri}, \citenamefont {Mizuno}, \citenamefont {Moiseev},
  \citenamefont {Monte}, \citenamefont {Monzani}, \citenamefont {Morselli},
  \citenamefont {Moskalenko}, \citenamefont {Murgia}, \citenamefont
  {Naumann-Godo}, \citenamefont {Norris}, \citenamefont {Nuss}, \citenamefont
  {Ohsugi}, \citenamefont {Orienti}, \citenamefont {Orlando}, \citenamefont
  {Ormes}, \citenamefont {Paneque}, \citenamefont {Panetta}, \citenamefont
  {Pesce-Rollins}, \citenamefont {Pierbattista}, \citenamefont {Piron},
  \citenamefont {Pivato}, \citenamefont {Poon}, \citenamefont {Rainò},
  \citenamefont {Rando}, \citenamefont {Razzano}, \citenamefont {Razzaque},
  \citenamefont {Reimer}, \citenamefont {Reimer}, \citenamefont {Romoli},
  \citenamefont {Sbarra}, \citenamefont {Scargle}, \citenamefont {Sgrò},
  \citenamefont {Siskind}, \citenamefont {Spandre}, \citenamefont {Spinelli},
  \citenamefont {?ukasz Stawarz}, \citenamefont {Strong}, \citenamefont
  {Suson}, \citenamefont {Tajima}, \citenamefont {Takahashi}, \citenamefont
  {Tanaka}, \citenamefont {Thayer}, \citenamefont {Thayer}, \citenamefont
  {Tibaldo}, \citenamefont {Tinivella}, \citenamefont {Tosti}, \citenamefont
  {Troja}, \citenamefont {Usher}, \citenamefont {Vandenbroucke}, \citenamefont
  {Vasileiou}, \citenamefont {Vianello}, \citenamefont {Vitale}, \citenamefont
  {Waite}, \citenamefont {Wallace}, \citenamefont {Wood}, \citenamefont {Wood},
  \citenamefont {Yang}, \citenamefont {Zaharijas},\ and\ \citenamefont
  {Zimmer}}]{0004-637X-761-2-91}%
  \BibitemOpen
  \bibfield  {author} {\bibinfo {author} {\bibnamefont {Ackermann},
  \bibfnamefont {M.}}, \bibinfo {author} {\bibnamefont {Ajello}, \bibfnamefont
  {M.}}, \bibinfo {author} {\bibnamefont {Atwood}, \bibfnamefont {W.~B.}},
  \emph {et~al.},\ }\href {http://stacks.iop.org/0004-637X/761/i=2/a=91}
  {\bibfield  {journal} {\bibinfo  {journal} {The Astrophysical Journal}\
  }\textbf {\bibinfo {volume} {761}},\ \bibinfo {pages} {91} (\bibinfo {year}
  {2012})}\BibitemShut {NoStop}%
\bibitem [{\citenamefont {{Ackermann}}\ \emph
  {et~al.}(2014{\natexlab{b}})\citenamefont {{Ackermann}}, \citenamefont
  {{Albert}}, \citenamefont {{Anderson}}, \citenamefont {{Baldini}},
  \citenamefont {{Ballet}}, \citenamefont {{Barbiellini}}, \citenamefont
  {{Bastieri}}, \citenamefont {{Bechtol}}, \citenamefont {{Bellazzini}},
  \citenamefont {{Bissaldi}}, \citenamefont {{Bloom}}, \citenamefont
  {{Bonamente}}, \citenamefont {{Bouvier}}, \citenamefont {{Brandt}},
  \citenamefont {{Bregeon}}, \citenamefont {{Brigida}}, \citenamefont
  {{Bruel}}, \citenamefont {{Buehler}}, \citenamefont {{Buson}}, \citenamefont
  {{Caliandro}}, \citenamefont {{Cameron}}, \citenamefont {{Caragiulo}},
  \citenamefont {{Caraveo}}, \citenamefont {{Cecchi}}, \citenamefont
  {{Charles}}, \citenamefont {{Chekhtman}}, \citenamefont {{Chiang}},
  \citenamefont {{Ciprini}}, \citenamefont {{Claus}}, \citenamefont
  {{Cohen-Tanugi}}, \citenamefont {{Conrad}}, \citenamefont {{D'Ammando}},
  \citenamefont {{de Angelis}}, \citenamefont {{Dermer}}, \citenamefont
  {{Digel}}, \citenamefont {{do Couto e Silva}}, \citenamefont {{Drell}},
  \citenamefont {{Drlica-Wagner}}, \citenamefont {{Essig}}, \citenamefont
  {{Favuzzi}}, \citenamefont {{Ferrara}}, \citenamefont {{Franckowiak}},
  \citenamefont {{Fukazawa}}, \citenamefont {{Funk}}, \citenamefont {{Fusco}},
  \citenamefont {{Gargano}}, \citenamefont {{Gasparrini}}, \citenamefont
  {{Giglietto}}, \citenamefont {{Giroletti}}, \citenamefont {{Godfrey}},
  \citenamefont {{Gomez-Vargas}}, \citenamefont {{Grenier}}, \citenamefont
  {{Guiriec}}, \citenamefont {{Gustafsson}}, \citenamefont {{Hayashida}},
  \citenamefont {{Hays}}, \citenamefont {{Hewitt}}, \citenamefont {{Hughes}},
  \citenamefont {{Jogler}}, \citenamefont {{Kamae}}, \citenamefont
  {{Kn{\"o}dlseder}}, \citenamefont {{Kocevski}}, \citenamefont {{Kuss}},
  \citenamefont {{Larsson}}, \citenamefont {{Latronico}}, \citenamefont {{Llena
  Garde}}, \citenamefont {{Longo}}, \citenamefont {{Loparco}}, \citenamefont
  {{Lovellette}}, \citenamefont {{Lubrano}}, \citenamefont {{Martinez}},
  \citenamefont {{Mayer}}, \citenamefont {{Mazziotta}}, \citenamefont
  {{Michelson}}, \citenamefont {{Mitthumsiri}}, \citenamefont {{Mizuno}},
  \citenamefont {{Moiseev}}, \citenamefont {{Monzani}}, \citenamefont
  {{Morselli}}, \citenamefont {{Moskalenko}}, \citenamefont {{Murgia}},
  \citenamefont {{Nemmen}}, \citenamefont {{Nuss}}, \citenamefont {{Ohsugi}},
  \citenamefont {{Orlando}}, \citenamefont {{Ormes}}, \citenamefont
  {{Perkins}}, \citenamefont {{Piron}}, \citenamefont {{Pivato}}, \citenamefont
  {{Porter}}, \citenamefont {{Rain{\`o}}}, \citenamefont {{Rando}},
  \citenamefont {{Razzano}}, \citenamefont {{Razzaque}}, \citenamefont
  {{Reimer}}, \citenamefont {{Reimer}}, \citenamefont {{Ritz}}, \citenamefont
  {{S{\'a}nchez-Conde}}, \citenamefont {{Sehgal}}, \citenamefont {{Sgr{\`o}}},
  \citenamefont {{Siskind}}, \citenamefont {{Spinelli}}, \citenamefont
  {{Strigari}}, \citenamefont {{Suson}}, \citenamefont {{Tajima}},
  \citenamefont {{Takahashi}}, \citenamefont {{Thayer}}, \citenamefont
  {{Tibaldo}}, \citenamefont {{Tinivella}}, \citenamefont {{Torres}},
  \citenamefont {{Uchiyama}}, \citenamefont {{Usher}}, \citenamefont
  {{Vandenbroucke}}, \citenamefont {{Vianello}}, \citenamefont {{Vitale}},
  \citenamefont {{Werner}}, \citenamefont {{Winer}}, \citenamefont {{Wood}},
  \citenamefont {{Wood}}, \citenamefont {{Zaharijas}}, \citenamefont
  {{Zimmer}},\ and\ \citenamefont {{Fermi-LAT
  Collaboration}}}]{2014PhRvD..89d2001A}%
  \BibitemOpen
  \bibfield  {author} {\bibinfo {author} {\bibnamefont {{Ackermann}},
  \bibfnamefont {M.}}, \bibinfo {author} {\bibnamefont {{Albert}},
  \bibfnamefont {A.}}, \bibinfo {author} {\bibnamefont {{Anderson}},
  \bibfnamefont {B.}},  \emph {et~al.} (\bibinfo {collaboration} {Fermi LAT
  Collaboration}),\ }\href {\doibase 10.1103/PhysRevD.89.042001} {\bibfield
  {journal} {\bibinfo  {journal} {\prd}\ }\textbf {\bibinfo {volume} {89}},\
  \bibinfo {eid} {042001} (\bibinfo {year} {2014}{\natexlab{b}})}\BibitemShut
  {NoStop}%
\bibitem [{\citenamefont {{Ackermann}}\ \emph {et~al.}(2012)\citenamefont
  {{Ackermann}}, \citenamefont {{Albert}}, \citenamefont {{Baldini}},
  \citenamefont {{Ballet}}, \citenamefont {{Barbiellini}}, \citenamefont
  {{Bastieri}}, \citenamefont {{Bechtol}}, \citenamefont {{Bellazzini}},
  \citenamefont {{Blandford}}, \citenamefont {{Bloom}}, \citenamefont
  {{Bonamente}}, \citenamefont {{Borgland}}, \citenamefont {{Bottacini}},
  \citenamefont {{Brandt}}, \citenamefont {{Bregeon}}, \citenamefont
  {{Brigida}}, \citenamefont {{Bruel}}, \citenamefont {{Buehler}},
  \citenamefont {{Burnett}}, \citenamefont {{Caliandro}}, \citenamefont
  {{Cameron}}, \citenamefont {{Caraveo}}, \citenamefont {{Casandjian}},
  \citenamefont {{Cecchi}}, \citenamefont {{Charles}}, \citenamefont
  {{Chiang}}, \citenamefont {{Ciprini}}, \citenamefont {{Claus}}, \citenamefont
  {{Cohen-Tanugi}}, \citenamefont {{Conrad}}, \citenamefont {{Cutini}},
  \citenamefont {{de Palma}}, \citenamefont {{Dermer}}, \citenamefont
  {{Digel}}, \citenamefont {{Silva}}, \citenamefont {{Drell}}, \citenamefont
  {{Drlica-Wagner}}, \citenamefont {{Essig}}, \citenamefont {{Falletti}},
  \citenamefont {{Favuzzi}}, \citenamefont {{Fegan}}, \citenamefont {{Focke}},
  \citenamefont {{Fukazawa}}, \citenamefont {{Funk}}, \citenamefont {{Fusco}},
  \citenamefont {{Gargano}}, \citenamefont {{Germani}}, \citenamefont
  {{Giglietto}}, \citenamefont {{Giordano}}, \citenamefont {{Giroletti}},
  \citenamefont {{Glanzman}}, \citenamefont {{Godfrey}}, \citenamefont
  {{Grenier}}, \citenamefont {{Guiriec}}, \citenamefont {{Gustafsson}},
  \citenamefont {{Hadasch}}, \citenamefont {{Hayashida}}, \citenamefont
  {{Hou}}, \citenamefont {{Hughes}}, \citenamefont {{Johnson}}, \citenamefont
  {{Johnson}}, \citenamefont {{Kamae}}, \citenamefont {{Katagiri}},
  \citenamefont {{Kataoka}}, \citenamefont {{Kn{\"o}dlseder}}, \citenamefont
  {{Kuss}}, \citenamefont {{Lande}}, \citenamefont {{Latronico}}, \citenamefont
  {{Lee}}, \citenamefont {{Lionetto}}, \citenamefont {{Llena Garde}},
  \citenamefont {{Longo}}, \citenamefont {{Loparco}}, \citenamefont
  {{Lovellette}}, \citenamefont {{Lubrano}}, \citenamefont {{Mazziotta}},
  \citenamefont {{McEnery}}, \citenamefont {{Michelson}}, \citenamefont
  {{Mitthumsiri}}, \citenamefont {{Mizuno}}, \citenamefont {{Moiseev}},
  \citenamefont {{Monte}}, \citenamefont {{Monzani}}, \citenamefont
  {{Morselli}}, \citenamefont {{Moskalenko}}, \citenamefont {{Murgia}},
  \citenamefont {{Naumann-Godo}}, \citenamefont {{Norris}}, \citenamefont
  {{Nuss}}, \citenamefont {{Ohsugi}}, \citenamefont {{Okumura}}, \citenamefont
  {{Orlando}}, \citenamefont {{Ormes}}, \citenamefont {{Ozaki}}, \citenamefont
  {{Paneque}}, \citenamefont {{Pelassa}}, \citenamefont {{Pierbattista}},
  \citenamefont {{Piron}}, \citenamefont {{Pivato}}, \citenamefont {{Porter}},
  \citenamefont {{Rain{\`o}}}, \citenamefont {{Rando}}, \citenamefont
  {{Razzano}}, \citenamefont {{Reimer}}, \citenamefont {{Reimer}},
  \citenamefont {{Ritz}}, \citenamefont {{Sadrozinski}}, \citenamefont
  {{Sehgal}}, \citenamefont {{Sgr{\`o}}}, \citenamefont {{Siskind}},
  \citenamefont {{Spinelli}}, \citenamefont {{Strigari}}, \citenamefont
  {{Suson}}, \citenamefont {{Tajima}}, \citenamefont {{Takahashi}},
  \citenamefont {{Tanaka}}, \citenamefont {{Thayer}}, \citenamefont {{Thayer}},
  \citenamefont {{Tibaldo}}, \citenamefont {{Tinivella}}, \citenamefont
  {{Torres}}, \citenamefont {{Troja}}, \citenamefont {{Uchiyama}},
  \citenamefont {{Usher}}, \citenamefont {{Vandenbroucke}}, \citenamefont
  {{Vasileiou}}, \citenamefont {{Vianello}}, \citenamefont {{Vitale}},
  \citenamefont {{Waite}}, \citenamefont {{Wang}}, \citenamefont {{Winer}},
  \citenamefont {{Wood}}, \citenamefont {{Yang}}, \citenamefont {{Zalewski}},\
  and\ \citenamefont {{Zimmer}}}]{2012ApJ...747..121A}%
  \BibitemOpen
  \bibfield  {author} {\bibinfo {author} {\bibnamefont {{Ackermann}},
  \bibfnamefont {M.}}, \bibinfo {author} {\bibnamefont {{Albert}},
  \bibfnamefont {A.}}, \bibinfo {author} {\bibnamefont {{Baldini}},
  \bibfnamefont {L.}},  \emph {et~al.} (\bibinfo {collaboration} {Fermi LAT
  Collaboration}),\ }\href {\doibase 10.1088/0004-637X/747/2/121} {\bibfield
  {journal} {\bibinfo  {journal} {\apj}\ }\textbf {\bibinfo {volume} {747}},\
  \bibinfo {eid} {121} (\bibinfo {year} {2012})},\ \Eprint
  {http://arxiv.org/abs/1201.2691} {arXiv:1201.2691 [astro-ph.HE]} \BibitemShut
  {NoStop}%
\bibitem [{\citenamefont {Ackermann}\ \emph {et~al.}(2011)\citenamefont
  {Ackermann} \emph {et~al.}}]{Ackermann:2011wa}%
  \BibitemOpen
  \bibfield  {author} {\bibinfo {author} {\bibnamefont {Ackermann},
  \bibfnamefont {M.}} \emph {et~al.} (\bibinfo {collaboration} {Fermi LAT
  Collaboration}),\ }\href {\doibase 10.1103/PhysRevLett.107.241302} {\bibfield
   {journal} {\bibinfo  {journal} {Phys.Rev.Lett.}\ }\textbf {\bibinfo {volume}
  {107}},\ \bibinfo {pages} {241302} (\bibinfo {year} {2011})},\ \Eprint
  {http://arxiv.org/abs/1108.3546} {arXiv:1108.3546 [astro-ph.HE]} \BibitemShut
  {NoStop}%
%%CITATION = ARXIV:1108.3546;%%
\bibitem [{\citenamefont {Ackermann}\ \emph
  {et~al.}(2012{\natexlab{a}})\citenamefont {Ackermann} \emph
  {et~al.}}]{Ackermann:2012rg}%
  \BibitemOpen
  \bibfield  {author} {\bibinfo {author} {\bibnamefont {Ackermann},
  \bibfnamefont {M.}} \emph {et~al.} (\bibinfo {collaboration} {Fermi LAT
  Collaboration}),\ }\href {\doibase 10.1088/0004-637X/761/2/91} {\bibfield
  {journal} {\bibinfo  {journal} {Astrophys.J.}\ }\textbf {\bibinfo {volume}
  {761}},\ \bibinfo {pages} {91} (\bibinfo {year} {2012}{\natexlab{a}})},\
  \Eprint {http://arxiv.org/abs/1205.6474} {arXiv:1205.6474 [astro-ph.CO]}
  \BibitemShut {NoStop}%
%%CITATION = ARXIV:1205.6474;%%
\bibitem [{\citenamefont {Ackermann}\ \emph
  {et~al.}(2012{\natexlab{b}})\citenamefont {Ackermann} \emph
  {et~al.}}]{Ackermann:2012qk}%
  \BibitemOpen
  \bibfield  {author} {\bibinfo {author} {\bibnamefont {Ackermann},
  \bibfnamefont {M.}} \emph {et~al.} (\bibinfo {collaboration} {Fermi LAT
  Collaboration}),\ }\href {\doibase 10.1103/PhysRevD.86.022002} {\bibfield
  {journal} {\bibinfo  {journal} {Phys.Rev.}\ }\textbf {\bibinfo {volume}
  {D86}},\ \bibinfo {pages} {022002} (\bibinfo {year} {2012}{\natexlab{b}})},\
  \Eprint {http://arxiv.org/abs/1205.2739} {arXiv:1205.2739 [astro-ph.HE]}
  \BibitemShut {NoStop}%
%%CITATION = ARXIV:1205.2739;%%
\bibitem [{\citenamefont {Ackermann}\ \emph
  {et~al.}(2012{\natexlab{c}})\citenamefont {Ackermann} \emph
  {et~al.}}]{Ackermann:2012nb}%
  \BibitemOpen
  \bibfield  {author} {\bibinfo {author} {\bibnamefont {Ackermann},
  \bibfnamefont {M.}} \emph {et~al.} (\bibinfo {collaboration} {Fermi LAT
  Collaboration}),\ }\href {\doibase 10.1088/0004-637X/747/2/121} {\bibfield
  {journal} {\bibinfo  {journal} {Astrophys.J.}\ }\textbf {\bibinfo {volume}
  {747}},\ \bibinfo {pages} {121} (\bibinfo {year} {2012}{\natexlab{c}})},\
  \Eprint {http://arxiv.org/abs/1201.2691} {arXiv:1201.2691 [astro-ph.HE]}
  \BibitemShut {NoStop}%
%%CITATION = ARXIV:1201.2691;%%
\bibitem [{\citenamefont {Ackermann}\ \emph {et~al.}(2013)\citenamefont
  {Ackermann} \emph {et~al.}}]{Ackermann:2013uma}%
  \BibitemOpen
  \bibfield  {author} {\bibinfo {author} {\bibnamefont {Ackermann},
  \bibfnamefont {M.}} \emph {et~al.} (\bibinfo {collaboration} {Fermi LAT
  Collaboration}),\ }\href {\doibase 10.1103/PhysRevD.88.082002} {\bibfield
  {journal} {\bibinfo  {journal} {Phys.Rev.}\ }\textbf {\bibinfo {volume}
  {D88}},\ \bibinfo {pages} {082002} (\bibinfo {year} {2013})},\ \Eprint
  {http://arxiv.org/abs/1305.5597} {arXiv:1305.5597 [astro-ph.HE]} \BibitemShut
  {NoStop}%
%%CITATION = ARXIV:1305.5597;%%
\bibitem [{\citenamefont {Ackermann}\ \emph {et~al.}(2015)\citenamefont
  {Ackermann} \emph {et~al.}}]{2015arXiv150302641F}%
  \BibitemOpen
  \bibfield  {author} {\bibinfo {author} {\bibnamefont {Ackermann},
  \bibfnamefont {M.}} \emph {et~al.},\ }\href@noop {} {\enquote {\bibinfo
  {title} {Searching for dark matter annihilation from milky way dwarf
  spheroidal galaxies with six years of fermi-lat dataxs},}\ } (\bibinfo {year}
  {2015}),\ \bibinfo {note} {submitted to PRL. 17 pages, 7 figures, 2 tables.
  Includes Supplementary Materials.},\ \Eprint
  {http://arxiv.org/abs/1503.02641} {arXiv:1503.02641 [astro-ph.HE]}
  \BibitemShut {NoStop}%
\bibitem [{\citenamefont {Actis}\ \emph {et~al.}(2011)\citenamefont {Actis}
  \emph {et~al.}}]{Consortium:2010bc}%
  \BibitemOpen
  \bibfield  {author} {\bibinfo {author} {\bibnamefont {Actis}, \bibfnamefont
  {M.}} \emph {et~al.} (\bibinfo {collaboration} {CTA Consortium}),\ }\href
  {\doibase 10.1007/s10686-011-9247-0} {\bibfield  {journal} {\bibinfo
  {journal} {Exper.Astron.}\ }\textbf {\bibinfo {volume} {32}},\ \bibinfo
  {pages} {193} (\bibinfo {year} {2011})},\ \Eprint
  {http://arxiv.org/abs/1008.3703} {arXiv:1008.3703 [astro-ph.IM]} \BibitemShut
  {NoStop}%
%%CITATION = ARXIV:1008.3703;%%
\bibitem [{\citenamefont {Ade}\ \emph {et~al.}(2015)\citenamefont {Ade} \emph
  {et~al.}}]{Planck:2015xua}%
  \BibitemOpen
  \bibfield  {author} {\bibinfo {author} {\bibnamefont {Ade}, \bibfnamefont
  {P.}} \emph {et~al.} (\bibinfo {collaboration} {Planck}),\ }\href@noop {}
  {\enquote {\bibinfo {title} {{Planck 2015 results. XIII. Cosmological
  parameters}},}\ } (\bibinfo {year} {2015}),\ \bibinfo {note} {submitted to
  A\&A},\ \Eprint {http://arxiv.org/abs/1502.01589} {arXiv:1502.01589
  [astro-ph.CO]} \BibitemShut {NoStop}%
%%CITATION = ARXIV:1502.01589;%%
\bibitem [{\citenamefont {{Adriani}}\ \emph {et~al.}(2009)\citenamefont
  {{Adriani}}, \citenamefont {{Barbarino}}, \citenamefont {{Bazilevskaya}},
  \citenamefont {{Bellotti}}, \citenamefont {{Boezio}}, \citenamefont
  {{Bogomolov}}, \citenamefont {{Bonechi}}, \citenamefont {{Bongi}},
  \citenamefont {{Bonvicini}}, \citenamefont {{Bottai}}, \citenamefont
  {{Bruno}}, \citenamefont {{Cafagna}}, \citenamefont {{Campana}},
  \citenamefont {{Carlson}}, \citenamefont {{Casolino}}, \citenamefont
  {{Castellini}}, \citenamefont {{de Pascale}}, \citenamefont {{de Rosa}},
  \citenamefont {{de Simone}}, \citenamefont {{di Felice}}, \citenamefont
  {{Galper}}, \citenamefont {{Grishantseva}}, \citenamefont {{Hofverberg}},
  \citenamefont {{Koldashov}}, \citenamefont {{Krutkov}}, \citenamefont
  {{Kvashnin}}, \citenamefont {{Leonov}}, \citenamefont {{Malvezzi}},
  \citenamefont {{Marcelli}}, \citenamefont {{Menn}}, \citenamefont
  {{Mikhailov}}, \citenamefont {{Mocchiutti}}, \citenamefont {{Orsi}},
  \citenamefont {{Osteria}}, \citenamefont {{Papini}}, \citenamefont
  {{Pearce}}, \citenamefont {{Picozza}}, \citenamefont {{Ricci}}, \citenamefont
  {{Ricciarini}}, \citenamefont {{Simon}}, \citenamefont {{Sparvoli}},
  \citenamefont {{Spillantini}}, \citenamefont {{Stozhkov}}, \citenamefont
  {{Vacchi}}, \citenamefont {{Vannuccini}}, \citenamefont {{Vasilyev}},
  \citenamefont {{Voronov}}, \citenamefont {{Yurkin}}, \citenamefont {{Zampa}},
  \citenamefont {{Zampa}},\ and\ \citenamefont
  {{Zverev}}}]{2009Natur.458..607A}%
  \BibitemOpen
  \bibfield  {author} {\bibinfo {author} {\bibnamefont {{Adriani}},
  \bibfnamefont {O.}}, \bibinfo {author} {\bibnamefont {{Barbarino}},
  \bibfnamefont {G.~C.}}, \bibinfo {author} {\bibnamefont {{Bazilevskaya}},
  \bibfnamefont {G.~A.}},  \emph {et~al.},\ }\href {\doibase
  10.1038/nature07942} {\bibfield  {journal} {\bibinfo  {journal} {\nat}\
  }\textbf {\bibinfo {volume} {458}},\ \bibinfo {pages} {607} (\bibinfo {year}
  {2009})},\ \Eprint {http://arxiv.org/abs/0810.4995} {arXiv:0810.4995}
  \BibitemShut {NoStop}%
\bibitem [{\citenamefont {{Agnello}}\ and\ \citenamefont
  {{Evans}}(2012)}]{Agnello:2012uc}%
  \BibitemOpen
  \bibfield  {author} {\bibinfo {author} {\bibnamefont {{Agnello}},
  \bibfnamefont {A.}}\ and\ \bibinfo {author} {\bibnamefont {{Evans}},
  \bibfnamefont {N.~W.}},\ }\href {\doibase 10.1088/2041-8205/754/2/L39}
  {\bibfield  {journal} {\bibinfo  {journal} {\apjl}\ }\textbf {\bibinfo
  {volume} {754}},\ \bibinfo {eid} {L39} (\bibinfo {year} {2012})},\ \Eprint
  {http://arxiv.org/abs/1205.6673} {arXiv:1205.6673 [astro-ph.GA]} \BibitemShut
  {NoStop}%
\bibitem [{\citenamefont {{Agnese}}\ \emph {et~al.}(2014)\citenamefont
  {{Agnese}}, \citenamefont {{Anderson}}, \citenamefont {{Asai}}, \citenamefont
  {{Balakishiyeva}}, \citenamefont {{Basu Thakur}}, \citenamefont {{Bauer}},
  \citenamefont {{Beaty}}, \citenamefont {{Billard}}, \citenamefont
  {{Borgland}}, \citenamefont {{Bowles}}, \citenamefont {{Brandt}},
  \citenamefont {{Brink}}, \citenamefont {{Bunker}}, \citenamefont {{Cabrera}},
  \citenamefont {{Caldwell}}, \citenamefont {{Cerdeno}}, \citenamefont
  {{Chagani}}, \citenamefont {{Chen}}, \citenamefont {{Cherry}}, \citenamefont
  {{Cooley}}, \citenamefont {{Cornell}}, \citenamefont {{Crewdson}},
  \citenamefont {{Cushman}}, \citenamefont {{Daal}}, \citenamefont {{DeVaney}},
  \citenamefont {{Di Stefano}}, \citenamefont {{Silva}}, \citenamefont
  {{Doughty}}, \citenamefont {{Esteban}}, \citenamefont {{Fallows}},
  \citenamefont {{Figueroa-Feliciano}}, \citenamefont {{Godfrey}},
  \citenamefont {{Golwala}}, \citenamefont {{Hall}}, \citenamefont {{Hansen}},
  \citenamefont {{Harris}}, \citenamefont {{Hertel}}, \citenamefont {{Hines}},
  \citenamefont {{Hofer}}, \citenamefont {{Holmgren}}, \citenamefont {{Hsu}},
  \citenamefont {{Huber}}, \citenamefont {{Jastram}}, \citenamefont {{Kamaev}},
  \citenamefont {{Kara}}, \citenamefont {{Kelsey}}, \citenamefont {{Kenany}},
  \citenamefont {{Kennedy}}, \citenamefont {{Kiveni}}, \citenamefont {{Koch}},
  \citenamefont {{Leder}}, \citenamefont {{Loer}}, \citenamefont {{Lopez
  Asamar}}, \citenamefont {{Mahapatra}}, \citenamefont {{Mandic}},
  \citenamefont {{Martinez}}, \citenamefont {{McCarthy}}, \citenamefont
  {{Mirabolfathi}}, \citenamefont {{Moffatt}}, \citenamefont {{Nelson}},
  \citenamefont {{Novak}}, \citenamefont {{Page}}, \citenamefont {{Partridge}},
  \citenamefont {{Pepin}}, \citenamefont {{Phipps}}, \citenamefont {{Platt}},
  \citenamefont {{Prasad}}, \citenamefont {{Pyle}}, \citenamefont {{Qiu}},
  \citenamefont {{Rau}}, \citenamefont {{Redl}}, \citenamefont {{Reisetter}},
  \citenamefont {{Resch}}, \citenamefont {{Ricci}}, \citenamefont {{Ruschman}},
  \citenamefont {{Saab}}, \citenamefont {{Sadoulet}}, \citenamefont {{Sander}},
  \citenamefont {{Schmitt}}, \citenamefont {{Schneck}}, \citenamefont
  {{Schnee}}, \citenamefont {{Scorza}}, \citenamefont {{Seitz}}, \citenamefont
  {{Serfass}}, \citenamefont {{Shank}}, \citenamefont {{Speller}},
  \citenamefont {{Tomada}}, \citenamefont {{Upadhyayula}}, \citenamefont
  {{Villano}}, \citenamefont {{Welliver}}, \citenamefont {{Wright}},
  \citenamefont {{Yellin}}, \citenamefont {{Yen}}, \citenamefont {{Young}},
  \citenamefont {{Zhang}},\ and\ \citenamefont {{SuperCDMS
  Collaboration}}}]{Agnese:2014aze}%
  \BibitemOpen
  \bibfield  {author} {\bibinfo {author} {\bibnamefont {{Agnese}},
  \bibfnamefont {R.}}, \bibinfo {author} {\bibnamefont {{Anderson}},
  \bibfnamefont {A.~J.}}, \bibinfo {author} {\bibnamefont {{Asai}},
  \bibfnamefont {M.}},  \emph {et~al.} (\bibinfo {collaboration} {SuperCDMS
  Collaboration}),\ }\href {\doibase 10.1103/PhysRevLett.112.241302} {\bibfield
   {journal} {\bibinfo  {journal} {Phys.Rev.Lett.}\ }\textbf {\bibinfo {volume}
  {112}},\ \bibinfo {pages} {241302} (\bibinfo {year} {2014})},\ \Eprint
  {http://arxiv.org/abs/1402.7137} {arXiv:1402.7137 [hep-ex]} \BibitemShut
  {NoStop}%
%%CITATION = ARXIV:1402.7137;%%
\bibitem [{\citenamefont {{Aharonian}}\ \emph
  {et~al.}(2009{\natexlab{a}})\citenamefont {{Aharonian}}, \citenamefont
  {{Akhperjanian}}, \citenamefont {{Anton}}, \citenamefont {{Barres de
  Almeida}}, \citenamefont {{Bazer-Bachi}}, \citenamefont {{Becherini}},
  \citenamefont {{Behera}}, \citenamefont {{Bernl{\"o}hr}}, \citenamefont
  {{Boisson}}, \citenamefont {{Bochow}}, \citenamefont {{Borrel}},
  \citenamefont {{Braun}}, \citenamefont {{Brion}}, \citenamefont {{Brucker}},
  \citenamefont {{Brun}}, \citenamefont {{B{\"u}hler}}, \citenamefont
  {{Bulik}}, \citenamefont {{B{\"u}sching}}, \citenamefont {{Boutelier}},
  \citenamefont {{Chadwick}}, \citenamefont {{Charbonnier}}, \citenamefont
  {{Chaves}}, \citenamefont {{Cheesebrough}}, \citenamefont {{Chounet}},
  \citenamefont {{Clapson}}, \citenamefont {{Coignet}}, \citenamefont
  {{Dalton}}, \citenamefont {{Daniel}}, \citenamefont {{Degrange}},
  \citenamefont {{Deil}}, \citenamefont {{Dickinson}}, \citenamefont
  {{Djannati-Ata{\"i}}}, \citenamefont {{Domainko}}, \citenamefont
  {{O'C.~Drury}}, \citenamefont {{Dubois}}, \citenamefont {{Dubus}},
  \citenamefont {{Dyks}}, \citenamefont {{Dyrda}}, \citenamefont {{Egberts}},
  \citenamefont {{Emmanoulopoulos}}, \citenamefont {{Espigat}}, \citenamefont
  {{Farnier}}, \citenamefont {{Feinstein}}, \citenamefont {{Fiasson}},
  \citenamefont {{F{\"o}rster}}, \citenamefont {{Fontaine}}, \citenamefont
  {{F{\"u}{\ss}ling}}, \citenamefont {{Gabici}}, \citenamefont {{Gallant}},
  \citenamefont {{G{\'e}rard}}, \citenamefont {{Giebels}}, \citenamefont
  {{Glicenstein}}, \citenamefont {{Gl{\"u}ck}}, \citenamefont {{Goret}},
  \citenamefont {{Hauser}}, \citenamefont {{Hauser}}, \citenamefont {{Heinz}},
  \citenamefont {{Heinzelmann}}, \citenamefont {{Henri}}, \citenamefont
  {{Hermann}}, \citenamefont {{Hinton}}, \citenamefont {{Hoffmann}},
  \citenamefont {{Hofmann}}, \citenamefont {{Holleran}}, \citenamefont
  {{Hoppe}}, \citenamefont {{Horns}}, \citenamefont {{Jacholkowska}},
  \citenamefont {{de Jager}}, \citenamefont {{Jung}}, \citenamefont
  {{Katarzy{\'n}ski}}, \citenamefont {{Katz}}, \citenamefont {{Kaufmann}},
  \citenamefont {{Kendziorra}}, \citenamefont {{Kerschhaggl}}, \citenamefont
  {{Khangulyan}}, \citenamefont {{Kh{\'e}lifi}}, \citenamefont {{Keogh}},
  \citenamefont {{Komin}}, \citenamefont {{Kosack}}, \citenamefont {{Lamanna}},
  \citenamefont {{Lenain}}, \citenamefont {{Lohse}}, \citenamefont
  {{Marandon}}, \citenamefont {{Martin}}, \citenamefont {{Martineau-Huynh}},
  \citenamefont {{Marcowith}}, \citenamefont {{Maurin}}, \citenamefont
  {{McComb}}, \citenamefont {{Medina}}, \citenamefont {{Moderski}},
  \citenamefont {{Moulin}}, \citenamefont {{Naumann-Godo}}, \citenamefont {{de
  Naurois}}, \citenamefont {{Nedbal}}, \citenamefont {{Nekrassov}},
  \citenamefont {{Niemiec}}, \citenamefont {{Nolan}}, \citenamefont {{Ohm}},
  \citenamefont {{Olive}}, \citenamefont {{de O{\~n}a Wilhelmi}}, \citenamefont
  {{Orford}}, \citenamefont {{Ostrowski}}, \citenamefont {{Panter}},
  \citenamefont {{Paz Arribas}}, \citenamefont {{Pedaletti}}, \citenamefont
  {{Pelletier}}, \citenamefont {{Petrucci}}, \citenamefont {{Pita}},
  \citenamefont {{P{\"u}hlhofer}}, \citenamefont {{Punch}}, \citenamefont
  {{Quirrenbach}}, \citenamefont {{Raubenheimer}}, \citenamefont {{Raue}},
  \citenamefont {{Rayner}}, \citenamefont {{Renaud}}, \citenamefont {{Rieger}},
  \citenamefont {{Ripken}}, \citenamefont {{Rob}}, \citenamefont
  {{Rosier-Lees}}, \citenamefont {{Rowell}}, \citenamefont {{Rudak}},
  \citenamefont {{Rulten}}, \citenamefont {{Ruppel}}, \citenamefont
  {{Sahakian}}, \citenamefont {{Santangelo}}, \citenamefont {{Schlickeiser}},
  \citenamefont {{Sch{\"o}ck}}, \citenamefont {{Schr{\"o}der}}, \citenamefont
  {{Schwanke}}, \citenamefont {{Schwarzburg}}, \citenamefont {{Schwemmer}},
  \citenamefont {{Shalchi}}, \citenamefont {{Skilton}}, \citenamefont {{Sol}},
  \citenamefont {{Spangler}}, \citenamefont {{Stawarz}}, \citenamefont
  {{Steenkamp}}, \citenamefont {{Stegmann}}, \citenamefont {{Superina}},
  \citenamefont {{Szostek}}, \citenamefont {{Tam}}, \citenamefont {{Tavernet}},
  \citenamefont {{Terrier}}, \citenamefont {{Tibolla}}, \citenamefont {{van
  Eldik}}, \citenamefont {{Vasileiadis}}, \citenamefont {{Venter}},
  \citenamefont {{Venter}}, \citenamefont {{Vialle}}, \citenamefont
  {{Vincent}}, \citenamefont {{Vivier}}, \citenamefont {{V{\"o}lk}},
  \citenamefont {{Volpe}}, \citenamefont {{Wagner}}, \citenamefont {{Ward}},
  \citenamefont {{Zdziarski}},\ and\ \citenamefont
  {{Zech}}}]{2009A&A...495...27A}%
  \BibitemOpen
  \bibfield  {author} {\bibinfo {author} {\bibnamefont {{Aharonian}},
  \bibfnamefont {F.}}, \bibinfo {author} {\bibnamefont {{Akhperjanian}},
  \bibfnamefont {A.~G.}}, \bibinfo {author} {\bibnamefont {{Anton}},
  \bibfnamefont {G.}},  \emph {et~al.},\ }\href {\doibase
  10.1051/0004-6361:200811372} {\bibfield  {journal} {\bibinfo  {journal}
  {\aap}\ }\textbf {\bibinfo {volume} {495}},\ \bibinfo {pages} {27} (\bibinfo
  {year} {2009}{\natexlab{a}})},\ \Eprint {http://arxiv.org/abs/0812.1638}
  {arXiv:0812.1638} \BibitemShut {NoStop}%
\bibitem [{\citenamefont {{Aharonian}}\ \emph
  {et~al.}(2009{\natexlab{b}})\citenamefont {{Aharonian}}, \citenamefont
  {{Akhperjanian}}, \citenamefont {{Anton}}, \citenamefont {{Barres de
  Almeida}}, \citenamefont {{Bazer-Bachi}}, \citenamefont {{Becherini}},
  \citenamefont {{Behera}}, \citenamefont {{Bernl{\"o}hr}}, \citenamefont
  {{Boisson}}, \citenamefont {{Bochow}}, \citenamefont {{Borrel}},
  \citenamefont {{Brion}}, \citenamefont {{Brucker}}, \citenamefont {{Brun}},
  \citenamefont {{B{\"u}hler}}, \citenamefont {{Bulik}}, \citenamefont
  {{B{\"u}sching}}, \citenamefont {{Boutelier}}, \citenamefont {{Chadwick}},
  \citenamefont {{Charbonnier}}, \citenamefont {{Chaves}}, \citenamefont
  {{Cheesebrough}}, \citenamefont {{Chounet}}, \citenamefont {{Clapson}},
  \citenamefont {{Coignet}}, \citenamefont {{Dalton}}, \citenamefont
  {{Daniel}}, \citenamefont {{Davids}}, \citenamefont {{Degrange}},
  \citenamefont {{Deil}}, \citenamefont {{Dickinson}}, \citenamefont
  {{Djannati-Ata{\"i}}}, \citenamefont {{Domainko}}, \citenamefont
  {{O'C.~Drury}}, \citenamefont {{Dubois}}, \citenamefont {{Dubus}},
  \citenamefont {{Dyks}}, \citenamefont {{Dyrda}}, \citenamefont {{Egberts}},
  \citenamefont {{Emmanoulopoulos}}, \citenamefont {{Espigat}}, \citenamefont
  {{Farnier}}, \citenamefont {{Feinstein}}, \citenamefont {{Fiasson}},
  \citenamefont {{F{\"o}rster}}, \citenamefont {{Fontaine}}, \citenamefont
  {{F{\"u}{\ss}ling}}, \citenamefont {{Gabici}}, \citenamefont {{Gallant}},
  \citenamefont {{G{\'e}rard}}, \citenamefont {{Giebels}}, \citenamefont
  {{Glicenstein}}, \citenamefont {{Gl{\"u}ck}}, \citenamefont {{Goret}},
  \citenamefont {{G{\"o}hring}}, \citenamefont {{Hauser}}, \citenamefont
  {{Hauser}}, \citenamefont {{Heinz}}, \citenamefont {{Heinzelmann}},
  \citenamefont {{Henri}}, \citenamefont {{Hermann}}, \citenamefont {{Hinton}},
  \citenamefont {{Hoffmann}}, \citenamefont {{Hofmann}}, \citenamefont
  {{Holleran}}, \citenamefont {{Hoppe}}, \citenamefont {{Horns}}, \citenamefont
  {{Inoue}}, \citenamefont {{Jacholkowska}}, \citenamefont {{de Jager}},
  \citenamefont {{Jahn}}, \citenamefont {{Jung}}, \citenamefont
  {{Katarzy{\'n}ski}}, \citenamefont {{Katz}}, \citenamefont {{Kaufmann}},
  \citenamefont {{Kendziorra}}, \citenamefont {{Kerschhaggl}}, \citenamefont
  {{Khangulyan}}, \citenamefont {{Kh{\'e}lifi}}, \citenamefont {{Keogh}},
  \citenamefont {{Klu{\'z}niak}}, \citenamefont {{Kneiske}}, \citenamefont
  {{Komin}}, \citenamefont {{Kosack}}, \citenamefont {{Lamanna}}, \citenamefont
  {{Lenain}}, \citenamefont {{Lohse}}, \citenamefont {{Marandon}},
  \citenamefont {{Martin}}, \citenamefont {{Martineau-Huynh}}, \citenamefont
  {{Marcowith}}, \citenamefont {{Maurin}}, \citenamefont {{McComb}},
  \citenamefont {{Medina}}, \citenamefont {{Moderski}}, \citenamefont
  {{Moulin}}, \citenamefont {{Naumann-Godo}}, \citenamefont {{de Naurois}},
  \citenamefont {{Nedbal}}, \citenamefont {{Nekrassov}}, \citenamefont
  {{Niemiec}}, \citenamefont {{Nolan}}, \citenamefont {{Ohm}}, \citenamefont
  {{Olive}}, \citenamefont {{de O{\~n}a Wilhelmi}}, \citenamefont {{Orford}},
  \citenamefont {{Ostrowski}}, \citenamefont {{Panter}}, \citenamefont {{Paz
  Arribas}}, \citenamefont {{Pedaletti}}, \citenamefont {{Pelletier}},
  \citenamefont {{Petrucci}}, \citenamefont {{Pita}}, \citenamefont
  {{P{\"u}hlhofer}}, \citenamefont {{Punch}}, \citenamefont {{Quirrenbach}},
  \citenamefont {{Raubenheimer}}, \citenamefont {{Raue}}, \citenamefont
  {{Rayner}}, \citenamefont {{Renaud}}, \citenamefont {{Reimer}}, \citenamefont
  {{Rieger}}, \citenamefont {{Ripken}}, \citenamefont {{Rob}}, \citenamefont
  {{Rosier-Lees}}, \citenamefont {{Rowell}}, \citenamefont {{Rudak}},
  \citenamefont {{Rulten}}, \citenamefont {{Ruppel}}, \citenamefont
  {{Sahakian}}, \citenamefont {{Santangelo}}, \citenamefont {{Schlickeiser}},
  \citenamefont {{Sch{\"o}ck}}, \citenamefont {{Schr{\"o}der}}, \citenamefont
  {{Schwanke}}, \citenamefont {{Schwarzburg}}, \citenamefont {{Schwemmer}},
  \citenamefont {{Shalchi}}, \citenamefont {{Sikora}}, \citenamefont
  {{Skilton}}, \citenamefont {{Sol}}, \citenamefont {{Spanglfoer}},
  \citenamefont {{Stawarz}}, \citenamefont {{Steenkamp}}, \citenamefont
  {{Stegmann}}, \citenamefont {{Superina}}, \citenamefont {{Szostek}},
  \citenamefont {{Tam}}, \citenamefont {{Tavernet}}, \citenamefont {{Terrier}},
  \citenamefont {{Tibolla}}, \citenamefont {{Tluczykont}}, \citenamefont {{van
  Eldik}}, \citenamefont {{Vasileiadis}}, \citenamefont {{Venter}},
  \citenamefont {{Venter}}, \citenamefont {{Vialle}}, \citenamefont
  {{Vincent}}, \citenamefont {{Vivier}}, \citenamefont {{V{\"o}lk}},
  \citenamefont {{Volpe}}, \citenamefont {{Wagner}}, \citenamefont {{Ward}},
  \citenamefont {{Zdziarski}},\ and\ \citenamefont
  {{Zech}}}]{2009A&A...502..437A}%
  \BibitemOpen
  \bibfield  {author} {\bibinfo {author} {\bibnamefont {{Aharonian}},
  \bibfnamefont {F.}}, \bibinfo {author} {\bibnamefont {{Akhperjanian}},
  \bibfnamefont {A.~G.}}, \bibinfo {author} {\bibnamefont {{Anton}},
  \bibfnamefont {G.}},  \emph {et~al.},\ }\href {\doibase
  10.1051/0004-6361/200912086} {\bibfield  {journal} {\bibinfo  {journal}
  {\aap}\ }\textbf {\bibinfo {volume} {502}},\ \bibinfo {pages} {437} (\bibinfo
  {year} {2009}{\natexlab{b}})},\ \Eprint {http://arxiv.org/abs/0907.0727}
  {arXiv:0907.0727 [astro-ph.CO]} \BibitemShut {NoStop}%
\bibitem [{\citenamefont {{Aharonian}}\ \emph {et~al.}(2004)\citenamefont
  {{Aharonian}}, \citenamefont {{Akhperjanian}}, \citenamefont {{Aye}},
  \citenamefont {{Bazer-Bachi}}, \citenamefont {{Beilicke}}, \citenamefont
  {{Benbow}}, \citenamefont {{Berge}}, \citenamefont {{Berghaus}},
  \citenamefont {{Bernl{\"o}hr}}, \citenamefont {{Bolz}}, \citenamefont
  {{Boisson}}, \citenamefont {{Borgmeier}}, \citenamefont {{Breitling}},
  \citenamefont {{Brown}}, \citenamefont {{Bussons Gordo}}, \citenamefont
  {{Chadwick}}, \citenamefont {{Chitnis}}, \citenamefont {{Chounet}},
  \citenamefont {{Cornils}}, \citenamefont {{Costamante}}, \citenamefont
  {{Degrange}}, \citenamefont {{Djannati-Ata{\"i}}}, \citenamefont
  {{O'C.~Drury}}, \citenamefont {{Ergin}}, \citenamefont {{Espigat}},
  \citenamefont {{Feinstein}}, \citenamefont {{Fleury}}, \citenamefont
  {{Fontaine}}, \citenamefont {{Funk}}, \citenamefont {{Gallant}},
  \citenamefont {{Giebels}}, \citenamefont {{Gillessen}}, \citenamefont
  {{Goret}}, \citenamefont {{Guy}}, \citenamefont {{Hadjichristidis}},
  \citenamefont {{Hauser}}, \citenamefont {{Heinzelmann}}, \citenamefont
  {{Henri}}, \citenamefont {{Hermann}}, \citenamefont {{Hinton}}, \citenamefont
  {{Hofmann}}, \citenamefont {{Holleran}}, \citenamefont {{Horns}},
  \citenamefont {{de Jager}}, \citenamefont {{Jung}}, \citenamefont
  {{Kh{\'e}lifi}}, \citenamefont {{Komin}}, \citenamefont {{Konopelko}},
  \citenamefont {{Latham}}, \citenamefont {{Le Gallou}}, \citenamefont
  {{Lemoine}}, \citenamefont {{Lemi{\`e}re}}, \citenamefont {{Leroy}},
  \citenamefont {{Lohse}}, \citenamefont {{Marcowith}}, \citenamefont
  {{Masterson}}, \citenamefont {{McComb}}, \citenamefont {{de Naurois}},
  \citenamefont {{Nolan}}, \citenamefont {{Noutsos}}, \citenamefont {{Orford}},
  \citenamefont {{Osborne}}, \citenamefont {{Ouchrif}}, \citenamefont
  {{Panter}}, \citenamefont {{Pelletier}}, \citenamefont {{Pita}},
  \citenamefont {{Pohl}}, \citenamefont {{P{\"u}hlhofer}}, \citenamefont
  {{Punch}}, \citenamefont {{Raubenheimer}}, \citenamefont {{Raue}},
  \citenamefont {{Raux}}, \citenamefont {{Rayner}}, \citenamefont {{Redondo}},
  \citenamefont {{Reimer}}, \citenamefont {{Reimer}}, \citenamefont {{Ripken}},
  \citenamefont {{Rivoal}}, \citenamefont {{Rob}}, \citenamefont {{Rolland}},
  \citenamefont {{Rowell}}, \citenamefont {{Sahakian}}, \citenamefont
  {{Saug{\'e}}}, \citenamefont {{Schlenker}}, \citenamefont {{Schlickeiser}},
  \citenamefont {{Schuster}}, \citenamefont {{Schwanke}}, \citenamefont
  {{Siewert}}, \citenamefont {{Sol}}, \citenamefont {{Steenkamp}},
  \citenamefont {{Stegmann}}, \citenamefont {{Tavernet}}, \citenamefont
  {{Th{\'e}oret}}, \citenamefont {{Tluczykont}}, \citenamefont {{van der
  Walt}}, \citenamefont {{Vasileiadis}}, \citenamefont {{Vincent}},
  \citenamefont {{Visser}}, \citenamefont {{V{\"o}lk}},\ and\ \citenamefont
  {{Wagner}}}]{2004A&A...425L..13A}%
  \BibitemOpen
  \bibfield  {author} {\bibinfo {author} {\bibnamefont {{Aharonian}},
  \bibfnamefont {F.}}, \bibinfo {author} {\bibnamefont {{Akhperjanian}},
  \bibfnamefont {A.~G.}}, \bibinfo {author} {\bibnamefont {{Aye}},
  \bibfnamefont {K.-M.}},  \emph {et~al.},\ }\href {\doibase
  10.1051/0004-6361:200400055} {\bibfield  {journal} {\bibinfo  {journal}
  {\aap}\ }\textbf {\bibinfo {volume} {425}},\ \bibinfo {pages} {L13} (\bibinfo
  {year} {2004})},\ \Eprint {http://arxiv.org/abs/astro-ph/0406658}
  {astro-ph/0406658} \BibitemShut {NoStop}%
\bibitem [{\citenamefont {{Aharonian}}\ \emph {et~al.}(2006)\citenamefont
  {{Aharonian}}, \citenamefont {{Akhperjanian}}, \citenamefont {{Bazer-Bachi}},
  \citenamefont {{Beilicke}}, \citenamefont {{Benbow}}, \citenamefont
  {{Berge}}, \citenamefont {{Bernl{\"o}hr}}, \citenamefont {{Boisson}},
  \citenamefont {{Bolz}}, \citenamefont {{Borrel}}, \citenamefont {{Braun}},
  \citenamefont {{Breitling}}, \citenamefont {{Brown}}, \citenamefont
  {{B{\"u}hler}}, \citenamefont {{B{\"u}sching}}, \citenamefont {{Carrigan}},
  \citenamefont {{Chadwick}}, \citenamefont {{Chounet}}, \citenamefont
  {{Cornils}}, \citenamefont {{Costamante}}, \citenamefont {{Degrange}},
  \citenamefont {{Dickinson}}, \citenamefont {{Djannati-Ata{\"i}}},
  \citenamefont {{Drury}}, \citenamefont {{Dubus}}, \citenamefont {{Egberts}},
  \citenamefont {{Emmanoulopoulos}}, \citenamefont {{Espigat}}, \citenamefont
  {{Feinstein}}, \citenamefont {{Ferrero}}, \citenamefont {{Fiasson}},
  \citenamefont {{Fontaine}}, \citenamefont {{Funk}}, \citenamefont {{Funk}},
  \citenamefont {{Gallant}}, \citenamefont {{Giebels}}, \citenamefont
  {{Glicenstein}}, \citenamefont {{Goret}}, \citenamefont {{Hadjichristidis}},
  \citenamefont {{Hauser}}, \citenamefont {{Hauser}}, \citenamefont
  {{Heinzelmann}}, \citenamefont {{Henri}}, \citenamefont {{Hermann}},
  \citenamefont {{Hinton}}, \citenamefont {{Hofmann}}, \citenamefont
  {{Holleran}}, \citenamefont {{Horns}}, \citenamefont {{Jacholkowska}},
  \citenamefont {{de Jager}}, \citenamefont {{Kh{\'e}lifi}}, \citenamefont
  {{Komin}}, \citenamefont {{Konopelko}}, \citenamefont {{Kosack}},
  \citenamefont {{Latham}}, \citenamefont {{Le Gallou}}, \citenamefont
  {{Lemi{\`e}re}}, \citenamefont {{Lemoine-Goumard}}, \citenamefont {{Lohse}},
  \citenamefont {{Martin}}, \citenamefont {{Martineau-Huynh}}, \citenamefont
  {{Marcowith}}, \citenamefont {{Masterson}}, \citenamefont {{McComb}},
  \citenamefont {{de Naurois}}, \citenamefont {{Nedbal}}, \citenamefont
  {{Nolan}}, \citenamefont {{Noutsos}}, \citenamefont {{Orford}}, \citenamefont
  {{Osborne}}, \citenamefont {{Ouchrif}}, \citenamefont {{Panter}},
  \citenamefont {{Pelletier}}, \citenamefont {{Pita}}, \citenamefont
  {{P{\"u}hlhofer}}, \citenamefont {{Punch}}, \citenamefont {{Raubenheimer}},
  \citenamefont {{Raue}}, \citenamefont {{Rayner}}, \citenamefont {{Reimer}},
  \citenamefont {{Reimer}}, \citenamefont {{Ripken}}, \citenamefont {{Rob}},
  \citenamefont {{Rolland}}, \citenamefont {{Rowell}}, \citenamefont
  {{Sahakian}}, \citenamefont {{Saug{\'e}}}, \citenamefont {{Schlenker}},
  \citenamefont {{Schlickeiser}}, \citenamefont {{Schwanke}}, \citenamefont
  {{Sol}}, \citenamefont {{Spangler}}, \citenamefont {{Spanier}}, \citenamefont
  {{Steenkamp}}, \citenamefont {{Stegmann}}, \citenamefont {{Superina}},
  \citenamefont {{Tavernet}}, \citenamefont {{Terrier}}, \citenamefont
  {{Th{\'e}oret}}, \citenamefont {{Tluczykont}}, \citenamefont {{van Eldik}},
  \citenamefont {{Vasileiadis}}, \citenamefont {{Venter}}, \citenamefont
  {{Vincent}}, \citenamefont {{V{\"o}lk}}, \citenamefont {{Wagner}},\ and\
  \citenamefont {{Ward}}}]{2006PhRvL..97v1102A}%
  \BibitemOpen
  \bibfield  {author} {\bibinfo {author} {\bibnamefont {{Aharonian}},
  \bibfnamefont {F.}}, \bibinfo {author} {\bibnamefont {{Akhperjanian}},
  \bibfnamefont {A.~G.}}, \bibinfo {author} {\bibnamefont {{Bazer-Bachi}},
  \bibfnamefont {A.~R.}},  \emph {et~al.},\ }\href {\doibase
  10.1103/PhysRevLett.97.221102} {\bibfield  {journal} {\bibinfo  {journal}
  {Physical Review Letters}\ }\textbf {\bibinfo {volume} {97}},\ \bibinfo {eid}
  {221102} (\bibinfo {year} {2006})},\ \Eprint
  {http://arxiv.org/abs/astro-ph/0610509} {astro-ph/0610509} \BibitemShut
  {NoStop}%
\bibitem [{\citenamefont {{Aharonian}}\ \emph {et~al.}(2008)\citenamefont
  {{Aharonian}}, \citenamefont {{Akhperjanian}}, \citenamefont {{Bazer-Bachi}},
  \citenamefont {{Beilicke}}, \citenamefont {{Benbow}}, \citenamefont
  {{Berge}}, \citenamefont {{Bernl{\"o}hr}}, \citenamefont {{Boisson}},
  \citenamefont {{Bolz}}, \citenamefont {{Borrel}}, \citenamefont {{Braun}},
  \citenamefont {{Brion}}, \citenamefont {{Brown}}, \citenamefont
  {{B{\"u}hler}}, \citenamefont {{B{\"u}sching}}, \citenamefont {{Boutelier}},
  \citenamefont {{Carrigan}}, \citenamefont {{Chadwick}}, \citenamefont
  {{Chounet}}, \citenamefont {{Coignet}}, \citenamefont {{Cornils}},
  \citenamefont {{Costamante}}, \citenamefont {{Degrange}}, \citenamefont
  {{Dickinson}}, \citenamefont {{Djannati-Ata{\"i}}}, \citenamefont {{Drury}},
  \citenamefont {{Dubus}}, \citenamefont {{Egberts}}, \citenamefont
  {{Emmanoulopoulos}}, \citenamefont {{Espigat}}, \citenamefont {{Farnier}},
  \citenamefont {{Feinstein}}, \citenamefont {{Ferrero}}, \citenamefont
  {{Fiasson}}, \citenamefont {{Fontaine}}, \citenamefont {{Funk}},
  \citenamefont {{Funk}}, \citenamefont {{F{\"u}{\ss}ling}}, \citenamefont
  {{Gallant}}, \citenamefont {{Giebels}}, \citenamefont {{Glicenstein}},
  \citenamefont {{Gl{\"u}ck}}, \citenamefont {{Goret}}, \citenamefont
  {{Hadjichristidis}}, \citenamefont {{Hauser}}, \citenamefont {{Hauser}},
  \citenamefont {{Heinzelmann}}, \citenamefont {{Henri}}, \citenamefont
  {{Hermann}}, \citenamefont {{Hinton}}, \citenamefont {{Hoffmann}},
  \citenamefont {{Hofmann}}, \citenamefont {{Holleran}}, \citenamefont
  {{Hoppe}}, \citenamefont {{Horns}}, \citenamefont {{Jacholkowska}},
  \citenamefont {{de Jager}}, \citenamefont {{Kendziorra}}, \citenamefont
  {{Kerschhaggl}}, \citenamefont {{Kh{\'e}lifi}}, \citenamefont {{Komin}},
  \citenamefont {{Kosack}}, \citenamefont {{Lamanna}}, \citenamefont
  {{Latham}}, \citenamefont {{Le Gallou}}, \citenamefont {{Lemi{\`e}re}},
  \citenamefont {{Lemoine-Goumard}}, \citenamefont {{Lohse}}, \citenamefont
  {{Martin}}, \citenamefont {{Martineau-Huynh}}, \citenamefont {{Marcowith}},
  \citenamefont {{Masterson}}, \citenamefont {{Maurin}}, \citenamefont
  {{McComb}}, \citenamefont {{Moulin}}, \citenamefont {{de Naurois}},
  \citenamefont {{Nedbal}}, \citenamefont {{Nolan}}, \citenamefont {{Noutsos}},
  \citenamefont {{Nuss}}, \citenamefont {{Olive}}, \citenamefont {{Orford}},
  \citenamefont {{Osborne}}, \citenamefont {{Panter}}, \citenamefont
  {{Pelletier}}, \citenamefont {{Petrucci}}, \citenamefont {{Pita}},
  \citenamefont {{P{\"u}hlhofer}}, \citenamefont {{Punch}}, \citenamefont
  {{Ranchon}}, \citenamefont {{Raubenheimer}}, \citenamefont {{Raue}},
  \citenamefont {{Rayner}}, \citenamefont {{Ripken}}, \citenamefont {{Rob}},
  \citenamefont {{Rolland}}, \citenamefont {{Rosier-Lees}}, \citenamefont
  {{Rowell}}, \citenamefont {{Sahakian}}, \citenamefont {{Santangelo}},
  \citenamefont {{Saug{\'e}}}, \citenamefont {{Schlenker}}, \citenamefont
  {{Schlickeiser}}, \citenamefont {{Schr{\"o}der}}, \citenamefont {{Schwanke}},
  \citenamefont {{Schwarzburg}}, \citenamefont {{Schwemmer}}, \citenamefont
  {{Shalchi}}, \citenamefont {{Sol}}, \citenamefont {{Spangler}}, \citenamefont
  {{Spanier}}, \citenamefont {{Steenkamp}}, \citenamefont {{Stegmann}},
  \citenamefont {{Superina}}, \citenamefont {{Tam}}, \citenamefont
  {{Tavernet}}, \citenamefont {{Terrier}}, \citenamefont {{Tluczykont}},
  \citenamefont {{van Eldik}}, \citenamefont {{Vasileiadis}}, \citenamefont
  {{Venter}}, \citenamefont {{Vialle}}, \citenamefont {{Vincent}},
  \citenamefont {{Vivier}}, \citenamefont {{V{\"o}lk}}, \citenamefont
  {{Wagner}},\ and\ \citenamefont {{Ward}}}]{2008APh....29...55A}%
  \BibitemOpen
  \bibfield  {author} {\bibinfo {author} {\bibnamefont {{Aharonian}},
  \bibfnamefont {F.}}, \bibinfo {author} {\bibnamefont {{Akhperjanian}},
  \bibfnamefont {A.~G.}}, \bibinfo {author} {\bibnamefont {{Bazer-Bachi}},
  \bibfnamefont {A.~R.}},  \emph {et~al.},\ }\href {\doibase
  10.1016/j.astropartphys.2007.11.007} {\bibfield  {journal} {\bibinfo
  {journal} {Astroparticle Physics}\ }\textbf {\bibinfo {volume} {29}},\
  \bibinfo {pages} {55} (\bibinfo {year} {2008})},\ \Eprint
  {http://arxiv.org/abs/0711.2369} {arXiv:0711.2369} \BibitemShut {NoStop}%
\bibitem [{\citenamefont {{Aharonian}}\ \emph {et~al.}(2010)\citenamefont
  {{Aharonian}}, \citenamefont {{Akhperjanian}}, \citenamefont {{Bazer-Bachi}},
  \citenamefont {{Beilicke}}, \citenamefont {{Benbow}}, \citenamefont
  {{Berge}}, \citenamefont {{Bernl{\"o}hr}}, \citenamefont {{Boisson}},
  \citenamefont {{Bolz}}, \citenamefont {{Borrel}}, \citenamefont {{Braun}},
  \citenamefont {{Brion}}, \citenamefont {{Brown}}, \citenamefont
  {{B{\"u}hler}}, \citenamefont {{B{\"u}sching}}, \citenamefont {{Boutelier}},
  \citenamefont {{Carrigan}}, \citenamefont {{Chadwick}}, \citenamefont
  {{Chounet}}, \citenamefont {{Coignet}}, \citenamefont {{Cornils}},
  \citenamefont {{Costamante}}, \citenamefont {{Degrange}}, \citenamefont
  {{Dickinson}}, \citenamefont {{Djannati-Ata{\"i}}}, \citenamefont {{Drury}},
  \citenamefont {{Dubus}}, \citenamefont {{Egberts}}, \citenamefont
  {{Emmanoulopoulos}}, \citenamefont {{Espigat}}, \citenamefont {{Farnier}},
  \citenamefont {{Feinstein}}, \citenamefont {{Ferrero}}, \citenamefont
  {{Fiasson}}, \citenamefont {{Fontaine}}, \citenamefont {{Funk}},
  \citenamefont {{Funk}}, \citenamefont {{F{\"u}{\ss}ling}}, \citenamefont
  {{Gallant}}, \citenamefont {{Giebels}}, \citenamefont {{Glicenstein}},
  \citenamefont {{Gl{\"u}ck}}, \citenamefont {{Goret}}, \citenamefont
  {{Hadjichristidis}}, \citenamefont {{Hauser}}, \citenamefont {{Hauser}},
  \citenamefont {{Heinzelmann}}, \citenamefont {{Henri}}, \citenamefont
  {{Hermann}}, \citenamefont {{Hinton}}, \citenamefont {{Hoffmann}},
  \citenamefont {{Hofmann}}, \citenamefont {{Holleran}}, \citenamefont
  {{Hoppe}}, \citenamefont {{Horns}}, \citenamefont {{Jacholkowska}},
  \citenamefont {{de Jager}}, \citenamefont {{Kendziorra}}, \citenamefont
  {{Kerschhaggl}}, \citenamefont {{Kh{\'e}lifi}}, \citenamefont {{Komin}},
  \citenamefont {{Kosack}}, \citenamefont {{Lamanna}}, \citenamefont
  {{Latham}}, \citenamefont {{Le Gallou}}, \citenamefont {{Lemi{\`e}re}},
  \citenamefont {{Lemoine-Goumard}}, \citenamefont {{Lohse}}, \citenamefont
  {{Martin}}, \citenamefont {{Martineau-Huynh}}, \citenamefont {{Marcowith}},
  \citenamefont {{Masterson}}, \citenamefont {{Maurin}}, \citenamefont
  {{McComb}}, \citenamefont {{Moulin}}, \citenamefont {{de Naurois}},
  \citenamefont {{Nedbal}}, \citenamefont {{Nolan}}, \citenamefont {{Noutsos}},
  \citenamefont {{Nuss}}, \citenamefont {{Olive}}, \citenamefont {{Orford}},
  \citenamefont {{Osborne}}, \citenamefont {{Panter}}, \citenamefont
  {{Pelletier}}, \citenamefont {{Petrucci}}, \citenamefont {{Pita}},
  \citenamefont {{P{\"u}hlhofer}}, \citenamefont {{Punch}}, \citenamefont
  {{Ranchon}}, \citenamefont {{Raubenheimer}}, \citenamefont {{Raue}},
  \citenamefont {{Rayner}}, \citenamefont {{Ripken}}, \citenamefont {{Rob}},
  \citenamefont {{Rolland}}, \citenamefont {{Rosier-Lees}}, \citenamefont
  {{Rowell}}, \citenamefont {{Sahakian}}, \citenamefont {{Santangelo}},
  \citenamefont {{Saug{\'e}}}, \citenamefont {{Schlenker}}, \citenamefont
  {{Schlickeiser}}, \citenamefont {{Schr{\"o}der}}, \citenamefont {{Schwanke}},
  \citenamefont {{Schwarzburg}}, \citenamefont {{Schwemmer}}, \citenamefont
  {{Shalchi}}, \citenamefont {{Sol}}, \citenamefont {{Spangler}}, \citenamefont
  {{Spanier}}, \citenamefont {{Steenkamp}}, \citenamefont {{Stegmann}},
  \citenamefont {{Superina}}, \citenamefont {{Tam}}, \citenamefont
  {{Tavernet}}, \citenamefont {{Terrier}}, \citenamefont {{Tluczykont}},
  \citenamefont {{van Eldik}}, \citenamefont {{Vasileiadis}}, \citenamefont
  {{Venter}}, \citenamefont {{Vialle}}, \citenamefont {{Vincent}},
  \citenamefont {{Vivier}}, \citenamefont {{V{\"o}lk}}, \citenamefont
  {{Wagner}},\ and\ \citenamefont {{Ward}}}]{2010APh....33..274A}%
  \BibitemOpen
  \bibfield  {author} {\bibinfo {author} {\bibnamefont {{Aharonian}},
  \bibfnamefont {F.}}, \bibinfo {author} {\bibnamefont {{Akhperjanian}},
  \bibfnamefont {A.~G.}}, \bibinfo {author} {\bibnamefont {{Bazer-Bachi}},
  \bibfnamefont {A.~R.}},  \emph {et~al.},\ }\href {\doibase
  10.1016/j.astropartphys.2010.01.007} {\bibfield  {journal} {\bibinfo
  {journal} {Astroparticle Physics}\ }\textbf {\bibinfo {volume} {33}},\
  \bibinfo {pages} {274} (\bibinfo {year} {2010})}\BibitemShut {NoStop}%
\bibitem [{\citenamefont {{Aharonian}}\ \emph
  {et~al.}(2009{\natexlab{c}})\citenamefont {{Aharonian}}, \citenamefont
  {{Akhperjanian}}, \citenamefont {{de Almeida}}, \citenamefont
  {{Bazer-Bachi}}, \citenamefont {{Behera}}, \citenamefont {{Benbow}},
  \citenamefont {{Bernl{\"o}hr}}, \citenamefont {{Boisson}}, \citenamefont
  {{Bochow}}, \citenamefont {{Borrel}}, \citenamefont {{Braun}}, \citenamefont
  {{Brion}}, \citenamefont {{Brucker}}, \citenamefont {{Brun}}, \citenamefont
  {{B{\"u}hler}}, \citenamefont {{Bulik}}, \citenamefont {{B{\"u}sching}},
  \citenamefont {{Boutelier}}, \citenamefont {{Carrigan}}, \citenamefont
  {{Chadwick}}, \citenamefont {{Charbonnier}}, \citenamefont {{Chaves}},
  \citenamefont {{Cheesebrough}}, \citenamefont {{Chounet}}, \citenamefont
  {{Clapson}}, \citenamefont {{Coignet}}, \citenamefont {{Costamante}},
  \citenamefont {{Dalton}}, \citenamefont {{Degrange}}, \citenamefont {{Deil}},
  \citenamefont {{Dickinson}}, \citenamefont {{Djannati-Ata{\"i}}},
  \citenamefont {{Domainko}}, \citenamefont {{Drury}}, \citenamefont
  {{Dubois}}, \citenamefont {{Dubus}}, \citenamefont {{Dyks}}, \citenamefont
  {{Dyrda}}, \citenamefont {{Egberts}}, \citenamefont {{Emmanoulopoulos}},
  \citenamefont {{Espigat}}, \citenamefont {{Farnier}}, \citenamefont
  {{Feinstein}}, \citenamefont {{Fiasson}}, \citenamefont {{F{\"o}rster}},
  \citenamefont {{Fontaine}}, \citenamefont {{F{\"u}ssling}}, \citenamefont
  {{Gabici}}, \citenamefont {{Gallant}}, \citenamefont {{G{\'e}rard}},
  \citenamefont {{Giebels}}, \citenamefont {{Glicenstein}}, \citenamefont
  {{Gl{\"u}ck}}, \citenamefont {{Goret}}, \citenamefont {{Hadjichristidis}},
  \citenamefont {{Hauser}}, \citenamefont {{Hauser}}, \citenamefont {{Heinz}},
  \citenamefont {{Heinzelmann}}, \citenamefont {{Henri}}, \citenamefont
  {{Hermann}}, \citenamefont {{Hinton}}, \citenamefont {{Hoffmann}},
  \citenamefont {{Hofmann}}, \citenamefont {{Holleran}}, \citenamefont
  {{Hoppe}}, \citenamefont {{Horns}}, \citenamefont {{Jacholkowska}},
  \citenamefont {{de Jager}}, \citenamefont {{Jung}}, \citenamefont
  {{Katarzy{\'n}ski}}, \citenamefont {{Kaufmann}}, \citenamefont
  {{Kendziorra}}, \citenamefont {{Kerschhaggl}}, \citenamefont {{Khangulyan}},
  \citenamefont {{Kh{\'e}lifi}}, \citenamefont {{Keogh}}, \citenamefont
  {{Komin}}, \citenamefont {{Kosack}}, \citenamefont {{Lamanna}}, \citenamefont
  {{Lenain}}, \citenamefont {{Lohse}}, \citenamefont {{Marandon}},
  \citenamefont {{Martin}}, \citenamefont {{Martineau-Huynh}}, \citenamefont
  {{Marcowith}}, \citenamefont {{Maurin}}, \citenamefont {{McComb}},
  \citenamefont {{Medina}}, \citenamefont {{Moderski}}, \citenamefont
  {{Moulin}}, \citenamefont {{Naumann-Godo}}, \citenamefont {{de Naurois}},
  \citenamefont {{Nedbal}}, \citenamefont {{Nekrassov}}, \citenamefont
  {{Niemiec}}, \citenamefont {{Nolan}}, \citenamefont {{Ohm}}, \citenamefont
  {{Olive}}, \citenamefont {{de O{\~n}a Wilhelmi}}, \citenamefont {{Orford}},
  \citenamefont {{Osborne}}, \citenamefont {{Ostrowski}}, \citenamefont
  {{Panter}}, \citenamefont {{Pedaletti}}, \citenamefont {{Pelletier}},
  \citenamefont {{Petrucci}}, \citenamefont {{Pita}}, \citenamefont
  {{P{\"u}hlhofer}}, \citenamefont {{Punch}}, \citenamefont {{Quirrenbach}},
  \citenamefont {{Raubenheimer}}, \citenamefont {{Raue}}, \citenamefont
  {{Rayner}}, \citenamefont {{Renaud}}, \citenamefont {{Rieger}}, \citenamefont
  {{Ripken}}, \citenamefont {{Rob}}, \citenamefont {{Rosier-Lees}},
  \citenamefont {{Rowell}}, \citenamefont {{Rudak}}, \citenamefont {{Rulten}},
  \citenamefont {{Ruppel}}, \citenamefont {{Sahakian}}, \citenamefont
  {{Santangelo}}, \citenamefont {{Schlickeiser}}, \citenamefont {{Sch{\"o}ck}},
  \citenamefont {{Schr{\"o}der}}, \citenamefont {{Schwanke}}, \citenamefont
  {{Schwarzburg}}, \citenamefont {{Schwemmer}}, \citenamefont {{Shalchi}},
  \citenamefont {{Skilton}}, \citenamefont {{Sol}}, \citenamefont {{Spangler}},
  \citenamefont {{Stawarz}}, \citenamefont {{Steenkamp}}, \citenamefont
  {{Stegmann}}, \citenamefont {{Superina}}, \citenamefont {{Tam}},
  \citenamefont {{Tavernet}}, \citenamefont {{Terrier}}, \citenamefont
  {{Tibolla}}, \citenamefont {{van Eldik}}, \citenamefont {{Vasileiadis}},
  \citenamefont {{Venter}}, \citenamefont {{Vialle}}, \citenamefont
  {{Vincent}}, \citenamefont {{Vivier}}, \citenamefont {{V{\"o}lk}},
  \citenamefont {{Volpe}}, \citenamefont {{Wagner}}, \citenamefont {{Ward}},
  \citenamefont {{Zdziarski}},\ and\ \citenamefont
  {{Zech}}}]{2009ApJ...691..175A}%
  \BibitemOpen
  \bibfield  {author} {\bibinfo {author} {\bibnamefont {{Aharonian}},
  \bibfnamefont {F.}}, \bibinfo {author} {\bibnamefont {{Akhperjanian}},
  \bibfnamefont {A.~G.}}, \bibinfo {author} {\bibnamefont {{de Almeida}},
  \bibfnamefont {U.~B.}},  \emph {et~al.},\ }\href {\doibase
  10.1088/0004-637X/691/1/175} {\bibfield  {journal} {\bibinfo  {journal}
  {\apj}\ }\textbf {\bibinfo {volume} {691}},\ \bibinfo {pages} {175} (\bibinfo
  {year} {2009}{\natexlab{c}})},\ \Eprint {http://arxiv.org/abs/0809.3894}
  {arXiv:0809.3894} \BibitemShut {NoStop}%
\bibitem [{\citenamefont {Ajello}\ \emph {et~al.}(2015)\citenamefont {Ajello},
  \citenamefont {Gasparrini}, \citenamefont {S\'anchez-Conde}, \citenamefont
  {Zaharijas}, \citenamefont {Gustafsson}, \citenamefont {Cohen-Tanugi},
  \citenamefont {Dermer}, \citenamefont {Inoue}, \citenamefont {Hartmann},
  \citenamefont {Ackermann}, \citenamefont {Bechtol}, \citenamefont
  {Franckowiak}, \citenamefont {Reimer}, \citenamefont {Romani},\ and\
  \citenamefont {Strong}}]{Ajello2015}%
  \BibitemOpen
  \bibfield  {author} {\bibinfo {author} {\bibnamefont {Ajello}, \bibfnamefont
  {M.}}, \bibinfo {author} {\bibnamefont {Gasparrini}, \bibfnamefont {D.}},
  \bibinfo {author} {\bibnamefont {S\'anchez-Conde}, \bibfnamefont {M.}},
  \emph {et~al.},\ }\href {http://adsabs.harvard.edu/abs/2015arXiv150105301A}
  {\enquote {\bibinfo {title} {The origin of the extragalactic gamma-ray
  background and implications for dark-matter annihilation},}\ } (\bibinfo
  {year} {2015}),\ \bibinfo {note} {accepted for publication in
  ApJL}\BibitemShut {NoStop}%
\bibitem [{\citenamefont {{Akerib}}\ \emph {et~al.}(2014)\citenamefont
  {{Akerib}}, \citenamefont {{Araujo}}, \citenamefont {{Bai}}, \citenamefont
  {{Bailey}}, \citenamefont {{Balajthy}}, \citenamefont {{Bedikian}},
  \citenamefont {{Bernard}}, \citenamefont {{Bernstein}}, \citenamefont
  {{Bolozdynya}}, \citenamefont {{Bradley}}, \citenamefont {{Byram}},
  \citenamefont {{Cahn}}, \citenamefont {{Carmona-Benitez}}, \citenamefont
  {{Chan}}, \citenamefont {{Chapman}}, \citenamefont {{Chiller}}, \citenamefont
  {{Chiller}}, \citenamefont {{Clark}}, \citenamefont {{Coffey}}, \citenamefont
  {{Currie}}, \citenamefont {{Curioni}}, \citenamefont {{Dazeley}},
  \citenamefont {{de Viveiros}}, \citenamefont {{Dobi}}, \citenamefont
  {{Dobson}}, \citenamefont {{Dragowsky}}, \citenamefont {{Druszkiewicz}},
  \citenamefont {{Edwards}}, \citenamefont {{Faham}}, \citenamefont
  {{Fiorucci}}, \citenamefont {{Flores}}, \citenamefont {{Gaitskell}},
  \citenamefont {{Gehman}}, \citenamefont {{Ghag}}, \citenamefont {{Gibson}},
  \citenamefont {{Gilchriese}}, \citenamefont {{Hall}}, \citenamefont
  {{Hanhardt}}, \citenamefont {{Hertel}}, \citenamefont {{Horn}}, \citenamefont
  {{Huang}}, \citenamefont {{Ihm}}, \citenamefont {{Jacobsen}}, \citenamefont
  {{Kastens}}, \citenamefont {{Kazkaz}}, \citenamefont {{Knoche}},
  \citenamefont {{Kyre}}, \citenamefont {{Lander}}, \citenamefont {{Larsen}},
  \citenamefont {{Lee}}, \citenamefont {{Leonard}}, \citenamefont {{Lesko}},
  \citenamefont {{Lindote}}, \citenamefont {{Lopes}}, \citenamefont
  {{Lyashenko}}, \citenamefont {{Malling}}, \citenamefont {{Mannino}},
  \citenamefont {{McKinsey}}, \citenamefont {{Mei}}, \citenamefont {{Mock}},
  \citenamefont {{Moongweluwan}}, \citenamefont {{Morad}}, \citenamefont
  {{Morii}}, \citenamefont {{Murphy}}, \citenamefont {{Nehrkorn}},
  \citenamefont {{Nelson}}, \citenamefont {{Neves}}, \citenamefont {{Nikkel}},
  \citenamefont {{Ott}}, \citenamefont {{Pangilinan}}, \citenamefont
  {{Parker}}, \citenamefont {{Pease}}, \citenamefont {{Pech}}, \citenamefont
  {{Phelps}}, \citenamefont {{Reichhart}}, \citenamefont {{Shutt}},
  \citenamefont {{Silva}}, \citenamefont {{Skulski}}, \citenamefont {{Sofka}},
  \citenamefont {{Solovov}}, \citenamefont {{Sorensen}}, \citenamefont
  {{Stiegler}}, \citenamefont {{O`Sullivan}}, \citenamefont {{Sumner}},
  \citenamefont {{Svoboda}}, \citenamefont {{Sweany}}, \citenamefont
  {{Szydagis}}, \citenamefont {{Taylor}}, \citenamefont {{Tennyson}},
  \citenamefont {{Tiedt}}, \citenamefont {{Tripathi}}, \citenamefont
  {{Uvarov}}, \citenamefont {{Verbus}}, \citenamefont {{Walsh}}, \citenamefont
  {{Webb}}, \citenamefont {{White}}, \citenamefont {{White}}, \citenamefont
  {{Witherell}}, \citenamefont {{Wlasenko}}, \citenamefont {{Wolfs}},
  \citenamefont {{Woods}},\ and\ \citenamefont {{Zhang}}}]{Akerib:2013tjd}%
  \BibitemOpen
  \bibfield  {author} {\bibinfo {author} {\bibnamefont {{Akerib}},
  \bibfnamefont {D.~S.}}, \bibinfo {author} {\bibnamefont {{Araujo}},
  \bibfnamefont {H.~M.}}, \bibinfo {author} {\bibnamefont {{Bai}},
  \bibfnamefont {X.}},  \emph {et~al.} (\bibinfo {collaboration} {LUX
  Collaboration}),\ }\href {\doibase 10.1103/PhysRevLett.112.091303} {\bibfield
   {journal} {\bibinfo  {journal} {Phys.Rev.Lett.}\ }\textbf {\bibinfo {volume}
  {112}},\ \bibinfo {pages} {091303} (\bibinfo {year} {2014})},\ \Eprint
  {http://arxiv.org/abs/1310.8214} {arXiv:1310.8214 [astro-ph.CO]} \BibitemShut
  {NoStop}%
%%CITATION = ARXIV:1310.8214;%%
\bibitem [{\citenamefont {{Albert}}\ \emph {et~al.}(2014)\citenamefont
  {{Albert}}, \citenamefont {{G{\'o}mez-Vargas}}, \citenamefont {{Grefe}},
  \citenamefont {{Mu{\~n}oz}}, \citenamefont {{Weniger}}, \citenamefont
  {{Bloom}}, \citenamefont {{Charles}}, \citenamefont {{Mazziotta}},\ and\
  \citenamefont {{Morselli}}}]{Albert:2014hwa}%
  \BibitemOpen
  \bibfield  {author} {\bibinfo {author} {\bibnamefont {{Albert}},
  \bibfnamefont {A.}}, \bibinfo {author} {\bibnamefont {{G{\'o}mez-Vargas}},
  \bibfnamefont {G.~A.}}, \bibinfo {author} {\bibnamefont {{Grefe}},
  \bibfnamefont {M.}},  \emph {et~al.},\ }\href {\doibase
  10.1088/1475-7516/2014/10/023} {\bibfield  {journal} {\bibinfo  {journal}
  {\jcap}\ }\textbf {\bibinfo {volume} {10}},\ \bibinfo {eid} {023} (\bibinfo
  {year} {2014})},\ \Eprint {http://arxiv.org/abs/1406.3430} {arXiv:1406.3430
  [astro-ph.HE]} \BibitemShut {NoStop}%
\bibitem [{\citenamefont {{Albert}}\ \emph {et~al.}(2006)\citenamefont
  {{Albert}}, \citenamefont {{Aliu}}, \citenamefont {{Anderhub}}, \citenamefont
  {{Antoranz}}, \citenamefont {{Armada}}, \citenamefont {{Asensio}},
  \citenamefont {{Baixeras}}, \citenamefont {{Barrio}}, \citenamefont
  {{Bartelt}}, \citenamefont {{Bartko}}, \citenamefont {{Bastieri}},
  \citenamefont {{Bavikadi}}, \citenamefont {{Bednarek}}, \citenamefont
  {{Berger}}, \citenamefont {{Bigongiari}}, \citenamefont {{Biland}},
  \citenamefont {{Bisesi}}, \citenamefont {{Bock}}, \citenamefont {{Bretz}},
  \citenamefont {{Britvitch}}, \citenamefont {{Camara}}, \citenamefont
  {{Chilingarian}}, \citenamefont {{Ciprini}}, \citenamefont {{Coarasa}},
  \citenamefont {{Commichau}}, \citenamefont {{Contreras}}, \citenamefont
  {{Cortina}}, \citenamefont {{Curtef}}, \citenamefont {{Danielyan}},
  \citenamefont {{Dazzi}}, \citenamefont {{De Angelis}}, \citenamefont {{de los
  Reyes}}, \citenamefont {{De Lotto}}, \citenamefont
  {{Domingo-Santamar{\'{\i}}a}}, \citenamefont {{Dorner}}, \citenamefont
  {{Doro}}, \citenamefont {{Errando}}, \citenamefont {{Fagiolini}},
  \citenamefont {{Ferenc}}, \citenamefont {{Fern{\'a}ndez}}, \citenamefont
  {{Firpo}}, \citenamefont {{Flix}}, \citenamefont {{Fonseca}}, \citenamefont
  {{Font}}, \citenamefont {{Galante}}, \citenamefont {{Garczarczyk}},
  \citenamefont {{Gaug}}, \citenamefont {{Giller}}, \citenamefont {{Goebel}},
  \citenamefont {{Hakobyan}}, \citenamefont {{Hayashida}}, \citenamefont
  {{Hengstebeck}}, \citenamefont {{H{\"o}hne}}, \citenamefont {{Hose}},
  \citenamefont {{Jacon}}, \citenamefont {{Kalekin}}, \citenamefont
  {{Kranich}}, \citenamefont {{Laille}}, \citenamefont {{Lenisa}},
  \citenamefont {{Liebing}}, \citenamefont {{Lindfors}}, \citenamefont
  {{Longo}}, \citenamefont {{L{\'o}pez}}, \citenamefont {{L{\'o}pez}},
  \citenamefont {{Lorenz}}, \citenamefont {{Lucarelli}}, \citenamefont
  {{Majumdar}}, \citenamefont {{Maneva}}, \citenamefont {{Mannheim}},
  \citenamefont {{Mariotti}}, \citenamefont {{Mart{\'{\i}}nez}}, \citenamefont
  {{Mase}}, \citenamefont {{Mazin}}, \citenamefont {{Merck}}, \citenamefont
  {{Meucci}}, \citenamefont {{Meyer}}, \citenamefont {{Miranda}}, \citenamefont
  {{Mirzoyan}}, \citenamefont {{Mizobuchi}}, \citenamefont {{Moralejo}},
  \citenamefont {{Nilsson}}, \citenamefont {{O{\~n}a-Wilhelmi}}, \citenamefont
  {{Ordu{\~n}a}}, \citenamefont {{Otte}}, \citenamefont {{Oya}}, \citenamefont
  {{Paneque}}, \citenamefont {{Paoletti}}, \citenamefont {{Pasanen}},
  \citenamefont {{Pascoli}}, \citenamefont {{Pauss}}, \citenamefont {{Pavel}},
  \citenamefont {{Pegna}}, \citenamefont {{Peruzzo}}, \citenamefont
  {{Piccioli}}, \citenamefont {{Prandini}}, \citenamefont {{Rico}},
  \citenamefont {{Rhode}}, \citenamefont {{Riegel}}, \citenamefont {{Rissi}},
  \citenamefont {{Robert}}, \citenamefont {{R{\"u}gamer}}, \citenamefont
  {{Saggion}}, \citenamefont {{S{\'a}nchez}}, \citenamefont {{Sartori}},
  \citenamefont {{Scalzotto}}, \citenamefont {{Schmitt}}, \citenamefont
  {{Schweizer}}, \citenamefont {{Shayduk}}, \citenamefont {{Shinozaki}},
  \citenamefont {{Shore}}, \citenamefont {{Sidro}}, \citenamefont
  {{Sillanp{\"a}{\"a}}}, \citenamefont {{Sobczynska}}, \citenamefont
  {{Stamerra}}, \citenamefont {{Stepanian}}, \citenamefont {{Stark}},
  \citenamefont {{Takalo}}, \citenamefont {{Temnikov}}, \citenamefont
  {{Tescaro}}, \citenamefont {{Teshima}}, \citenamefont {{Tonello}},
  \citenamefont {{Torres}}, \citenamefont {{Torres}}, \citenamefont {{Turini}},
  \citenamefont {{Vankov}}, \citenamefont {{Vardanyan}}, \citenamefont
  {{Vitale}}, \citenamefont {{Wagner}}, \citenamefont {{Wibig}}, \citenamefont
  {{Wittek}},\ and\ \citenamefont {{Zapatero}}}]{2006ApJ...638L.101A}%
  \BibitemOpen
  \bibfield  {author} {\bibinfo {author} {\bibnamefont {{Albert}},
  \bibfnamefont {J.}}, \bibinfo {author} {\bibnamefont {{Aliu}}, \bibfnamefont
  {E.}}, \bibinfo {author} {\bibnamefont {{Anderhub}}, \bibfnamefont {H.}},
  \emph {et~al.},\ }\href {\doibase 10.1086/501164} {\bibfield  {journal}
  {\bibinfo  {journal} {\apjl}\ }\textbf {\bibinfo {volume} {638}},\ \bibinfo
  {pages} {L101} (\bibinfo {year} {2006})},\ \Eprint
  {http://arxiv.org/abs/astro-ph/0512469} {astro-ph/0512469} \BibitemShut
  {NoStop}%
\bibitem [{\citenamefont {Albert}\ \emph {et~al.}(2008)\citenamefont {Albert}
  \emph {et~al.}}]{Albert:2007xg}%
  \BibitemOpen
  \bibfield  {author} {\bibinfo {author} {\bibnamefont {Albert}, \bibfnamefont
  {J.}} \emph {et~al.} (\bibinfo {collaboration} {MAGIC Collaboration}),\
  }\href {\doibase 10.1086/529135} {\bibfield  {journal} {\bibinfo  {journal}
  {Astrophys.J.}\ }\textbf {\bibinfo {volume} {679}},\ \bibinfo {pages} {428}
  (\bibinfo {year} {2008})},\ \Eprint {http://arxiv.org/abs/0711.2574}
  {arXiv:0711.2574 [astro-ph]} \BibitemShut {NoStop}%
%%CITATION = ARXIV:0711.2574;%%
\bibitem [{\citenamefont {{Aleksi{\'c}}}\ \emph {et~al.}(2012)\citenamefont
  {{Aleksi{\'c}}}, \citenamefont {{Alvarez}}, \citenamefont {{Antonelli}},
  \citenamefont {{Antoranz}}, \citenamefont {{Asensio}}, \citenamefont
  {{Backes}}, \citenamefont {{Barres de Almeida}}, \citenamefont {{Barrio}},
  \citenamefont {{Bastieri}}, \citenamefont {{Becerra Gonz{\'a}lez}},
  \citenamefont {{Bednarek}}, \citenamefont {{Berdyugin}}, \citenamefont
  {{Berger}}, \citenamefont {{Bernardini}}, \citenamefont {{Biland}},
  \citenamefont {{Blanch}}, \citenamefont {{Bock}}, \citenamefont {{Boller}},
  \citenamefont {{Bonnoli}}, \citenamefont {{Borla Tridon}}, \citenamefont
  {{Braun}}, \citenamefont {{Bretz}}, \citenamefont {{Ca{\~n}ellas}},
  \citenamefont {{Carmona}}, \citenamefont {{Carosi}}, \citenamefont {{Colin}},
  \citenamefont {{Colombo}}, \citenamefont {{Contreras}}, \citenamefont
  {{Cortina}}, \citenamefont {{Cossio}}, \citenamefont {{Covino}},
  \citenamefont {{Dazzi}}, \citenamefont {{de Angelis}}, \citenamefont {{de
  Caneva}}, \citenamefont {{de Cea Del Pozo}}, \citenamefont {{de Lotto}},
  \citenamefont {{Delgado Mendez}}, \citenamefont {{Diago Ortega}},
  \citenamefont {{Doert}}, \citenamefont {{Dom{\'{\i}}nguez}}, \citenamefont
  {{Dominis Prester}}, \citenamefont {{Dorner}}, \citenamefont {{Doro}},
  \citenamefont {{Eisenacher}}, \citenamefont {{Elsaesser}}, \citenamefont
  {{Ferenc}}, \citenamefont {{Fonseca}}, \citenamefont {{Font}}, \citenamefont
  {{Fruck}}, \citenamefont {{Garc{\'{\i}}a L{\'o}pez}}, \citenamefont
  {{Garczarczyk}}, \citenamefont {{Garrido}}, \citenamefont {{Giavitto}},
  \citenamefont {{Godinovi{\'c}}}, \citenamefont {{Gozzini}}, \citenamefont
  {{Hadasch}}, \citenamefont {{H{\"a}fner}}, \citenamefont {{Herrero}},
  \citenamefont {{Hildebrand}}, \citenamefont {{H{\"o}hne-M{\"o}nch}},
  \citenamefont {{Hose}}, \citenamefont {{Hrupec}}, \citenamefont {{Jogler}},
  \citenamefont {{Kellermann}}, \citenamefont {{Klepser}}, \citenamefont
  {{Kr{\"a}henb{\"u}hl}}, \citenamefont {{Krause}}, \citenamefont {{Kushida}},
  \citenamefont {{La Barbera}}, \citenamefont {{Lelas}}, \citenamefont
  {{Leonardo}}, \citenamefont {{Lewandowska}}, \citenamefont {{Lindfors}},
  \citenamefont {{Lombardi}}, \citenamefont {{L{\'o}pez}}, \citenamefont
  {{L{\'o}pez}}, \citenamefont {{L{\'o}pez-Oramas}}, \citenamefont {{Lorenz}},
  \citenamefont {{Makariev}}, \citenamefont {{Maneva}}, \citenamefont
  {{Mankuzhiyil}}, \citenamefont {{Mannheim}}, \citenamefont {{Maraschi}},
  \citenamefont {{Mariotti}}, \citenamefont {{Mart{\'{\i}}nez}}, \citenamefont
  {{Mazin}}, \citenamefont {{Meucci}}, \citenamefont {{Miranda}}, \citenamefont
  {{Mirzoyan}}, \citenamefont {{Mold{\'o}n}}, \citenamefont {{Moralejo}},
  \citenamefont {{Munar-Adrover}}, \citenamefont {{Niedzwiecki}}, \citenamefont
  {{Nieto}}, \citenamefont {{Nilsson}}, \citenamefont {{Nowak}}, \citenamefont
  {{Orito}}, \citenamefont {{Paiano}}, \citenamefont {{Paneque}}, \citenamefont
  {{Paoletti}}, \citenamefont {{Pardo}}, \citenamefont {{Paredes}},
  \citenamefont {{Partini}}, \citenamefont {{Perez-Torres}}, \citenamefont
  {{Persic}}, \citenamefont {{Peruzzo}}, \citenamefont {{Pilia}}, \citenamefont
  {{Pochon}}, \citenamefont {{Prada}}, \citenamefont {{Prada Moroni}},
  \citenamefont {{Prandini}}, \citenamefont {{Puerto Gimenez}}, \citenamefont
  {{Puljak}}, \citenamefont {{Reichardt}}, \citenamefont {{Reinthal}},
  \citenamefont {{Rhode}}, \citenamefont {{Rib{\'o}}}, \citenamefont {{Rico}},
  \citenamefont {{R{\"u}gamer}}, \citenamefont {{Saggion}}, \citenamefont
  {{Saito}}, \citenamefont {{Saito}}, \citenamefont {{Salvati}}, \citenamefont
  {{Satalecka}}, \citenamefont {{Scalzotto}}, \citenamefont {{Scapin}},
  \citenamefont {{Schultz}}, \citenamefont {{Schweizer}}, \citenamefont
  {{Shayduk}}, \citenamefont {{Shore}}, \citenamefont {{Sillanp{\"a}{\"a}}},
  \citenamefont {{Sitarek}}, \citenamefont {{Snidaric}}, \citenamefont
  {{Sobczynska}}, \citenamefont {{Spanier}}, \citenamefont {{Spiro}},
  \citenamefont {{Stamatescu}}, \citenamefont {{Stamerra}}, \citenamefont
  {{Steinke}}, \citenamefont {{Storz}}, \citenamefont {{Strah}}, \citenamefont
  {{Sun}}, \citenamefont {{Suri{\'c}}}, \citenamefont {{Takalo}}, \citenamefont
  {{Takami}}, \citenamefont {{Tavecchio}}, \citenamefont {{Temnikov}},
  \citenamefont {{Terzi{\'c}}}, \citenamefont {{Tescaro}}, \citenamefont
  {{Teshima}}, \citenamefont {{Tibolla}}, \citenamefont {{Torres}},
  \citenamefont {{Treves}}, \citenamefont {{Uellenbeck}}, \citenamefont
  {{Vankov}}, \citenamefont {{Vogler}}, \citenamefont {{Wagner}}, \citenamefont
  {{Weitzel}}, \citenamefont {{Zabalza}}, \citenamefont {{Zandanel}},
  \citenamefont {{Zanin}}, \citenamefont {{MAGIC Collaboration}}, \citenamefont
  {{Pfrommer}},\ and\ \citenamefont {{Pinzke}}}]{2012A&A...541A..99A}%
  \BibitemOpen
  \bibfield  {author} {\bibinfo {author} {\bibnamefont {{Aleksi{\'c}}},
  \bibfnamefont {J.}}, \bibinfo {author} {\bibnamefont {{Alvarez}},
  \bibfnamefont {E.~A.}}, \bibinfo {author} {\bibnamefont {{Antonelli}},
  \bibfnamefont {L.~A.}},  \emph {et~al.},\ }\href {\doibase
  10.1051/0004-6361/201118502} {\bibfield  {journal} {\bibinfo  {journal}
  {\aap}\ }\textbf {\bibinfo {volume} {541}},\ \bibinfo {eid} {A99} (\bibinfo
  {year} {2012})},\ \Eprint {http://arxiv.org/abs/1111.5544} {arXiv:1111.5544
  [astro-ph.HE]} \BibitemShut {NoStop}%
\bibitem [{\citenamefont {{Aleksi{\'c}}}\ \emph {et~al.}(2011)\citenamefont
  {{Aleksi{\'c}}}, \citenamefont {{Alvarez}}, \citenamefont {{Antonelli}},
  \citenamefont {{Antoranz}}, \citenamefont {{Asensio}}, \citenamefont
  {{Backes}}, \citenamefont {{Barrio}}, \citenamefont {{Bastieri}},
  \citenamefont {{Becerra Gonz{\'a}lez}}, \citenamefont {{Bednarek}},
  \citenamefont {{Berdyugin}}, \citenamefont {{Berger}}, \citenamefont
  {{Bernardini}}, \citenamefont {{Biland}}, \citenamefont {{Blanch}},
  \citenamefont {{Bock}}, \citenamefont {{Boller}}, \citenamefont {{Bonnoli}},
  \citenamefont {{Borla Tridon}}, \citenamefont {{Braun}}, \citenamefont
  {{Bretz}}, \citenamefont {{Ca{\~n}ellas}}, \citenamefont {{Carmona}},
  \citenamefont {{Carosi}}, \citenamefont {{Colin}}, \citenamefont {{Colombo}},
  \citenamefont {{Contreras}}, \citenamefont {{Cortina}}, \citenamefont
  {{Cossio}}, \citenamefont {{Covino}}, \citenamefont {{Dazzi}}, \citenamefont
  {{De Angelis}}, \citenamefont {{De Cea del Pozo}}, \citenamefont {{De
  Lotto}}, \citenamefont {{Delgado Mendez}}, \citenamefont {{Diago Ortega}},
  \citenamefont {{Doert}}, \citenamefont {{Dom{\'{\i}}nguez}}, \citenamefont
  {{Dominis Prester}}, \citenamefont {{Dorner}}, \citenamefont {{Doro}},
  \citenamefont {{Elsaesser}}, \citenamefont {{Ferenc}}, \citenamefont
  {{Fonseca}}, \citenamefont {{Font}}, \citenamefont {{Fruck}}, \citenamefont
  {{Garc{\'{\i}}a L{\'o}pez}}, \citenamefont {{Garczarczyk}}, \citenamefont
  {{Garrido}}, \citenamefont {{Giavitto}}, \citenamefont {{Godinovi{\'c}}},
  \citenamefont {{Hadasch}}, \citenamefont {{H{\"a}fner}}, \citenamefont
  {{Herrero}}, \citenamefont {{Hildebrand}}, \citenamefont
  {{H{\"o}hne-M{\"o}nch}}, \citenamefont {{Hose}}, \citenamefont {{Hrupec}},
  \citenamefont {{Huber}}, \citenamefont {{Jogler}}, \citenamefont {{Klepser}},
  \citenamefont {{Kr{\"a}henb{\"u}hl}}, \citenamefont {{Krause}}, \citenamefont
  {{La Barbera}}, \citenamefont {{Lelas}}, \citenamefont {{Leonardo}},
  \citenamefont {{Lindfors}}, \citenamefont {{Lombardi}}, \citenamefont
  {{L{\'o}pez}}, \citenamefont {{Lorenz}}, \citenamefont {{Makariev}},
  \citenamefont {{Maneva}}, \citenamefont {{Mankuzhiyil}}, \citenamefont
  {{Mannheim}}, \citenamefont {{Maraschi}}, \citenamefont {{Mariotti}},
  \citenamefont {{Mart{\'{\i}}nez}}, \citenamefont {{Mazin}}, \citenamefont
  {{Meucci}}, \citenamefont {{Miranda}}, \citenamefont {{Mirzoyan}},
  \citenamefont {{Miyamoto}}, \citenamefont {{Mold{\'o}n}}, \citenamefont
  {{Moralejo}}, \citenamefont {{Munar-Androver}}, \citenamefont {{Nieto}},
  \citenamefont {{Nilsson}}, \citenamefont {{Orito}}, \citenamefont {{Oya}},
  \citenamefont {{Paiano}}, \citenamefont {{Paneque}}, \citenamefont
  {{Paoletti}}, \citenamefont {{Pardo}}, \citenamefont {{Paredes}},
  \citenamefont {{Partini}}, \citenamefont {{Pasanen}}, \citenamefont
  {{Pauss}}, \citenamefont {{Perez-Torres}}, \citenamefont {{Persic}},
  \citenamefont {{Peruzzo}}, \citenamefont {{Pilia}}, \citenamefont {{Pochon}},
  \citenamefont {{Prada}}, \citenamefont {{Prada Moroni}}, \citenamefont
  {{Prandini}}, \citenamefont {{Puljak}}, \citenamefont {{Reichardt}},
  \citenamefont {{Reinthal}}, \citenamefont {{Rhode}}, \citenamefont
  {{Rib{\'o}}}, \citenamefont {{Rico}}, \citenamefont {{R{\"u}gamer}},
  \citenamefont {{Saggion}}, \citenamefont {{Saito}}, \citenamefont {{Saito}},
  \citenamefont {{Salvati}}, \citenamefont {{Satalecka}}, \citenamefont
  {{Scalzotto}}, \citenamefont {{Scapin}}, \citenamefont {{Schultz}},
  \citenamefont {{Schweizer}}, \citenamefont {{Shayduk}}, \citenamefont
  {{Shore}}, \citenamefont {{Sillanp{\"a}{\"a}}}, \citenamefont {{Sitarek}},
  \citenamefont {{Sobczynska}}, \citenamefont {{Spanier}}, \citenamefont
  {{Spiro}}, \citenamefont {{Stamerra}}, \citenamefont {{Steinke}},
  \citenamefont {{Storz}}, \citenamefont {{Strah}}, \citenamefont
  {{Suri{\'c}}}, \citenamefont {{Takalo}}, \citenamefont {{Takami}},
  \citenamefont {{Tavecchio}}, \citenamefont {{Temnikov}}, \citenamefont
  {{Terzi{\'c}}}, \citenamefont {{Tescaro}}, \citenamefont {{Teshima}},
  \citenamefont {{Thom}}, \citenamefont {{Tibolla}}, \citenamefont {{Torres}},
  \citenamefont {{Treves}}, \citenamefont {{Vankov}}, \citenamefont {{Vogler}},
  \citenamefont {{Wagner}}, \citenamefont {{Weitzel}}, \citenamefont
  {{Zabalza}}, \citenamefont {{Zandanel}}, \citenamefont {{Zanin}},
  \citenamefont {{Fornasa}}, \citenamefont {{Essig}}, \citenamefont
  {{Sehgal}},\ and\ \citenamefont {{Strigari}}}]{2011JCAP...06..035A}%
  \BibitemOpen
  \bibfield  {author} {\bibinfo {author} {\bibnamefont {{Aleksi{\'c}}},
  \bibfnamefont {J.}}, \bibinfo {author} {\bibnamefont {{Alvarez}},
  \bibfnamefont {E.~A.}}, \bibinfo {author} {\bibnamefont {{Antonelli}},
  \bibfnamefont {L.~A.}},  \emph {et~al.},\ }\href {\doibase
  10.1088/1475-7516/2011/06/035} {\bibfield  {journal} {\bibinfo  {journal}
  {\jcap}\ }\textbf {\bibinfo {volume} {6}},\ \bibinfo {eid} {035} (\bibinfo
  {year} {2011})},\ \Eprint {http://arxiv.org/abs/1103.0477} {arXiv:1103.0477
  [astro-ph.HE]} \BibitemShut {NoStop}%
\bibitem [{\citenamefont {{Aleksi{\'c}}}\ \emph {et~al.}(2014)\citenamefont
  {{Aleksi{\'c}}}, \citenamefont {{Ansoldi}}, \citenamefont {{Antonelli}},
  \citenamefont {{Antoranz}}, \citenamefont {{Babic}}, \citenamefont
  {{Bangale}}, \citenamefont {{Barres de Almeida}}, \citenamefont {{Barrio}},
  \citenamefont {{Becerra Gonz{\'a}lez}}, \citenamefont {{Bednarek}},
  \citenamefont {{Berger}}, \citenamefont {{Bernardini}}, \citenamefont
  {{Biland}}, \citenamefont {{Blanch}}, \citenamefont {{Bock}}, \citenamefont
  {{Bonnefoy}}, \citenamefont {{Bonnoli}}, \citenamefont {{Borracci}},
  \citenamefont {{Bretz}}, \citenamefont {{Carmona}}, \citenamefont {{Carosi}},
  \citenamefont {{Carreto Fidalgo}}, \citenamefont {{Colin}}, \citenamefont
  {{Colombo}}, \citenamefont {{Contreras}}, \citenamefont {{Cortina}},
  \citenamefont {{Covino}}, \citenamefont {{Da Vela}}, \citenamefont {{Dazzi}},
  \citenamefont {{De Angelis}}, \citenamefont {{De Caneva}}, \citenamefont {{De
  Lotto}}, \citenamefont {{Delgado Mendez}}, \citenamefont {{Doert}},
  \citenamefont {{Dom{\'{\i}}nguez}}, \citenamefont {{Dominis Prester}},
  \citenamefont {{Dorner}}, \citenamefont {{Doro}}, \citenamefont {{Einecke}},
  \citenamefont {{Eisenacher}}, \citenamefont {{Elsaesser}}, \citenamefont
  {{Farina}}, \citenamefont {{Ferenc}}, \citenamefont {{Fonseca}},
  \citenamefont {{Font}}, \citenamefont {{Frantzen}}, \citenamefont {{Fruck}},
  \citenamefont {{Garc{\'{\i}}a L{\'o}pez}}, \citenamefont {{Garczarczyk}},
  \citenamefont {{Garrido Terrats}}, \citenamefont {{Gaug}}, \citenamefont
  {{Giavitto}}, \citenamefont {{Godinovi{\'c}}}, \citenamefont {{Gonz{\'a}lez
  Mu{\~n}oz}}, \citenamefont {{Gozzini}}, \citenamefont {{Hadasch}},
  \citenamefont {{Hayashida}}, \citenamefont {{Herrero}}, \citenamefont
  {{Hildebrand}}, \citenamefont {{Hose}}, \citenamefont {{Hrupec}},
  \citenamefont {{Idec}}, \citenamefont {{Kadenius}}, \citenamefont
  {{Kellermann}}, \citenamefont {{Kodani}}, \citenamefont {{Konno}},
  \citenamefont {{Krause}}, \citenamefont {{Kubo}}, \citenamefont {{Kushida}},
  \citenamefont {{La Barbera}}, \citenamefont {{Lelas}}, \citenamefont
  {{Lewandowska}}, \citenamefont {{Lindfors}}, \citenamefont {{Lombardi}},
  \citenamefont {{L{\'o}pez}}, \citenamefont {{L{\'o}pez-Coto}}, \citenamefont
  {{L{\'o}pez-Oramas}}, \citenamefont {{Lorenz}}, \citenamefont {{Lozano}},
  \citenamefont {{Makariev}}, \citenamefont {{Mallot}}, \citenamefont
  {{Maneva}}, \citenamefont {{Mankuzhiyil}}, \citenamefont {{Mannheim}},
  \citenamefont {{Maraschi}}, \citenamefont {{Marcote}}, \citenamefont
  {{Mariotti}}, \citenamefont {{Mart{\'{\i}}nez}}, \citenamefont {{Mazin}},
  \citenamefont {{Menzel}}, \citenamefont {{Meucci}}, \citenamefont
  {{Miranda}}, \citenamefont {{Mirzoyan}}, \citenamefont {{Moralejo}},
  \citenamefont {{Munar-Adrover}}, \citenamefont {{Nakajima}}, \citenamefont
  {{Niedzwiecki}}, \citenamefont {{Nilsson}}, \citenamefont {{Nishijima}},
  \citenamefont {{Nowak}}, \citenamefont {{Orito}}, \citenamefont
  {{Overkemping}}, \citenamefont {{Paiano}}, \citenamefont {{Palatiello}},
  \citenamefont {{Paneque}}, \citenamefont {{Paoletti}}, \citenamefont
  {{Paredes}}, \citenamefont {{Paredes-Fortuny}}, \citenamefont {{Partini}},
  \citenamefont {{Persic}}, \citenamefont {{Prada}}, \citenamefont {{Prada
  Moroni}}, \citenamefont {{Prandini}}, \citenamefont {{Preziuso}},
  \citenamefont {{Puljak}}, \citenamefont {{Reinthal}}, \citenamefont
  {{Rhode}}, \citenamefont {{Rib{\'o}}}, \citenamefont {{Rico}}, \citenamefont
  {{Rodriguez Garcia}}, \citenamefont {{R{\"u}gamer}}, \citenamefont
  {{Saggion}}, \citenamefont {{Saito}}, \citenamefont {{Saito}}, \citenamefont
  {{Salvati}}, \citenamefont {{Satalecka}}, \citenamefont {{Scalzotto}},
  \citenamefont {{Scapin}}, \citenamefont {{Schultz}}, \citenamefont
  {{Schweizer}}, \citenamefont {{Sillanp{\"a}{\"a}}}, \citenamefont
  {{Sitarek}}, \citenamefont {{Snidaric}}, \citenamefont {{Sobczynska}},
  \citenamefont {{Spanier}}, \citenamefont {{Stamatescu}}, \citenamefont
  {{Stamerra}}, \citenamefont {{Steinbring}}, \citenamefont {{Storz}},
  \citenamefont {{Sun}}, \citenamefont {{Suri{\'c}}}, \citenamefont {{Takalo}},
  \citenamefont {{Takami}}, \citenamefont {{Tavecchio}}, \citenamefont
  {{Temnikov}}, \citenamefont {{Terzi{\'c}}}, \citenamefont {{Tescaro}},
  \citenamefont {{Teshima}}, \citenamefont {{Thaele}}, \citenamefont
  {{Tibolla}}, \citenamefont {{Torres}}, \citenamefont {{Toyama}},
  \citenamefont {{Treves}}, \citenamefont {{Uellenbeck}}, \citenamefont
  {{Vogler}}, \citenamefont {{Wagner}}, \citenamefont {{Zandanel}},
  \citenamefont {{Zanin}},\ and\ \citenamefont
  {{Ibarra}}}]{2014JCAP...02..008A}%
  \BibitemOpen
  \bibfield  {author} {\bibinfo {author} {\bibnamefont {{Aleksi{\'c}}},
  \bibfnamefont {J.}}, \bibinfo {author} {\bibnamefont {{Ansoldi}},
  \bibfnamefont {S.}}, \bibinfo {author} {\bibnamefont {{Antonelli}},
  \bibfnamefont {L.~A.}},  \emph {et~al.},\ }\href {\doibase
  10.1088/1475-7516/2014/02/008} {\bibfield  {journal} {\bibinfo  {journal}
  {\jcap}\ }\textbf {\bibinfo {volume} {2}},\ \bibinfo {eid} {008} (\bibinfo
  {year} {2014})},\ \Eprint {http://arxiv.org/abs/1312.1535} {arXiv:1312.1535
  [hep-ph]} \BibitemShut {NoStop}%
\bibitem [{\citenamefont {{Aleksi{\'c}}}\ \emph {et~al.}(2010)\citenamefont
  {{Aleksi{\'c}}}, \citenamefont {{Antonelli}}, \citenamefont {{Antoranz}},
  \citenamefont {{Backes}}, \citenamefont {{Baixeras}}, \citenamefont
  {{Balestra}}, \citenamefont {{Barrio}}, \citenamefont {{Bastieri}},
  \citenamefont {{Becerra Gonz{\'a}lez}}, \citenamefont {{Bednarek}},
  \citenamefont {{Berdyugin}}, \citenamefont {{Berger}}, \citenamefont
  {{Bernardini}}, \citenamefont {{Biland}}, \citenamefont {{Bock}},
  \citenamefont {{Bonnoli}}, \citenamefont {{Bordas}}, \citenamefont {{Borla
  Tridon}}, \citenamefont {{Bosch-Ramon}}, \citenamefont {{Bose}},
  \citenamefont {{Braun}}, \citenamefont {{Bretz}}, \citenamefont {{Britzger}},
  \citenamefont {{Camara}}, \citenamefont {{Carmona}}, \citenamefont
  {{Carosi}}, \citenamefont {{Colin}}, \citenamefont {{Commichau}},
  \citenamefont {{Contreras}}, \citenamefont {{Cortina}}, \citenamefont
  {{Costado}}, \citenamefont {{Covino}}, \citenamefont {{Dazzi}}, \citenamefont
  {{De Angelis}}, \citenamefont {{De Cea del Pozo}}, \citenamefont {{De los
  Reyes}}, \citenamefont {{De Lotto}}, \citenamefont {{De Maria}},
  \citenamefont {{De Sabata}}, \citenamefont {{Delgado Mendez}}, \citenamefont
  {{Doert}}, \citenamefont {{Dom{\'{\i}}nguez}}, \citenamefont {{Dominis
  Prester}}, \citenamefont {{Dorner}}, \citenamefont {{Doro}}, \citenamefont
  {{Elsaesser}}, \citenamefont {{Errando}}, \citenamefont {{Ferenc}},
  \citenamefont {{Fonseca}}, \citenamefont {{Font}}, \citenamefont {{Galante}},
  \citenamefont {{Garc{\'{\i}}a L{\'o}pez}}, \citenamefont {{Garczarczyk}},
  \citenamefont {{Gaug}}, \citenamefont {{Godinovic}}, \citenamefont
  {{Hadasch}}, \citenamefont {{Herrero}}, \citenamefont {{Hildebrand}},
  \citenamefont {{H{\"o}hne-M{\"o}nch}}, \citenamefont {{Hose}}, \citenamefont
  {{Hrupec}}, \citenamefont {{Hsu}}, \citenamefont {{Jogler}}, \citenamefont
  {{Klepser}}, \citenamefont {{Kr{\"a}henb{\"u}hl}}, \citenamefont {{Kranich}},
  \citenamefont {{La Barbera}}, \citenamefont {{Laille}}, \citenamefont
  {{Leonardo}}, \citenamefont {{Lindfors}}, \citenamefont {{Lombardi}},
  \citenamefont {{Longo}}, \citenamefont {{L{\'o}pez}}, \citenamefont
  {{Lorenz}}, \citenamefont {{Majumdar}}, \citenamefont {{Maneva}},
  \citenamefont {{Mankuzhiyil}}, \citenamefont {{Mannheim}}, \citenamefont
  {{Maraschi}}, \citenamefont {{Mariotti}}, \citenamefont {{Mart{\'{\i}}nez}},
  \citenamefont {{Mazin}}, \citenamefont {{Meucci}}, \citenamefont {{Miranda}},
  \citenamefont {{Mirzoyan}}, \citenamefont {{Miyamoto}}, \citenamefont
  {{Mold{\'o}n}}, \citenamefont {{Moles}}, \citenamefont {{Moralejo}},
  \citenamefont {{Nieto}}, \citenamefont {{Nilsson}}, \citenamefont
  {{Ninkovic}}, \citenamefont {{Orito}}, \citenamefont {{Oya}}, \citenamefont
  {{Paiano}}, \citenamefont {{Paoletti}}, \citenamefont {{Paredes}},
  \citenamefont {{Partini}}, \citenamefont {{Pasanen}}, \citenamefont
  {{Pascoli}}, \citenamefont {{Pauss}}, \citenamefont {{Pegna}}, \citenamefont
  {{Perez-Torres}}, \citenamefont {{Persic}}, \citenamefont {{Peruzzo}},
  \citenamefont {{Prada}}, \citenamefont {{Prandini}}, \citenamefont
  {{Puchades}}, \citenamefont {{Puljak}}, \citenamefont {{Reichardt}},
  \citenamefont {{Rhode}}, \citenamefont {{Rib{\'o}}}, \citenamefont {{Rico}},
  \citenamefont {{Rissi}}, \citenamefont {{R{\"u}gamer}}, \citenamefont
  {{Saggion}}, \citenamefont {{Saito}}, \citenamefont {{Salvati}},
  \citenamefont {{S{\'a}nchez-Conde}}, \citenamefont {{Satalecka}},
  \citenamefont {{Scalzotto}}, \citenamefont {{Scapin}}, \citenamefont
  {{Schultz}}, \citenamefont {{Schweizer}}, \citenamefont {{Shayduk}},
  \citenamefont {{Shore}}, \citenamefont {{Sierpowska-Bartosik}}, \citenamefont
  {{Sillanp{\"a}{\"a}}}, \citenamefont {{Sitarek}}, \citenamefont
  {{Sobczynska}}, \citenamefont {{Spanier}}, \citenamefont {{Spiro}},
  \citenamefont {{Stamerra}}, \citenamefont {{Steinke}}, \citenamefont
  {{Struebig}}, \citenamefont {{Suric}}, \citenamefont {{Takalo}},
  \citenamefont {{Tavecchio}}, \citenamefont {{Temnikov}}, \citenamefont
  {{Terzic}}, \citenamefont {{Tescaro}}, \citenamefont {{Teshima}},
  \citenamefont {{Torres}}, \citenamefont {{Vankov}}, \citenamefont {{Wagner}},
  \citenamefont {{Zabalza}}, \citenamefont {{Zandanel}}, \citenamefont
  {{Zanin}}, \citenamefont {{Zapatero}}, \citenamefont {{Pfrommer}},
  \citenamefont {{Pinzke}}, \citenamefont {{En{\ss}lin}}, \citenamefont
  {{Inoue}}, \citenamefont {{Ghisellini}},\ and\ \citenamefont {{MAGIC
  Collaboration}}}]{2010ApJ...710..634A}%
  \BibitemOpen
  \bibfield  {author} {\bibinfo {author} {\bibnamefont {{Aleksi{\'c}}},
  \bibfnamefont {J.}}, \bibinfo {author} {\bibnamefont {{Antonelli}},
  \bibfnamefont {L.~A.}}, \bibinfo {author} {\bibnamefont {{Antoranz}},
  \bibfnamefont {P.}},  \emph {et~al.},\ }\href {\doibase
  10.1088/0004-637X/710/1/634} {\bibfield  {journal} {\bibinfo  {journal}
  {\apj}\ }\textbf {\bibinfo {volume} {710}},\ \bibinfo {pages} {634} (\bibinfo
  {year} {2010})},\ \Eprint {http://arxiv.org/abs/0909.3267} {arXiv:0909.3267
  [astro-ph.HE]} \BibitemShut {NoStop}%
\bibitem [{\citenamefont {{Aleksi{\'c}}}, \citenamefont {{Rico}},\ and\
  \citenamefont {{Martinez}}(2012)}]{2012JCAP...10..032A}%
  \BibitemOpen
  \bibfield  {author} {\bibinfo {author} {\bibnamefont {{Aleksi{\'c}}},
  \bibfnamefont {J.}}, \bibinfo {author} {\bibnamefont {{Rico}}, \bibfnamefont
  {J.}}, \ and\ \bibinfo {author} {\bibnamefont {{Martinez}}, \bibfnamefont
  {M.}},\ }\href {\doibase 10.1088/1475-7516/2012/10/032} {\bibfield  {journal}
  {\bibinfo  {journal} {\jcap}\ }\textbf {\bibinfo {volume} {10}},\ \bibinfo
  {eid} {032} (\bibinfo {year} {2012})},\ \Eprint
  {http://arxiv.org/abs/1209.5589} {arXiv:1209.5589 [astro-ph.HE]} \BibitemShut
  {NoStop}%
\bibitem [{\citenamefont {Aliu}\ \emph {et~al.}(2012)\citenamefont {Aliu},
  \citenamefont {Archambault}, \citenamefont {Arlen}, \citenamefont {Aune},
  \citenamefont {Beilicke}, \citenamefont {Benbow}, \citenamefont {Bouvier},
  \citenamefont {Bradbury}, \citenamefont {Buckley}, \citenamefont {Bugaev},
  \citenamefont {Byrum}, \citenamefont {Cannon}, \citenamefont {Cesarini},
  \citenamefont {Christiansen}, \citenamefont {Ciupik}, \citenamefont
  {Collins-Hughes}, \citenamefont {Connolly}, \citenamefont {Cui},
  \citenamefont {Decerprit}, \citenamefont {Dickherber}, \citenamefont {Dumm},
  \citenamefont {Errando}, \citenamefont {Falcone}, \citenamefont {Feng},
  \citenamefont {Ferrer}, \citenamefont {Finley}, \citenamefont {Finnegan},
  \citenamefont {Fortson}, \citenamefont {Furniss}, \citenamefont {Galante},
  \citenamefont {Gall}, \citenamefont {Godambe}, \citenamefont {Griffin},
  \citenamefont {Grube}, \citenamefont {Gyuk}, \citenamefont {Hanna},
  \citenamefont {Holder}, \citenamefont {Huan}, \citenamefont {Hughes},
  \citenamefont {Humensky}, \citenamefont {Kaaret}, \citenamefont {Karlsson},
  \citenamefont {Kertzman}, \citenamefont {Khassen}, \citenamefont {Kieda},
  \citenamefont {Krawczynski}, \citenamefont {Krennrich}, \citenamefont {Lee},
  \citenamefont {Madhavan}, \citenamefont {Maier}, \citenamefont {Majumdar},
  \citenamefont {McArthur}, \citenamefont {McCann}, \citenamefont {Moriarty},
  \citenamefont {Mukherjee}, \citenamefont {Ong}, \citenamefont {Orr},
  \citenamefont {Otte}, \citenamefont {Park}, \citenamefont {Perkins},
  \citenamefont {Pohl}, \citenamefont {Prokoph}, \citenamefont {Quinn},
  \citenamefont {Ragan}, \citenamefont {Reyes}, \citenamefont {Reynolds},
  \citenamefont {Roache}, \citenamefont {Rose}, \citenamefont {Ruppel},
  \citenamefont {Saxon}, \citenamefont {Schroedter}, \citenamefont {Sembroski},
  \citenamefont {Senturk}, \citenamefont {Skole}, \citenamefont {Smith},
  \citenamefont {Staszak}, \citenamefont {Telezhinsky}, \citenamefont {Tesic},
  \citenamefont {Theiling}, \citenamefont {Thibadeau}, \citenamefont
  {Tsurusaki}, \citenamefont {Varlotta}, \citenamefont {Vassiliev},
  \citenamefont {Vincent}, \citenamefont {Vivier}, \citenamefont {Wagner},
  \citenamefont {Wakely}, \citenamefont {Ward}, \citenamefont {Weekes},
  \citenamefont {Weinstein}, \citenamefont {Weisgarber}, \citenamefont
  {Williams}, \citenamefont {Zitzer},\ and\ \citenamefont
  {Collaboration}}]{1202.2144v1}%
  \BibitemOpen
  \bibfield  {author} {\bibinfo {author} {\bibnamefont {Aliu}, \bibfnamefont
  {E.}}, \bibinfo {author} {\bibnamefont {Archambault}, \bibfnamefont {S.}},
  \bibinfo {author} {\bibnamefont {Arlen}, \bibfnamefont {T.}},  \emph
  {et~al.},\ }\href {\doibase 10.1103/PhysRevD.85.062001} {\bibfield  {journal}
  {\bibinfo  {journal} {Phys. Rev. D 85, 062001 (2012)}\ } (\bibinfo {year}
  {2012}),\ 10.1103/PhysRevD.85.062001},\ \Eprint
  {http://arxiv.org/abs/1202.2144v1} {arXiv:1202.2144v1 [astro-ph.HE]}
  \BibitemShut {NoStop}%
\bibitem [{\citenamefont {Aliu}\ \emph {et~al.}(2009)\citenamefont {Aliu} \emph
  {et~al.}}]{Aliu:2008ny}%
  \BibitemOpen
  \bibfield  {author} {\bibinfo {author} {\bibnamefont {Aliu}, \bibfnamefont
  {E.}} \emph {et~al.} (\bibinfo {collaboration} {MAGIC Collaboration}),\
  }\href {\doibase 10.1088/0004-637X/697/2/1299} {\bibfield  {journal}
  {\bibinfo  {journal} {Astrophys.J.}\ }\textbf {\bibinfo {volume} {697}},\
  \bibinfo {pages} {1299} (\bibinfo {year} {2009})},\ \Eprint
  {http://arxiv.org/abs/0810.3561} {arXiv:0810.3561 [astro-ph]} \BibitemShut
  {NoStop}%
%%CITATION = ARXIV:0810.3561;%%
\bibitem [{\citenamefont {Allen}, \citenamefont {Evrard},\ and\ \citenamefont
  {Mantz}(2011)}]{Allen:2011zs}%
  \BibitemOpen
  \bibfield  {author} {\bibinfo {author} {\bibnamefont {Allen}, \bibfnamefont
  {S.~W.}}, \bibinfo {author} {\bibnamefont {Evrard}, \bibfnamefont {A.~E.}}, \
  and\ \bibinfo {author} {\bibnamefont {Mantz}, \bibfnamefont {A.~B.}},\ }\href
  {\doibase 10.1146/annurev-astro-081710-102514} {\bibfield  {journal}
  {\bibinfo  {journal} {Ann.Rev.Astron.Astrophys.}\ }\textbf {\bibinfo {volume}
  {49}},\ \bibinfo {pages} {409} (\bibinfo {year} {2011})},\ \Eprint
  {http://arxiv.org/abs/1103.4829} {arXiv:1103.4829 [astro-ph.CO]} \BibitemShut
  {NoStop}%
%%CITATION = ARXIV:1103.4829;%%
\bibitem [{\citenamefont {Allgood}\ \emph {et~al.}(2006)\citenamefont
  {Allgood}, \citenamefont {Flores}, \citenamefont {Primack}, \citenamefont
  {Kravtsov}, \citenamefont {Wechsler} \emph {et~al.}}]{Allgood:2005eu}%
  \BibitemOpen
  \bibfield  {author} {\bibinfo {author} {\bibnamefont {Allgood}, \bibfnamefont
  {B.}}, \bibinfo {author} {\bibnamefont {Flores}, \bibfnamefont {R.~A.}},
  \bibinfo {author} {\bibnamefont {Primack}, \bibfnamefont {J.~R.}},  \emph
  {et~al.},\ }\href {\doibase 10.1111/j.1365-2966.2006.10094.x} {\bibfield
  {journal} {\bibinfo  {journal} {Mon.Not.Roy.Astron.Soc.}\ }\textbf {\bibinfo
  {volume} {367}},\ \bibinfo {pages} {1781} (\bibinfo {year} {2006})},\ \Eprint
  {http://arxiv.org/abs/astro-ph/0508497} {arXiv:astro-ph/0508497 [astro-ph]}
  \BibitemShut {NoStop}%
%%CITATION = ASTRO-PH/0508497;%%
\bibitem [{\citenamefont {{Amorisco}}\ and\ \citenamefont
  {{Evans}}(2012)}]{Amorisco:2011hb}%
  \BibitemOpen
  \bibfield  {author} {\bibinfo {author} {\bibnamefont {{Amorisco}},
  \bibfnamefont {N.~C.}}\ and\ \bibinfo {author} {\bibnamefont {{Evans}},
  \bibfnamefont {N.~W.}},\ }\href {\doibase 10.1111/j.1365-2966.2011.19684.x}
  {\bibfield  {journal} {\bibinfo  {journal} {\mnras}\ }\textbf {\bibinfo
  {volume} {419}},\ \bibinfo {pages} {184} (\bibinfo {year} {2012})},\ \Eprint
  {http://arxiv.org/abs/1106.1062} {arXiv:1106.1062 [astro-ph.CO]} \BibitemShut
  {NoStop}%
\bibitem [{\citenamefont {Anderhalden}\ and\ \citenamefont
  {Diemand}(2013)}]{Anderhalden:2013wd}%
  \BibitemOpen
  \bibfield  {author} {\bibinfo {author} {\bibnamefont {Anderhalden},
  \bibfnamefont {D.}}\ and\ \bibinfo {author} {\bibnamefont {Diemand},
  \bibfnamefont {J.}},\ }\href {\doibase 10.1088/1475-7516/2013/04/009,
  10.1088/1475-7516/2013/08/E02} {\bibfield  {journal} {\bibinfo  {journal}
  {JCAP}\ }\textbf {\bibinfo {volume} {1304}},\ \bibinfo {pages} {009}
  (\bibinfo {year} {2013})},\ \Eprint {http://arxiv.org/abs/1302.0003}
  {arXiv:1302.0003 [astro-ph.CO]} \BibitemShut {NoStop}%
%%CITATION = ARXIV:1302.0003;%%
\bibitem [{\citenamefont {Ando}\ and\ \citenamefont
  {Nagai}(2012)}]{Ando:2012vu}%
  \BibitemOpen
  \bibfield  {author} {\bibinfo {author} {\bibnamefont {Ando}, \bibfnamefont
  {S.}}\ and\ \bibinfo {author} {\bibnamefont {Nagai}, \bibfnamefont {D.}},\
  }\href {\doibase 10.1088/1475-7516/2012/07/017} {\bibfield  {journal}
  {\bibinfo  {journal} {JCAP}\ }\textbf {\bibinfo {volume} {1207}},\ \bibinfo
  {pages} {017} (\bibinfo {year} {2012})},\ \Eprint
  {http://arxiv.org/abs/1201.0753} {arXiv:1201.0753 [astro-ph.HE]} \BibitemShut
  {NoStop}%
%%CITATION = ARXIV:1201.0753;%%
\bibitem [{\citenamefont {{Aprile}}\ \emph {et~al.}(2012)\citenamefont
  {{Aprile}}, \citenamefont {{Alfonsi}}, \citenamefont {{Arisaka}},
  \citenamefont {{Arneodo}}, \citenamefont {{Balan}}, \citenamefont {{Baudis}},
  \citenamefont {{Bauermeister}}, \citenamefont {{Behrens}}, \citenamefont
  {{Beltrame}}, \citenamefont {{Bokeloh}}, \citenamefont {{Brown}},
  \citenamefont {{Bruno}}, \citenamefont {{Budnik}}, \citenamefont {{Cardoso}},
  \citenamefont {{Chen}}, \citenamefont {{Choi}}, \citenamefont {{Cline}},
  \citenamefont {{Colijn}}, \citenamefont {{Contreras}}, \citenamefont
  {{Cussonneau}}, \citenamefont {{Decowski}}, \citenamefont {{Duchovni}},
  \citenamefont {{Fattori}}, \citenamefont {{Ferella}}, \citenamefont
  {{Fulgione}}, \citenamefont {{Gao}}, \citenamefont {{Garbini}}, \citenamefont
  {{Ghag}}, \citenamefont {{Giboni}}, \citenamefont {{Goetzke}}, \citenamefont
  {{Grignon}}, \citenamefont {{Gross}}, \citenamefont {{Hampel}}, \citenamefont
  {{Kaether}}, \citenamefont {{Kish}}, \citenamefont {{Lamblin}}, \citenamefont
  {{Landsman}}, \citenamefont {{Lang}}, \citenamefont {{Le Calloch}},
  \citenamefont {{Levy}}, \citenamefont {{Lim}}, \citenamefont {{Lin}},
  \citenamefont {{Lindemann}}, \citenamefont {{Lindner}}, \citenamefont
  {{Lopes}}, \citenamefont {{Lung}}, \citenamefont {{Marrod{\'a}n Undagoitia}},
  \citenamefont {{Massoli}}, \citenamefont {{Melgarejo Fernandez}},
  \citenamefont {{Meng}}, \citenamefont {{Molinario}}, \citenamefont {{Nativ}},
  \citenamefont {{Ni}}, \citenamefont {{Oberlack}}, \citenamefont {{Orrigo}},
  \citenamefont {{Pantic}}, \citenamefont {{Persiani}}, \citenamefont
  {{Plante}}, \citenamefont {{Priel}}, \citenamefont {{Rizzo}}, \citenamefont
  {{Rosendahl}}, \citenamefont {{dos Santos}}, \citenamefont {{Sartorelli}},
  \citenamefont {{Schreiner}}, \citenamefont {{Schumann}}, \citenamefont
  {{Scotto Lavina}}, \citenamefont {{Scovell}}, \citenamefont {{Selvi}},
  \citenamefont {{Shagin}}, \citenamefont {{Simgen}}, \citenamefont
  {{Teymourian}}, \citenamefont {{Thers}}, \citenamefont {{Vitells}},
  \citenamefont {{Wang}}, \citenamefont {{Weber}},\ and\ \citenamefont
  {{Weinheimer}}}]{Aprile:2012nq}%
  \BibitemOpen
  \bibfield  {author} {\bibinfo {author} {\bibnamefont {{Aprile}},
  \bibfnamefont {E.}}, \bibinfo {author} {\bibnamefont {{Alfonsi}},
  \bibfnamefont {M.}}, \bibinfo {author} {\bibnamefont {{Arisaka}},
  \bibfnamefont {K.}},  \emph {et~al.},\ }\href {\doibase
  10.1103/PhysRevLett.109.181301} {\bibfield  {journal} {\bibinfo  {journal}
  {Physical Review Letters}\ }\textbf {\bibinfo {volume} {109}},\ \bibinfo
  {eid} {181301} (\bibinfo {year} {2012})},\ \Eprint
  {http://arxiv.org/abs/1207.5988} {arXiv:1207.5988 [astro-ph.CO]} \BibitemShut
  {NoStop}%
\bibitem [{\citenamefont {{Arlen}}\ \emph {et~al.}(2012)\citenamefont
  {{Arlen}}, \citenamefont {{Aune}}, \citenamefont {{Beilicke}}, \citenamefont
  {{Benbow}}, \citenamefont {{Bouvier}}, \citenamefont {{Buckley}},
  \citenamefont {{Bugaev}}, \citenamefont {{Byrum}}, \citenamefont {{Cannon}},
  \citenamefont {{Cesarini}}, \citenamefont {{Ciupik}}, \citenamefont
  {{Collins-Hughes}}, \citenamefont {{Connolly}}, \citenamefont {{Cui}},
  \citenamefont {{Dickherber}}, \citenamefont {{Dumm}}, \citenamefont
  {{Falcone}}, \citenamefont {{Federici}}, \citenamefont {{Feng}},
  \citenamefont {{Finley}}, \citenamefont {{Finnegan}}, \citenamefont
  {{Fortson}}, \citenamefont {{Furniss}}, \citenamefont {{Galante}},
  \citenamefont {{Gall}}, \citenamefont {{Godambe}}, \citenamefont {{Griffin}},
  \citenamefont {{Grube}}, \citenamefont {{Gyuk}}, \citenamefont {{Holder}},
  \citenamefont {{Huan}}, \citenamefont {{Hughes}}, \citenamefont {{Humensky}},
  \citenamefont {{Imran}}, \citenamefont {{Kaaret}}, \citenamefont
  {{Karlsson}}, \citenamefont {{Kertzman}}, \citenamefont {{Khassen}},
  \citenamefont {{Kieda}}, \citenamefont {{Krawczynski}}, \citenamefont
  {{Krennrich}}, \citenamefont {{Lee}}, \citenamefont {{Madhavan}},
  \citenamefont {{Maier}}, \citenamefont {{Majumdar}}, \citenamefont
  {{McArthur}}, \citenamefont {{McCann}}, \citenamefont {{Moriarty}},
  \citenamefont {{Mukherjee}}, \citenamefont {{Nelson}}, \citenamefont
  {{O'Faol{\'a}in de Bhr{\'o}ithe}}, \citenamefont {{Ong}}, \citenamefont
  {{Orr}}, \citenamefont {{Otte}}, \citenamefont {{Park}}, \citenamefont
  {{Perkins}}, \citenamefont {{Pohl}}, \citenamefont {{Prokoph}}, \citenamefont
  {{Quinn}}, \citenamefont {{Ragan}}, \citenamefont {{Reyes}}, \citenamefont
  {{Reynolds}}, \citenamefont {{Roache}}, \citenamefont {{Ruppel}},
  \citenamefont {{Saxon}}, \citenamefont {{Schroedter}}, \citenamefont
  {{Sembroski}}, \citenamefont {{Skole}}, \citenamefont {{Smith}},
  \citenamefont {{Telezhinsky}}, \citenamefont {{Te{\v s}i{\'c}}},
  \citenamefont {{Theiling}}, \citenamefont {{Thibadeau}}, \citenamefont
  {{Tsurusaki}}, \citenamefont {{Varlotta}}, \citenamefont {{Vivier}},
  \citenamefont {{Wakely}}, \citenamefont {{Ward}}, \citenamefont
  {{Weinstein}}, \citenamefont {{Welsing}}, \citenamefont {{Williams}},
  \citenamefont {{Zitzer}}, \citenamefont {{Pfrommer}},\ and\ \citenamefont
  {{Pinzke}}}]{2012ApJ...757..123A}%
  \BibitemOpen
  \bibfield  {author} {\bibinfo {author} {\bibnamefont {{Arlen}}, \bibfnamefont
  {T.}}, \bibinfo {author} {\bibnamefont {{Aune}}, \bibfnamefont {T.}},
  \bibinfo {author} {\bibnamefont {{Beilicke}}, \bibfnamefont {M.}},  \emph
  {et~al.},\ }\href {\doibase 10.1088/0004-637X/757/2/123} {\bibfield
  {journal} {\bibinfo  {journal} {\apj}\ }\textbf {\bibinfo {volume} {757}},\
  \bibinfo {eid} {123} (\bibinfo {year} {2012})},\ \Eprint
  {http://arxiv.org/abs/1208.0676} {arXiv:1208.0676 [astro-ph.HE]} \BibitemShut
  {NoStop}%
\bibitem [{\citenamefont {{Atkins}}\ \emph {et~al.}(2003)\citenamefont
  {{Atkins}}, \citenamefont {{Benbow}}, \citenamefont {{Berley}}, \citenamefont
  {{Blaufuss}}, \citenamefont {{Bussons}}, \citenamefont {{Coyne}},
  \citenamefont {{Delay}}, \citenamefont {{De Young}}, \citenamefont
  {{Dingus}}, \citenamefont {{Dorfan}}, \citenamefont {{Ellsworth}},
  \citenamefont {{Falcone}}, \citenamefont {{Fleysher}}, \citenamefont
  {{Fleysher}}, \citenamefont {{Gisler}}, \citenamefont {{Gonzalez}},
  \citenamefont {{Goodman}}, \citenamefont {{Haines}}, \citenamefont {{Hays}},
  \citenamefont {{Hoffman}}, \citenamefont {{Kelley}}, \citenamefont {{Laird}},
  \citenamefont {{McCullough}}, \citenamefont {{McEnery}}, \citenamefont
  {{Miller}}, \citenamefont {{Mincer}}, \citenamefont {{Morales}},
  \citenamefont {{Nemethy}}, \citenamefont {{Noyes}}, \citenamefont {{Ryan}},
  \citenamefont {{Samuelson}}, \citenamefont {{Schneider}}, \citenamefont
  {{Shen}}, \citenamefont {{Shoup}}, \citenamefont {{Sinnis}}, \citenamefont
  {{Smith}}, \citenamefont {{Sullivan}}, \citenamefont {{Tumer}}, \citenamefont
  {{Wang}}, \citenamefont {{Wascko}}, \citenamefont {{Williams}}, \citenamefont
  {{Westerhoff}}, \citenamefont {{Wilson}}, \citenamefont {{Xu}},\ and\
  \citenamefont {{Yodh}}}]{2003ApJ...595..803A}%
  \BibitemOpen
  \bibfield  {author} {\bibinfo {author} {\bibnamefont {{Atkins}},
  \bibfnamefont {R.}}, \bibinfo {author} {\bibnamefont {{Benbow}},
  \bibfnamefont {W.}}, \bibinfo {author} {\bibnamefont {{Berley}},
  \bibfnamefont {D.}},  \emph {et~al.},\ }\href {\doibase 10.1086/377498}
  {\bibfield  {journal} {\bibinfo  {journal} {\apj}\ }\textbf {\bibinfo
  {volume} {595}},\ \bibinfo {pages} {803} (\bibinfo {year} {2003})},\ \Eprint
  {http://arxiv.org/abs/astro-ph/0305308} {astro-ph/0305308} \BibitemShut
  {NoStop}%
\bibitem [{\citenamefont {Atwood}\ \emph {et~al.}(2012)\citenamefont {Atwood},
  \citenamefont {Albert}, \citenamefont {Baldini}, \citenamefont {Tinivella},
  \citenamefont {Bregeon}, \citenamefont {Pesce-Rollins}, \citenamefont {Sgrò},
  \citenamefont {Bruel}, \citenamefont {Charles}, \citenamefont
  {Drlica-Wagner}, \citenamefont {Franckowiak}, \citenamefont {Jogler},
  \citenamefont {Rochester}, \citenamefont {Usher}, \citenamefont {Wood},
  \citenamefont {Cohen-Tanugi},\ and\ \citenamefont {{S. Zimmer for the
  Fermi-LAT Collaboration}}}]{2013arXiv1303.3514A}%
  \BibitemOpen
  \bibfield  {author} {\bibinfo {author} {\bibnamefont {Atwood}, \bibfnamefont
  {W.}}, \bibinfo {author} {\bibnamefont {Albert}, \bibfnamefont {A.}},
  \bibinfo {author} {\bibnamefont {Baldini}, \bibfnamefont {L.}},  \emph
  {et~al.},\ }in\ \href
  {http://fermi.gsfc.nasa.gov/science/mtgs/symposia/2012/} {\emph {\bibinfo
  {booktitle} {Proceedings of the 4th Fermi Symposium 2012}}}\ (\bibinfo
  {address} {Monterey, California},\ \bibinfo {year} {2012})\BibitemShut
  {NoStop}%
\bibitem [{\citenamefont {Atwood}\ \emph {et~al.}(2009)\citenamefont {Atwood}
  \emph {et~al.}}]{Atwood:2009ez}%
  \BibitemOpen
  \bibfield  {author} {\bibinfo {author} {\bibnamefont {Atwood}, \bibfnamefont
  {W.}} \emph {et~al.} (\bibinfo {collaboration} {Fermi LAT Collaboration}),\
  }\href {\doibase 10.1088/0004-637X/697/2/1071} {\bibfield  {journal}
  {\bibinfo  {journal} {\apj}\ }\textbf {\bibinfo {volume} {697}},\ \bibinfo
  {pages} {1071} (\bibinfo {year} {2009})},\ \Eprint
  {http://arxiv.org/abs/0902.1089} {arXiv:0902.1089 [astro-ph.IM]} \BibitemShut
  {NoStop}%
\bibitem [{\citenamefont {{Atwood}}\ \emph {et~al.}(2013)\citenamefont
  {{Atwood}}, \citenamefont {{Baldini}}, \citenamefont {{Bregeon}},
  \citenamefont {{Bruel}}, \citenamefont {{Chekhtman}}, \citenamefont
  {{Cohen-Tanugi}}, \citenamefont {{Drlica-Wagner}}, \citenamefont {{Granot}},
  \citenamefont {{Longo}}, \citenamefont {{Omodei}}, \citenamefont
  {{Pesce-Rollins}}, \citenamefont {{Razzaque}}, \citenamefont {{Rochester}},
  \citenamefont {{Sgr{\`o}}}, \citenamefont {{Tinivella}}, \citenamefont
  {{Usher}},\ and\ \citenamefont {{Zimmer}}}]{2013ApJ...774...76A}%
  \BibitemOpen
  \bibfield  {author} {\bibinfo {author} {\bibnamefont {{Atwood}},
  \bibfnamefont {W.~B.}}, \bibinfo {author} {\bibnamefont {{Baldini}},
  \bibfnamefont {L.}}, \bibinfo {author} {\bibnamefont {{Bregeon}},
  \bibfnamefont {J.}},  \emph {et~al.},\ }\href {\doibase
  10.1088/0004-637X/774/1/76} {\bibfield  {journal} {\bibinfo  {journal}
  {\apj}\ }\textbf {\bibinfo {volume} {774}},\ \bibinfo {eid} {76} (\bibinfo
  {year} {2013})},\ \Eprint {http://arxiv.org/abs/1307.3037} {arXiv:1307.3037
  [astro-ph.HE]} \BibitemShut {NoStop}%
\bibitem [{\citenamefont {Baer}\ \emph {et~al.}(2014)\citenamefont {Baer},
  \citenamefont {Choi}, \citenamefont {Kim},\ and\ \citenamefont
  {Roszkowski}}]{Baer:2014eja}%
  \BibitemOpen
  \bibfield  {author} {\bibinfo {author} {\bibnamefont {Baer}, \bibfnamefont
  {H.}}, \bibinfo {author} {\bibnamefont {Choi}, \bibfnamefont {K.-Y.}},
  \bibinfo {author} {\bibnamefont {Kim}, \bibfnamefont {J.~E.}}, \ and\
  \bibinfo {author} {\bibnamefont {Roszkowski}, \bibfnamefont {L.}},\ }\href
  {\doibase 10.1016/j.physrep.2014.10.002} {\bibfield  {journal} {\bibinfo
  {journal} {Phys.Rept.}\ }\textbf {\bibinfo {volume} {555}},\ \bibinfo {pages}
  {1} (\bibinfo {year} {2014})},\ \Eprint {http://arxiv.org/abs/1407.0017}
  {arXiv:1407.0017 [hep-ph]} \BibitemShut {NoStop}%
%%CITATION = ARXIV:1407.0017;%%
\bibitem [{\citenamefont {Baltz}\ \emph {et~al.}(2008)\citenamefont {Baltz},
  \citenamefont {Berenji}, \citenamefont {Bertone}, \citenamefont {Bergstrom},
  \citenamefont {Bloom} \emph {et~al.}}]{Baltz:2008wd}%
  \BibitemOpen
  \bibfield  {author} {\bibinfo {author} {\bibnamefont {Baltz}, \bibfnamefont
  {E.}}, \bibinfo {author} {\bibnamefont {Berenji}, \bibfnamefont {B.}},
  \bibinfo {author} {\bibnamefont {Bertone}, \bibfnamefont {G.}},  \emph
  {et~al.},\ }\href {\doibase 10.1088/1475-7516/2008/07/013} {\bibfield
  {journal} {\bibinfo  {journal} {\jcap}\ }\textbf {\bibinfo {volume} {0807}},\
  \bibinfo {pages} {013} (\bibinfo {year} {2008})},\ \Eprint
  {http://arxiv.org/abs/0806.2911} {arXiv:0806.2911 [astro-ph]} \BibitemShut
  {NoStop}%
\bibitem [{\citenamefont {Baltz}\ \emph {et~al.}(2006)\citenamefont {Baltz},
  \citenamefont {Battaglia}, \citenamefont {Peskin},\ and\ \citenamefont
  {Wizansky}}]{Baltz:2006fm}%
  \BibitemOpen
  \bibfield  {author} {\bibinfo {author} {\bibnamefont {Baltz}, \bibfnamefont
  {E.~A.}}, \bibinfo {author} {\bibnamefont {Battaglia}, \bibfnamefont {M.}},
  \bibinfo {author} {\bibnamefont {Peskin}, \bibfnamefont {M.~E.}}, \ and\
  \bibinfo {author} {\bibnamefont {Wizansky}, \bibfnamefont {T.}},\ }\href
  {\doibase 10.1103/PhysRevD.74.103521} {\bibfield  {journal} {\bibinfo
  {journal} {Phys.Rev.}\ }\textbf {\bibinfo {volume} {D74}},\ \bibinfo {pages}
  {103521} (\bibinfo {year} {2006})},\ \Eprint
  {http://arxiv.org/abs/hep-ph/0602187} {arXiv:hep-ph/0602187 [hep-ph]}
  \BibitemShut {NoStop}%
%%CITATION = HEP-PH/0602187;%%
\bibitem [{\citenamefont {Baltz}, \citenamefont {Taylor},\ and\ \citenamefont
  {Wai}(2007)}]{Baltz:2006sv}%
  \BibitemOpen
  \bibfield  {author} {\bibinfo {author} {\bibnamefont {Baltz}, \bibfnamefont
  {E.~A.}}, \bibinfo {author} {\bibnamefont {Taylor}, \bibfnamefont {J.~E.}}, \
  and\ \bibinfo {author} {\bibnamefont {Wai}, \bibfnamefont {L.~L.}},\ }\href
  {\doibase 10.1086/517882} {\bibfield  {journal} {\bibinfo  {journal}
  {Astrophys.J.}\ }\textbf {\bibinfo {volume} {659}},\ \bibinfo {pages} {L125}
  (\bibinfo {year} {2007})},\ \Eprint {http://arxiv.org/abs/astro-ph/0610731}
  {arXiv:astro-ph/0610731 [astro-ph]} \BibitemShut {NoStop}%
%%CITATION = ASTRO-PH/0610731;%%
\bibitem [{\citenamefont {{Battaglia}}\ \emph {et~al.}(2005)\citenamefont
  {{Battaglia}}, \citenamefont {{Helmi}}, \citenamefont {{Morrison}},
  \citenamefont {{Harding}}, \citenamefont {{Olszewski}}, \citenamefont
  {{Mateo}}, \citenamefont {{Freeman}}, \citenamefont {{Norris}},\ and\
  \citenamefont {{Shectman}}}]{2005MNRAS.364..433B}%
  \BibitemOpen
  \bibfield  {author} {\bibinfo {author} {\bibnamefont {{Battaglia}},
  \bibfnamefont {G.}}, \bibinfo {author} {\bibnamefont {{Helmi}}, \bibfnamefont
  {A.}}, \bibinfo {author} {\bibnamefont {{Morrison}}, \bibfnamefont {H.}},
  \emph {et~al.},\ }\href {\doibase 10.1111/j.1365-2966.2005.09367.x}
  {\bibfield  {journal} {\bibinfo  {journal} {\mnras}\ }\textbf {\bibinfo
  {volume} {364}},\ \bibinfo {pages} {433} (\bibinfo {year} {2005})},\ \Eprint
  {http://arxiv.org/abs/astro-ph/0506102} {astro-ph/0506102} \BibitemShut
  {NoStop}%
\bibitem [{\citenamefont {{Battaglia}}\ \emph {et~al.}(2008)\citenamefont
  {{Battaglia}}, \citenamefont {{Helmi}}, \citenamefont {{Tolstoy}},
  \citenamefont {{Irwin}}, \citenamefont {{Hill}},\ and\ \citenamefont
  {{Jablonka}}}]{Battaglia:2008jz}%
  \BibitemOpen
  \bibfield  {author} {\bibinfo {author} {\bibnamefont {{Battaglia}},
  \bibfnamefont {G.}}, \bibinfo {author} {\bibnamefont {{Helmi}}, \bibfnamefont
  {A.}}, \bibinfo {author} {\bibnamefont {{Tolstoy}}, \bibfnamefont {E.}},
  \emph {et~al.},\ }\href {\doibase 10.1086/590179} {\bibfield  {journal}
  {\bibinfo  {journal} {\apjl}\ }\textbf {\bibinfo {volume} {681}},\ \bibinfo
  {pages} {L13} (\bibinfo {year} {2008})},\ \Eprint
  {http://arxiv.org/abs/0802.4220} {arXiv:0802.4220} \BibitemShut {NoStop}%
\bibitem [{\citenamefont {{Bechtol}}\ \emph {et~al.}(2015)\citenamefont
  {{Bechtol}}, \citenamefont {{Drlica-Wagner}}, \citenamefont {{Balbinot}},
  \citenamefont {{Pieres}}, \citenamefont {{Simon}}, \citenamefont {{Yanny}},
  \citenamefont {{Santiago}}, \citenamefont {{Wechsler}}, \citenamefont
  {{Frieman}}, \citenamefont {{Walker}}, \citenamefont {{Williams}},
  \citenamefont {{Rozo}}, \citenamefont {{Rykoff}}, \citenamefont {{Queiroz}},
  \citenamefont {{Luque}}, \citenamefont {{Benoit-Levy}}, \citenamefont
  {{Bernstein}}, \citenamefont {{Tucker}}, \citenamefont {{Sevilla}},
  \citenamefont {{Gruendl}}, \citenamefont {{da Costa}}, \citenamefont {{Fausti
  Neto}}, \citenamefont {{Maia}}, \citenamefont {{Abbott}}, \citenamefont
  {{Allam}}, \citenamefont {{Armstrong}}, \citenamefont {{Bauer}},
  \citenamefont {{Bernstein}}, \citenamefont {{Bertin}}, \citenamefont
  {{Brooks}}, \citenamefont {{Buckley-Geer}}, \citenamefont {{Burke}},
  \citenamefont {{Carnero Rosell}}, \citenamefont {{Castander}}, \citenamefont
  {{D'Andrea}}, \citenamefont {{DePoy}}, \citenamefont {{Desai}}, \citenamefont
  {{Diehl}}, \citenamefont {{Eifler}}, \citenamefont {{Estrada}}, \citenamefont
  {{Evrard}}, \citenamefont {{Fernandez}}, \citenamefont {{Finley}},
  \citenamefont {{Flaugher}}, \citenamefont {{Gaztanaga}}, \citenamefont
  {{Gerdes}}, \citenamefont {{Girardi}}, \citenamefont {{Gladders}},
  \citenamefont {{Gruen}}, \citenamefont {{Gutierrez}}, \citenamefont {{Hao}},
  \citenamefont {{Honscheid}}, \citenamefont {{Jain}}, \citenamefont {{James}},
  \citenamefont {{Kent}}, \citenamefont {{Kron}}, \citenamefont {{Kuehn}},
  \citenamefont {{Kuropatkin}}, \citenamefont {{Lahav}}, \citenamefont {{Li}},
  \citenamefont {{Lin}}, \citenamefont {{Makler}}, \citenamefont {{March}},
  \citenamefont {{Marshall}}, \citenamefont {{Martini}}, \citenamefont
  {{Merritt}}, \citenamefont {{Miller}}, \citenamefont {{Miquel}},
  \citenamefont {{Mohr}}, \citenamefont {{Neilsen}}, \citenamefont {{Nichol}},
  \citenamefont {{Nord}}, \citenamefont {{Ogando}}, \citenamefont {{Peoples}},
  \citenamefont {{Petravick}}, \citenamefont {{Plazas}}, \citenamefont
  {{Romer}}, \citenamefont {{Roodman}}, \citenamefont {{Sako}}, \citenamefont
  {{Sanchez}}, \citenamefont {{Scarpine}}, \citenamefont {{Schubnell}},
  \citenamefont {{Smith}}, \citenamefont {{Soares-Santos}}, \citenamefont
  {{Sobreira}}, \citenamefont {{Suchyta}}, \citenamefont {{Swanson}},
  \citenamefont {{Tarle}}, \citenamefont {{Thaler}}, \citenamefont {{Thomas}},
  \citenamefont {{Wester}},\ and\ \citenamefont
  {{Zuntz}}}]{2015arXiv150302584T}%
  \BibitemOpen
  \bibfield  {author} {\bibinfo {author} {\bibnamefont {{Bechtol}},
  \bibfnamefont {K.}}, \bibinfo {author} {\bibnamefont {{Drlica-Wagner}},
  \bibfnamefont {A.}}, \bibinfo {author} {\bibnamefont {{Balbinot}},
  \bibfnamefont {E.}},  \emph {et~al.} (\bibinfo {collaboration} {DES
  collaboration}),\ }\href {http://adsabs.harvard.edu/abs/2015arXiv150302584T}
  {\enquote {\bibinfo {title} {{Eight New Milky Way Companions Discovered in
  First-Year Dark Energy Survey Data}},}\ } (\bibinfo {year} {2015}),\ \bibinfo
  {note} {submitted to ApJ},\ \Eprint {http://arxiv.org/abs/1503.02584}
  {arXiv:1503.02584} \BibitemShut {NoStop}%
\bibitem [{\citenamefont {Behroozi}, \citenamefont {Conroy},\ and\
  \citenamefont {Wechsler}(2010)}]{Behroozi:2010rx}%
  \BibitemOpen
  \bibfield  {author} {\bibinfo {author} {\bibnamefont {Behroozi},
  \bibfnamefont {P.~S.}}, \bibinfo {author} {\bibnamefont {Conroy},
  \bibfnamefont {C.}}, \ and\ \bibinfo {author} {\bibnamefont {Wechsler},
  \bibfnamefont {R.~H.}},\ }\href {\doibase 10.1088/0004-637X/717/1/379}
  {\bibfield  {journal} {\bibinfo  {journal} {Astrophys.J.}\ }\textbf {\bibinfo
  {volume} {717}},\ \bibinfo {pages} {379} (\bibinfo {year} {2010})},\ \Eprint
  {http://arxiv.org/abs/1001.0015} {arXiv:1001.0015 [astro-ph.CO]} \BibitemShut
  {NoStop}%
%%CITATION = ARXIV:1001.0015;%%
\bibitem [{\citenamefont {Bergstrom}(2012)}]{Bergstrom:2012fi}%
  \BibitemOpen
  \bibfield  {author} {\bibinfo {author} {\bibnamefont {Bergstrom},
  \bibfnamefont {L.}},\ }\href {\doibase 10.1002/andp.201200116} {\bibfield
  {journal} {\bibinfo  {journal} {Annalen Phys.}\ }\textbf {\bibinfo {volume}
  {524}},\ \bibinfo {pages} {479} (\bibinfo {year} {2012})},\ \Eprint
  {http://arxiv.org/abs/1205.4882} {arXiv:1205.4882 [astro-ph.HE]} \BibitemShut
  {NoStop}%
%%CITATION = ARXIV:1205.4882;%%
\bibitem [{\citenamefont {Bergstrom}\ \emph {et~al.}(2012)\citenamefont
  {Bergstrom}, \citenamefont {Bertone}, \citenamefont {Conrad}, \citenamefont
  {Farnier},\ and\ \citenamefont {Weniger}}]{Bergstrom:2012vd}%
  \BibitemOpen
  \bibfield  {author} {\bibinfo {author} {\bibnamefont {Bergstrom},
  \bibfnamefont {L.}}, \bibinfo {author} {\bibnamefont {Bertone}, \bibfnamefont
  {G.}}, \bibinfo {author} {\bibnamefont {Conrad}, \bibfnamefont {J.}},
  \bibinfo {author} {\bibnamefont {Farnier}, \bibfnamefont {C.}}, \ and\
  \bibinfo {author} {\bibnamefont {Weniger}, \bibfnamefont {C.}},\ }\href
  {\doibase 10.1088/1475-7516/2012/11/025} {\bibfield  {journal} {\bibinfo
  {journal} {JCAP}\ }\textbf {\bibinfo {volume} {1211}},\ \bibinfo {pages}
  {025} (\bibinfo {year} {2012})},\ \Eprint {http://arxiv.org/abs/1207.6773}
  {arXiv:1207.6773 [hep-ph]} \BibitemShut {NoStop}%
%%CITATION = ARXIV:1207.6773;%%
\bibitem [{\citenamefont {Bergstrom}, \citenamefont {Bringmann},\ and\
  \citenamefont {Edsjo}(2011)}]{Bergstrom:2010gh}%
  \BibitemOpen
  \bibfield  {author} {\bibinfo {author} {\bibnamefont {Bergstrom},
  \bibfnamefont {L.}}, \bibinfo {author} {\bibnamefont {Bringmann},
  \bibfnamefont {T.}}, \ and\ \bibinfo {author} {\bibnamefont {Edsjo},
  \bibfnamefont {J.}},\ }\href {\doibase 10.1103/PhysRevD.83.045024} {\bibfield
   {journal} {\bibinfo  {journal} {Phys.Rev.}\ }\textbf {\bibinfo {volume}
  {D83}},\ \bibinfo {pages} {045024} (\bibinfo {year} {2011})},\ \Eprint
  {http://arxiv.org/abs/1011.4514} {arXiv:1011.4514 [hep-ph]} \BibitemShut
  {NoStop}%
%%CITATION = ARXIV:1011.4514;%%
\bibitem [{\citenamefont {Bergstrom}\ and\ \citenamefont
  {Snellman}(1988)}]{Bergstrom:1988fp}%
  \BibitemOpen
  \bibfield  {author} {\bibinfo {author} {\bibnamefont {Bergstrom},
  \bibfnamefont {L.}}\ and\ \bibinfo {author} {\bibnamefont {Snellman},
  \bibfnamefont {H.}},\ }\href {\doibase 10.1103/PhysRevD.37.3737} {\bibfield
  {journal} {\bibinfo  {journal} {Phys.Rev.}\ }\textbf {\bibinfo {volume}
  {D37}},\ \bibinfo {pages} {3737} (\bibinfo {year} {1988})}\BibitemShut
  {NoStop}%
%%CITATION = PHRVA,D37,3737;%%
\bibitem [{\citenamefont {Bertone}\ \emph {et~al.}(2012)\citenamefont
  {Bertone}, \citenamefont {Cerdeno}, \citenamefont {Fornasa}, \citenamefont
  {Pieri}, \citenamefont {Ruiz~de Austri} \emph {et~al.}}]{Bertone:2011pq}%
  \BibitemOpen
  \bibfield  {author} {\bibinfo {author} {\bibnamefont {Bertone}, \bibfnamefont
  {G.}}, \bibinfo {author} {\bibnamefont {Cerdeno}, \bibfnamefont {D.}},
  \bibinfo {author} {\bibnamefont {Fornasa}, \bibfnamefont {M.}},  \emph
  {et~al.},\ }\href {\doibase 10.1103/PhysRevD.85.055014} {\bibfield  {journal}
  {\bibinfo  {journal} {Phys.Rev.}\ }\textbf {\bibinfo {volume} {D85}},\
  \bibinfo {pages} {055014} (\bibinfo {year} {2012})},\ \Eprint
  {http://arxiv.org/abs/1111.2607} {arXiv:1111.2607 [astro-ph.HE]} \BibitemShut
  {NoStop}%
%%CITATION = ARXIV:1111.2607;%%
\bibitem [{\citenamefont {Bertone}, \citenamefont {Hooper},\ and\ \citenamefont
  {Silk}(2005)}]{Bertone:2004pz}%
  \BibitemOpen
  \bibfield  {author} {\bibinfo {author} {\bibnamefont {Bertone}, \bibfnamefont
  {G.}}, \bibinfo {author} {\bibnamefont {Hooper}, \bibfnamefont {D.}}, \ and\
  \bibinfo {author} {\bibnamefont {Silk}, \bibfnamefont {J.}},\ }\href
  {\doibase 10.1016/j.physrep.2004.08.031} {\bibfield  {journal} {\bibinfo
  {journal} {\physrep}\ }\textbf {\bibinfo {volume} {405}},\ \bibinfo {pages}
  {279} (\bibinfo {year} {2005})},\ \Eprint
  {http://arxiv.org/abs/hep-ph/0404175} {arXiv:hep-ph/0404175 [hep-ph]}
  \BibitemShut {NoStop}%
\bibitem [{\citenamefont {Bertone}\ and\ \citenamefont
  {Merritt}(2005)}]{Bertone:2005xv}%
  \BibitemOpen
  \bibfield  {author} {\bibinfo {author} {\bibnamefont {Bertone}, \bibfnamefont
  {G.}}\ and\ \bibinfo {author} {\bibnamefont {Merritt}, \bibfnamefont {D.}},\
  }\href {\doibase 10.1142/S0217732305017391} {\bibfield  {journal} {\bibinfo
  {journal} {Mod.Phys.Lett.}\ }\textbf {\bibinfo {volume} {A20}},\ \bibinfo
  {pages} {1021} (\bibinfo {year} {2005})},\ \Eprint
  {http://arxiv.org/abs/astro-ph/0504422} {arXiv:astro-ph/0504422 [astro-ph]}
  \BibitemShut {NoStop}%
%%CITATION = ASTRO-PH/0504422;%%
\bibitem [{\citenamefont {{Blitz}}\ \emph {et~al.}(1999)\citenamefont
  {{Blitz}}, \citenamefont {{Spergel}}, \citenamefont {{Teuben}}, \citenamefont
  {{Hartmann}},\ and\ \citenamefont {{Burton}}}]{1999ApJ...514..818B}%
  \BibitemOpen
  \bibfield  {author} {\bibinfo {author} {\bibnamefont {{Blitz}}, \bibfnamefont
  {L.}}, \bibinfo {author} {\bibnamefont {{Spergel}}, \bibfnamefont {D.~N.}},
  \bibinfo {author} {\bibnamefont {{Teuben}}, \bibfnamefont {P.~J.}}, \bibinfo
  {author} {\bibnamefont {{Hartmann}}, \bibfnamefont {D.}}, \ and\ \bibinfo
  {author} {\bibnamefont {{Burton}}, \bibfnamefont {W.~B.}},\ }\href {\doibase
  10.1086/306963} {\bibfield  {journal} {\bibinfo  {journal} {\apj}\ }\textbf
  {\bibinfo {volume} {514}},\ \bibinfo {pages} {818} (\bibinfo {year}
  {1999})},\ \Eprint {http://arxiv.org/abs/astro-ph/9803251} {astro-ph/9803251}
  \BibitemShut {NoStop}%
\bibitem [{\citenamefont {{Bovy}}\ and\ \citenamefont
  {{Rix}}(2013)}]{2013ApJ...779..115B}%
  \BibitemOpen
  \bibfield  {author} {\bibinfo {author} {\bibnamefont {{Bovy}}, \bibfnamefont
  {J.}}\ and\ \bibinfo {author} {\bibnamefont {{Rix}}, \bibfnamefont {H.-W.}},\
  }\href {\doibase 10.1088/0004-637X/779/2/115} {\bibfield  {journal} {\bibinfo
   {journal} {\apj}\ }\textbf {\bibinfo {volume} {779}},\ \bibinfo {eid} {115}
  (\bibinfo {year} {2013})},\ \Eprint {http://arxiv.org/abs/1309.0809}
  {arXiv:1309.0809 [astro-ph.GA]} \BibitemShut {NoStop}%
\bibitem [{\citenamefont {Bovy}\ and\ \citenamefont
  {Rix}(2013)}]{Bovy:2013raa}%
  \BibitemOpen
  \bibfield  {author} {\bibinfo {author} {\bibnamefont {Bovy}, \bibfnamefont
  {J.}}\ and\ \bibinfo {author} {\bibnamefont {Rix}, \bibfnamefont {H.-W.}},\
  }\href {\doibase 10.1088/0004-637X/779/2/115} {\bibfield  {journal} {\bibinfo
   {journal} {Astrophys.J.}\ }\textbf {\bibinfo {volume} {779}},\ \bibinfo
  {pages} {115} (\bibinfo {year} {2013})},\ \Eprint
  {http://arxiv.org/abs/1309.0809} {arXiv:1309.0809 [astro-ph.GA]} \BibitemShut
  {NoStop}%
%%CITATION = ARXIV:1309.0809;%%
\bibitem [{\citenamefont {Boylan-Kolchin}, \citenamefont {Bullock},\ and\
  \citenamefont {Kaplinghat}(2011)}]{BoylanKolchin:2011de}%
  \BibitemOpen
  \bibfield  {author} {\bibinfo {author} {\bibnamefont {Boylan-Kolchin},
  \bibfnamefont {M.}}, \bibinfo {author} {\bibnamefont {Bullock}, \bibfnamefont
  {J.~S.}}, \ and\ \bibinfo {author} {\bibnamefont {Kaplinghat}, \bibfnamefont
  {M.}},\ }\href@noop {} {\bibfield  {journal} {\bibinfo  {journal}
  {Mon.Not.Roy.Astron.Soc.}\ }\textbf {\bibinfo {volume} {415}},\ \bibinfo
  {pages} {L40} (\bibinfo {year} {2011})},\ \Eprint
  {http://arxiv.org/abs/1103.0007} {arXiv:1103.0007 [astro-ph.CO]} \BibitemShut
  {NoStop}%
%%CITATION = ARXIV:1103.0007;%%
\bibitem [{\citenamefont {{Boylan-Kolchin}}\ \emph {et~al.}(2013)\citenamefont
  {{Boylan-Kolchin}}, \citenamefont {{Bullock}}, \citenamefont {{Sohn}},
  \citenamefont {{Besla}},\ and\ \citenamefont {{van der
  Marel}}}]{BoylanKolchin:2012xy}%
  \BibitemOpen
  \bibfield  {author} {\bibinfo {author} {\bibnamefont {{Boylan-Kolchin}},
  \bibfnamefont {M.}}, \bibinfo {author} {\bibnamefont {{Bullock}},
  \bibfnamefont {J.~S.}}, \bibinfo {author} {\bibnamefont {{Sohn}},
  \bibfnamefont {S.~T.}}, \bibinfo {author} {\bibnamefont {{Besla}},
  \bibfnamefont {G.}}, \ and\ \bibinfo {author} {\bibnamefont {{van der
  Marel}}, \bibfnamefont {R.~P.}},\ }\href {\doibase
  10.1088/0004-637X/768/2/140} {\bibfield  {journal} {\bibinfo  {journal}
  {\apj}\ }\textbf {\bibinfo {volume} {768}},\ \bibinfo {eid} {140} (\bibinfo
  {year} {2013})},\ \Eprint {http://arxiv.org/abs/1210.6046} {arXiv:1210.6046
  [astro-ph.CO]} \BibitemShut {NoStop}%
\bibitem [{\citenamefont {{Boylan-Kolchin}}\ \emph {et~al.}(2009)\citenamefont
  {{Boylan-Kolchin}}, \citenamefont {{Springel}}, \citenamefont {{White}},
  \citenamefont {{Jenkins}},\ and\ \citenamefont
  {{Lemson}}}]{2009MNRAS.398.1150B}%
  \BibitemOpen
  \bibfield  {author} {\bibinfo {author} {\bibnamefont {{Boylan-Kolchin}},
  \bibfnamefont {M.}}, \bibinfo {author} {\bibnamefont {{Springel}},
  \bibfnamefont {V.}}, \bibinfo {author} {\bibnamefont {{White}}, \bibfnamefont
  {S.~D.~M.}}, \bibinfo {author} {\bibnamefont {{Jenkins}}, \bibfnamefont
  {A.}}, \ and\ \bibinfo {author} {\bibnamefont {{Lemson}}, \bibfnamefont
  {G.}},\ }\href {\doibase 10.1111/j.1365-2966.2009.15191.x} {\bibfield
  {journal} {\bibinfo  {journal} {\mnras}\ }\textbf {\bibinfo {volume} {398}},\
  \bibinfo {pages} {1150} (\bibinfo {year} {2009})},\ \Eprint
  {http://arxiv.org/abs/0903.3041} {arXiv:0903.3041 [astro-ph.CO]} \BibitemShut
  {NoStop}%
\bibitem [{\citenamefont {{Breddels}}\ \emph {et~al.}(2013)\citenamefont
  {{Breddels}}, \citenamefont {{Helmi}}, \citenamefont {{van den Bosch}},
  \citenamefont {{van de Ven}},\ and\ \citenamefont
  {{Battaglia}}}]{Breddels:2012cq}%
  \BibitemOpen
  \bibfield  {author} {\bibinfo {author} {\bibnamefont {{Breddels}},
  \bibfnamefont {M.~A.}}, \bibinfo {author} {\bibnamefont {{Helmi}},
  \bibfnamefont {A.}}, \bibinfo {author} {\bibnamefont {{van den Bosch}},
  \bibfnamefont {R.~C.~E.}}, \bibinfo {author} {\bibnamefont {{van de Ven}},
  \bibfnamefont {G.}}, \ and\ \bibinfo {author} {\bibnamefont {{Battaglia}},
  \bibfnamefont {G.}},\ }\href {\doibase 10.1093/mnras/stt956} {\bibfield
  {journal} {\bibinfo  {journal} {\mnras}\ }\textbf {\bibinfo {volume} {433}},\
  \bibinfo {pages} {3173} (\bibinfo {year} {2013})},\ \Eprint
  {http://arxiv.org/abs/1205.4712} {arXiv:1205.4712 [astro-ph.CO]} \BibitemShut
  {NoStop}%
\bibitem [{\citenamefont {Bringmann}, \citenamefont {Bergstrom},\ and\
  \citenamefont {Edsjo}(2008)}]{Bringmann:2007nk}%
  \BibitemOpen
  \bibfield  {author} {\bibinfo {author} {\bibnamefont {Bringmann},
  \bibfnamefont {T.}}, \bibinfo {author} {\bibnamefont {Bergstrom},
  \bibfnamefont {L.}}, \ and\ \bibinfo {author} {\bibnamefont {Edsjo},
  \bibfnamefont {J.}},\ }\href {\doibase 10.1088/1126-6708/2008/01/049}
  {\bibfield  {journal} {\bibinfo  {journal} {JHEP}\ }\textbf {\bibinfo
  {volume} {0801}},\ \bibinfo {pages} {049} (\bibinfo {year} {2008})},\ \Eprint
  {http://arxiv.org/abs/0710.3169} {arXiv:0710.3169 [hep-ph]} \BibitemShut
  {NoStop}%
%%CITATION = ARXIV:0710.3169;%%
\bibitem [{\citenamefont {{Bringmann}}\ \emph {et~al.}(2014)\citenamefont
  {{Bringmann}}, \citenamefont {{Calore}}, \citenamefont {{Di Mauro}},\ and\
  \citenamefont {{Donato}}}]{2014PhRvD..89b3012B}%
  \BibitemOpen
  \bibfield  {author} {\bibinfo {author} {\bibnamefont {{Bringmann}},
  \bibfnamefont {T.}}, \bibinfo {author} {\bibnamefont {{Calore}},
  \bibfnamefont {F.}}, \bibinfo {author} {\bibnamefont {{Di Mauro}},
  \bibfnamefont {M.}}, \ and\ \bibinfo {author} {\bibnamefont {{Donato}},
  \bibfnamefont {F.}},\ }\href {\doibase 10.1103/PhysRevD.89.023012} {\bibfield
   {journal} {\bibinfo  {journal} {\prd}\ }\textbf {\bibinfo {volume} {89}},\
  \bibinfo {eid} {023012} (\bibinfo {year} {2014})},\ \Eprint
  {http://arxiv.org/abs/1303.3284} {arXiv:1303.3284 [astro-ph.CO]} \BibitemShut
  {NoStop}%
\bibitem [{\citenamefont {Bringmann}\ and\ \citenamefont
  {Weniger}(2012)}]{Bringmann:2012ez}%
  \BibitemOpen
  \bibfield  {author} {\bibinfo {author} {\bibnamefont {Bringmann},
  \bibfnamefont {T.}}\ and\ \bibinfo {author} {\bibnamefont {Weniger},
  \bibfnamefont {C.}},\ }\href {\doibase 10.1016/j.dark.2012.10.005} {\bibfield
   {journal} {\bibinfo  {journal} {Phys.Dark Univ.}\ }\textbf {\bibinfo
  {volume} {1}},\ \bibinfo {pages} {194} (\bibinfo {year} {2012})},\ \Eprint
  {http://arxiv.org/abs/1208.5481} {arXiv:1208.5481 [hep-ph]} \BibitemShut
  {NoStop}%
%%CITATION = ARXIV:1208.5481;%%
\bibitem [{\citenamefont {{Buckley}}\ \emph {et~al.}(2013)\citenamefont
  {{Buckley}}, \citenamefont {{Cowen}}, \citenamefont {{Profumo}},
  \citenamefont {{Archer}}, \citenamefont {{Cahill-Rowley}}, \citenamefont
  {{Cotta}}, \citenamefont {{Digel}}, \citenamefont {{Drlica-Wagner}},
  \citenamefont {{Ferrer}}, \citenamefont {{Funk}}, \citenamefont {{Hewett}},
  \citenamefont {{Holder}}, \citenamefont {{Humensky}}, \citenamefont
  {{Ismail}}, \citenamefont {{Israel}}, \citenamefont {{Jeltema}},
  \citenamefont {{Olinto}}, \citenamefont {{Peter}}, \citenamefont {{Pretz}},
  \citenamefont {{Rizzo}}, \citenamefont {{Siegal-Gaskins}}, \citenamefont
  {{Smith}}, \citenamefont {{Staszak}}, \citenamefont {{Vandenbroucke}},\ and\
  \citenamefont {{Wood}}}]{2013arXiv1310.7040B}%
  \BibitemOpen
  \bibfield  {author} {\bibinfo {author} {\bibnamefont {{Buckley}},
  \bibfnamefont {J.}}, \bibinfo {author} {\bibnamefont {{Cowen}}, \bibfnamefont
  {D.~F.}}, \bibinfo {author} {\bibnamefont {{Profumo}}, \bibfnamefont {S.}},
  \emph {et~al.},\ }\href@noop {} {\enquote {\bibinfo {title} {{Cosmic Frontier
  Indirect Dark Matter Detection Working Group Summary}},}\ } (\bibinfo {year}
  {2013}),\ \bibinfo {note} {{Snowmass Indirect Dark Matter Detection CF2
  Working Group Summary}},\ \Eprint {http://arxiv.org/abs/1310.7040}
  {arXiv:1310.7040 [astro-ph.HE]} \BibitemShut {NoStop}%
\bibitem [{\citenamefont {Buckley}\ \emph {et~al.}(2015)\citenamefont
  {Buckley}, \citenamefont {Charles}, \citenamefont {Gaskins}, \citenamefont
  {Brooks}, \citenamefont {Drlica-Wagner} \emph {et~al.}}]{Buckley:2015doa}%
  \BibitemOpen
  \bibfield  {author} {\bibinfo {author} {\bibnamefont {Buckley}, \bibfnamefont
  {M.~R.}}, \bibinfo {author} {\bibnamefont {Charles}, \bibfnamefont {E.}},
  \bibinfo {author} {\bibnamefont {Gaskins}, \bibfnamefont {J.~M.}},  \emph
  {et~al.},\ }\href@noop {} {\bibfield  {journal} {\bibinfo  {journal}
  {Phys.Rev.D}\ } (\bibinfo {year} {2015})},\ \Eprint
  {http://arxiv.org/abs/1502.01020} {arXiv:1502.01020 [astro-ph.HE]}
  \BibitemShut {NoStop}%
%%CITATION = ARXIV:1502.01020;%%
\bibitem [{\citenamefont {Buckley}\ and\ \citenamefont
  {Hooper}(2010)}]{Buckley:2010vg}%
  \BibitemOpen
  \bibfield  {author} {\bibinfo {author} {\bibnamefont {Buckley}, \bibfnamefont
  {M.~R.}}\ and\ \bibinfo {author} {\bibnamefont {Hooper}, \bibfnamefont
  {D.}},\ }\href {\doibase 10.1103/PhysRevD.82.063501} {\bibfield  {journal}
  {\bibinfo  {journal} {Phys.Rev.}\ }\textbf {\bibinfo {volume} {D82}},\
  \bibinfo {pages} {063501} (\bibinfo {year} {2010})},\ \Eprint
  {http://arxiv.org/abs/1004.1644} {arXiv:1004.1644 [hep-ph]} \BibitemShut
  {NoStop}%
%%CITATION = ARXIV:1004.1644;%%
\bibitem [{\citenamefont {{Burkert}}(1995)}]{1995ApJ...447L..25B}%
  \BibitemOpen
  \bibfield  {author} {\bibinfo {author} {\bibnamefont {{Burkert}},
  \bibfnamefont {A.}},\ }\href {\doibase 10.1086/309560} {\bibfield  {journal}
  {\bibinfo  {journal} {\apjl}\ }\textbf {\bibinfo {volume} {447}},\ \bibinfo
  {pages} {L25} (\bibinfo {year} {1995})},\ \Eprint
  {http://arxiv.org/abs/astro-ph/9504041} {astro-ph/9504041} \BibitemShut
  {NoStop}%
\bibitem [{\citenamefont {Busha}\ \emph {et~al.}(2011)\citenamefont {Busha},
  \citenamefont {Marshall}, \citenamefont {Wechsler}, \citenamefont {Klypin},\
  and\ \citenamefont {Primack}}]{Busha:2010sg}%
  \BibitemOpen
  \bibfield  {author} {\bibinfo {author} {\bibnamefont {Busha}, \bibfnamefont
  {M.~T.}}, \bibinfo {author} {\bibnamefont {Marshall}, \bibfnamefont {P.~J.}},
  \bibinfo {author} {\bibnamefont {Wechsler}, \bibfnamefont {R.~H.}}, \bibinfo
  {author} {\bibnamefont {Klypin}, \bibfnamefont {A.}}, \ and\ \bibinfo
  {author} {\bibnamefont {Primack}, \bibfnamefont {J.}},\ }\href {\doibase
  10.1088/0004-637X/743/1/40} {\bibfield  {journal} {\bibinfo  {journal}
  {Astrophys.J.}\ }\textbf {\bibinfo {volume} {743}},\ \bibinfo {pages} {40}
  (\bibinfo {year} {2011})},\ \Eprint {http://arxiv.org/abs/1011.2203}
  {arXiv:1011.2203 [astro-ph.GA]} \BibitemShut {NoStop}%
%%CITATION = ARXIV:1011.2203;%%
\bibitem [{\citenamefont {Calore}\ \emph {et~al.}(2014)\citenamefont {Calore},
  \citenamefont {Cholis}, \citenamefont {McCabe},\ and\ \citenamefont
  {Weniger}}]{Calore:2014nla}%
  \BibitemOpen
  \bibfield  {author} {\bibinfo {author} {\bibnamefont {Calore}, \bibfnamefont
  {F.}}, \bibinfo {author} {\bibnamefont {Cholis}, \bibfnamefont {I.}},
  \bibinfo {author} {\bibnamefont {McCabe}, \bibfnamefont {C.}}, \ and\
  \bibinfo {author} {\bibnamefont {Weniger}, \bibfnamefont {C.}},\ }\href
  {http://www.arxiv.org/abs/1411.4647} {\enquote {\bibinfo {title} {{A Tale of
  Tails: Dark Matter Interpretations of the Fermi GeV Excess in Light of
  Background Model Systematics}},}\ } (\bibinfo {year} {2014}),\ \Eprint
  {http://arxiv.org/abs/1411.4647} {arXiv:1411.4647 [hep-ph]} \BibitemShut
  {NoStop}%
%%CITATION = ARXIV:1411.4647;%%
\bibitem [{\citenamefont {Calore}, \citenamefont {Cholis},\ and\ \citenamefont
  {Weniger}(2014)}]{Calore:2014xka}%
  \BibitemOpen
  \bibfield  {author} {\bibinfo {author} {\bibnamefont {Calore}, \bibfnamefont
  {F.}}, \bibinfo {author} {\bibnamefont {Cholis}, \bibfnamefont {I.}}, \ and\
  \bibinfo {author} {\bibnamefont {Weniger}, \bibfnamefont {C.}},\ }\href
  {http://www.arxiv.org/abs/1409.0042} {\enquote {\bibinfo {title} {{Background
  model systematics for the Fermi GeV excess}},}\ } (\bibinfo {year} {2014}),\
  \Eprint {http://arxiv.org/abs/1409.0042} {arXiv:1409.0042 [astro-ph.CO]}
  \BibitemShut {NoStop}%
%%CITATION = ARXIV:1409.0042;%%
\bibitem [{\citenamefont {{Campbell}}\ and\ \citenamefont
  {{Beacom}}(2013)}]{2013arXiv1312.3945C}%
  \BibitemOpen
  \bibfield  {author} {\bibinfo {author} {\bibnamefont {{Campbell}},
  \bibfnamefont {S.~S.}}\ and\ \bibinfo {author} {\bibnamefont {{Beacom}},
  \bibfnamefont {J.~F.}},\ }\href@noop {} {\bibfield  {journal} {\bibinfo
  {journal} {ArXiv e-prints}\ } (\bibinfo {year} {2013})},\ \Eprint
  {http://arxiv.org/abs/1312.3945} {arXiv:1312.3945 [astro-ph.HE]} \BibitemShut
  {NoStop}%
\bibitem [{\citenamefont {Carlson}\ and\ \citenamefont
  {Profumo}(2014)}]{Carlson:2014cwa}%
  \BibitemOpen
  \bibfield  {author} {\bibinfo {author} {\bibnamefont {Carlson}, \bibfnamefont
  {E.}}\ and\ \bibinfo {author} {\bibnamefont {Profumo}, \bibfnamefont {S.}},\
  }\href {\doibase 10.1103/PhysRevD.90.023015} {\bibfield  {journal} {\bibinfo
  {journal} {Phys.Rev.}\ }\textbf {\bibinfo {volume} {D90}},\ \bibinfo {pages}
  {023015} (\bibinfo {year} {2014})},\ \Eprint {http://arxiv.org/abs/1405.7685}
  {arXiv:1405.7685 [astro-ph.HE]} \BibitemShut {NoStop}%
%%CITATION = ARXIV:1405.7685;%%
\bibitem [{\citenamefont {{Cesarini}}\ \emph {et~al.}(2004)\citenamefont
  {{Cesarini}}, \citenamefont {{Fucito}}, \citenamefont {{Lionetto}},
  \citenamefont {{Morselli}},\ and\ \citenamefont
  {{Ullio}}}]{2004APh....21..267C}%
  \BibitemOpen
  \bibfield  {author} {\bibinfo {author} {\bibnamefont {{Cesarini}},
  \bibfnamefont {A.}}, \bibinfo {author} {\bibnamefont {{Fucito}},
  \bibfnamefont {F.}}, \bibinfo {author} {\bibnamefont {{Lionetto}},
  \bibfnamefont {A.}}, \bibinfo {author} {\bibnamefont {{Morselli}},
  \bibfnamefont {A.}}, \ and\ \bibinfo {author} {\bibnamefont {{Ullio}},
  \bibfnamefont {P.}},\ }\href {\doibase 10.1016/j.astropartphys.2004.02.001}
  {\bibfield  {journal} {\bibinfo  {journal} {Astroparticle Physics}\ }\textbf
  {\bibinfo {volume} {21}},\ \bibinfo {pages} {267} (\bibinfo {year} {2004})},\
  \Eprint {http://arxiv.org/abs/astro-ph/0305075} {astro-ph/0305075}
  \BibitemShut {NoStop}%
\bibitem [{\citenamefont {Charbonnier}\ \emph {et~al.}(2011)\citenamefont
  {Charbonnier}, \citenamefont {Combet}, \citenamefont {Daniel}, \citenamefont
  {Funk}, \citenamefont {Hinton} \emph {et~al.}}]{Charbonnier:2011ft}%
  \BibitemOpen
  \bibfield  {author} {\bibinfo {author} {\bibnamefont {Charbonnier},
  \bibfnamefont {A.}}, \bibinfo {author} {\bibnamefont {Combet}, \bibfnamefont
  {C.}}, \bibinfo {author} {\bibnamefont {Daniel}, \bibfnamefont {M.}},  \emph
  {et~al.},\ }\href {\doibase 10.1111/j.1365-2966.2011.19387.x} {\bibfield
  {journal} {\bibinfo  {journal} {Mon.Not.Roy.Astron.Soc.}\ }\textbf {\bibinfo
  {volume} {418}},\ \bibinfo {pages} {1526} (\bibinfo {year} {2011})},\ \Eprint
  {http://arxiv.org/abs/1104.0412} {arXiv:1104.0412 [astro-ph.HE]} \BibitemShut
  {NoStop}%
%%CITATION = ARXIV:1104.0412;%%
\bibitem [{\citenamefont {Chernoff}(1954)}]{Chernoff:1954}%
  \BibitemOpen
  \bibfield  {author} {\bibinfo {author} {\bibnamefont {Chernoff},
  \bibfnamefont {H.}},\ }\href@noop {} {\bibfield  {journal} {\bibinfo
  {journal} {Ann. Math. Statist.}\ }\textbf {\bibinfo {volume} {25}},\ \bibinfo
  {pages} {573} (\bibinfo {year} {1954})}\BibitemShut {NoStop}%
\bibitem [{\citenamefont {Cohen-Tanugi}(2009)}]{cohen-tanugi_fermi-lat_2009}%
  \BibitemOpen
  \bibfield  {author} {\bibinfo {author} {\bibnamefont {Cohen-Tanugi},
  \bibfnamefont {J.}},\ }in\ \href
  {http://www.slac.stanford.edu/econf/C0911022/} {\emph {\bibinfo {booktitle}
  {2009 Fermi Symposium. {eConf} Proceedings C0911022}}}\ (\bibinfo {address}
  {Washington, {D.C.}},\ \bibinfo {year} {2009})\BibitemShut {NoStop}%
\bibitem [{\citenamefont {Cohen-Tanugi}\ \emph {et~al.}(2009)\citenamefont
  {Cohen-Tanugi}, \citenamefont {Pohl}, \citenamefont {Tibolla},\ and\
  \citenamefont {Nuss}}]{cohen-tanugi_gev-band_2009}%
  \BibitemOpen
  \bibfield  {author} {\bibinfo {author} {\bibnamefont {Cohen-Tanugi},
  \bibfnamefont {J.}}, \bibinfo {author} {\bibnamefont {Pohl}, \bibfnamefont
  {M.}}, \bibinfo {author} {\bibnamefont {Tibolla}, \bibfnamefont {O.}}, \ and\
  \bibinfo {author} {\bibnamefont {Nuss}, \bibfnamefont {E.}},\ }in\ \href
  {http://icrc2009.uni.lodz.pl/proc/pdf/icrc0645.pdf?} {\emph {\bibinfo
  {booktitle} {Proceedings of the 31st International Cosmic-Ray Conference}}}\
  (\bibinfo {address} {Lodz, Poland},\ \bibinfo {year} {2009})\BibitemShut
  {NoStop}%
\bibitem [{\citenamefont {{Conrad}}(2014)}]{2014arXiv1411.1925C}%
  \BibitemOpen
  \bibfield  {author} {\bibinfo {author} {\bibnamefont {{Conrad}},
  \bibfnamefont {J.}},\ }\href@noop {} {\enquote {\bibinfo {title} {{Indirect
  Detection of WIMP Dark Matter: a compact review}},}\ } (\bibinfo {year}
  {2014}),\ \bibinfo {note} {{invited contribution to ``Interplay between
  Particle and Astroparticle Physics'', Queen Mary University of London (UK),
  2014}},\ \Eprint {http://arxiv.org/abs/1411.1925} {arXiv:1411.1925 [hep-ph]}
  \BibitemShut {NoStop}%
\bibitem [{\citenamefont {Conrad}(2014)}]{Conrad:2014nna}%
  \BibitemOpen
  \bibfield  {author} {\bibinfo {author} {\bibnamefont {Conrad}, \bibfnamefont
  {J.}},\ }\href {\doibase 10.1016/j.astropartphys.2014.09.003} {\bibfield
  {journal} {\bibinfo  {journal} {Astropart.Phys.}\ }\textbf {\bibinfo {volume}
  {62}},\ \bibinfo {pages} {165} (\bibinfo {year} {2014})},\ \Eprint
  {http://arxiv.org/abs/1407.6617} {arXiv:1407.6617 [astro-ph.CO]} \BibitemShut
  {NoStop}%
%%CITATION = ARXIV:1407.6617;%%
\bibitem [{\citenamefont {Cuoco}\ \emph {et~al.}(2008)\citenamefont {Cuoco},
  \citenamefont {Brandbyge}, \citenamefont {Hannestad}, \citenamefont
  {Haugboelle},\ and\ \citenamefont {Miele}}]{Cuoco:2007sh}%
  \BibitemOpen
  \bibfield  {author} {\bibinfo {author} {\bibnamefont {Cuoco}, \bibfnamefont
  {A.}}, \bibinfo {author} {\bibnamefont {Brandbyge}, \bibfnamefont {J.}},
  \bibinfo {author} {\bibnamefont {Hannestad}, \bibfnamefont {S.}}, \bibinfo
  {author} {\bibnamefont {Haugboelle}, \bibfnamefont {T.}}, \ and\ \bibinfo
  {author} {\bibnamefont {Miele}, \bibfnamefont {G.}},\ }\href {\doibase
  10.1103/PhysRevD.77.123518} {\bibfield  {journal} {\bibinfo  {journal}
  {Phys.Rev.}\ }\textbf {\bibinfo {volume} {D77}},\ \bibinfo {pages} {123518}
  (\bibinfo {year} {2008})},\ \Eprint {http://arxiv.org/abs/0710.4136}
  {arXiv:0710.4136 [astro-ph]} \BibitemShut {NoStop}%
%%CITATION = ARXIV:0710.4136;%%
\bibitem [{\citenamefont {{Cuoco}}\ \emph {et~al.}(2011)\citenamefont
  {{Cuoco}}, \citenamefont {{Sellerholm}}, \citenamefont {{Conrad}},\ and\
  \citenamefont {{Hannestad}}}]{2011MNRAS.414.2040C}%
  \BibitemOpen
  \bibfield  {author} {\bibinfo {author} {\bibnamefont {{Cuoco}}, \bibfnamefont
  {A.}}, \bibinfo {author} {\bibnamefont {{Sellerholm}}, \bibfnamefont {A.}},
  \bibinfo {author} {\bibnamefont {{Conrad}}, \bibfnamefont {J.}}, \ and\
  \bibinfo {author} {\bibnamefont {{Hannestad}}, \bibfnamefont {S.}},\ }\href
  {\doibase 10.1111/j.1365-2966.2011.18525.x} {\bibfield  {journal} {\bibinfo
  {journal} {\mnras}\ }\textbf {\bibinfo {volume} {414}},\ \bibinfo {pages}
  {2040} (\bibinfo {year} {2011})},\ \Eprint {http://arxiv.org/abs/1005.0843}
  {arXiv:1005.0843 [astro-ph.HE]} \BibitemShut {NoStop}%
\bibitem [{\citenamefont {Cushman}\ \emph {et~al.}(2013)\citenamefont
  {Cushman}, \citenamefont {Galbiati}, \citenamefont {McKinsey}, \citenamefont
  {Robertson}, \citenamefont {Tait} \emph {et~al.}}]{Cushman:2013zza}%
  \BibitemOpen
  \bibfield  {author} {\bibinfo {author} {\bibnamefont {Cushman}, \bibfnamefont
  {P.}}, \bibinfo {author} {\bibnamefont {Galbiati}, \bibfnamefont {C.}},
  \bibinfo {author} {\bibnamefont {McKinsey}, \bibfnamefont {D.}},  \emph
  {et~al.},\ }\href@noop {} {\enquote {\bibinfo {title} {{Working Group Report:
  WIMP Dark Matter Direct Detection}},}\ } (\bibinfo {year} {2013}),\ \bibinfo
  {note} {snowmass CF1 Final Summary Report: 47 pages and 28 figures with a 5
  page appendix on instrumentation R\&D},\ \Eprint
  {http://arxiv.org/abs/1310.8327} {arXiv:1310.8327 [hep-ex]} \BibitemShut
  {NoStop}%
%%CITATION = ARXIV:1310.8327;%%
\bibitem [{\citenamefont {Daylan}\ \emph {et~al.}(2014)\citenamefont {Daylan},
  \citenamefont {Finkbeiner}, \citenamefont {Hooper}, \citenamefont {Linden},
  \citenamefont {Portillo} \emph {et~al.}}]{Daylan:2014rsa}%
  \BibitemOpen
  \bibfield  {author} {\bibinfo {author} {\bibnamefont {Daylan}, \bibfnamefont
  {T.}}, \bibinfo {author} {\bibnamefont {Finkbeiner}, \bibfnamefont {D.~P.}},
  \bibinfo {author} {\bibnamefont {Hooper}, \bibfnamefont {D.}},  \emph
  {et~al.},\ }\href {http://www.arxiv.org/abs/1402.6703} {\enquote {\bibinfo
  {title} {{The Characterization of the Gamma-Ray Signal from the Central Milky
  Way: A Compelling Case for Annihilating Dark Matter}},}\ } (\bibinfo {year}
  {2014}),\ \Eprint {http://arxiv.org/abs/1402.6703} {arXiv:1402.6703
  [astro-ph.HE]} \BibitemShut {NoStop}%
%%CITATION = ARXIV:1402.6703;%%
\bibitem [{\citenamefont {{Deason}}\ \emph {et~al.}(2012)\citenamefont
  {{Deason}}, \citenamefont {{Belokurov}}, \citenamefont {{Evans}},\ and\
  \citenamefont {{An}}}]{Deason:2012wm}%
  \BibitemOpen
  \bibfield  {author} {\bibinfo {author} {\bibnamefont {{Deason}},
  \bibfnamefont {A.~J.}}, \bibinfo {author} {\bibnamefont {{Belokurov}},
  \bibfnamefont {V.}}, \bibinfo {author} {\bibnamefont {{Evans}}, \bibfnamefont
  {N.~W.}}, \ and\ \bibinfo {author} {\bibnamefont {{An}}, \bibfnamefont
  {J.}},\ }\href {\doibase 10.1111/j.1745-3933.2012.01283.x} {\bibfield
  {journal} {\bibinfo  {journal} {\mnras}\ }\textbf {\bibinfo {volume} {424}},\
  \bibinfo {pages} {L44} (\bibinfo {year} {2012})},\ \Eprint
  {http://arxiv.org/abs/1204.5189} {arXiv:1204.5189 [astro-ph.GA]} \BibitemShut
  {NoStop}%
\bibitem [{\citenamefont {{Di Mauro}}\ and\ \citenamefont
  {{Donato}}(2015)}]{2015arXiv150105316D}%
  \BibitemOpen
  \bibfield  {author} {\bibinfo {author} {\bibnamefont {{Di Mauro}},
  \bibfnamefont {M.}}\ and\ \bibinfo {author} {\bibnamefont {{Donato}},
  \bibfnamefont {F.}},\ }\href
  {http://adsabs.harvard.edu/abs/2015arXiv150105316D} {\enquote {\bibinfo
  {title} {{The composition of the Fermi-LAT IGRB intensity: emission from
  extragalactic point sources and dark matter annihilations}},}\ } (\bibinfo
  {year} {2015}),\ \bibinfo {note} {submitted to \prd},\ \Eprint
  {http://arxiv.org/abs/1501.05316} {arXiv:1501.05316 [astro-ph.HE]}
  \BibitemShut {NoStop}%
\bibitem [{\citenamefont {Dickinson}\ and\ \citenamefont
  {Conrad}(2013)}]{Dickinson:2012wp}%
  \BibitemOpen
  \bibfield  {author} {\bibinfo {author} {\bibnamefont {Dickinson},
  \bibfnamefont {H.}}\ and\ \bibinfo {author} {\bibnamefont {Conrad},
  \bibfnamefont {J.}},\ }\href {\doibase 10.1016/j.astropartphys.2012.10.004}
  {\bibfield  {journal} {\bibinfo  {journal} {Astropart.Phys.}\ }\textbf
  {\bibinfo {volume} {41}},\ \bibinfo {pages} {17} (\bibinfo {year} {2013})},\
  \Eprint {http://arxiv.org/abs/1203.5643} {arXiv:1203.5643 [astro-ph.IM]}
  \BibitemShut {NoStop}%
%%CITATION = ARXIV:1203.5643;%%
\bibitem [{\citenamefont {Diemand}\ \emph {et~al.}(2008)\citenamefont
  {Diemand}, \citenamefont {Kuhlen}, \citenamefont {Madau}, \citenamefont
  {Zemp}, \citenamefont {Moore} \emph {et~al.}}]{Diemand:2008in}%
  \BibitemOpen
  \bibfield  {author} {\bibinfo {author} {\bibnamefont {Diemand}, \bibfnamefont
  {J.}}, \bibinfo {author} {\bibnamefont {Kuhlen}, \bibfnamefont {M.}},
  \bibinfo {author} {\bibnamefont {Madau}, \bibfnamefont {P.}},  \emph
  {et~al.},\ }\href {\doibase 10.1038/nature07153} {\bibfield  {journal}
  {\bibinfo  {journal} {Nature}\ }\textbf {\bibinfo {volume} {454}},\ \bibinfo
  {pages} {735} (\bibinfo {year} {2008})},\ \Eprint
  {http://arxiv.org/abs/0805.1244} {arXiv:0805.1244 [astro-ph]} \BibitemShut
  {NoStop}%
%%CITATION = ARXIV:0805.1244;%%
\bibitem [{\citenamefont {Doro}\ \emph {et~al.}(2013)\citenamefont {Doro} \emph
  {et~al.}}]{Doro:2012xx}%
  \BibitemOpen
  \bibfield  {author} {\bibinfo {author} {\bibnamefont {Doro}, \bibfnamefont
  {M.}} \emph {et~al.} (\bibinfo {collaboration} {CTA Collaboration}),\ }\href
  {\doibase 10.1016/j.astropartphys.2012.08.002} {\bibfield  {journal}
  {\bibinfo  {journal} {Astropart.Phys.}\ }\textbf {\bibinfo {volume} {43}},\
  \bibinfo {pages} {189} (\bibinfo {year} {2013})},\ \Eprint
  {http://arxiv.org/abs/1208.5356} {arXiv:1208.5356 [astro-ph.IM]} \BibitemShut
  {NoStop}%
%%CITATION = ARXIV:1208.5356;%%
\bibitem [{\citenamefont {{Drlica-Wagner}}\ \emph {et~al.}(2015)\citenamefont
  {{Drlica-Wagner}}, \citenamefont {{Albert}}, \citenamefont {{Bechtol}},
  \citenamefont {{Wood}}, \citenamefont {{Strigari}}, \citenamefont
  {{Sanchez-Conde}}, \citenamefont {{Baldini}}, \citenamefont {{Essig}},
  \citenamefont {{Cohen-Tanugi}}, \citenamefont {{Anderson}}, \citenamefont
  {{Bellazzini}}, \citenamefont {{Bloom}}, \citenamefont {{Caputo}},
  \citenamefont {{Cecchi}}, \citenamefont {{Charles}}, \citenamefont
  {{Chiang}}, \citenamefont {{Conrad}}, \citenamefont {{de Angelis}},
  \citenamefont {{Funk}}, \citenamefont {{Fusco}}, \citenamefont {{Gargano}},
  \citenamefont {{Giglietto}}, \citenamefont {{Giordano}}, \citenamefont
  {{Guiriec}}, \citenamefont {{Gustafsson}}, \citenamefont {{Kuss}},
  \citenamefont {{Loparco}}, \citenamefont {{Lubrano}}, \citenamefont
  {{Mirabal}}, \citenamefont {{Mizuno}}, \citenamefont {{Morselli}},
  \citenamefont {{Ohsugi}}, \citenamefont {{Orlando}}, \citenamefont
  {{Persic}}, \citenamefont {{Raino}}, \citenamefont {{Spada}}, \citenamefont
  {{Suson}}, \citenamefont {{Zaharijas}}, \citenamefont {{Zimmer}},
  \citenamefont {{Abbott}}, \citenamefont {{Allam}}, \citenamefont
  {{Balbinot}}, \citenamefont {{Bauer}}, \citenamefont {{Benoit-Levy}},
  \citenamefont {{Bernstein}}, \citenamefont {{Bernstein}}, \citenamefont
  {{Bertin}}, \citenamefont {{Brooks}}, \citenamefont {{Buckley-Geer}},
  \citenamefont {{Burke}}, \citenamefont {{Carnero Rosell}}, \citenamefont
  {{Castander}}, \citenamefont {{Covarrubias}}, \citenamefont {{D'Andrea}},
  \citenamefont {{da Costa}}, \citenamefont {{DePoy}}, \citenamefont {{Desai}},
  \citenamefont {{Diehl}}, \citenamefont {{Cunha}}, \citenamefont {{Eifler}},
  \citenamefont {{Estrada}}, \citenamefont {{Evrard}}, \citenamefont {{Fausti
  Neto}}, \citenamefont {{Fernandez}}, \citenamefont {{Finley}}, \citenamefont
  {{Flaugher}}, \citenamefont {{Frieman}}, \citenamefont {{Gaztanaga}},
  \citenamefont {{Gerdes}}, \citenamefont {{Gruen}}, \citenamefont {{Gruendl}},
  \citenamefont {{Gutierrez}}, \citenamefont {{Honscheid}}, \citenamefont
  {{Jain}}, \citenamefont {{James}}, \citenamefont {{Jeltema}}, \citenamefont
  {{Kent}}, \citenamefont {{Kron}}, \citenamefont {{Kuropatkin}}, \citenamefont
  {{Lahav}}, \citenamefont {{Li}}, \citenamefont {{Luque}}, \citenamefont
  {{Maia}}, \citenamefont {{Makler}}, \citenamefont {{March}}, \citenamefont
  {{Marshall}}, \citenamefont {{Martini}}, \citenamefont {{Merritt}},
  \citenamefont {{Miller}}, \citenamefont {{Miquel}}, \citenamefont {{Mohr}},
  \citenamefont {{Neilsen}}, \citenamefont {{Nord}}, \citenamefont {{Ogando}},
  \citenamefont {{Peoples}}, \citenamefont {{Petravick}}, \citenamefont
  {{Pieres}}, \citenamefont {{Plazas}}, \citenamefont {{Queiroz}},
  \citenamefont {{Romer}}, \citenamefont {{Roodman}}, \citenamefont {{Rykoff}},
  \citenamefont {{Sako}}, \citenamefont {{Sanchez}}, \citenamefont
  {{Santiago}}, \citenamefont {{Scarpine}}, \citenamefont {{Schubnell}},
  \citenamefont {{Sevilla}}, \citenamefont {{Smith}}, \citenamefont
  {{Soares-Santos}}, \citenamefont {{Sobreira}}, \citenamefont {{Suchyta}},
  \citenamefont {{Swanson}}, \citenamefont {{Tarle}}, \citenamefont {{Thaler}},
  \citenamefont {{Thomas}}, \citenamefont {{Tucker}}, \citenamefont {{Walker}},
  \citenamefont {{Wechsler}}, \citenamefont {{Wester}}, \citenamefont
  {{Williams}}, \citenamefont {{Yanny}},\ and\ \citenamefont
  {{Zuntz}}}]{2015arXiv150302632T}%
  \BibitemOpen
  \bibfield  {author} {\bibinfo {author} {\bibnamefont {{Drlica-Wagner}},
  \bibfnamefont {A.}}, \bibinfo {author} {\bibnamefont {{Albert}},
  \bibfnamefont {A.}}, \bibinfo {author} {\bibnamefont {{Bechtol}},
  \bibfnamefont {K.}},  \emph {et~al.} (\bibinfo {collaboration} {The
  {Fermi-LAT Collaboration} and {DES Collaboration}}),\ }\href
  {http://adsabs.harvard.edu/abs/2015arXiv150302632T} {\enquote {\bibinfo
  {title} {{Search for Gamma-Ray Emission from DES Dwarf Spheroidal Galaxy
  Candidates with Fermi-LAT Data}},}\ } (\bibinfo {year} {2015}),\ \bibinfo
  {note} {submitted to ApJ Lett.},\ \Eprint {http://arxiv.org/abs/1503.02632}
  {arXiv:1503.02632 [astro-ph.HE]} \BibitemShut {NoStop}%
\bibitem [{\citenamefont {Drlica-Wagner}\ \emph {et~al.}(2014)\citenamefont
  {Drlica-Wagner}, \citenamefont {Gómez-Vargas}, \citenamefont {Hewitt},
  \citenamefont {Linden},\ and\ \citenamefont {Tibaldo}}]{0004-637X-790-1-24}%
  \BibitemOpen
  \bibfield  {author} {\bibinfo {author} {\bibnamefont {Drlica-Wagner},
  \bibfnamefont {A.}}, \bibinfo {author} {\bibnamefont {Gómez-Vargas},
  \bibfnamefont {G.~A.}}, \bibinfo {author} {\bibnamefont {Hewitt},
  \bibfnamefont {J.~W.}}, \bibinfo {author} {\bibnamefont {Linden},
  \bibfnamefont {T.}}, \ and\ \bibinfo {author} {\bibnamefont {Tibaldo},
  \bibfnamefont {L.}},\ }\href {http://stacks.iop.org/0004-637X/790/i=1/a=24}
  {\bibfield  {journal} {\bibinfo  {journal} {The Astrophysical Journal}\
  }\textbf {\bibinfo {volume} {790}},\ \bibinfo {pages} {24} (\bibinfo {year}
  {2014})}\BibitemShut {NoStop}%
\bibitem [{\citenamefont {Dutta}(2014)}]{Dutta:2014mya}%
  \BibitemOpen
  \bibfield  {author} {\bibinfo {author} {\bibnamefont {Dutta}, \bibfnamefont
  {B.}},\ }\href@noop {} {\enquote {\bibinfo {title} {{Dark Matter Searches at
  Accelerator Facilities}},}\ } (\bibinfo {year} {2014}),\ \bibinfo {note}
  {symposium on Cosmology and Particle Astrophysics 2013 Honolulu, HI, November
  12- 15, 2013},\ \Eprint {http://arxiv.org/abs/1403.6217} {arXiv:1403.6217
  [hep-ph]} \BibitemShut {NoStop}%
%%CITATION = ARXIV:1403.6217;%%
\bibitem [{\citenamefont {{Essig}}\ \emph {et~al.}(2013)\citenamefont
  {{Essig}}, \citenamefont {{Kuflik}}, \citenamefont {{McDermott}},
  \citenamefont {{Volansky}},\ and\ \citenamefont
  {{Zurek}}}]{2013JHEP...11..193E}%
  \BibitemOpen
  \bibfield  {author} {\bibinfo {author} {\bibnamefont {{Essig}}, \bibfnamefont
  {R.}}, \bibinfo {author} {\bibnamefont {{Kuflik}}, \bibfnamefont {E.}},
  \bibinfo {author} {\bibnamefont {{McDermott}}, \bibfnamefont {S.~D.}},
  \bibinfo {author} {\bibnamefont {{Volansky}}, \bibfnamefont {T.}}, \ and\
  \bibinfo {author} {\bibnamefont {{Zurek}}, \bibfnamefont {K.~M.}},\ }\href
  {\doibase 10.1007/JHEP11(2013)193} {\bibfield  {journal} {\bibinfo  {journal}
  {Journal of High Energy Physics}\ }\textbf {\bibinfo {volume} {11}},\
  \bibinfo {eid} {193} (\bibinfo {year} {2013})},\ \Eprint
  {http://arxiv.org/abs/1309.4091} {arXiv:1309.4091 [hep-ph]} \BibitemShut
  {NoStop}%
\bibitem [{\citenamefont {Essig}, \citenamefont {Sehgal},\ and\ \citenamefont
  {Strigari}(2009)}]{Essig:2009jx}%
  \BibitemOpen
  \bibfield  {author} {\bibinfo {author} {\bibnamefont {Essig}, \bibfnamefont
  {R.}}, \bibinfo {author} {\bibnamefont {Sehgal}, \bibfnamefont {N.}}, \ and\
  \bibinfo {author} {\bibnamefont {Strigari}, \bibfnamefont {L.~E.}},\ }\href
  {\doibase 10.1103/PhysRevD.80.023506} {\bibfield  {journal} {\bibinfo
  {journal} {\prd}\ }\textbf {\bibinfo {volume} {80}},\ \bibinfo {pages}
  {023506} (\bibinfo {year} {2009})},\ \Eprint {http://arxiv.org/abs/0902.4750}
  {arXiv:0902.4750 [hep-ph]} \BibitemShut {NoStop}%
\bibitem [{\citenamefont {Essig}\ \emph {et~al.}(2010)\citenamefont {Essig},
  \citenamefont {Sehgal}, \citenamefont {Strigari}, \citenamefont {Geha},\ and\
  \citenamefont {Simon}}]{Essig:2010em}%
  \BibitemOpen
  \bibfield  {author} {\bibinfo {author} {\bibnamefont {Essig}, \bibfnamefont
  {R.}}, \bibinfo {author} {\bibnamefont {Sehgal}, \bibfnamefont {N.}},
  \bibinfo {author} {\bibnamefont {Strigari}, \bibfnamefont {L.~E.}}, \bibinfo
  {author} {\bibnamefont {Geha}, \bibfnamefont {M.}}, \ and\ \bibinfo {author}
  {\bibnamefont {Simon}, \bibfnamefont {J.~D.}},\ }\href {\doibase
  10.1103/PhysRevD.82.123503} {\bibfield  {journal} {\bibinfo  {journal}
  {Phys.Rev.}\ }\textbf {\bibinfo {volume} {D82}},\ \bibinfo {pages} {123503}
  (\bibinfo {year} {2010})},\ \Eprint {http://arxiv.org/abs/1007.4199}
  {arXiv:1007.4199 [astro-ph.CO]} \BibitemShut {NoStop}%
%%CITATION = ARXIV:1007.4199;%%
\bibitem [{\citenamefont {{Faerman}}, \citenamefont {{Sternberg}},\ and\
  \citenamefont {{McKee}}(2013)}]{2013ApJ...777..119F}%
  \BibitemOpen
  \bibfield  {author} {\bibinfo {author} {\bibnamefont {{Faerman}},
  \bibfnamefont {Y.}}, \bibinfo {author} {\bibnamefont {{Sternberg}},
  \bibfnamefont {A.}}, \ and\ \bibinfo {author} {\bibnamefont {{McKee}},
  \bibfnamefont {C.~F.}},\ }\href {\doibase 10.1088/0004-637X/777/2/119}
  {\bibfield  {journal} {\bibinfo  {journal} {Astrophys. J.}\ }\textbf
  {\bibinfo {volume} {777}},\ \bibinfo {eid} {119} (\bibinfo {year} {2013})},\
  \Eprint {http://arxiv.org/abs/1309.0815} {arXiv:1309.0815 [astro-ph.CO]}
  \BibitemShut {NoStop}%
\bibitem [{\citenamefont {Fargion}\ \emph {et~al.}(2000)\citenamefont
  {Fargion}, \citenamefont {Konoplich}, \citenamefont {Grossi},\ and\
  \citenamefont {Khlopov}}]{Fargion:1998ya}%
  \BibitemOpen
  \bibfield  {author} {\bibinfo {author} {\bibnamefont {Fargion}, \bibfnamefont
  {D.}}, \bibinfo {author} {\bibnamefont {Konoplich}, \bibfnamefont {R.}},
  \bibinfo {author} {\bibnamefont {Grossi}, \bibfnamefont {M.}}, \ and\
  \bibinfo {author} {\bibnamefont {Khlopov}, \bibfnamefont {M.}},\ }\href
  {\doibase 10.1016/S0927-6505(99)00094-8} {\bibfield  {journal} {\bibinfo
  {journal} {Astropart.Phys.}\ }\textbf {\bibinfo {volume} {12}},\ \bibinfo
  {pages} {307} (\bibinfo {year} {2000})},\ \Eprint
  {http://arxiv.org/abs/astro-ph/9809260} {arXiv:astro-ph/9809260 [astro-ph]}
  \BibitemShut {NoStop}%
%%CITATION = ASTRO-PH/9809260;%%
\bibitem [{\citenamefont {Feldman}\ and\ \citenamefont
  {Cousins}(1998)}]{Feldman:1997qc}%
  \BibitemOpen
  \bibfield  {author} {\bibinfo {author} {\bibnamefont {Feldman}, \bibfnamefont
  {G.~J.}}\ and\ \bibinfo {author} {\bibnamefont {Cousins}, \bibfnamefont
  {R.~D.}},\ }\href {\doibase 10.1103/PhysRevD.57.3873} {\bibfield  {journal}
  {\bibinfo  {journal} {Phys.Rev.}\ }\textbf {\bibinfo {volume} {D57}},\
  \bibinfo {pages} {3873} (\bibinfo {year} {1998})},\ \Eprint
  {http://arxiv.org/abs/physics/9711021} {arXiv:physics/9711021
  [physics.data-an]} \BibitemShut {NoStop}%
%%CITATION = PHYSICS/9711021;%%
\bibitem [{\citenamefont {Fornasa}\ and\ \citenamefont
  {Sanchez-Conde}(2015)}]{Fornasa:2015qua}%
  \BibitemOpen
  \bibfield  {author} {\bibinfo {author} {\bibnamefont {Fornasa}, \bibfnamefont
  {M.}}\ and\ \bibinfo {author} {\bibnamefont {Sanchez-Conde}, \bibfnamefont
  {M.~A.}},\ }\href@noop {} {\enquote {\bibinfo {title} {{The nature of the
  Diffuse Gamma-Ray Background}},}\ } (\bibinfo {year} {2015}),\ \bibinfo
  {note} {submitted to Physics Reports},\ \Eprint
  {http://arxiv.org/abs/1502.02866} {arXiv:1502.02866 [astro-ph.CO]}
  \BibitemShut {NoStop}%
%%CITATION = ARXIV:1502.02866;%%
\bibitem [{\citenamefont {Frenk}\ and\ \citenamefont
  {White}(2012)}]{Frenk:2012ph}%
  \BibitemOpen
  \bibfield  {author} {\bibinfo {author} {\bibnamefont {Frenk}, \bibfnamefont
  {C.}}\ and\ \bibinfo {author} {\bibnamefont {White}, \bibfnamefont {S.~D.}},\
  }\href {\doibase 10.1002/andp.201200212} {\bibfield  {journal} {\bibinfo
  {journal} {Annalen Phys.}\ }\textbf {\bibinfo {volume} {524}},\ \bibinfo
  {pages} {507} (\bibinfo {year} {2012})},\ \Eprint
  {http://arxiv.org/abs/1210.0544} {arXiv:1210.0544 [astro-ph.CO]} \BibitemShut
  {NoStop}%
%%CITATION = ARXIV:1210.0544;%%
\bibitem [{\citenamefont {{Galper}}\ \emph
  {et~al.}(2013{\natexlab{a}})\citenamefont {{Galper}}, \citenamefont
  {{Adriani}}, \citenamefont {{Aptekar}}, \citenamefont {{Arkhangelskaja}},
  \citenamefont {{Arkhangelskiy}}, \citenamefont {{Boezio}}, \citenamefont
  {{Bonvicini}}, \citenamefont {{Boyarchuk}}, \citenamefont {{Fradkin}},
  \citenamefont {{Gusakov}}, \citenamefont {{Kaplin}}, \citenamefont
  {{Kachanov}}, \citenamefont {{Kheymits}}, \citenamefont {{Leonov}},
  \citenamefont {{Longo}}, \citenamefont {{Mazets}}, \citenamefont {{Maestro}},
  \citenamefont {{Marrocchesi}}, \citenamefont {{Mereminskiy}}, \citenamefont
  {{Mikhailov}}, \citenamefont {{Moiseev}}, \citenamefont {{Mocchiutti}},
  \citenamefont {{Mori}}, \citenamefont {{Moskalenko}}, \citenamefont
  {{Naumov}}, \citenamefont {{Papini}}, \citenamefont {{Picozza}},
  \citenamefont {{Rodin}}, \citenamefont {{Runtso}}, \citenamefont
  {{Sparvoli}}, \citenamefont {{Spillantini}}, \citenamefont {{Suchkov}},
  \citenamefont {{Tavani}}, \citenamefont {{Topchiev}}, \citenamefont
  {{Vacchi}}, \citenamefont {{Vannuccini}}, \citenamefont {{Yurkin}},
  \citenamefont {{Zampa}}, \citenamefont {{Zverev}},\ and\ \citenamefont
  {{Zirakashvili}}}]{2013AIPC.1516..288G}%
  \BibitemOpen
  \bibfield  {author} {\bibinfo {author} {\bibnamefont {{Galper}},
  \bibfnamefont {A.~M.}}, \bibinfo {author} {\bibnamefont {{Adriani}},
  \bibfnamefont {O.}}, \bibinfo {author} {\bibnamefont {{Aptekar}},
  \bibfnamefont {R.~L.}},  \emph {et~al.},\ }in\ \href {\doibase
  10.1063/1.4792586} {\emph {\bibinfo {booktitle} {American Institute of
  Physics Conference Series}}},\ \bibinfo {series} {American Institute of
  Physics Conference Series}, Vol.\ \bibinfo {volume} {1516},\ \bibinfo
  {editor} {edited by\ \bibinfo {editor} {\bibfnamefont {J.~F.}\ \bibnamefont
  {{Ormes}}}}\ (\bibinfo {year} {2013})\ pp.\ \bibinfo {pages} {288--292},\
  \Eprint {http://arxiv.org/abs/1210.1457} {arXiv:1210.1457 [astro-ph.IM]}
  \BibitemShut {NoStop}%
\bibitem [{\citenamefont {{Galper}}\ \emph
  {et~al.}(2013{\natexlab{b}})\citenamefont {{Galper}}, \citenamefont
  {{Adriani}}, \citenamefont {{Aptekar}}, \citenamefont {{Arkhangelskaja}},
  \citenamefont {{Arkhangelskiy}}, \citenamefont {{Boezio}}, \citenamefont
  {{Bonvicini}}, \citenamefont {{Boyarchuk}}, \citenamefont {{Gusakov}},
  \citenamefont {{Farber}}, \citenamefont {{Fradkin}}, \citenamefont
  {{Kachanov}}, \citenamefont {{Kaplin}}, \citenamefont {{Kheymits}},
  \citenamefont {{Leonov}}, \citenamefont {{Longo}}, \citenamefont {{Maestro}},
  \citenamefont {{Marrocchesi}}, \citenamefont {{Mazets}}, \citenamefont
  {{Mocchiutti}}, \citenamefont {{Moiseev}}, \citenamefont {{Mori}},
  \citenamefont {{Moskalenko}}, \citenamefont {{Naumov}}, \citenamefont
  {{Papini}}, \citenamefont {{Picozza}}, \citenamefont {{Rodin}}, \citenamefont
  {{Runtso}}, \citenamefont {{Sparvoli}}, \citenamefont {{Spillantini}},
  \citenamefont {{Suchkov}}, \citenamefont {{Tavani}}, \citenamefont
  {{Topchiev}}, \citenamefont {{Vacchi}}, \citenamefont {{Vannuccini}},
  \citenamefont {{Yurkin}}, \citenamefont {{Zampa}},\ and\ \citenamefont
  {{Zverev}}}]{2013AdSpR..51..297G}%
  \BibitemOpen
  \bibfield  {author} {\bibinfo {author} {\bibnamefont {{Galper}},
  \bibfnamefont {A.~M.}}, \bibinfo {author} {\bibnamefont {{Adriani}},
  \bibfnamefont {O.}}, \bibinfo {author} {\bibnamefont {{Aptekar}},
  \bibfnamefont {R.~L.}},  \emph {et~al.},\ }\href {\doibase
  10.1016/j.asr.2012.01.019} {\bibfield  {journal} {\bibinfo  {journal}
  {Advances in Space Research}\ }\textbf {\bibinfo {volume} {51}},\ \bibinfo
  {pages} {297} (\bibinfo {year} {2013}{\natexlab{b}})},\ \Eprint
  {http://arxiv.org/abs/1201.2490} {arXiv:1201.2490 [astro-ph.IM]} \BibitemShut
  {NoStop}%
\bibitem [{\citenamefont {Gao}\ \emph {et~al.}(2012)\citenamefont {Gao},
  \citenamefont {Frenk}, \citenamefont {Jenkins}, \citenamefont {Springel},\
  and\ \citenamefont {White}}]{Gao:2011rf}%
  \BibitemOpen
  \bibfield  {author} {\bibinfo {author} {\bibnamefont {Gao}, \bibfnamefont
  {L.}}, \bibinfo {author} {\bibnamefont {Frenk}, \bibfnamefont {C.}}, \bibinfo
  {author} {\bibnamefont {Jenkins}, \bibfnamefont {A.}}, \bibinfo {author}
  {\bibnamefont {Springel}, \bibfnamefont {V.}}, \ and\ \bibinfo {author}
  {\bibnamefont {White}, \bibfnamefont {S.}},\ }\href@noop {} {\bibfield
  {journal} {\bibinfo  {journal} {Mon.Not.Roy.Astron.Soc.}\ }\textbf {\bibinfo
  {volume} {419}},\ \bibinfo {pages} {1721} (\bibinfo {year} {2012})},\ \Eprint
  {http://arxiv.org/abs/1107.1916} {arXiv:1107.1916 [astro-ph.CO]} \BibitemShut
  {NoStop}%
%%CITATION = ARXIV:1107.1916;%%
\bibitem [{\citenamefont {Geringer-Sameth}\ and\ \citenamefont
  {Koushiappas}(2011)}]{PhysRevLett.107.241303}%
  \BibitemOpen
  \bibfield  {author} {\bibinfo {author} {\bibnamefont {Geringer-Sameth},
  \bibfnamefont {A.}}\ and\ \bibinfo {author} {\bibnamefont {Koushiappas},
  \bibfnamefont {S.~M.}},\ }\href {\doibase 10.1103/PhysRevLett.107.241303}
  {\bibfield  {journal} {\bibinfo  {journal} {Phys. Rev. Lett.}\ }\textbf
  {\bibinfo {volume} {107}},\ \bibinfo {pages} {241303} (\bibinfo {year}
  {2011})}\BibitemShut {NoStop}%
\bibitem [{\citenamefont {Geringer-Sameth}\ and\ \citenamefont
  {Koushiappas}(2012)}]{GeringerSameth:2012sr}%
  \BibitemOpen
  \bibfield  {author} {\bibinfo {author} {\bibnamefont {Geringer-Sameth},
  \bibfnamefont {A.}}\ and\ \bibinfo {author} {\bibnamefont {Koushiappas},
  \bibfnamefont {S.~M.}},\ }\href {\doibase 10.1103/PhysRevD.86.021302}
  {\bibfield  {journal} {\bibinfo  {journal} {Phys.Rev.}\ }\textbf {\bibinfo
  {volume} {D86}},\ \bibinfo {pages} {021302} (\bibinfo {year} {2012})},\
  \Eprint {http://arxiv.org/abs/1206.0796} {arXiv:1206.0796 [astro-ph.HE]}
  \BibitemShut {NoStop}%
%%CITATION = ARXIV:1206.0796;%%
\bibitem [{\citenamefont {{Geringer-Sameth}}, \citenamefont {{Koushiappas}},\
  and\ \citenamefont {{Walker}}(2015)}]{Geringer-Sameth:2014yza}%
  \BibitemOpen
  \bibfield  {author} {\bibinfo {author} {\bibnamefont {{Geringer-Sameth}},
  \bibfnamefont {A.}}, \bibinfo {author} {\bibnamefont {{Koushiappas}},
  \bibfnamefont {S.~M.}}, \ and\ \bibinfo {author} {\bibnamefont {{Walker}},
  \bibfnamefont {M.}},\ }\href {\doibase 10.1088/0004-637X/801/2/74} {\bibfield
   {journal} {\bibinfo  {journal} {\apj}\ }\textbf {\bibinfo {volume} {801}},\
  \bibinfo {eid} {74} (\bibinfo {year} {2015})},\ \Eprint
  {http://arxiv.org/abs/1408.0002} {arXiv:1408.0002} \BibitemShut {NoStop}%
\bibitem [{\citenamefont {Geringer-Sameth}, \citenamefont {Koushiappas},\ and\
  \citenamefont {Walker}(2014)}]{Geringer-Sameth:2014qqa}%
  \BibitemOpen
  \bibfield  {author} {\bibinfo {author} {\bibnamefont {Geringer-Sameth},
  \bibfnamefont {A.}}, \bibinfo {author} {\bibnamefont {Koushiappas},
  \bibfnamefont {S.~M.}}, \ and\ \bibinfo {author} {\bibnamefont {Walker},
  \bibfnamefont {M.~G.}},\ }\href@noop {} {\  (\bibinfo {year} {2014})},\
  \Eprint {http://arxiv.org/abs/1410.2242} {arXiv:1410.2242 [astro-ph.CO]}
  \BibitemShut {NoStop}%
%%CITATION = ARXIV:1410.2242;%%
\bibitem [{\citenamefont {Geringer-Sameth}\ \emph {et~al.}(2015)\citenamefont
  {Geringer-Sameth}, \citenamefont {Walker}, \citenamefont {Koushiappas},
  \citenamefont {Koposov}, \citenamefont {Belokurov} \emph
  {et~al.}}]{Geringer-Sameth:2015lua}%
  \BibitemOpen
  \bibfield  {author} {\bibinfo {author} {\bibnamefont {Geringer-Sameth},
  \bibfnamefont {A.}}, \bibinfo {author} {\bibnamefont {Walker}, \bibfnamefont
  {M.~G.}}, \bibinfo {author} {\bibnamefont {Koushiappas}, \bibfnamefont
  {S.~M.}},  \emph {et~al.},\ }\href@noop {} {\  (\bibinfo {year} {2015})},\
  \Eprint {http://arxiv.org/abs/1503.02320} {arXiv:1503.02320 [astro-ph.HE]}
  \BibitemShut {NoStop}%
%%CITATION = ARXIV:1503.02320;%%
\bibitem [{\citenamefont {Gibbons}, \citenamefont {Belokurov},\ and\
  \citenamefont {Evans}(2014)}]{Gibbons:2014ewa}%
  \BibitemOpen
  \bibfield  {author} {\bibinfo {author} {\bibnamefont {Gibbons}, \bibfnamefont
  {S.}}, \bibinfo {author} {\bibnamefont {Belokurov}, \bibfnamefont {V.}}, \
  and\ \bibinfo {author} {\bibnamefont {Evans}, \bibfnamefont {N.}},\ }\href
  {\doibase 10.1093/mnras/stu1986} {\bibfield  {journal} {\bibinfo  {journal}
  {Mon.Not.Roy.Astron.Soc.}\ }\textbf {\bibinfo {volume} {445}},\ \bibinfo
  {pages} {3788} (\bibinfo {year} {2014})},\ \Eprint
  {http://arxiv.org/abs/1406.2243} {arXiv:1406.2243 [astro-ph.GA]} \BibitemShut
  {NoStop}%
%%CITATION = ARXIV:1406.2243;%%
\bibitem [{\citenamefont {{Gilmore}}\ \emph {et~al.}(2007)\citenamefont
  {{Gilmore}}, \citenamefont {{Wilkinson}}, \citenamefont {{Wyse}},
  \citenamefont {{Kleyna}}, \citenamefont {{Koch}}, \citenamefont {{Evans}},\
  and\ \citenamefont {{Grebel}}}]{Gilmore:2007}%
  \BibitemOpen
  \bibfield  {author} {\bibinfo {author} {\bibnamefont {{Gilmore}},
  \bibfnamefont {G.}}, \bibinfo {author} {\bibnamefont {{Wilkinson}},
  \bibfnamefont {M.~I.}}, \bibinfo {author} {\bibnamefont {{Wyse}},
  \bibfnamefont {R.~F.~G.}},  \emph {et~al.},\ }\href {\doibase 10.1086/518025}
  {\bibfield  {journal} {\bibinfo  {journal} {\apj}\ }\textbf {\bibinfo
  {volume} {663}},\ \bibinfo {pages} {948} (\bibinfo {year} {2007})},\ \Eprint
  {http://arxiv.org/abs/arXiv:astro-ph/0703308} {arXiv:astro-ph/0703308}
  \BibitemShut {NoStop}%
\bibitem [{\citenamefont {{Gnedin}}\ \emph {et~al.}(2010)\citenamefont
  {{Gnedin}}, \citenamefont {{Brown}}, \citenamefont {{Geller}},\ and\
  \citenamefont {{Kenyon}}}]{2010ApJ...720L.108G}%
  \BibitemOpen
  \bibfield  {author} {\bibinfo {author} {\bibnamefont {{Gnedin}},
  \bibfnamefont {O.~Y.}}, \bibinfo {author} {\bibnamefont {{Brown}},
  \bibfnamefont {W.~R.}}, \bibinfo {author} {\bibnamefont {{Geller}},
  \bibfnamefont {M.~J.}}, \ and\ \bibinfo {author} {\bibnamefont {{Kenyon}},
  \bibfnamefont {S.~J.}},\ }\href {\doibase 10.1088/2041-8205/720/1/L108}
  {\bibfield  {journal} {\bibinfo  {journal} {\apjl}\ }\textbf {\bibinfo
  {volume} {720}},\ \bibinfo {pages} {L108} (\bibinfo {year} {2010})},\ \Eprint
  {http://arxiv.org/abs/1005.2619} {arXiv:1005.2619 [astro-ph.GA]} \BibitemShut
  {NoStop}%
\bibitem [{\citenamefont {Gnedin}\ and\ \citenamefont
  {Primack}(2004)}]{Gnedin:2003rj}%
  \BibitemOpen
  \bibfield  {author} {\bibinfo {author} {\bibnamefont {Gnedin}, \bibfnamefont
  {O.~Y.}}\ and\ \bibinfo {author} {\bibnamefont {Primack}, \bibfnamefont
  {J.~R.}},\ }\href {\doibase 10.1103/PhysRevLett.93.061302} {\bibfield
  {journal} {\bibinfo  {journal} {Phys.Rev.Lett.}\ }\textbf {\bibinfo {volume}
  {93}},\ \bibinfo {pages} {061302} (\bibinfo {year} {2004})},\ \Eprint
  {http://arxiv.org/abs/astro-ph/0308385} {arXiv:astro-ph/0308385 [astro-ph]}
  \BibitemShut {NoStop}%
%%CITATION = ASTRO-PH/0308385;%%
\bibitem [{\citenamefont {{G{\'o}mez-Vargas}}\ \emph
  {et~al.}(2013)\citenamefont {{G{\'o}mez-Vargas}}, \citenamefont
  {{S{\'a}nchez-Conde}}, \citenamefont {{Huh}}, \citenamefont {{Peir{\'o}}},
  \citenamefont {{Prada}}, \citenamefont {{Morselli}}, \citenamefont
  {{Klypin}}, \citenamefont {{Cerde{\~n}o}}, \citenamefont {{Mambrini}},\ and\
  \citenamefont {{Mu{\~n}oz}}}]{2013JCAP...10..029G}%
  \BibitemOpen
  \bibfield  {author} {\bibinfo {author} {\bibnamefont {{G{\'o}mez-Vargas}},
  \bibfnamefont {G.~A.}}, \bibinfo {author} {\bibnamefont
  {{S{\'a}nchez-Conde}}, \bibfnamefont {M.~A.}}, \bibinfo {author}
  {\bibnamefont {{Huh}}, \bibfnamefont {J.-H.}},  \emph {et~al.},\ }\href
  {\doibase 10.1088/1475-7516/2013/10/029} {\bibfield  {journal} {\bibinfo
  {journal} {\jcap}\ }\textbf {\bibinfo {volume} {10}},\ \bibinfo {eid} {029}
  (\bibinfo {year} {2013})},\ \Eprint {http://arxiv.org/abs/1308.3515}
  {arXiv:1308.3515 [astro-ph.HE]} \BibitemShut {NoStop}%
\bibitem [{\citenamefont {Gondolo}\ and\ \citenamefont
  {Silk}(1999)}]{Gondolo:1999ef}%
  \BibitemOpen
  \bibfield  {author} {\bibinfo {author} {\bibnamefont {Gondolo}, \bibfnamefont
  {P.}}\ and\ \bibinfo {author} {\bibnamefont {Silk}, \bibfnamefont {J.}},\
  }\href {\doibase 10.1103/PhysRevLett.83.1719} {\bibfield  {journal} {\bibinfo
   {journal} {Phys.Rev.Lett.}\ }\textbf {\bibinfo {volume} {83}},\ \bibinfo
  {pages} {1719} (\bibinfo {year} {1999})},\ \Eprint
  {http://arxiv.org/abs/astro-ph/9906391} {arXiv:astro-ph/9906391 [astro-ph]}
  \BibitemShut {NoStop}%
%%CITATION = ASTRO-PH/9906391;%%
\bibitem [{\citenamefont {Green}, \citenamefont {Hofmann},\ and\ \citenamefont
  {Schwarz}(2004)}]{Green:2003un}%
  \BibitemOpen
  \bibfield  {author} {\bibinfo {author} {\bibnamefont {Green}, \bibfnamefont
  {A.~M.}}, \bibinfo {author} {\bibnamefont {Hofmann}, \bibfnamefont {S.}}, \
  and\ \bibinfo {author} {\bibnamefont {Schwarz}, \bibfnamefont {D.~J.}},\
  }\href {\doibase 10.1111/j.1365-2966.2004.08232.x} {\bibfield  {journal}
  {\bibinfo  {journal} {Mon.Not.Roy.Astron.Soc.}\ }\textbf {\bibinfo {volume}
  {353}},\ \bibinfo {pages} {L23} (\bibinfo {year} {2004})},\ \Eprint
  {http://arxiv.org/abs/astro-ph/0309621} {arXiv:astro-ph/0309621 [astro-ph]}
  \BibitemShut {NoStop}%
%%CITATION = ASTRO-PH/0309621;%%
\bibitem [{\citenamefont {Gross}\ and\ \citenamefont
  {Vitells}(2010)}]{Gross:2010bma}%
  \BibitemOpen
  \bibfield  {author} {\bibinfo {author} {\bibnamefont {Gross}, \bibfnamefont
  {E.}}\ and\ \bibinfo {author} {\bibnamefont {Vitells}, \bibfnamefont {O.}},\
  }\href {\doibase 10.1140/epjc/s10052-010-1470-8} {\bibfield  {journal}
  {\bibinfo  {journal} {Eur.Phys.J.}\ }\textbf {\bibinfo {volume} {C70}},\
  \bibinfo {pages} {525} (\bibinfo {year} {2010})},\ \Eprint
  {http://arxiv.org/abs/1005.1891} {arXiv:1005.1891 [physics.data-an]}
  \BibitemShut {NoStop}%
%%CITATION = ARXIV:1005.1891;%%
\bibitem [{\citenamefont {{Guo}}\ \emph {et~al.}(2010)\citenamefont {{Guo}},
  \citenamefont {{White}}, \citenamefont {{Li}},\ and\ \citenamefont
  {{Boylan-Kolchin}}}]{2010MNRAS.404.1111G}%
  \BibitemOpen
  \bibfield  {author} {\bibinfo {author} {\bibnamefont {{Guo}}, \bibfnamefont
  {Q.}}, \bibinfo {author} {\bibnamefont {{White}}, \bibfnamefont {S.}},
  \bibinfo {author} {\bibnamefont {{Li}}, \bibfnamefont {C.}}, \ and\ \bibinfo
  {author} {\bibnamefont {{Boylan-Kolchin}}, \bibfnamefont {M.}},\ }\href
  {\doibase 10.1111/j.1365-2966.2010.16341.x} {\bibfield  {journal} {\bibinfo
  {journal} {\mnras}\ }\textbf {\bibinfo {volume} {404}},\ \bibinfo {pages}
  {1111} (\bibinfo {year} {2010})},\ \Eprint {http://arxiv.org/abs/0909.4305}
  {arXiv:0909.4305 [astro-ph.CO]} \BibitemShut {NoStop}%
\bibitem [{\citenamefont {Han}\ \emph {et~al.}(2012)\citenamefont {Han},
  \citenamefont {Frenk}, \citenamefont {Eke}, \citenamefont {Gao},
  \citenamefont {White} \emph {et~al.}}]{Han:2012uw}%
  \BibitemOpen
  \bibfield  {author} {\bibinfo {author} {\bibnamefont {Han}, \bibfnamefont
  {J.}}, \bibinfo {author} {\bibnamefont {Frenk}, \bibfnamefont {C.~S.}},
  \bibinfo {author} {\bibnamefont {Eke}, \bibfnamefont {V.~R.}},  \emph
  {et~al.},\ }\href {\doibase 10.1111/j.1365-2966.2012.22080.x} {\bibfield
  {journal} {\bibinfo  {journal} {Mon.Not.Roy.Astron.Soc.}\ }\textbf {\bibinfo
  {volume} {427}},\ \bibinfo {pages} {1651} (\bibinfo {year} {2012})},\ \Eprint
  {http://arxiv.org/abs/1207.6749} {arXiv:1207.6749 [astro-ph.CO]} \BibitemShut
  {NoStop}%
%%CITATION = ARXIV:1207.6749;%%
\bibitem [{\citenamefont {{Han}}\ \emph {et~al.}(2012)\citenamefont {{Han}},
  \citenamefont {{Frenk}}, \citenamefont {{Eke}}, \citenamefont {{Gao}},\ and\
  \citenamefont {{White}}}]{2012arXiv1201.1003H}%
  \BibitemOpen
  \bibfield  {author} {\bibinfo {author} {\bibnamefont {{Han}}, \bibfnamefont
  {J.}}, \bibinfo {author} {\bibnamefont {{Frenk}}, \bibfnamefont {C.~S.}},
  \bibinfo {author} {\bibnamefont {{Eke}}, \bibfnamefont {V.~R.}}, \bibinfo
  {author} {\bibnamefont {{Gao}}, \bibfnamefont {L.}}, \ and\ \bibinfo {author}
  {\bibnamefont {{White}}, \bibfnamefont {S.~D.~M.}},\ }\href@noop {}
  {\bibfield  {journal} {\bibinfo  {journal} {ArXiv e-prints}\ } (\bibinfo
  {year} {2012})},\ \Eprint {http://arxiv.org/abs/1201.1003} {arXiv:1201.1003
  [astro-ph.HE]} \BibitemShut {NoStop}%
\bibitem [{\citenamefont {Hayashi}\ and\ \citenamefont
  {Chiba}(2012)}]{Hayashi:2012si}%
  \BibitemOpen
  \bibfield  {author} {\bibinfo {author} {\bibnamefont {Hayashi}, \bibfnamefont
  {K.}}\ and\ \bibinfo {author} {\bibnamefont {Chiba}, \bibfnamefont {M.}},\
  }\href {\doibase 10.1088/0004-637X/755/2/145} {\bibfield  {journal} {\bibinfo
   {journal} {Astrophys.J.}\ }\textbf {\bibinfo {volume} {755}},\ \bibinfo
  {pages} {145} (\bibinfo {year} {2012})},\ \Eprint
  {http://arxiv.org/abs/1206.3888} {arXiv:1206.3888 [astro-ph.CO]} \BibitemShut
  {NoStop}%
%%CITATION = ARXIV:1206.3888;%%
\bibitem [{\citenamefont {{Hensley}}, \citenamefont {{Siegal-Gaskins}},\ and\
  \citenamefont {{Pavlidou}}(2010)}]{2010ApJ...723..277H}%
  \BibitemOpen
  \bibfield  {author} {\bibinfo {author} {\bibnamefont {{Hensley}},
  \bibfnamefont {B.~S.}}, \bibinfo {author} {\bibnamefont {{Siegal-Gaskins}},
  \bibfnamefont {J.~M.}}, \ and\ \bibinfo {author} {\bibnamefont {{Pavlidou}},
  \bibfnamefont {V.}},\ }\href {\doibase 10.1088/0004-637X/723/1/277}
  {\bibfield  {journal} {\bibinfo  {journal} {\apj}\ }\textbf {\bibinfo
  {volume} {723}},\ \bibinfo {pages} {277} (\bibinfo {year} {2010})},\ \Eprint
  {http://arxiv.org/abs/0912.1854} {arXiv:0912.1854 [astro-ph.CO]} \BibitemShut
  {NoStop}%
\bibitem [{\citenamefont {{Hewitt}}\ \emph {et~al.}(2013)\citenamefont
  {{Hewitt}}, \citenamefont {{Acero}}, \citenamefont {{Brandt}}, \citenamefont
  {{Cohen}}, \citenamefont {{de Palma}}, \citenamefont {{Giordano}},\ and\
  \citenamefont {{for the Fermi LAT Collaboration}}}]{2013arXiv1307.6570H}%
  \BibitemOpen
  \bibfield  {author} {\bibinfo {author} {\bibnamefont {{Hewitt}},
  \bibfnamefont {J.~W.}}, \bibinfo {author} {\bibnamefont {{Acero}},
  \bibfnamefont {F.}}, \bibinfo {author} {\bibnamefont {{Brandt}},
  \bibfnamefont {T.~J.}},  \emph {et~al.},\ }\href@noop {} {\bibfield
  {journal} {\bibinfo  {journal} {ArXiv e-prints}\ } (\bibinfo {year}
  {2013})},\ \Eprint {http://arxiv.org/abs/1307.6570} {arXiv:1307.6570
  [astro-ph.HE]} \BibitemShut {NoStop}%
\bibitem [{\citenamefont {Hinton}(2004)}]{Hinton:2004eu}%
  \BibitemOpen
  \bibfield  {author} {\bibinfo {author} {\bibnamefont {Hinton}, \bibfnamefont
  {J.}} (\bibinfo {collaboration} {HESS Collaboration}),\ }\href {\doibase
  10.1016/j.newar.2003.12.004} {\bibfield  {journal} {\bibinfo  {journal} {New
  Astron.Rev.}\ }\textbf {\bibinfo {volume} {48}},\ \bibinfo {pages} {331}
  (\bibinfo {year} {2004})},\ \Eprint {http://arxiv.org/abs/astro-ph/0403052}
  {arXiv:astro-ph/0403052 [astro-ph]} \BibitemShut {NoStop}%
%%CITATION = ASTRO-PH/0403052;%%
\bibitem [{\citenamefont {{Hofmann}}\ \emph {et~al.}(2003)\citenamefont
  {{Hofmann}}, \citenamefont {{Horns}}, \citenamefont {{Lampeitl}},\ and\
  \citenamefont {{HEGRA Collaboration}}}]{2003ICRC....3.1685H}%
  \BibitemOpen
  \bibfield  {author} {\bibinfo {author} {\bibnamefont {{Hofmann}},
  \bibfnamefont {W.}}, \bibinfo {author} {\bibnamefont {{Horns}}, \bibfnamefont
  {D.}}, \bibinfo {author} {\bibnamefont {{Lampeitl}}, \bibfnamefont {H.}}, \
  and\ \bibinfo {author} {\bibnamefont {{HEGRA Collaboration}},},\ }\href@noop
  {} {\bibfield  {journal} {\bibinfo  {journal} {International Cosmic Ray
  Conference}\ }\textbf {\bibinfo {volume} {3}},\ \bibinfo {pages} {1685}
  (\bibinfo {year} {2003})}\BibitemShut {NoStop}%
\bibitem [{\citenamefont {{Holder}}\ \emph {et~al.}(2006)\citenamefont
  {{Holder}}, \citenamefont {{Atkins}}, \citenamefont {{Badran}}, \citenamefont
  {{Blaylock}}, \citenamefont {{Bradbury}}, \citenamefont {{Buckley}},
  \citenamefont {{Byrum}}, \citenamefont {{Carter-Lewis}}, \citenamefont
  {{Celik}}, \citenamefont {{Chow}}, \citenamefont {{Cogan}}, \citenamefont
  {{Cui}}, \citenamefont {{Daniel}}, \citenamefont {{de la Calle Perez}},
  \citenamefont {{Dowdall}}, \citenamefont {{Dowkontt}}, \citenamefont
  {{Duke}}, \citenamefont {{Falcone}}, \citenamefont {{Fegan}}, \citenamefont
  {{Finley}}, \citenamefont {{Fortin}}, \citenamefont {{Fortson}},
  \citenamefont {{Gibbs}}, \citenamefont {{Gillanders}}, \citenamefont
  {{Glidewell}}, \citenamefont {{Grube}}, \citenamefont {{Gutierrez}},
  \citenamefont {{Gyuk}}, \citenamefont {{Hall}}, \citenamefont {{Hanna}},
  \citenamefont {{Hays}}, \citenamefont {{Horan}}, \citenamefont {{Hughes}},
  \citenamefont {{Humensky}}, \citenamefont {{Imran}}, \citenamefont {{Jung}},
  \citenamefont {{Kaaret}}, \citenamefont {{Kenny}}, \citenamefont {{Kieda}},
  \citenamefont {{Kildea}}, \citenamefont {{Knapp}}, \citenamefont
  {{Krawczynski}}, \citenamefont {{Krennrich}}, \citenamefont {{Lang}},
  \citenamefont {{LeBohec}}, \citenamefont {{Linton}}, \citenamefont
  {{Little}}, \citenamefont {{Maier}}, \citenamefont {{Manseri}}, \citenamefont
  {{Milovanovic}}, \citenamefont {{Moriarty}}, \citenamefont {{Mukherjee}},
  \citenamefont {{Ogden}}, \citenamefont {{Ong}}, \citenamefont {{Petry}},
  \citenamefont {{Perkins}}, \citenamefont {{Pizlo}}, \citenamefont {{Pohl}},
  \citenamefont {{Quinn}}, \citenamefont {{Ragan}}, \citenamefont {{Reynolds}},
  \citenamefont {{Roache}}, \citenamefont {{Rose}}, \citenamefont
  {{Schroedter}}, \citenamefont {{Sembroski}}, \citenamefont {{Sleege}},
  \citenamefont {{Steele}}, \citenamefont {{Swordy}}, \citenamefont {{Syson}},
  \citenamefont {{Toner}}, \citenamefont {{Valcarcel}}, \citenamefont
  {{Vassiliev}}, \citenamefont {{Wakely}}, \citenamefont {{Weekes}},
  \citenamefont {{White}}, \citenamefont {{Williams}},\ and\ \citenamefont
  {{Wagner}}}]{2006APh....25..391H}%
  \BibitemOpen
  \bibfield  {author} {\bibinfo {author} {\bibnamefont {{Holder}},
  \bibfnamefont {J.}}, \bibinfo {author} {\bibnamefont {{Atkins}},
  \bibfnamefont {R.~W.}}, \bibinfo {author} {\bibnamefont {{Badran}},
  \bibfnamefont {H.~M.}},  \emph {et~al.},\ }\href {\doibase
  10.1016/j.astropartphys.2006.04.002} {\bibfield  {journal} {\bibinfo
  {journal} {Astroparticle Physics}\ }\textbf {\bibinfo {volume} {25}},\
  \bibinfo {pages} {391} (\bibinfo {year} {2006})},\ \Eprint
  {http://arxiv.org/abs/astro-ph/0604119} {astro-ph/0604119} \BibitemShut
  {NoStop}%
\bibitem [{\citenamefont {{Hooper}}\ and\ \citenamefont
  {{Goodenough}}(2011)}]{2011PhLB..697..412H}%
  \BibitemOpen
  \bibfield  {author} {\bibinfo {author} {\bibnamefont {{Hooper}},
  \bibfnamefont {D.}}\ and\ \bibinfo {author} {\bibnamefont {{Goodenough}},
  \bibfnamefont {L.}},\ }\href {\doibase 10.1016/j.physletb.2011.02.029}
  {\bibfield  {journal} {\bibinfo  {journal} {Physics Letters B}\ }\textbf
  {\bibinfo {volume} {697}},\ \bibinfo {pages} {412} (\bibinfo {year}
  {2011})},\ \Eprint {http://arxiv.org/abs/1010.2752} {arXiv:1010.2752
  [hep-ph]} \BibitemShut {NoStop}%
\bibitem [{\citenamefont {Hooper}\ and\ \citenamefont
  {Linden}(2011)}]{Hooper:2011ti}%
  \BibitemOpen
  \bibfield  {author} {\bibinfo {author} {\bibnamefont {Hooper}, \bibfnamefont
  {D.}}\ and\ \bibinfo {author} {\bibnamefont {Linden}, \bibfnamefont {T.}},\
  }\href {\doibase 10.1103/PhysRevD.84.123005} {\bibfield  {journal} {\bibinfo
  {journal} {Phys.Rev.}\ }\textbf {\bibinfo {volume} {D84}},\ \bibinfo {pages}
  {123005} (\bibinfo {year} {2011})},\ \Eprint {http://arxiv.org/abs/1110.0006}
  {arXiv:1110.0006 [astro-ph.HE]} \BibitemShut {NoStop}%
%%CITATION = ARXIV:1110.0006;%%
\bibitem [{\citenamefont {{Iocco}}, \citenamefont {{Pato}},\ and\ \citenamefont
  {{Bertone}}(2015)}]{Iocco:2015xga}%
  \BibitemOpen
  \bibfield  {author} {\bibinfo {author} {\bibnamefont {{Iocco}}, \bibfnamefont
  {F.}}, \bibinfo {author} {\bibnamefont {{Pato}}, \bibfnamefont {M.}}, \ and\
  \bibinfo {author} {\bibnamefont {{Bertone}}, \bibfnamefont {G.}},\ }\href
  {\doibase 10.1038/nphys3237} {\bibfield  {journal} {\bibinfo  {journal}
  {Nature Physics}\ }\textbf {\bibinfo {volume} {11}},\ \bibinfo {pages} {245}
  (\bibinfo {year} {2015})},\ \Eprint {http://arxiv.org/abs/1502.03821}
  {arXiv:1502.03821} \BibitemShut {NoStop}%
\bibitem [{\citenamefont {Iocco}\ \emph {et~al.}(2011)\citenamefont {Iocco},
  \citenamefont {Pato}, \citenamefont {Bertone},\ and\ \citenamefont
  {Jetzer}}]{Iocco:2011jz}%
  \BibitemOpen
  \bibfield  {author} {\bibinfo {author} {\bibnamefont {Iocco}, \bibfnamefont
  {F.}}, \bibinfo {author} {\bibnamefont {Pato}, \bibfnamefont {M.}}, \bibinfo
  {author} {\bibnamefont {Bertone}, \bibfnamefont {G.}}, \ and\ \bibinfo
  {author} {\bibnamefont {Jetzer}, \bibfnamefont {P.}},\ }\href {\doibase
  10.1088/1475-7516/2011/11/029} {\bibfield  {journal} {\bibinfo  {journal}
  {JCAP}\ }\textbf {\bibinfo {volume} {1111}},\ \bibinfo {pages} {029}
  (\bibinfo {year} {2011})},\ \Eprint {http://arxiv.org/abs/1107.5810}
  {arXiv:1107.5810 [astro-ph.GA]} \BibitemShut {NoStop}%
%%CITATION = ARXIV:1107.5810;%%
\bibitem [{\citenamefont {Ishiyama}, \citenamefont {Makino},\ and\
  \citenamefont {Ebisuzaki}(2010)}]{Ishiyama:2010es}%
  \BibitemOpen
  \bibfield  {author} {\bibinfo {author} {\bibnamefont {Ishiyama},
  \bibfnamefont {T.}}, \bibinfo {author} {\bibnamefont {Makino}, \bibfnamefont
  {J.}}, \ and\ \bibinfo {author} {\bibnamefont {Ebisuzaki}, \bibfnamefont
  {T.}},\ }\href {\doibase 10.1088/2041-8205/723/2/L195} {\bibfield  {journal}
  {\bibinfo  {journal} {ApJ}\ }\textbf {\bibinfo {volume} {723}},\ \bibinfo
  {pages} {L195} (\bibinfo {year} {2010})},\ \Eprint
  {http://arxiv.org/abs/1006.3392} {arXiv:1006.3392 [astro-ph.CO]} \BibitemShut
  {NoStop}%
%%CITATION = ARXIV:1006.3392;%%
\bibitem [{\citenamefont {James}(2006)}]{James:2006zz}%
  \BibitemOpen
  \bibfield  {author} {\bibinfo {author} {\bibnamefont {James}, \bibfnamefont
  {F.}},\ }\href@noop {} {\emph {\bibinfo {title} {{Statistical methods in
  experimental physics}}}}\ (\bibinfo  {publisher} {Publisher: World Scientific
  Publishing Company; 2 edition},\ \bibinfo {year} {2006})\BibitemShut
  {NoStop}%
%%CITATION = ISBN-9789812567956 ETC.;%%
\bibitem [{\citenamefont {Jardel}\ \emph {et~al.}(2013)\citenamefont {Jardel},
  \citenamefont {Gebhardt}, \citenamefont {Fabricius}, \citenamefont {Drory},\
  and\ \citenamefont {Williams}}]{Jardel:2012am}%
  \BibitemOpen
  \bibfield  {author} {\bibinfo {author} {\bibnamefont {Jardel}, \bibfnamefont
  {J.~R.}}, \bibinfo {author} {\bibnamefont {Gebhardt}, \bibfnamefont {K.}},
  \bibinfo {author} {\bibnamefont {Fabricius}, \bibfnamefont {M.~H.}}, \bibinfo
  {author} {\bibnamefont {Drory}, \bibfnamefont {N.}}, \ and\ \bibinfo {author}
  {\bibnamefont {Williams}, \bibfnamefont {M.~J.}},\ }\href {\doibase
  10.1088/0004-637X/763/2/91} {\bibfield  {journal} {\bibinfo  {journal}
  {Astrophys.J.}\ }\textbf {\bibinfo {volume} {763}},\ \bibinfo {pages} {91}
  (\bibinfo {year} {2013})},\ \Eprint {http://arxiv.org/abs/1211.5376}
  {arXiv:1211.5376 [astro-ph.CO]} \BibitemShut {NoStop}%
%%CITATION = ARXIV:1211.5376;%%
\bibitem [{\citenamefont {Jungman}, \citenamefont {Kamionkowski},\ and\
  \citenamefont {Griest}(1996)}]{Jungman:1995df}%
  \BibitemOpen
  \bibfield  {author} {\bibinfo {author} {\bibnamefont {Jungman}, \bibfnamefont
  {G.}}, \bibinfo {author} {\bibnamefont {Kamionkowski}, \bibfnamefont {M.}}, \
  and\ \bibinfo {author} {\bibnamefont {Griest}, \bibfnamefont {K.}},\ }\href
  {\doibase 10.1016/0370-1573(95)00058-5} {\bibfield  {journal} {\bibinfo
  {journal} {\physrep}\ }\textbf {\bibinfo {volume} {267}},\ \bibinfo {pages}
  {195} (\bibinfo {year} {1996})},\ \Eprint
  {http://arxiv.org/abs/hep-ph/9506380} {arXiv:hep-ph/9506380 [hep-ph]}
  \BibitemShut {NoStop}%
\bibitem [{\citenamefont {Kauffmann}, \citenamefont {White},\ and\
  \citenamefont {Guiderdoni}(1993)}]{Kauffmann:1993gv}%
  \BibitemOpen
  \bibfield  {author} {\bibinfo {author} {\bibnamefont {Kauffmann},
  \bibfnamefont {G.}}, \bibinfo {author} {\bibnamefont {White}, \bibfnamefont
  {S.~D.}}, \ and\ \bibinfo {author} {\bibnamefont {Guiderdoni}, \bibfnamefont
  {B.}},\ }\href@noop {} {\bibfield  {journal} {\bibinfo  {journal}
  {Mon.Not.Roy.Astron.Soc.}\ }\textbf {\bibinfo {volume} {264}},\ \bibinfo
  {pages} {201} (\bibinfo {year} {1993})}\BibitemShut {NoStop}%
%%CITATION = MNRAA,264,201;%%
\bibitem [{\citenamefont {Kissmann}(2014)}]{kissmann_picard:_2014}%
  \BibitemOpen
  \bibfield  {author} {\bibinfo {author} {\bibnamefont {Kissmann},
  \bibfnamefont {R.}},\ }\href {\doibase 10.1016/j.astropartphys.2014.02.002}
  {\bibfield  {journal} {\bibinfo  {journal} {Astroparticle Physics}\ }\textbf
  {\bibinfo {volume} {55}},\ \bibinfo {pages} {37} (\bibinfo {year}
  {2014})}\BibitemShut {NoStop}%
\bibitem [{\citenamefont {{Klypin}}, \citenamefont {{Zhao}},\ and\
  \citenamefont {{Somerville}}(2002)}]{2002ApJ...573..597K}%
  \BibitemOpen
  \bibfield  {author} {\bibinfo {author} {\bibnamefont {{Klypin}},
  \bibfnamefont {A.}}, \bibinfo {author} {\bibnamefont {{Zhao}}, \bibfnamefont
  {H.}}, \ and\ \bibinfo {author} {\bibnamefont {{Somerville}}, \bibfnamefont
  {R.~S.}},\ }\href {\doibase 10.1086/340656} {\bibfield  {journal} {\bibinfo
  {journal} {\apj}\ }\textbf {\bibinfo {volume} {573}},\ \bibinfo {pages} {597}
  (\bibinfo {year} {2002})},\ \Eprint {http://arxiv.org/abs/astro-ph/0110390}
  {astro-ph/0110390} \BibitemShut {NoStop}%
\bibitem [{\citenamefont {Klypin}\ \emph {et~al.}(1999)\citenamefont {Klypin},
  \citenamefont {Kravtsov}, \citenamefont {Valenzuela},\ and\ \citenamefont
  {Prada}}]{Klypin:1999uc}%
  \BibitemOpen
  \bibfield  {author} {\bibinfo {author} {\bibnamefont {Klypin}, \bibfnamefont
  {A.~A.}}, \bibinfo {author} {\bibnamefont {Kravtsov}, \bibfnamefont {A.~V.}},
  \bibinfo {author} {\bibnamefont {Valenzuela}, \bibfnamefont {O.}}, \ and\
  \bibinfo {author} {\bibnamefont {Prada}, \bibfnamefont {F.}},\ }\href
  {\doibase 10.1086/307643} {\bibfield  {journal} {\bibinfo  {journal}
  {Astrophys.J.}\ }\textbf {\bibinfo {volume} {522}},\ \bibinfo {pages} {82}
  (\bibinfo {year} {1999})},\ \Eprint {http://arxiv.org/abs/astro-ph/9901240}
  {arXiv:astro-ph/9901240 [astro-ph]} \BibitemShut {NoStop}%
%%CITATION = ASTRO-PH/9901240;%%
\bibitem [{\citenamefont {{Klypin}}, \citenamefont {{Trujillo-Gomez}},\ and\
  \citenamefont {{Primack}}(2011)}]{2011ApJ...740..102K}%
  \BibitemOpen
  \bibfield  {author} {\bibinfo {author} {\bibnamefont {{Klypin}},
  \bibfnamefont {A.~A.}}, \bibinfo {author} {\bibnamefont {{Trujillo-Gomez}},
  \bibfnamefont {S.}}, \ and\ \bibinfo {author} {\bibnamefont {{Primack}},
  \bibfnamefont {J.}},\ }\href {\doibase 10.1088/0004-637X/740/2/102}
  {\bibfield  {journal} {\bibinfo  {journal} {\apj}\ }\textbf {\bibinfo
  {volume} {740}},\ \bibinfo {eid} {102} (\bibinfo {year} {2011})},\ \Eprint
  {http://arxiv.org/abs/1002.3660} {arXiv:1002.3660 [astro-ph.CO]} \BibitemShut
  {NoStop}%
\bibitem [{\citenamefont {Koposov}\ \emph {et~al.}(2015)\citenamefont
  {Koposov}, \citenamefont {Belokurov}, \citenamefont {Torrealba},\ and\
  \citenamefont {Evans}}]{Koposov:2015cua}%
  \BibitemOpen
  \bibfield  {author} {\bibinfo {author} {\bibnamefont {Koposov}, \bibfnamefont
  {S.~E.}}, \bibinfo {author} {\bibnamefont {Belokurov}, \bibfnamefont {V.}},
  \bibinfo {author} {\bibnamefont {Torrealba}, \bibfnamefont {G.}}, \ and\
  \bibinfo {author} {\bibnamefont {Evans}, \bibfnamefont {N.~W.}},\ }\href@noop
  {} {\  (\bibinfo {year} {2015})},\ \Eprint {http://arxiv.org/abs/1503.02079}
  {arXiv:1503.02079 [astro-ph.GA]} \BibitemShut {NoStop}%
%%CITATION = ARXIV:1503.02079;%%
\bibitem [{\citenamefont {Koushiappas}, \citenamefont {Zentner},\ and\
  \citenamefont {Walker}(2004)}]{Koushiappas:2003bn}%
  \BibitemOpen
  \bibfield  {author} {\bibinfo {author} {\bibnamefont {Koushiappas},
  \bibfnamefont {S.~M.}}, \bibinfo {author} {\bibnamefont {Zentner},
  \bibfnamefont {A.~R.}}, \ and\ \bibinfo {author} {\bibnamefont {Walker},
  \bibfnamefont {T.~P.}},\ }\href {\doibase 10.1103/PhysRevD.69.043501}
  {\bibfield  {journal} {\bibinfo  {journal} {Phys.Rev.}\ }\textbf {\bibinfo
  {volume} {D69}},\ \bibinfo {pages} {043501} (\bibinfo {year} {2004})},\
  \Eprint {http://arxiv.org/abs/astro-ph/0309464} {arXiv:astro-ph/0309464
  [astro-ph]} \BibitemShut {NoStop}%
\bibitem [{\citenamefont {{Kravtsov}}, \citenamefont {{Vikhlinin}},\ and\
  \citenamefont {{Meshscheryakov}}(2014)}]{2014arXiv1401.7329K}%
  \BibitemOpen
  \bibfield  {author} {\bibinfo {author} {\bibnamefont {{Kravtsov}},
  \bibfnamefont {A.}}, \bibinfo {author} {\bibnamefont {{Vikhlinin}},
  \bibfnamefont {A.}}, \ and\ \bibinfo {author} {\bibnamefont
  {{Meshscheryakov}}, \bibfnamefont {A.}},\ }\href@noop {} {\bibfield
  {journal} {\bibinfo  {journal} {ArXiv e-prints}\ } (\bibinfo {year}
  {2014})},\ \Eprint {http://arxiv.org/abs/1401.7329} {arXiv:1401.7329
  [astro-ph.CO]} \BibitemShut {NoStop}%
\bibitem [{\citenamefont {{Laevens}}\ \emph {et~al.}(2015)\citenamefont
  {{Laevens}}, \citenamefont {{Martin}}, \citenamefont {{Ibata}}, \citenamefont
  {{Rix}}, \citenamefont {{Bernard}}, \citenamefont {{Bell}}, \citenamefont
  {{Sesar}}, \citenamefont {{Ferguson}}, \citenamefont {{Schlafly}},
  \citenamefont {{Slater}}, \citenamefont {{Burgett}}, \citenamefont
  {{Chambers}}, \citenamefont {{Flewelling}}, \citenamefont {{Hodapp}},
  \citenamefont {{Kaiser}}, \citenamefont {{Kudritzki}}, \citenamefont
  {{Lupton}}, \citenamefont {{Magnier}}, \citenamefont {{Metcalfe}},
  \citenamefont {{Morgan}}, \citenamefont {{Price}}, \citenamefont {{Tonry}},
  \citenamefont {{Wainscoat}},\ and\ \citenamefont
  {{Waters}}}]{2015arXiv150305554L}%
  \BibitemOpen
  \bibfield  {author} {\bibinfo {author} {\bibnamefont {{Laevens}},
  \bibfnamefont {B.~P.~M.}}, \bibinfo {author} {\bibnamefont {{Martin}},
  \bibfnamefont {N.~F.}}, \bibinfo {author} {\bibnamefont {{Ibata}},
  \bibfnamefont {R.~A.}},  \emph {et~al.},\ }\href@noop {} {\bibfield
  {journal} {\bibinfo  {journal} {ArXiv e-prints}\ } (\bibinfo {year}
  {2015})},\ \Eprint {http://arxiv.org/abs/1503.05554} {arXiv:1503.05554}
  \BibitemShut {NoStop}%
\bibitem [{\citenamefont {{Lamanna}}\ \emph {et~al.}(2013)\citenamefont
  {{Lamanna}}, \citenamefont {{Farnier}}, \citenamefont {{Jacholkowska}},
  \citenamefont {{Kieffer}},\ and\ \citenamefont
  {{Trichard}}}]{2013arXiv1307.4918L}%
  \BibitemOpen
  \bibfield  {author} {\bibinfo {author} {\bibnamefont {{Lamanna}},
  \bibfnamefont {G.}}, \bibinfo {author} {\bibnamefont {{Farnier}},
  \bibfnamefont {C.}}, \bibinfo {author} {\bibnamefont {{Jacholkowska}},
  \bibfnamefont {A.}}, \bibinfo {author} {\bibnamefont {{Kieffer}},
  \bibfnamefont {M.}}, \ and\ \bibinfo {author} {\bibnamefont {{Trichard}},
  \bibfnamefont {C.}} (\bibinfo {collaboration} {H.~E.~S.~S.~Collaboration}),\
  }in\ \href@noop {} {\emph {\bibinfo {booktitle} {Proceedings of the 33rd
  International Cosmic Ray Conference ({ICRC}2013), Rio de Janeiro (Brazil)}}}\
  (\bibinfo {year} {2013})\ p.~\bibinfo {pages} {5},\ \Eprint
  {http://arxiv.org/abs/1307.4918} {arXiv:1307.4918 [astro-ph.HE]} \BibitemShut
  {NoStop}%
\bibitem [{\citenamefont {{Lavalle}}\ \emph {et~al.}(2006)\citenamefont
  {{Lavalle}}, \citenamefont {{Manseri}}, \citenamefont {{Jacholkowska}},
  \citenamefont {{Brion}}, \citenamefont {{Britto}}, \citenamefont {{Bruel}},
  \citenamefont {{Bussons Gordo}}, \citenamefont {{Dumora}}, \citenamefont
  {{Durand}}, \citenamefont {{Giraud}}, \citenamefont {{Lott}}, \citenamefont
  {{M{\"u}nz}}, \citenamefont {{Nuss}}, \citenamefont {{Piron}}, \citenamefont
  {{Reposeur}},\ and\ \citenamefont {{Smith}}}]{2006A&A...450....1L}%
  \BibitemOpen
  \bibfield  {author} {\bibinfo {author} {\bibnamefont {{Lavalle}},
  \bibfnamefont {J.}}, \bibinfo {author} {\bibnamefont {{Manseri}},
  \bibfnamefont {H.}}, \bibinfo {author} {\bibnamefont {{Jacholkowska}},
  \bibfnamefont {A.}},  \emph {et~al.},\ }\href {\doibase
  10.1051/0004-6361:20054340} {\bibfield  {journal} {\bibinfo  {journal}
  {\aap}\ }\textbf {\bibinfo {volume} {450}},\ \bibinfo {pages} {1} (\bibinfo
  {year} {2006})},\ \Eprint {http://arxiv.org/abs/astro-ph/0601298}
  {astro-ph/0601298} \BibitemShut {NoStop}%
\bibitem [{\citenamefont {{Lefranc}}\ \emph {et~al.}(2015)\citenamefont
  {{Lefranc}}, \citenamefont {{Moulin}}, \citenamefont {{Panci}},\ and\
  \citenamefont {{Silk}}}]{2015arXiv150205064L}%
  \BibitemOpen
  \bibfield  {author} {\bibinfo {author} {\bibnamefont {{Lefranc}},
  \bibfnamefont {V.}}, \bibinfo {author} {\bibnamefont {{Moulin}},
  \bibfnamefont {E.}}, \bibinfo {author} {\bibnamefont {{Panci}}, \bibfnamefont
  {P.}}, \ and\ \bibinfo {author} {\bibnamefont {{Silk}}, \bibfnamefont {J.}},\
  }\href@noop {} {\enquote {\bibinfo {title} {{Prospects for Annihilating Dark
  Matter in the inner Galactic halo by the Cherenkov Telescope Array}},}\ }
  (\bibinfo {year} {2015}),\ \bibinfo {note} {provided by the SAO/NASA
  Astrophysics Data System},\ \Eprint {http://arxiv.org/abs/1502.05064}
  {arXiv:1502.05064 [astro-ph.HE]} \BibitemShut {NoStop}%
\bibitem [{\citenamefont {Li}\ and\ \citenamefont {Ma}(1983)}]{Li:1983fv}%
  \BibitemOpen
  \bibfield  {author} {\bibinfo {author} {\bibnamefont {Li}, \bibfnamefont
  {T.-P.}}\ and\ \bibinfo {author} {\bibnamefont {Ma}, \bibfnamefont {Y.-Q.}},\
  }\href {\doibase 10.1086/161295} {\bibfield  {journal} {\bibinfo  {journal}
  {Astrophys.J.}\ }\textbf {\bibinfo {volume} {272}},\ \bibinfo {pages} {317}
  (\bibinfo {year} {1983})}\BibitemShut {NoStop}%
%%CITATION = ASJOA,272,317;%%
\bibitem [{\citenamefont {{Li}}\ and\ \citenamefont
  {{Yuan}}(2012)}]{2012PhLB..715...35L}%
  \BibitemOpen
  \bibfield  {author} {\bibinfo {author} {\bibnamefont {{Li}}, \bibfnamefont
  {Y.}}\ and\ \bibinfo {author} {\bibnamefont {{Yuan}}, \bibfnamefont {Q.}},\
  }\href {\doibase 10.1016/j.physletb.2012.07.057} {\bibfield  {journal}
  {\bibinfo  {journal} {Physics Letters B}\ }\textbf {\bibinfo {volume}
  {715}},\ \bibinfo {pages} {35} (\bibinfo {year} {2012})},\ \Eprint
  {http://arxiv.org/abs/1206.2241} {arXiv:1206.2241 [astro-ph.HE]} \BibitemShut
  {NoStop}%
\bibitem [{\citenamefont {Li}\ and\ \citenamefont {White}(2008)}]{Li:2007eg}%
  \BibitemOpen
  \bibfield  {author} {\bibinfo {author} {\bibnamefont {Li}, \bibfnamefont
  {Y.-S.}}\ and\ \bibinfo {author} {\bibnamefont {White}, \bibfnamefont
  {S.~D.}},\ }\href {\doibase 10.1111/j.1365-2966.2007.12748.x} {\bibfield
  {journal} {\bibinfo  {journal} {Mon.Not.Roy.Astron.Soc.}\ }\textbf {\bibinfo
  {volume} {384}},\ \bibinfo {pages} {1459} (\bibinfo {year} {2008})},\ \Eprint
  {http://arxiv.org/abs/0710.3740} {arXiv:0710.3740 [astro-ph]} \BibitemShut
  {NoStop}%
%%CITATION = ARXIV:0710.3740;%%
\bibitem [{\citenamefont {{Li}}, \citenamefont {{Yuan}},\ and\ \citenamefont
  {{Xu}}(2013)}]{2013arXiv1312.7609L}%
  \BibitemOpen
  \bibfield  {author} {\bibinfo {author} {\bibnamefont {{Li}}, \bibfnamefont
  {Z.}}, \bibinfo {author} {\bibnamefont {{Yuan}}, \bibfnamefont {Q.}}, \ and\
  \bibinfo {author} {\bibnamefont {{Xu}}, \bibfnamefont {Y.}},\ }\href@noop {}
  {\bibfield  {journal} {\bibinfo  {journal} {ArXiv e-prints}\ } (\bibinfo
  {year} {2013})},\ \Eprint {http://arxiv.org/abs/1312.7609} {arXiv:1312.7609
  [astro-ph.CO]} \BibitemShut {NoStop}%
\bibitem [{\citenamefont {{Little}}\ and\ \citenamefont
  {{Tremaine}}(1987)}]{Little:1987}%
  \BibitemOpen
  \bibfield  {author} {\bibinfo {author} {\bibnamefont {{Little}},
  \bibfnamefont {B.}}\ and\ \bibinfo {author} {\bibnamefont {{Tremaine}},
  \bibfnamefont {S.}},\ }\href {\doibase 10.1086/165567} {\bibfield  {journal}
  {\bibinfo  {journal} {\apj}\ }\textbf {\bibinfo {volume} {320}},\ \bibinfo
  {pages} {493} (\bibinfo {year} {1987})}\BibitemShut {NoStop}%
\bibitem [{\citenamefont {Loeb}\ and\ \citenamefont
  {Zaldarriaga}(2005)}]{Loeb:2005pm}%
  \BibitemOpen
  \bibfield  {author} {\bibinfo {author} {\bibnamefont {Loeb}, \bibfnamefont
  {A.}}\ and\ \bibinfo {author} {\bibnamefont {Zaldarriaga}, \bibfnamefont
  {M.}},\ }\href {\doibase 10.1103/PhysRevD.71.103520} {\bibfield  {journal}
  {\bibinfo  {journal} {Phys.Rev.}\ }\textbf {\bibinfo {volume} {D71}},\
  \bibinfo {pages} {103520} (\bibinfo {year} {2005})},\ \Eprint
  {http://arxiv.org/abs/astro-ph/0504112} {arXiv:astro-ph/0504112 [astro-ph]}
  \BibitemShut {NoStop}%
%%CITATION = ASTRO-PH/0504112;%%
\bibitem [{\citenamefont {Lorenz}(2004)}]{Lorenz:2004ah}%
  \BibitemOpen
  \bibfield  {author} {\bibinfo {author} {\bibnamefont {Lorenz}, \bibfnamefont
  {E.}} (\bibinfo {collaboration} {MAGIC Collaboration}),\ }\href {\doibase
  10.1016/j.newar.2003.12.059} {\bibfield  {journal} {\bibinfo  {journal} {New
  Astron.Rev.}\ }\textbf {\bibinfo {volume} {48}},\ \bibinfo {pages} {339}
  (\bibinfo {year} {2004})}\BibitemShut {NoStop}%
%%CITATION = ASTRE,48,339;%%
\bibitem [{\citenamefont {{Mac{\'{\i}}as-Ram{\'{\i}}rez}}\ \emph
  {et~al.}(2012)\citenamefont {{Mac{\'{\i}}as-Ram{\'{\i}}rez}}, \citenamefont
  {{Gordon}}, \citenamefont {{Brown}},\ and\ \citenamefont
  {{Adams}}}]{2012PhRvD..86g6004M}%
  \BibitemOpen
  \bibfield  {author} {\bibinfo {author} {\bibnamefont
  {{Mac{\'{\i}}as-Ram{\'{\i}}rez}}, \bibfnamefont {O.}}, \bibinfo {author}
  {\bibnamefont {{Gordon}}, \bibfnamefont {C.}}, \bibinfo {author}
  {\bibnamefont {{Brown}}, \bibfnamefont {A.~M.}}, \ and\ \bibinfo {author}
  {\bibnamefont {{Adams}}, \bibfnamefont {J.}},\ }\href {\doibase
  10.1103/PhysRevD.86.076004} {\bibfield  {journal} {\bibinfo  {journal}
  {\prd}\ }\textbf {\bibinfo {volume} {86}},\ \bibinfo {eid} {076004} (\bibinfo
  {year} {2012})},\ \Eprint {http://arxiv.org/abs/1207.6257} {arXiv:1207.6257
  [astro-ph.HE]} \BibitemShut {NoStop}%
\bibitem [{\citenamefont {Mack}\ \emph {et~al.}(2008)\citenamefont {Mack},
  \citenamefont {Jacques}, \citenamefont {Beacom}, \citenamefont {Bell},\ and\
  \citenamefont {Yuksel}}]{Mack:2008wu}%
  \BibitemOpen
  \bibfield  {author} {\bibinfo {author} {\bibnamefont {Mack}, \bibfnamefont
  {G.~D.}}, \bibinfo {author} {\bibnamefont {Jacques}, \bibfnamefont {T.~D.}},
  \bibinfo {author} {\bibnamefont {Beacom}, \bibfnamefont {J.~F.}}, \bibinfo
  {author} {\bibnamefont {Bell}, \bibfnamefont {N.~F.}}, \ and\ \bibinfo
  {author} {\bibnamefont {Yuksel}, \bibfnamefont {H.}},\ }\href {\doibase
  10.1103/PhysRevD.78.063542} {\bibfield  {journal} {\bibinfo  {journal}
  {Phys.Rev.}\ }\textbf {\bibinfo {volume} {D78}},\ \bibinfo {pages} {063542}
  (\bibinfo {year} {2008})},\ \Eprint {http://arxiv.org/abs/0803.0157}
  {arXiv:0803.0157 [astro-ph]} \BibitemShut {NoStop}%
%%CITATION = ARXIV:0803.0157;%%
\bibitem [{\citenamefont {Martinez}(2013)}]{Martinez:2013els}%
  \BibitemOpen
  \bibfield  {author} {\bibinfo {author} {\bibnamefont {Martinez},
  \bibfnamefont {G.~D.}},\ }\href@noop {} {\enquote {\bibinfo {title} {{A
  Robust Determination of Milky Way Satellite Properties using Hierarchical
  Mass Modeling}},}\ } (\bibinfo {year} {2013}),\ \Eprint
  {http://arxiv.org/abs/1309.2641} {arXiv:1309.2641 [astro-ph.GA]} \BibitemShut
  {NoStop}%
%%CITATION = ARXIV:1309.2641;%%
\bibitem [{\citenamefont {Martinez}\ \emph {et~al.}(2009)\citenamefont
  {Martinez}, \citenamefont {Bullock}, \citenamefont {Kaplinghat},
  \citenamefont {Strigari},\ and\ \citenamefont {Trotta}}]{Martinez:2009jh}%
  \BibitemOpen
  \bibfield  {author} {\bibinfo {author} {\bibnamefont {Martinez},
  \bibfnamefont {G.~D.}}, \bibinfo {author} {\bibnamefont {Bullock},
  \bibfnamefont {J.~S.}}, \bibinfo {author} {\bibnamefont {Kaplinghat},
  \bibfnamefont {M.}}, \bibinfo {author} {\bibnamefont {Strigari},
  \bibfnamefont {L.~E.}}, \ and\ \bibinfo {author} {\bibnamefont {Trotta},
  \bibfnamefont {R.}},\ }\href {\doibase 10.1088/1475-7516/2009/06/014}
  {\bibfield  {journal} {\bibinfo  {journal} {JCAP}\ }\textbf {\bibinfo
  {volume} {0906}},\ \bibinfo {pages} {014} (\bibinfo {year} {2009})},\ \Eprint
  {http://arxiv.org/abs/0902.4715} {arXiv:0902.4715 [astro-ph.HE]} \BibitemShut
  {NoStop}%
%%CITATION = ARXIV:0902.4715;%%
\bibitem [{\citenamefont {{Mayer-Hasselwander}}\ \emph
  {et~al.}(1998)\citenamefont {{Mayer-Hasselwander}}, \citenamefont
  {{Bertsch}}, \citenamefont {{Dingus}}, \citenamefont {{Eckart}},
  \citenamefont {{Esposito}}, \citenamefont {{Genzel}}, \citenamefont
  {{Hartman}}, \citenamefont {{Hunter}}, \citenamefont {{Kanbach}},
  \citenamefont {{Kniffen}}, \citenamefont {{Lin}}, \citenamefont
  {{Michelson}}, \citenamefont {{Muecke}}, \citenamefont {{von Montigny}},
  \citenamefont {{Mukherjee}}, \citenamefont {{Nolan}}, \citenamefont {{Pohl}},
  \citenamefont {{Reimer}}, \citenamefont {{Schneid}}, \citenamefont
  {{Sreekumar}},\ and\ \citenamefont {{Thompson}}}]{1998A&A...335..161M}%
  \BibitemOpen
  \bibfield  {author} {\bibinfo {author} {\bibnamefont {{Mayer-Hasselwander}},
  \bibfnamefont {H.~A.}}, \bibinfo {author} {\bibnamefont {{Bertsch}},
  \bibfnamefont {D.~L.}}, \bibinfo {author} {\bibnamefont {{Dingus}},
  \bibfnamefont {B.~L.}},  \emph {et~al.},\ }\href@noop {} {\bibfield
  {journal} {\bibinfo  {journal} {\aap}\ }\textbf {\bibinfo {volume} {335}},\
  \bibinfo {pages} {161} (\bibinfo {year} {1998})}\BibitemShut {NoStop}%
\bibitem [{\citenamefont
  {{McConnachie}}(2012{\natexlab{a}})}]{2012AJ....144....4M}%
  \BibitemOpen
  \bibfield  {author} {\bibinfo {author} {\bibnamefont {{McConnachie}},
  \bibfnamefont {A.~W.}},\ }\href {\doibase 10.1088/0004-6256/144/1/4}
  {\bibfield  {journal} {\bibinfo  {journal} {\aj}\ }\textbf {\bibinfo {volume}
  {144}},\ \bibinfo {eid} {4} (\bibinfo {year} {2012}{\natexlab{a}})},\ \Eprint
  {http://arxiv.org/abs/1204.1562} {arXiv:1204.1562 [astro-ph.CO]} \BibitemShut
  {NoStop}%
\bibitem [{\citenamefont
  {{McConnachie}}(2012{\natexlab{b}})}]{McConnachie:2012AJ....144....4M}%
  \BibitemOpen
  \bibfield  {author} {\bibinfo {author} {\bibnamefont {{McConnachie}},
  \bibfnamefont {A.~W.}},\ }\href {\doibase 10.1088/0004-6256/144/1/4}
  {\bibfield  {journal} {\bibinfo  {journal} {\aj}\ }\textbf {\bibinfo {volume}
  {144}},\ \bibinfo {eid} {4} (\bibinfo {year} {2012}{\natexlab{b}})},\ \Eprint
  {http://arxiv.org/abs/1204.1562} {arXiv:1204.1562 [astro-ph.CO]} \BibitemShut
  {NoStop}%
\bibitem [{\citenamefont {McGaugh}\ \emph {et~al.}(2015)\citenamefont
  {McGaugh}, \citenamefont {Lelli}, \citenamefont {Pawlowski}, \citenamefont
  {Angus}, \citenamefont {Bienaymé} \emph {et~al.}}]{McGaugh:2015tha}%
  \BibitemOpen
  \bibfield  {author} {\bibinfo {author} {\bibnamefont {McGaugh}, \bibfnamefont
  {S.}}, \bibinfo {author} {\bibnamefont {Lelli}, \bibfnamefont {F.}}, \bibinfo
  {author} {\bibnamefont {Pawlowski}, \bibfnamefont {M.}},  \emph {et~al.},\
  }\href@noop {} {\  (\bibinfo {year} {2015})},\ \Eprint
  {http://arxiv.org/abs/1503.07813} {arXiv:1503.07813 [astro-ph.GA]}
  \BibitemShut {NoStop}%
%%CITATION = ARXIV:1503.07813;%%
\bibitem [{\citenamefont {{Miville-Desch{\^e}nes}}\ and\ \citenamefont
  {{Lagache}}(2005)}]{2005ApJS..157..302M}%
  \BibitemOpen
  \bibfield  {author} {\bibinfo {author} {\bibnamefont
  {{Miville-Desch{\^e}nes}}, \bibfnamefont {M.-A.}}\ and\ \bibinfo {author}
  {\bibnamefont {{Lagache}}, \bibfnamefont {G.}},\ }\href {\doibase
  10.1086/427938} {\bibfield  {journal} {\bibinfo  {journal} {\apjs}\ }\textbf
  {\bibinfo {volume} {157}},\ \bibinfo {pages} {302} (\bibinfo {year}
  {2005})},\ \Eprint {http://arxiv.org/abs/astro-ph/0412216} {astro-ph/0412216}
  \BibitemShut {NoStop}%
\bibitem [{\citenamefont {{Moiseev}}\ \emph {et~al.}(2013)\citenamefont
  {{Moiseev}}, \citenamefont {{Galper}}, \citenamefont {{Adriani}},
  \citenamefont {{Aptekar}}, \citenamefont {{Arkhangelskaja}}, \citenamefont
  {{Arkhangelskiy}}, \citenamefont {{Avanesov}}, \citenamefont {{Bergstrom}},
  \citenamefont {{Boezio}}, \citenamefont {{Bonvicini}}, \citenamefont
  {{Boyarchuk}}, \citenamefont {{Dogiel}}, \citenamefont {{Gusakov}},
  \citenamefont {{Fradkin}}, \citenamefont {{Fuglesang}}, \citenamefont
  {{Hnatyk}}, \citenamefont {{Kachanov}}, \citenamefont {{Kaplin}},
  \citenamefont {{Kheymits}}, \citenamefont {{Korepanov}}, \citenamefont
  {{Larsson}}, \citenamefont {{Leonov}}, \citenamefont {{Longo}}, \citenamefont
  {{Maestro}}, \citenamefont {{Marrocchesi}}, \citenamefont {{Mazets}},
  \citenamefont {{Mikhailov}}, \citenamefont {{Mocchiutti}}, \citenamefont
  {{Mori}}, \citenamefont {{Moskalenko}}, \citenamefont {{Naumov}},
  \citenamefont {{Papini}}, \citenamefont {{Pearce}}, \citenamefont
  {{Picozza}}, \citenamefont {{Runtso}}, \citenamefont {{Ryde}}, \citenamefont
  {{Sparvoli}}, \citenamefont {{Spillantini}}, \citenamefont {{Suchkov}},
  \citenamefont {{Tavani}}, \citenamefont {{Topchiev}}, \citenamefont
  {{Vacchi}}, \citenamefont {{Vannuccini}}, \citenamefont {{Yurkin}},
  \citenamefont {{Zampa}}, \citenamefont {{Zarikashvili}},\ and\ \citenamefont
  {{Zverev}}}]{2013arXiv1307.2345M}%
  \BibitemOpen
  \bibfield  {author} {\bibinfo {author} {\bibnamefont {{Moiseev}},
  \bibfnamefont {A.~A.}}, \bibinfo {author} {\bibnamefont {{Galper}},
  \bibfnamefont {A.~M.}}, \bibinfo {author} {\bibnamefont {{Adriani}},
  \bibfnamefont {O.}},  \emph {et~al.},\ }\href@noop {} {\bibfield  {journal}
  {\bibinfo  {journal} {ArXiv e-prints}\ } (\bibinfo {year} {2013})},\ \Eprint
  {http://arxiv.org/abs/1307.2345} {arXiv:1307.2345 [astro-ph.IM]} \BibitemShut
  {NoStop}%
\bibitem [{\citenamefont {{Moore}}\ \emph {et~al.}(1999)\citenamefont
  {{Moore}}, \citenamefont {{Ghigna}}, \citenamefont {{Governato}},
  \citenamefont {{Lake}}, \citenamefont {{Quinn}}, \citenamefont {{Stadel}},\
  and\ \citenamefont {{Tozzi}}}]{Moore:1999}%
  \BibitemOpen
  \bibfield  {author} {\bibinfo {author} {\bibnamefont {{Moore}}, \bibfnamefont
  {B.}}, \bibinfo {author} {\bibnamefont {{Ghigna}}, \bibfnamefont {S.}},
  \bibinfo {author} {\bibnamefont {{Governato}}, \bibfnamefont {F.}},  \emph
  {et~al.},\ }\href {\doibase 10.1086/312287} {\bibfield  {journal} {\bibinfo
  {journal} {\apjl}\ }\textbf {\bibinfo {volume} {524}},\ \bibinfo {pages}
  {L19} (\bibinfo {year} {1999})},\ \Eprint
  {http://arxiv.org/abs/arXiv:astro-ph/9907411} {arXiv:astro-ph/9907411}
  \BibitemShut {NoStop}%
\bibitem [{\citenamefont {Moster}\ \emph {et~al.}(2010)\citenamefont {Moster},
  \citenamefont {Somerville}, \citenamefont {Maulbetsch}, \citenamefont
  {Bosch}, \citenamefont {Maccio'} \emph {et~al.}}]{Moster:2009fk}%
  \BibitemOpen
  \bibfield  {author} {\bibinfo {author} {\bibnamefont {Moster}, \bibfnamefont
  {B.~P.}}, \bibinfo {author} {\bibnamefont {Somerville}, \bibfnamefont
  {R.~S.}}, \bibinfo {author} {\bibnamefont {Maulbetsch}, \bibfnamefont {C.}},
  \emph {et~al.},\ }\href {\doibase 10.1088/0004-637X/710/2/903} {\bibfield
  {journal} {\bibinfo  {journal} {Astrophys.J.}\ }\textbf {\bibinfo {volume}
  {710}},\ \bibinfo {pages} {903} (\bibinfo {year} {2010})},\ \Eprint
  {http://arxiv.org/abs/0903.4682} {arXiv:0903.4682 [astro-ph.CO]} \BibitemShut
  {NoStop}%
%%CITATION = ARXIV:0903.4682;%%
\bibitem [{\citenamefont {Nagai}, \citenamefont {Vikhlinin},\ and\
  \citenamefont {Kravtsov}(2007)}]{Nagai:2006sz}%
  \BibitemOpen
  \bibfield  {author} {\bibinfo {author} {\bibnamefont {Nagai}, \bibfnamefont
  {D.}}, \bibinfo {author} {\bibnamefont {Vikhlinin}, \bibfnamefont {A.}}, \
  and\ \bibinfo {author} {\bibnamefont {Kravtsov}, \bibfnamefont {A.~V.}},\
  }\href {\doibase 10.1086/509868} {\bibfield  {journal} {\bibinfo  {journal}
  {Astrophys.J.}\ }\textbf {\bibinfo {volume} {655}},\ \bibinfo {pages} {98}
  (\bibinfo {year} {2007})},\ \Eprint {http://arxiv.org/abs/astro-ph/0609247}
  {arXiv:astro-ph/0609247 [astro-ph]} \BibitemShut {NoStop}%
%%CITATION = ASTRO-PH/0609247;%%
\bibitem [{\citenamefont {Navarro}, \citenamefont {Frenk},\ and\ \citenamefont
  {White}(1997)}]{Navarro:1996gj}%
  \BibitemOpen
  \bibfield  {author} {\bibinfo {author} {\bibnamefont {Navarro}, \bibfnamefont
  {J.~F.}}, \bibinfo {author} {\bibnamefont {Frenk}, \bibfnamefont {C.~S.}}, \
  and\ \bibinfo {author} {\bibnamefont {White}, \bibfnamefont {S.~D.}},\ }\href
  {\doibase 10.1086/304888} {\bibfield  {journal} {\bibinfo  {journal} {\apj}\
  }\textbf {\bibinfo {volume} {490}},\ \bibinfo {pages} {493} (\bibinfo {year}
  {1997})},\ \Eprint {http://arxiv.org/abs/astro-ph/9611107}
  {arXiv:astro-ph/9611107 [astro-ph]} \BibitemShut {NoStop}%
\bibitem [{\citenamefont {Newman}\ \emph {et~al.}(2013)\citenamefont {Newman},
  \citenamefont {Treu}, \citenamefont {Ellis},\ and\ \citenamefont
  {Sand}}]{Newman:2012nw}%
  \BibitemOpen
  \bibfield  {author} {\bibinfo {author} {\bibnamefont {Newman}, \bibfnamefont
  {A.~B.}}, \bibinfo {author} {\bibnamefont {Treu}, \bibfnamefont {T.}},
  \bibinfo {author} {\bibnamefont {Ellis}, \bibfnamefont {R.~S.}}, \ and\
  \bibinfo {author} {\bibnamefont {Sand}, \bibfnamefont {D.~J.}},\ }\href
  {\doibase 10.1088/0004-637X/765/1/25} {\bibfield  {journal} {\bibinfo
  {journal} {Astrophys.J.}\ }\textbf {\bibinfo {volume} {765}},\ \bibinfo
  {pages} {25} (\bibinfo {year} {2013})},\ \Eprint
  {http://arxiv.org/abs/1209.1392} {arXiv:1209.1392 [astro-ph.CO]} \BibitemShut
  {NoStop}%
%%CITATION = ARXIV:1209.1392;%%
\bibitem [{\citenamefont {Ng}\ \emph {et~al.}(2014)\citenamefont {Ng},
  \citenamefont {Laha}, \citenamefont {Campbell}, \citenamefont {Horiuchi},
  \citenamefont {Dasgupta} \emph {et~al.}}]{Ng:2013xha}%
  \BibitemOpen
  \bibfield  {author} {\bibinfo {author} {\bibnamefont {Ng}, \bibfnamefont
  {K.~C.~Y.}}, \bibinfo {author} {\bibnamefont {Laha}, \bibfnamefont {R.}},
  \bibinfo {author} {\bibnamefont {Campbell}, \bibfnamefont {S.}},  \emph
  {et~al.},\ }\href {\doibase 10.1103/PhysRevD.89.083001} {\bibfield  {journal}
  {\bibinfo  {journal} {Phys.Rev.}\ }\textbf {\bibinfo {volume} {D89}},\
  \bibinfo {pages} {083001} (\bibinfo {year} {2014})},\ \Eprint
  {http://arxiv.org/abs/1310.1915} {arXiv:1310.1915 [astro-ph.CO]} \BibitemShut
  {NoStop}%
%%CITATION = ARXIV:1310.1915;%%
\bibitem [{\citenamefont {{Perkins}}\ \emph {et~al.}(2006)\citenamefont
  {{Perkins}}, \citenamefont {{Badran}}, \citenamefont {{Blaylock}},
  \citenamefont {{Bradbury}}, \citenamefont {{Cogan}}, \citenamefont {{Chow}},
  \citenamefont {{Cui}}, \citenamefont {{Daniel}}, \citenamefont {{Falcone}},
  \citenamefont {{Fegan}}, \citenamefont {{Finley}}, \citenamefont {{Fortin}},
  \citenamefont {{Fortson}}, \citenamefont {{Gillanders}}, \citenamefont
  {{Gutierrez}}, \citenamefont {{Grube}}, \citenamefont {{Hall}}, \citenamefont
  {{Hanna}}, \citenamefont {{Holder}}, \citenamefont {{Horan}}, \citenamefont
  {{Hughes}}, \citenamefont {{Humensky}}, \citenamefont {{Kenny}},
  \citenamefont {{Kertzman}}, \citenamefont {{Kieda}}, \citenamefont
  {{Kildea}}, \citenamefont {{Kosack}}, \citenamefont {{Krawczynski}},
  \citenamefont {{Krennrich}}, \citenamefont {{Lang}}, \citenamefont
  {{LeBohec}}, \citenamefont {{Maier}}, \citenamefont {{Moriarty}},
  \citenamefont {{Ong}}, \citenamefont {{Pohl}}, \citenamefont {{Ragan}},
  \citenamefont {{Rebillot}}, \citenamefont {{Sembroski}}, \citenamefont
  {{Steele}}, \citenamefont {{Swordy}}, \citenamefont {{Valcarcel}},
  \citenamefont {{Vassiliev}}, \citenamefont {{Wakely}}, \citenamefont
  {{Weekes}}, \citenamefont {{Williams}},\ and\ \citenamefont {{VERITAS
  Collaboration}}}]{2006ApJ...644..148P}%
  \BibitemOpen
  \bibfield  {author} {\bibinfo {author} {\bibnamefont {{Perkins}},
  \bibfnamefont {J.~S.}}, \bibinfo {author} {\bibnamefont {{Badran}},
  \bibfnamefont {H.~M.}}, \bibinfo {author} {\bibnamefont {{Blaylock}},
  \bibfnamefont {G.}},  \emph {et~al.},\ }\href {\doibase 10.1086/503321}
  {\bibfield  {journal} {\bibinfo  {journal} {\apj}\ }\textbf {\bibinfo
  {volume} {644}},\ \bibinfo {pages} {148} (\bibinfo {year} {2006})},\ \Eprint
  {http://arxiv.org/abs/astro-ph/0602258} {astro-ph/0602258} \BibitemShut
  {NoStop}%
\bibitem [{\citenamefont {Perryman}\ \emph {et~al.}(2001)\citenamefont
  {Perryman}, \citenamefont {de~Boer}, \citenamefont {Gilmore}, \citenamefont
  {Hog}, \citenamefont {Lattanzi} \emph {et~al.}}]{Perryman:2001sp}%
  \BibitemOpen
  \bibfield  {author} {\bibinfo {author} {\bibnamefont {Perryman},
  \bibfnamefont {M.}}, \bibinfo {author} {\bibnamefont {de~Boer}, \bibfnamefont
  {K.~S.}}, \bibinfo {author} {\bibnamefont {Gilmore}, \bibfnamefont {G.}},
  \emph {et~al.},\ }\href {\doibase 10.1051/0004-6361:20010085} {\bibfield
  {journal} {\bibinfo  {journal} {Astron.Astrophys.}\ }\textbf {\bibinfo
  {volume} {369}},\ \bibinfo {pages} {339} (\bibinfo {year} {2001})},\ \Eprint
  {http://arxiv.org/abs/astro-ph/0101235} {arXiv:astro-ph/0101235 [astro-ph]}
  \BibitemShut {NoStop}%
%%CITATION = ASTRO-PH/0101235;%%
\bibitem [{\citenamefont {Petrovi{\'c}}, \citenamefont {Serpico},\ and\
  \citenamefont {Zaharija{\v s}}(2014)}]{Petrovic:2014uda}%
  \BibitemOpen
  \bibfield  {author} {\bibinfo {author} {\bibnamefont {Petrovi{\'c}},
  \bibfnamefont {J.}}, \bibinfo {author} {\bibnamefont {Serpico}, \bibfnamefont
  {P.}}, \ and\ \bibinfo {author} {\bibnamefont {Zaharija{\v s}}, \bibfnamefont
  {G.}},\ }\href {\doibase 10.1088/1475-7516/2014/10/052} {\bibfield  {journal}
  {\bibinfo  {journal} {\jcap}\ }\textbf {\bibinfo {volume} {10}},\ \bibinfo
  {eid} {052} (\bibinfo {year} {2014})},\ \Eprint
  {http://arxiv.org/abs/1405.7928} {arXiv:1405.7928 [astro-ph.HE]} \BibitemShut
  {NoStop}%
\bibitem [{\citenamefont {Pieri}, \citenamefont {Bertone},\ and\ \citenamefont
  {Branchini}(2008)}]{Pieri:2007ir}%
  \BibitemOpen
  \bibfield  {author} {\bibinfo {author} {\bibnamefont {Pieri}, \bibfnamefont
  {L.}}, \bibinfo {author} {\bibnamefont {Bertone}, \bibfnamefont {G.}}, \ and\
  \bibinfo {author} {\bibnamefont {Branchini}, \bibfnamefont {E.}},\ }\href
  {\doibase 10.1111/j.1365-2966.2007.12828.x} {\bibfield  {journal} {\bibinfo
  {journal} {Mon.Not.Roy.Astron.Soc.}\ }\textbf {\bibinfo {volume} {384}},\
  \bibinfo {pages} {1627} (\bibinfo {year} {2008})},\ \Eprint
  {http://arxiv.org/abs/0706.2101} {arXiv:0706.2101 [astro-ph]} \BibitemShut
  {NoStop}%
\bibitem [{\citenamefont {Pierre}, \citenamefont {Siegal-Gaskins},\ and\
  \citenamefont {Scott}(2014)}]{Pierre:2014tra}%
  \BibitemOpen
  \bibfield  {author} {\bibinfo {author} {\bibnamefont {Pierre}, \bibfnamefont
  {M.}}, \bibinfo {author} {\bibnamefont {Siegal-Gaskins}, \bibfnamefont
  {J.~M.}}, \ and\ \bibinfo {author} {\bibnamefont {Scott}, \bibfnamefont
  {P.}},\ }\href {\doibase 10.1088/1475-7516/2014/06/024} {\bibfield  {journal}
  {\bibinfo  {journal} {JCAP}\ }\textbf {\bibinfo {volume} {1406}},\ \bibinfo
  {pages} {024} (\bibinfo {year} {2014})},\ \Eprint
  {http://arxiv.org/abs/1401.7330} {arXiv:1401.7330 [astro-ph.HE]} \BibitemShut
  {NoStop}%
%%CITATION = ARXIV:1401.7330;%%
\bibitem [{\citenamefont {{Piffl}}\ \emph {et~al.}(2014)\citenamefont
  {{Piffl}}, \citenamefont {{Scannapieco}}, \citenamefont {{Binney}},
  \citenamefont {{Steinmetz}}, \citenamefont {{Scholz}}, \citenamefont
  {{Williams}}, \citenamefont {{de Jong}}, \citenamefont {{Kordopatis}},
  \citenamefont {{Matijevi{\v c}}}, \citenamefont {{Bienaym{\'e}}},
  \citenamefont {{Bland-Hawthorn}}, \citenamefont {{Boeche}}, \citenamefont
  {{Freeman}}, \citenamefont {{Gibson}}, \citenamefont {{Gilmore}},
  \citenamefont {{Grebel}}, \citenamefont {{Helmi}}, \citenamefont {{Munari}},
  \citenamefont {{Navarro}}, \citenamefont {{Parker}}, \citenamefont {{Reid}},
  \citenamefont {{Seabroke}}, \citenamefont {{Watson}}, \citenamefont
  {{Wyse}},\ and\ \citenamefont {{Zwitter}}}]{2014A&A...562A..91P}%
  \BibitemOpen
  \bibfield  {author} {\bibinfo {author} {\bibnamefont {{Piffl}}, \bibfnamefont
  {T.}}, \bibinfo {author} {\bibnamefont {{Scannapieco}}, \bibfnamefont {C.}},
  \bibinfo {author} {\bibnamefont {{Binney}}, \bibfnamefont {J.}},  \emph
  {et~al.},\ }\href {\doibase 10.1051/0004-6361/201322531} {\bibfield
  {journal} {\bibinfo  {journal} {\aap}\ }\textbf {\bibinfo {volume} {562}},\
  \bibinfo {eid} {A91} (\bibinfo {year} {2014})},\ \Eprint
  {http://arxiv.org/abs/1309.4293} {arXiv:1309.4293 [astro-ph.GA]} \BibitemShut
  {NoStop}%
\bibitem [{\citenamefont {Pinzke}, \citenamefont {Pfrommer},\ and\
  \citenamefont {Bergstrom}(2011)}]{Pinzke:2011ek}%
  \BibitemOpen
  \bibfield  {author} {\bibinfo {author} {\bibnamefont {Pinzke}, \bibfnamefont
  {A.}}, \bibinfo {author} {\bibnamefont {Pfrommer}, \bibfnamefont {C.}}, \
  and\ \bibinfo {author} {\bibnamefont {Bergstrom}, \bibfnamefont {L.}},\
  }\href {\doibase 10.1103/PhysRevD.84.123509} {\bibfield  {journal} {\bibinfo
  {journal} {Phys.Rev.}\ }\textbf {\bibinfo {volume} {D84}},\ \bibinfo {pages}
  {123509} (\bibinfo {year} {2011})},\ \Eprint {http://arxiv.org/abs/1105.3240}
  {arXiv:1105.3240 [astro-ph.HE]} \BibitemShut {NoStop}%
%%CITATION = ARXIV:1105.3240;%%
\bibitem [{\citenamefont {Pontzen}\ and\ \citenamefont
  {Governato}(2012)}]{Pontzen:2011ty}%
  \BibitemOpen
  \bibfield  {author} {\bibinfo {author} {\bibnamefont {Pontzen}, \bibfnamefont
  {A.}}\ and\ \bibinfo {author} {\bibnamefont {Governato}, \bibfnamefont
  {F.}},\ }\href {\doibase 10.1111/j.1365-2966.2012.20571.x} {\bibfield
  {journal} {\bibinfo  {journal} {Mon.Not.Roy.Astron.Soc.}\ }\textbf {\bibinfo
  {volume} {421}},\ \bibinfo {pages} {3464} (\bibinfo {year} {2012})},\ \Eprint
  {http://arxiv.org/abs/1106.0499} {arXiv:1106.0499 [astro-ph.CO]} \BibitemShut
  {NoStop}%
%%CITATION = ARXIV:1106.0499;%%
\bibitem [{\citenamefont {Porter}, \citenamefont {Johnson},\ and\ \citenamefont
  {Graham}(2011)}]{porter_dark_2011}%
  \BibitemOpen
  \bibfield  {author} {\bibinfo {author} {\bibnamefont {Porter}, \bibfnamefont
  {T.~A.}}, \bibinfo {author} {\bibnamefont {Johnson}, \bibfnamefont {R.~P.}},
  \ and\ \bibinfo {author} {\bibnamefont {Graham}, \bibfnamefont {P.~W.}},\
  }\href {\doibase 10.1146/annurev-astro-081710-102528} {\bibfield  {journal}
  {\bibinfo  {journal} {Annual Review of Astronomy and Astrophysics}\ }\textbf
  {\bibinfo {volume} {49}},\ \bibinfo {pages} {155} (\bibinfo {year}
  {2011})}\BibitemShut {NoStop}%
\bibitem [{\citenamefont {Profumo}, \citenamefont {Sigurdson},\ and\
  \citenamefont {Kamionkowski}(2006)}]{Profumo:2006bv}%
  \BibitemOpen
  \bibfield  {author} {\bibinfo {author} {\bibnamefont {Profumo}, \bibfnamefont
  {S.}}, \bibinfo {author} {\bibnamefont {Sigurdson}, \bibfnamefont {K.}}, \
  and\ \bibinfo {author} {\bibnamefont {Kamionkowski}, \bibfnamefont {M.}},\
  }\href {\doibase 10.1103/PhysRevLett.97.031301} {\bibfield  {journal}
  {\bibinfo  {journal} {Phys.Rev.Lett.}\ }\textbf {\bibinfo {volume} {97}},\
  \bibinfo {pages} {031301} (\bibinfo {year} {2006})},\ \Eprint
  {http://arxiv.org/abs/astro-ph/0603373} {arXiv:astro-ph/0603373 [astro-ph]}
  \BibitemShut {NoStop}%
%%CITATION = ASTRO-PH/0603373;%%
\bibitem [{\citenamefont {Pshirkov}, \citenamefont {Vasiliev},\ and\
  \citenamefont {Postnov}(2015)}]{Pshirkov:2015hda}%
  \BibitemOpen
  \bibfield  {author} {\bibinfo {author} {\bibnamefont {Pshirkov},
  \bibfnamefont {M.}}, \bibinfo {author} {\bibnamefont {Vasiliev},
  \bibfnamefont {V.}}, \ and\ \bibinfo {author} {\bibnamefont {Postnov},
  \bibfnamefont {K.}},\ }\href@noop {} {\  (\bibinfo {year} {2015})},\ \Eprint
  {http://arxiv.org/abs/1501.03460} {arXiv:1501.03460 [astro-ph.GA]}
  \BibitemShut {NoStop}%
%%CITATION = ARXIV:1501.03460;%%
\bibitem [{\citenamefont {Pullen}, \citenamefont {Chary},\ and\ \citenamefont
  {Kamionkowski}(2007)}]{Pullen:2006sy}%
  \BibitemOpen
  \bibfield  {author} {\bibinfo {author} {\bibnamefont {Pullen}, \bibfnamefont
  {A.~R.}}, \bibinfo {author} {\bibnamefont {Chary}, \bibfnamefont {R.-R.}}, \
  and\ \bibinfo {author} {\bibnamefont {Kamionkowski}, \bibfnamefont {M.}},\
  }\href {\doibase 10.1103/PhysRevD.76.063006, 10.1103/PhysRevD.83.029904}
  {\bibfield  {journal} {\bibinfo  {journal} {Phys.Rev.}\ }\textbf {\bibinfo
  {volume} {D76}},\ \bibinfo {pages} {063006} (\bibinfo {year} {2007})},\
  \Eprint {http://arxiv.org/abs/astro-ph/0610295} {arXiv:astro-ph/0610295
  [astro-ph]} \BibitemShut {NoStop}%
%%CITATION = ASTRO-PH/0610295;%%
\bibitem [{\citenamefont {Read}(2014)}]{Read:2014qva}%
  \BibitemOpen
  \bibfield  {author} {\bibinfo {author} {\bibnamefont {Read}, \bibfnamefont
  {J.}},\ }\href {\doibase 10.1088/0954-3899/41/6/063101} {\bibfield  {journal}
  {\bibinfo  {journal} {J.Phys.}\ }\textbf {\bibinfo {volume} {G41}},\ \bibinfo
  {pages} {063101} (\bibinfo {year} {2014})},\ \Eprint
  {http://arxiv.org/abs/1404.1938} {arXiv:1404.1938 [astro-ph.GA]} \BibitemShut
  {NoStop}%
%%CITATION = ARXIV:1404.1938;%%
\bibitem [{\citenamefont {{Reimer}}\ \emph {et~al.}(2003)\citenamefont
  {{Reimer}}, \citenamefont {{Pohl}}, \citenamefont {{Sreekumar}},\ and\
  \citenamefont {{Mattox}}}]{2003ApJ...588..155R}%
  \BibitemOpen
  \bibfield  {author} {\bibinfo {author} {\bibnamefont {{Reimer}},
  \bibfnamefont {O.}}, \bibinfo {author} {\bibnamefont {{Pohl}}, \bibfnamefont
  {M.}}, \bibinfo {author} {\bibnamefont {{Sreekumar}}, \bibfnamefont {P.}}, \
  and\ \bibinfo {author} {\bibnamefont {{Mattox}}, \bibfnamefont {J.~R.}},\
  }\href {\doibase 10.1086/374046} {\bibfield  {journal} {\bibinfo  {journal}
  {\apj}\ }\textbf {\bibinfo {volume} {588}},\ \bibinfo {pages} {155} (\bibinfo
  {year} {2003})},\ \Eprint {http://arxiv.org/abs/astro-ph/0301362}
  {astro-ph/0301362} \BibitemShut {NoStop}%
\bibitem [{\citenamefont {Richardson}\ and\ \citenamefont
  {Fairbairn}(2013)}]{Richardson:2013lja}%
  \BibitemOpen
  \bibfield  {author} {\bibinfo {author} {\bibnamefont {Richardson},
  \bibfnamefont {T.}}\ and\ \bibinfo {author} {\bibnamefont {Fairbairn},
  \bibfnamefont {M.}},\ }\href {http://www.arxiv.org/abs/1305.0670} {\enquote
  {\bibinfo {title} {{Cores in Classical Dwarf Spheroidal Galaxies? A
  Dispersion-Kurtosis Jeans Analysis Without Restricted Anisotropy}},}\ }
  (\bibinfo {year} {2013}),\ \bibinfo {note} {prepared for submission to
  \mnras.},\ \Eprint {http://arxiv.org/abs/1305.0670} {arXiv:1305.0670
  [astro-ph.GA]} \BibitemShut {NoStop}%
%%CITATION = ARXIV:1305.0670;%%
\bibitem [{\citenamefont {Ripken}\ \emph {et~al.}(2014)\citenamefont {Ripken},
  \citenamefont {Cuoco}, \citenamefont {Zechlin}, \citenamefont {Conrad},\ and\
  \citenamefont {Horns}}]{Ripken:2012db}%
  \BibitemOpen
  \bibfield  {author} {\bibinfo {author} {\bibnamefont {Ripken}, \bibfnamefont
  {J.}}, \bibinfo {author} {\bibnamefont {Cuoco}, \bibfnamefont {A.}}, \bibinfo
  {author} {\bibnamefont {Zechlin}, \bibfnamefont {H.-S.}}, \bibinfo {author}
  {\bibnamefont {Conrad}, \bibfnamefont {J.}}, \ and\ \bibinfo {author}
  {\bibnamefont {Horns}, \bibfnamefont {D.}},\ }\href {\doibase
  10.1088/1475-7516/2014/01/049} {\bibfield  {journal} {\bibinfo  {journal}
  {JCAP}\ }\textbf {\bibinfo {volume} {1401}},\ \bibinfo {pages} {049}
  (\bibinfo {year} {2014})},\ \Eprint {http://arxiv.org/abs/1211.6922}
  {arXiv:1211.6922 [astro-ph.HE]} \BibitemShut {NoStop}%
%%CITATION = ARXIV:1211.6922;%%
\bibitem [{\citenamefont {Rolke}, \citenamefont {Lopez},\ and\ \citenamefont
  {Conrad}(2005)}]{Rolke:2004mj}%
  \BibitemOpen
  \bibfield  {author} {\bibinfo {author} {\bibnamefont {Rolke}, \bibfnamefont
  {W.~A.}}, \bibinfo {author} {\bibnamefont {Lopez}, \bibfnamefont {A.~M.}}, \
  and\ \bibinfo {author} {\bibnamefont {Conrad}, \bibfnamefont {J.}},\ }\href
  {\doibase 10.1016/j.nima.2005.05.068} {\bibfield  {journal} {\bibinfo
  {journal} {Nucl.Instrum.Meth.}\ }\textbf {\bibinfo {volume} {A551}},\
  \bibinfo {pages} {493} (\bibinfo {year} {2005})},\ \Eprint
  {http://arxiv.org/abs/physics/0403059} {arXiv:physics/0403059 [physics]}
  \BibitemShut {NoStop}%
%%CITATION = PHYSICS/0403059;%%
\bibitem [{\citenamefont {Roszkowski}, \citenamefont {Sessolo},\ and\
  \citenamefont {Williams}(2015)}]{Roszkowski:2014iqa}%
  \BibitemOpen
  \bibfield  {author} {\bibinfo {author} {\bibnamefont {Roszkowski},
  \bibfnamefont {L.}}, \bibinfo {author} {\bibnamefont {Sessolo}, \bibfnamefont
  {E.~M.}}, \ and\ \bibinfo {author} {\bibnamefont {Williams}, \bibfnamefont
  {A.~J.}},\ }\href {\doibase 10.1007/JHEP02(2015)014} {\bibfield  {journal}
  {\bibinfo  {journal} {JHEP}\ }\textbf {\bibinfo {volume} {1502}},\ \bibinfo
  {pages} {014} (\bibinfo {year} {2015})},\ \Eprint
  {http://arxiv.org/abs/1411.5214} {arXiv:1411.5214 [hep-ph]} \BibitemShut
  {NoStop}%
%%CITATION = ARXIV:1411.5214;%%
\bibitem [{\citenamefont {{S{\'a}nchez-Conde}}\ and\ \citenamefont
  {{Prada}}(2014)}]{Sanchez-Conde:2013yxa}%
  \BibitemOpen
  \bibfield  {author} {\bibinfo {author} {\bibnamefont {{S{\'a}nchez-Conde}},
  \bibfnamefont {M.~A.}}\ and\ \bibinfo {author} {\bibnamefont {{Prada}},
  \bibfnamefont {F.}},\ }\href {\doibase 10.1093/mnras/stu1014} {\bibfield
  {journal} {\bibinfo  {journal} {\mnras}\ }\textbf {\bibinfo {volume} {442}},\
  \bibinfo {pages} {2271} (\bibinfo {year} {2014})},\ \Eprint
  {http://arxiv.org/abs/1312.1729} {arXiv:1312.1729} \BibitemShut {NoStop}%
\bibitem [{\citenamefont {{Sch{\"o}del}}, \citenamefont {{Merritt}},\ and\
  \citenamefont {{Eckart}}(2009)}]{Schodel2009}%
  \BibitemOpen
  \bibfield  {author} {\bibinfo {author} {\bibnamefont {{Sch{\"o}del}},
  \bibfnamefont {R.}}, \bibinfo {author} {\bibnamefont {{Merritt}},
  \bibfnamefont {D.}}, \ and\ \bibinfo {author} {\bibnamefont {{Eckart}},
  \bibfnamefont {A.}},\ }\href {\doibase 10.1051/0004-6361/200810922}
  {\bibfield  {journal} {\bibinfo  {journal} {\aap}\ }\textbf {\bibinfo
  {volume} {502}},\ \bibinfo {pages} {91} (\bibinfo {year} {2009})},\ \Eprint
  {http://arxiv.org/abs/0902.3892} {arXiv:0902.3892 [astro-ph.GA]} \BibitemShut
  {NoStop}%
\bibitem [{\citenamefont {Scott}\ \emph {et~al.}(2010)\citenamefont {Scott},
  \citenamefont {Conrad}, \citenamefont {Edsjo}, \citenamefont {Bergstrom},
  \citenamefont {Farnier} \emph {et~al.}}]{Scott:2009jn}%
  \BibitemOpen
  \bibfield  {author} {\bibinfo {author} {\bibnamefont {Scott}, \bibfnamefont
  {P.}}, \bibinfo {author} {\bibnamefont {Conrad}, \bibfnamefont {J.}},
  \bibinfo {author} {\bibnamefont {Edsjo}, \bibfnamefont {J.}},  \emph
  {et~al.},\ }\href {\doibase 10.1088/1475-7516/2010/01/031} {\bibfield
  {journal} {\bibinfo  {journal} {JCAP}\ }\textbf {\bibinfo {volume} {1001}},\
  \bibinfo {pages} {031} (\bibinfo {year} {2010})},\ \Eprint
  {http://arxiv.org/abs/0909.3300} {arXiv:0909.3300 [astro-ph.CO]} \BibitemShut
  {NoStop}%
%%CITATION = ARXIV:0909.3300;%%
\bibitem [{\citenamefont {Sefusatti}\ \emph {et~al.}(2014)\citenamefont
  {Sefusatti}, \citenamefont {Zaharijas}, \citenamefont {Serpico},
  \citenamefont {Theurel},\ and\ \citenamefont
  {Gustafsson}}]{Sefusatti:2014vha}%
  \BibitemOpen
  \bibfield  {author} {\bibinfo {author} {\bibnamefont {Sefusatti},
  \bibfnamefont {E.}}, \bibinfo {author} {\bibnamefont {Zaharijas},
  \bibfnamefont {G.}}, \bibinfo {author} {\bibnamefont {Serpico}, \bibfnamefont
  {P.~D.}}, \bibinfo {author} {\bibnamefont {Theurel}, \bibfnamefont {D.}}, \
  and\ \bibinfo {author} {\bibnamefont {Gustafsson}, \bibfnamefont {M.}},\
  }\href {\doibase 10.1093/mnras/stu686} {\bibfield  {journal} {\bibinfo
  {journal} {Mon.Not.Roy.Astron.Soc.}\ }\textbf {\bibinfo {volume} {441}},\
  \bibinfo {pages} {1861} (\bibinfo {year} {2014})},\ \Eprint
  {http://arxiv.org/abs/1401.2117} {arXiv:1401.2117 [astro-ph.CO]} \BibitemShut
  {NoStop}%
%%CITATION = ARXIV:1401.2117;%%
\bibitem [{\citenamefont {{Serpico}}\ and\ \citenamefont
  {{Zaharijas}}(2008)}]{2008APh....29..380S}%
  \BibitemOpen
  \bibfield  {author} {\bibinfo {author} {\bibnamefont {{Serpico}},
  \bibfnamefont {P.~D.}}\ and\ \bibinfo {author} {\bibnamefont {{Zaharijas}},
  \bibfnamefont {G.}},\ }\href {\doibase 10.1016/j.astropartphys.2008.04.001}
  {\bibfield  {journal} {\bibinfo  {journal} {Astroparticle Physics}\ }\textbf
  {\bibinfo {volume} {29}},\ \bibinfo {pages} {380} (\bibinfo {year} {2008})},\
  \Eprint {http://arxiv.org/abs/0802.3245} {arXiv:0802.3245} \BibitemShut
  {NoStop}%
\bibitem [{\citenamefont {Silverwood}\ \emph {et~al.}(2015)\citenamefont
  {Silverwood}, \citenamefont {Weniger}, \citenamefont {Scott},\ and\
  \citenamefont {Bertone}}]{Silverwood:2014yza}%
  \BibitemOpen
  \bibfield  {author} {\bibinfo {author} {\bibnamefont {Silverwood},
  \bibfnamefont {H.}}, \bibinfo {author} {\bibnamefont {Weniger}, \bibfnamefont
  {C.}}, \bibinfo {author} {\bibnamefont {Scott}, \bibfnamefont {P.}}, \ and\
  \bibinfo {author} {\bibnamefont {Bertone}, \bibfnamefont {G.}},\ }\href
  {http://stacks.iop.org/1475-7516/2015/i=03/a=055} {\bibfield  {journal}
  {\bibinfo  {journal} {\jcap}\ }\textbf {\bibinfo {volume} {2015}},\ \bibinfo
  {pages} {055} (\bibinfo {year} {2015})},\ \Eprint
  {http://arxiv.org/abs/1408.4131} {arXiv:1408.4131 [astro-ph.HE]} \BibitemShut
  {NoStop}%
\bibitem [{\citenamefont {{Smith}}\ \emph {et~al.}(2013)\citenamefont
  {{Smith}}, \citenamefont {{Bird}}, \citenamefont {{Buckley}}, \citenamefont
  {{Byrum}}, \citenamefont {{Finley}}, \citenamefont {{Galante}}, \citenamefont
  {{Geringer-Sameth}}, \citenamefont {{Hanna}}, \citenamefont {{Holder}},
  \citenamefont {{Kieda}}, \citenamefont {{Koushiappas}}, \citenamefont
  {{Ong}}, \citenamefont {{Staszak}},\ and\ \citenamefont
  {{Zitzer}}}]{2013arXiv1304.6367S}%
  \BibitemOpen
  \bibfield  {author} {\bibinfo {author} {\bibnamefont {{Smith}}, \bibfnamefont
  {A.~W.}}, \bibinfo {author} {\bibnamefont {{Bird}}, \bibfnamefont {R.}},
  \bibinfo {author} {\bibnamefont {{Buckley}}, \bibfnamefont {J.}},  \emph
  {et~al.},\ }\href@noop {} {\enquote {\bibinfo {title} {{CF2 White Paper:
  Status and Prospects of The VERITAS Indirect Dark Matter Detection
  Program}},}\ } (\bibinfo {year} {2013}),\ \bibinfo {note} {{Submitted to the
  Snowmass 2013 proceedings, Cosmic Frontier Subgroup 2}},\ \Eprint
  {http://arxiv.org/abs/1304.6367} {arXiv:1304.6367 [astro-ph.HE]} \BibitemShut
  {NoStop}%
\bibitem [{\citenamefont {Spengler}(2015)}]{Spengler:2015dda}%
  \BibitemOpen
  \bibfield  {author} {\bibinfo {author} {\bibnamefont {Spengler},
  \bibfnamefont {G.}},\ }\href@noop {} {\enquote {\bibinfo {title}
  {{Significance in Gamma Ray Astronomy with Systematic Errors}},}\ } (\bibinfo
  {year} {2015}),\ \bibinfo {note} {accepted by Astroparticle Physics},\
  \Eprint {http://arxiv.org/abs/1502.03249} {arXiv:1502.03249 [astro-ph.IM]}
  \BibitemShut {NoStop}%
%%CITATION = ARXIV:1502.03249;%%
\bibitem [{\citenamefont {{Springel}}(2005)}]{2005MNRAS.364.1105S}%
  \BibitemOpen
  \bibfield  {author} {\bibinfo {author} {\bibnamefont {{Springel}},
  \bibfnamefont {V.}},\ }\href {\doibase 10.1111/j.1365-2966.2005.09655.x}
  {\bibfield  {journal} {\bibinfo  {journal} {\mnras}\ }\textbf {\bibinfo
  {volume} {364}},\ \bibinfo {pages} {1105} (\bibinfo {year} {2005})},\ \Eprint
  {http://arxiv.org/abs/astro-ph/0505010} {astro-ph/0505010} \BibitemShut
  {NoStop}%
\bibitem [{\citenamefont {Springel}\ \emph
  {et~al.}(2008{\natexlab{a}})\citenamefont {Springel}, \citenamefont {Wang},
  \citenamefont {Vogelsberger}, \citenamefont {Ludlow}, \citenamefont {Jenkins}
  \emph {et~al.}}]{Springel:2008cc}%
  \BibitemOpen
  \bibfield  {author} {\bibinfo {author} {\bibnamefont {Springel},
  \bibfnamefont {V.}}, \bibinfo {author} {\bibnamefont {Wang}, \bibfnamefont
  {J.}}, \bibinfo {author} {\bibnamefont {Vogelsberger}, \bibfnamefont {M.}},
  \emph {et~al.},\ }\href {\doibase 10.1111/j.1365-2966.2008.14066.x}
  {\bibfield  {journal} {\bibinfo  {journal} {Mon.Not.Roy.Astron.Soc.}\
  }\textbf {\bibinfo {volume} {391}},\ \bibinfo {pages} {1685} (\bibinfo {year}
  {2008}{\natexlab{a}})},\ \Eprint {http://arxiv.org/abs/0809.0898}
  {arXiv:0809.0898 [astro-ph]} \BibitemShut {NoStop}%
%%CITATION = ARXIV:0809.0898;%%
\bibitem [{\citenamefont {Springel}\ \emph
  {et~al.}(2008{\natexlab{b}})\citenamefont {Springel}, \citenamefont {White},
  \citenamefont {Frenk}, \citenamefont {Navarro}, \citenamefont {Jenkins} \emph
  {et~al.}}]{Springel:2008zz}%
  \BibitemOpen
  \bibfield  {author} {\bibinfo {author} {\bibnamefont {Springel},
  \bibfnamefont {V.}}, \bibinfo {author} {\bibnamefont {White}, \bibfnamefont
  {S.}}, \bibinfo {author} {\bibnamefont {Frenk}, \bibfnamefont {C.}},  \emph
  {et~al.},\ }\href {\doibase 10.1038/nature07411} {\bibfield  {journal}
  {\bibinfo  {journal} {\nat}\ }\textbf {\bibinfo {volume} {456N7218}},\
  \bibinfo {pages} {73} (\bibinfo {year} {2008}{\natexlab{b}})}\BibitemShut
  {NoStop}%
\bibitem [{\citenamefont {{Sreekumar}}\ \emph {et~al.}(1998)\citenamefont
  {{Sreekumar}}, \citenamefont {{Bertsch}}, \citenamefont {{Dingus}},
  \citenamefont {{Esposito}}, \citenamefont {{Fichtel}}, \citenamefont
  {{Hartman}}, \citenamefont {{Hunter}}, \citenamefont {{Kanbach}},
  \citenamefont {{Kniffen}}, \citenamefont {{Lin}}, \citenamefont
  {{Mayer-Hasselwander}}, \citenamefont {{Michelson}}, \citenamefont {{von
  Montigny}}, \citenamefont {{M{\"u}cke}}, \citenamefont {{Mukherjee}},
  \citenamefont {{Nolan}}, \citenamefont {{Pohl}}, \citenamefont {{Reimer}},
  \citenamefont {{Schneid}}, \citenamefont {{Stacy}}, \citenamefont
  {{Stecker}}, \citenamefont {{Thompson}},\ and\ \citenamefont
  {{Willis}}}]{1998ApJ...494..523S}%
  \BibitemOpen
  \bibfield  {author} {\bibinfo {author} {\bibnamefont {{Sreekumar}},
  \bibfnamefont {P.}}, \bibinfo {author} {\bibnamefont {{Bertsch}},
  \bibfnamefont {D.~L.}}, \bibinfo {author} {\bibnamefont {{Dingus}},
  \bibfnamefont {B.~L.}},  \emph {et~al.},\ }\href {\doibase 10.1086/305222}
  {\bibfield  {journal} {\bibinfo  {journal} {\apj}\ }\textbf {\bibinfo
  {volume} {494}},\ \bibinfo {pages} {523} (\bibinfo {year} {1998})},\ \Eprint
  {http://arxiv.org/abs/astro-ph/9709257} {astro-ph/9709257} \BibitemShut
  {NoStop}%
\bibitem [{\citenamefont {Steigman}, \citenamefont {Dasgupta},\ and\
  \citenamefont {Beacom}(2012)}]{Steigman:2012nb}%
  \BibitemOpen
  \bibfield  {author} {\bibinfo {author} {\bibnamefont {Steigman},
  \bibfnamefont {G.}}, \bibinfo {author} {\bibnamefont {Dasgupta},
  \bibfnamefont {B.}}, \ and\ \bibinfo {author} {\bibnamefont {Beacom},
  \bibfnamefont {J.~F.}},\ }\href {\doibase 10.1103/PhysRevD.86.023506}
  {\bibfield  {journal} {\bibinfo  {journal} {Phys.Rev.}\ }\textbf {\bibinfo
  {volume} {D86}},\ \bibinfo {pages} {023506} (\bibinfo {year} {2012})},\
  \Eprint {http://arxiv.org/abs/1204.3622} {arXiv:1204.3622 [hep-ph]}
  \BibitemShut {NoStop}%
%%CITATION = ARXIV:1204.3622;%%
\bibitem [{\citenamefont {{Stoehr}}\ \emph {et~al.}(2003)\citenamefont
  {{Stoehr}}, \citenamefont {{White}}, \citenamefont {{Springel}},
  \citenamefont {{Tormen}},\ and\ \citenamefont
  {{Yoshida}}}]{2003MNRAS.345.1313S}%
  \BibitemOpen
  \bibfield  {author} {\bibinfo {author} {\bibnamefont {{Stoehr}},
  \bibfnamefont {F.}}, \bibinfo {author} {\bibnamefont {{White}}, \bibfnamefont
  {S.~D.~M.}}, \bibinfo {author} {\bibnamefont {{Springel}}, \bibfnamefont
  {V.}}, \bibinfo {author} {\bibnamefont {{Tormen}}, \bibfnamefont {G.}}, \
  and\ \bibinfo {author} {\bibnamefont {{Yoshida}}, \bibfnamefont {N.}},\
  }\href {\doibase 10.1046/j.1365-2966.2003.07052.x} {\bibfield  {journal}
  {\bibinfo  {journal} {\mnras}\ }\textbf {\bibinfo {volume} {345}},\ \bibinfo
  {pages} {1313} (\bibinfo {year} {2003})},\ \Eprint
  {http://arxiv.org/abs/astro-ph/0307026} {astro-ph/0307026} \BibitemShut
  {NoStop}%
\bibitem [{\citenamefont {Strigari}(2013)}]{Strigari:2013iaa}%
  \BibitemOpen
  \bibfield  {author} {\bibinfo {author} {\bibnamefont {Strigari},
  \bibfnamefont {L.~E.}},\ }\href {\doibase 10.1016/j.physrep.2013.05.004}
  {\bibfield  {journal} {\bibinfo  {journal} {Phys.Rept.}\ }\textbf {\bibinfo
  {volume} {531}},\ \bibinfo {pages} {1} (\bibinfo {year} {2013})},\ \Eprint
  {http://arxiv.org/abs/1211.7090} {arXiv:1211.7090 [astro-ph.CO]} \BibitemShut
  {NoStop}%
%%CITATION = ARXIV:1211.7090;%%
\bibitem [{\citenamefont {Strigari}, \citenamefont {Frenk},\ and\ \citenamefont
  {White}(2010)}]{Strigari:2010un}%
  \BibitemOpen
  \bibfield  {author} {\bibinfo {author} {\bibnamefont {Strigari},
  \bibfnamefont {L.~E.}}, \bibinfo {author} {\bibnamefont {Frenk},
  \bibfnamefont {C.~S.}}, \ and\ \bibinfo {author} {\bibnamefont {White},
  \bibfnamefont {S.~D.}},\ }\href {\doibase 10.1111/j.1365-2966.2010.17287.x}
  {\bibfield  {journal} {\bibinfo  {journal} {Mon.Not.Roy.Astron.Soc.}\
  }\textbf {\bibinfo {volume} {408}},\ \bibinfo {pages} {2364} (\bibinfo {year}
  {2010})},\ \Eprint {http://arxiv.org/abs/1003.4268} {arXiv:1003.4268
  [astro-ph.CO]} \BibitemShut {NoStop}%
%%CITATION = ARXIV:1003.4268;%%
\bibitem [{\citenamefont {Strigari}, \citenamefont {Frenk},\ and\ \citenamefont
  {White}(2014)}]{Strigari:2014yea}%
  \BibitemOpen
  \bibfield  {author} {\bibinfo {author} {\bibnamefont {Strigari},
  \bibfnamefont {L.~E.}}, \bibinfo {author} {\bibnamefont {Frenk},
  \bibfnamefont {C.~S.}}, \ and\ \bibinfo {author} {\bibnamefont {White},
  \bibfnamefont {S.~D.~M.}},\ }\href@noop {} {\enquote {\bibinfo {title}
  {{Dynamical models for the Sculptor dwarf spheroidal in a Lambda CDM
  universe}},}\ } (\bibinfo {year} {2014}),\ \bibinfo {note} {submitted to
  MNRAS},\ \Eprint {http://arxiv.org/abs/1406.6079} {arXiv:1406.6079
  [astro-ph.GA]} \BibitemShut {NoStop}%
%%CITATION = ARXIV:1406.6079;%%
\bibitem [{\citenamefont {Strigari}\ \emph {et~al.}(2007)\citenamefont
  {Strigari}, \citenamefont {Koushiappas}, \citenamefont {Bullock},\ and\
  \citenamefont {Kaplinghat}}]{Strigari:2006rd}%
  \BibitemOpen
  \bibfield  {author} {\bibinfo {author} {\bibnamefont {Strigari},
  \bibfnamefont {L.~E.}}, \bibinfo {author} {\bibnamefont {Koushiappas},
  \bibfnamefont {S.~M.}}, \bibinfo {author} {\bibnamefont {Bullock},
  \bibfnamefont {J.~S.}}, \ and\ \bibinfo {author} {\bibnamefont {Kaplinghat},
  \bibfnamefont {M.}},\ }\href {\doibase 10.1103/PhysRevD.75.083526} {\bibfield
   {journal} {\bibinfo  {journal} {\prd}\ }\textbf {\bibinfo {volume} {75}},\
  \bibinfo {pages} {083526} (\bibinfo {year} {2007})},\ \Eprint
  {http://arxiv.org/abs/astro-ph/0611925} {arXiv:astro-ph/0611925 [astro-ph]}
  \BibitemShut {NoStop}%
\bibitem [{\citenamefont {{Strigari}}\ \emph {et~al.}(2008)\citenamefont
  {{Strigari}}, \citenamefont {{Koushiappas}}, \citenamefont {{Bullock}},
  \citenamefont {{Kaplinghat}}, \citenamefont {{Simon}}, \citenamefont
  {{Geha}},\ and\ \citenamefont {{Willman}}}]{Strigari:2007at}%
  \BibitemOpen
  \bibfield  {author} {\bibinfo {author} {\bibnamefont {{Strigari}},
  \bibfnamefont {L.~E.}}, \bibinfo {author} {\bibnamefont {{Koushiappas}},
  \bibfnamefont {S.~M.}}, \bibinfo {author} {\bibnamefont {{Bullock}},
  \bibfnamefont {J.~S.}},  \emph {et~al.},\ }\href {\doibase 10.1086/529488}
  {\bibfield  {journal} {\bibinfo  {journal} {\apj}\ }\textbf {\bibinfo
  {volume} {678}},\ \bibinfo {pages} {614} (\bibinfo {year} {2008})},\ \Eprint
  {http://arxiv.org/abs/0709.1510} {arXiv:0709.1510} \BibitemShut {NoStop}%
\bibitem [{\citenamefont {Strong}, \citenamefont {Moskalenko},\ and\
  \citenamefont {Ptuskin}(2007)}]{strong_cosmic-ray_2007}%
  \BibitemOpen
  \bibfield  {author} {\bibinfo {author} {\bibnamefont {Strong}, \bibfnamefont
  {A.~W.}}, \bibinfo {author} {\bibnamefont {Moskalenko}, \bibfnamefont
  {I.~V.}}, \ and\ \bibinfo {author} {\bibnamefont {Ptuskin}, \bibfnamefont
  {V.~S.}},\ }\href {\doibase 10.1146/annurev.nucl.57.090506.123011} {\bibfield
   {journal} {\bibinfo  {journal} {Annual Review of Nuclear and Particle
  Science}\ }\textbf {\bibinfo {volume} {57}},\ \bibinfo {pages} {285}
  (\bibinfo {year} {2007})}\BibitemShut {NoStop}%
\bibitem [{\citenamefont {Su}\ and\ \citenamefont
  {Finkbeiner}(2012)}]{Su:2012ft}%
  \BibitemOpen
  \bibfield  {author} {\bibinfo {author} {\bibnamefont {Su}, \bibfnamefont
  {M.}}\ and\ \bibinfo {author} {\bibnamefont {Finkbeiner}, \bibfnamefont
  {D.~P.}},\ }\href@noop {} {\enquote {\bibinfo {title} {{Strong Evidence for
  Gamma-ray Line Emission from the Inner Galaxy}},}\ } (\bibinfo {year}
  {2012}),\ \Eprint {http://arxiv.org/abs/1206.1616} {arXiv:1206.1616
  [astro-ph.HE]} \BibitemShut {NoStop}%
%%CITATION = ARXIV:1206.1616;%%
\bibitem [{\citenamefont {Tasitsiomi}\ and\ \citenamefont
  {Olinto}(2002)}]{PhysRevD.66.083006}%
  \BibitemOpen
  \bibfield  {author} {\bibinfo {author} {\bibnamefont {Tasitsiomi},
  \bibfnamefont {A.}}\ and\ \bibinfo {author} {\bibnamefont {Olinto},
  \bibfnamefont {A.~V.}},\ }\href {\doibase 10.1103/PhysRevD.66.083006}
  {\bibfield  {journal} {\bibinfo  {journal} {Phys. Rev. D}\ }\textbf {\bibinfo
  {volume} {66}},\ \bibinfo {pages} {083006} (\bibinfo {year}
  {2002})}\BibitemShut {NoStop}%
\bibitem [{\citenamefont {Trotta}\ \emph {et~al.}(2011)\citenamefont {Trotta},
  \citenamefont {Johannesson}, \citenamefont {Moskalenko}, \citenamefont
  {Porter}, \citenamefont {de~Austri} \emph {et~al.}}]{Trotta:2010mx}%
  \BibitemOpen
  \bibfield  {author} {\bibinfo {author} {\bibnamefont {Trotta}, \bibfnamefont
  {R.}}, \bibinfo {author} {\bibnamefont {Johannesson}, \bibfnamefont {G.}},
  \bibinfo {author} {\bibnamefont {Moskalenko}, \bibfnamefont {I.}},  \emph
  {et~al.},\ }\href {\doibase 10.1088/0004-637X/729/2/106} {\bibfield
  {journal} {\bibinfo  {journal} {Astrophys.J.}\ }\textbf {\bibinfo {volume}
  {729}},\ \bibinfo {pages} {106} (\bibinfo {year} {2011})},\ \Eprint
  {http://arxiv.org/abs/1011.0037} {arXiv:1011.0037 [astro-ph.HE]} \BibitemShut
  {NoStop}%
%%CITATION = ARXIV:1011.0037;%%
\bibitem [{\citenamefont {Ullio}\ \emph {et~al.}(2002)\citenamefont {Ullio},
  \citenamefont {Bergstrom}, \citenamefont {Edsjo},\ and\ \citenamefont
  {Lacey}}]{Ullio:2002pj}%
  \BibitemOpen
  \bibfield  {author} {\bibinfo {author} {\bibnamefont {Ullio}, \bibfnamefont
  {P.}}, \bibinfo {author} {\bibnamefont {Bergstrom}, \bibfnamefont {L.}},
  \bibinfo {author} {\bibnamefont {Edsjo}, \bibfnamefont {J.}}, \ and\ \bibinfo
  {author} {\bibnamefont {Lacey}, \bibfnamefont {C.~G.}},\ }\href {\doibase
  10.1103/PhysRevD.66.123502} {\bibfield  {journal} {\bibinfo  {journal}
  {Phys.Rev.}\ }\textbf {\bibinfo {volume} {D66}},\ \bibinfo {pages} {123502}
  (\bibinfo {year} {2002})},\ \Eprint {http://arxiv.org/abs/astro-ph/0207125}
  {arXiv:astro-ph/0207125 [astro-ph]} \BibitemShut {NoStop}%
%%CITATION = ASTRO-PH/0207125;%%
\bibitem [{\citenamefont {Ullio}, \citenamefont {Zhao},\ and\ \citenamefont
  {Kamionkowski}(2001)}]{Ullio:2001fb}%
  \BibitemOpen
  \bibfield  {author} {\bibinfo {author} {\bibnamefont {Ullio}, \bibfnamefont
  {P.}}, \bibinfo {author} {\bibnamefont {Zhao}, \bibfnamefont {H.}}, \ and\
  \bibinfo {author} {\bibnamefont {Kamionkowski}, \bibfnamefont {M.}},\ }\href
  {\doibase 10.1103/PhysRevD.64.043504} {\bibfield  {journal} {\bibinfo
  {journal} {Phys.Rev.}\ }\textbf {\bibinfo {volume} {D64}},\ \bibinfo {pages}
  {043504} (\bibinfo {year} {2001})},\ \Eprint
  {http://arxiv.org/abs/astro-ph/0101481} {arXiv:astro-ph/0101481 [astro-ph]}
  \BibitemShut {NoStop}%
%%CITATION = ASTRO-PH/0101481;%%
\bibitem [{\citenamefont {{van den Bosch}}\ and\ \citenamefont
  {{Jiang}}(2014)}]{2014arXiv1403.6835V}%
  \BibitemOpen
  \bibfield  {author} {\bibinfo {author} {\bibnamefont {{van den Bosch}},
  \bibfnamefont {F.~C.}}\ and\ \bibinfo {author} {\bibnamefont {{Jiang}},
  \bibfnamefont {F.}},\ }\href@noop {} {\bibfield  {journal} {\bibinfo
  {journal} {ArXiv e-prints}\ } (\bibinfo {year} {2014})},\ \Eprint
  {http://arxiv.org/abs/1403.6835} {arXiv:1403.6835} \BibitemShut {NoStop}%
\bibitem [{\citenamefont {Vikhlinin}\ \emph {et~al.}(2009)\citenamefont
  {Vikhlinin}, \citenamefont {Burenin}, \citenamefont {Ebeling}, \citenamefont
  {Forman}, \citenamefont {Hornstrup} \emph {et~al.}}]{Vikhlinin:2008cd}%
  \BibitemOpen
  \bibfield  {author} {\bibinfo {author} {\bibnamefont {Vikhlinin},
  \bibfnamefont {A.}}, \bibinfo {author} {\bibnamefont {Burenin}, \bibfnamefont
  {R.}}, \bibinfo {author} {\bibnamefont {Ebeling}, \bibfnamefont {H.}},  \emph
  {et~al.},\ }\href {\doibase 10.1088/0004-637X/692/2/1033} {\bibfield
  {journal} {\bibinfo  {journal} {Astrophys.J.}\ }\textbf {\bibinfo {volume}
  {692}},\ \bibinfo {pages} {1033} (\bibinfo {year} {2009})},\ \Eprint
  {http://arxiv.org/abs/0805.2207} {arXiv:0805.2207 [astro-ph]} \BibitemShut
  {NoStop}%
%%CITATION = ARXIV:0805.2207;%%
\bibitem [{\citenamefont {Vladimirov}\ \emph {et~al.}(2011)\citenamefont
  {Vladimirov}, \citenamefont {Digel}, \citenamefont {Johannesson},
  \citenamefont {Michelson}, \citenamefont {Moskalenko} \emph
  {et~al.}}]{Vladimirov:2010aq}%
  \BibitemOpen
  \bibfield  {author} {\bibinfo {author} {\bibnamefont {Vladimirov},
  \bibfnamefont {A.~E.}}, \bibinfo {author} {\bibnamefont {Digel},
  \bibfnamefont {S.~W.}}, \bibinfo {author} {\bibnamefont {Johannesson},
  \bibfnamefont {G.}},  \emph {et~al.},\ }\href {\doibase
  10.1016/j.cpc.2011.01.017} {\bibfield  {journal} {\bibinfo  {journal}
  {Comput.Phys.Commun.}\ }\textbf {\bibinfo {volume} {182}},\ \bibinfo {pages}
  {1156} (\bibinfo {year} {2011})},\ \Eprint {http://arxiv.org/abs/1008.3642}
  {arXiv:1008.3642 [astro-ph.HE]} \BibitemShut {NoStop}%
%%CITATION = ARXIV:1008.3642;%%
\bibitem [{\citenamefont {Vogelsberger}\ \emph {et~al.}(2009)\citenamefont
  {Vogelsberger}, \citenamefont {Helmi}, \citenamefont {Springel},
  \citenamefont {White}, \citenamefont {Wang} \emph
  {et~al.}}]{Vogelsberger:2008qb}%
  \BibitemOpen
  \bibfield  {author} {\bibinfo {author} {\bibnamefont {Vogelsberger},
  \bibfnamefont {M.}}, \bibinfo {author} {\bibnamefont {Helmi}, \bibfnamefont
  {A.}}, \bibinfo {author} {\bibnamefont {Springel}, \bibfnamefont {V.}},
  \emph {et~al.},\ }\href {\doibase 10.1111/j.1365-2966.2009.14630.x}
  {\bibfield  {journal} {\bibinfo  {journal} {Mon.Not.Roy.Astron.Soc.}\
  }\textbf {\bibinfo {volume} {395}},\ \bibinfo {pages} {797} (\bibinfo {year}
  {2009})},\ \Eprint {http://arxiv.org/abs/0812.0362} {arXiv:0812.0362
  [astro-ph]} \BibitemShut {NoStop}%
%%CITATION = ARXIV:0812.0362;%%
\bibitem [{\citenamefont {{Walker}}(2013)}]{Walker:2012td}%
  \BibitemOpen
  \bibfield  {author} {\bibinfo {author} {\bibnamefont {{Walker}},
  \bibfnamefont {M.}},\ }\enquote {\bibinfo {title} {{Dark Matter in the
  Galactic Dwarf Spheroidal Satellites}},}\ in\ \href {\doibase
  10.1007/978-94-007-5612-0_20} {\emph {\bibinfo {booktitle} {Planets, Stars
  and Stellar Systems.~Volume 5: Galactic Structure and Stellar
  Populations}}},\ \bibinfo {editor} {edited by\ \bibinfo {editor}
  {\bibfnamefont {T.~D.}\ \bibnamefont {{Oswalt}}}\ and\ \bibinfo {editor}
  {\bibfnamefont {G.}~\bibnamefont {{Gilmore}}}}\ (\bibinfo  {publisher}
  {Springer Netherlands},\ \bibinfo {year} {2013})\ p.\ \bibinfo {pages}
  {1039}\BibitemShut {NoStop}%
\bibitem [{\citenamefont {{Walker}}, \citenamefont {{Mateo}},\ and\
  \citenamefont {{Olszewski}}(2009)}]{Walker:2008ax}%
  \BibitemOpen
  \bibfield  {author} {\bibinfo {author} {\bibnamefont {{Walker}},
  \bibfnamefont {M.~G.}}, \bibinfo {author} {\bibnamefont {{Mateo}},
  \bibfnamefont {M.}}, \ and\ \bibinfo {author} {\bibnamefont {{Olszewski}},
  \bibfnamefont {E.~W.}},\ }\href {\doibase 10.1088/0004-6256/137/2/3100}
  {\bibfield  {journal} {\bibinfo  {journal} {\aj}\ }\textbf {\bibinfo {volume}
  {137}},\ \bibinfo {pages} {3100} (\bibinfo {year} {2009})},\ \Eprint
  {http://arxiv.org/abs/0811.0118} {arXiv:0811.0118} \BibitemShut {NoStop}%
\bibitem [{\citenamefont {Walker}\ \emph {et~al.}(2009)\citenamefont {Walker},
  \citenamefont {Mateo}, \citenamefont {Olszewski}, \citenamefont {Penarrubia},
  \citenamefont {Evans} \emph {et~al.}}]{Walker:2009zp}%
  \BibitemOpen
  \bibfield  {author} {\bibinfo {author} {\bibnamefont {Walker}, \bibfnamefont
  {M.~G.}}, \bibinfo {author} {\bibnamefont {Mateo}, \bibfnamefont {M.}},
  \bibinfo {author} {\bibnamefont {Olszewski}, \bibfnamefont {E.~W.}},  \emph
  {et~al.},\ }\href {\doibase 10.1088/0004-637X/704/2/1274,
  10.1088/0004-637X/710/1/886} {\bibfield  {journal} {\bibinfo  {journal}
  {\apj}\ }\textbf {\bibinfo {volume} {704}},\ \bibinfo {pages} {1274}
  (\bibinfo {year} {2009})},\ \Eprint {http://arxiv.org/abs/0906.0341}
  {arXiv:0906.0341 [astro-ph.CO]} \BibitemShut {NoStop}%
\bibitem [{\citenamefont {Walker}\ and\ \citenamefont
  {Penarrubia}(2011)}]{Walker:2011zu}%
  \BibitemOpen
  \bibfield  {author} {\bibinfo {author} {\bibnamefont {Walker}, \bibfnamefont
  {M.~G.}}\ and\ \bibinfo {author} {\bibnamefont {Penarrubia}, \bibfnamefont
  {J.}},\ }\href {\doibase 10.1088/0004-637X/742/1/20} {\bibfield  {journal}
  {\bibinfo  {journal} {Astrophys.J.}\ }\textbf {\bibinfo {volume} {742}},\
  \bibinfo {pages} {20} (\bibinfo {year} {2011})},\ \Eprint
  {http://arxiv.org/abs/1108.2404} {arXiv:1108.2404 [astro-ph.CO]} \BibitemShut
  {NoStop}%
%%CITATION = ARXIV:1108.2404;%%
\bibitem [{\citenamefont {Weniger}(2012)}]{Weniger:2012tx}%
  \BibitemOpen
  \bibfield  {author} {\bibinfo {author} {\bibnamefont {Weniger}, \bibfnamefont
  {C.}},\ }\href {\doibase 10.1088/1475-7516/2012/08/007} {\bibfield  {journal}
  {\bibinfo  {journal} {JCAP}\ }\textbf {\bibinfo {volume} {1208}},\ \bibinfo
  {pages} {007} (\bibinfo {year} {2012})},\ \Eprint
  {http://arxiv.org/abs/1204.2797} {arXiv:1204.2797 [hep-ph]} \BibitemShut
  {NoStop}%
%%CITATION = ARXIV:1204.2797;%%
\bibitem [{\citenamefont {Weniger}(2013)}]{Weniger:2013dya}%
  \BibitemOpen
  \bibfield  {author} {\bibinfo {author} {\bibnamefont {Weniger}, \bibfnamefont
  {C.}},\ }in\ \href@noop {} {\emph {\bibinfo {booktitle} {2012 Fermi Symposium
  proceedings - eConf C121028}}}\ (\bibinfo {year} {2013})\ \Eprint
  {http://arxiv.org/abs/1303.1798} {arXiv:1303.1798 [astro-ph.HE]} \BibitemShut
  {NoStop}%
%%CITATION = ARXIV:1303.1798;%%
\bibitem [{\citenamefont {Wilks}(1938)}]{wilks_large-sample_1938}%
  \BibitemOpen
  \bibfield  {author} {\bibinfo {author} {\bibnamefont {Wilks}, \bibfnamefont
  {S.~S.}},\ }\href {\doibase 10.1214/aoms/1177732360} {\bibfield  {journal}
  {\bibinfo  {journal} {The Annals of Mathematical Statistics}\ }\textbf
  {\bibinfo {volume} {9}},\ \bibinfo {pages} {60} (\bibinfo {year}
  {1938})}\BibitemShut {NoStop}%
\bibitem [{\citenamefont {Wolf}\ \emph {et~al.}(2010)\citenamefont {Wolf},
  \citenamefont {Martinez}, \citenamefont {Bullock}, \citenamefont
  {Kaplinghat}, \citenamefont {Geha} \emph {et~al.}}]{Wolf:2009tu}%
  \BibitemOpen
  \bibfield  {author} {\bibinfo {author} {\bibnamefont {Wolf}, \bibfnamefont
  {J.}}, \bibinfo {author} {\bibnamefont {Martinez}, \bibfnamefont {G.~D.}},
  \bibinfo {author} {\bibnamefont {Bullock}, \bibfnamefont {J.~S.}},  \emph
  {et~al.},\ }\href@noop {} {\bibfield  {journal} {\bibinfo  {journal}
  {\mnras}\ }\textbf {\bibinfo {volume} {406}},\ \bibinfo {pages} {1220}
  (\bibinfo {year} {2010})},\ \Eprint {http://arxiv.org/abs/0908.2995}
  {arXiv:0908.2995 [astro-ph.CO]} \BibitemShut {NoStop}%
\bibitem [{\citenamefont {Wood}\ \emph {et~al.}(2013)\citenamefont {Wood},
  \citenamefont {Buckley}, \citenamefont {Digel}, \citenamefont {Funk},
  \citenamefont {Nieto} \emph {et~al.}}]{Wood:2013taa}%
  \BibitemOpen
  \bibfield  {author} {\bibinfo {author} {\bibnamefont {Wood}, \bibfnamefont
  {M.}}, \bibinfo {author} {\bibnamefont {Buckley}, \bibfnamefont {J.}},
  \bibinfo {author} {\bibnamefont {Digel}, \bibfnamefont {S.}},  \emph
  {et~al.},\ }\href@noop {} {\enquote {\bibinfo {title} {{Prospects for
  Indirect Detection of Dark Matter with CTA}},}\ } (\bibinfo {year} {2013}),\
  \bibinfo {note} {white paper contribution for Snowmass 2013 in the Cosmic
  Frontier Working Group ``CF2: WIMP Dark Matter Indirect Detection''},\
  \Eprint {http://arxiv.org/abs/1305.0302} {arXiv:1305.0302 [astro-ph.HE]}
  \BibitemShut {NoStop}%
%%CITATION = ARXIV:1305.0302;%%
\bibitem [{\citenamefont {{Wu}}\ and\ \citenamefont
  {{Chang}}(2013)}]{DAMPE_ICRC2013}%
  \BibitemOpen
  \bibfield  {author} {\bibinfo {author} {\bibnamefont {{Wu}}, \bibfnamefont
  {J.}}\ and\ \bibinfo {author} {\bibnamefont {{Chang}}, \bibfnamefont {J.}},\
  }in\ \href@noop {} {\emph {\bibinfo {booktitle} {ICRC 2013, Rio de Janeiro,
  Brazil}}}\ (\bibinfo {year} {2013})\BibitemShut {NoStop}%
\bibitem [{\citenamefont {Wu}\ \emph {et~al.}(2014)\citenamefont {Wu},
  \citenamefont {Su}, \citenamefont {Bravar}, \citenamefont {Chang},
  \citenamefont {Fan} \emph {et~al.}}]{Wu:2014tya}%
  \BibitemOpen
  \bibfield  {author} {\bibinfo {author} {\bibnamefont {Wu}, \bibfnamefont
  {X.}}, \bibinfo {author} {\bibnamefont {Su}, \bibfnamefont {M.}}, \bibinfo
  {author} {\bibnamefont {Bravar}, \bibfnamefont {A.}},  \emph {et~al.},\
  }\href {\doibase 10.1117/12.2057251} {\bibfield  {journal} {\bibinfo
  {journal} {Proc.SPIE Int.Soc.Opt.Eng.}\ }\textbf {\bibinfo {volume} {9144}},\
  \bibinfo {pages} {91440F} (\bibinfo {year} {2014})},\ \Eprint
  {http://arxiv.org/abs/1407.0710} {arXiv:1407.0710 [astro-ph.IM]} \BibitemShut
  {NoStop}%
%%CITATION = ARXIV:1407.0710;%%
\bibitem [{\citenamefont {Xue}\ \emph {et~al.}(2008)\citenamefont {Xue} \emph
  {et~al.}}]{Xue:2008se}%
  \BibitemOpen
  \bibfield  {author} {\bibinfo {author} {\bibnamefont {Xue}, \bibfnamefont
  {X.}} \emph {et~al.} (\bibinfo {collaboration} {SDSS Collaboration}),\ }\href
  {\doibase 10.1086/589500} {\bibfield  {journal} {\bibinfo  {journal}
  {Astrophys.J.}\ }\textbf {\bibinfo {volume} {684}},\ \bibinfo {pages} {1143}
  (\bibinfo {year} {2008})},\ \Eprint {http://arxiv.org/abs/0801.1232}
  {arXiv:0801.1232 [astro-ph]} \BibitemShut {NoStop}%
%%CITATION = ARXIV:0801.1232;%%
\bibitem [{\citenamefont {{Yoshida}}(2013)}]{CALET_ICRC2013}%
  \BibitemOpen
  \bibfield  {author} {\bibinfo {author} {\bibnamefont {{Yoshida}},
  \bibfnamefont {K.}} (\bibinfo {collaboration} {CALET collaboration}),\ }in\
  \href@noop {} {\emph {\bibinfo {booktitle} {ICRC 2013, Rio de Janeiro,
  Brazil}}}\ (\bibinfo {year} {2013})\BibitemShut {NoStop}%
\bibitem [{\citenamefont {Zeldovich}\ \emph {et~al.}(1980)\citenamefont
  {Zeldovich}, \citenamefont {Klypin}, \citenamefont {Khlopov},\ and\
  \citenamefont {Chechetkin}}]{Zeldovich:1980st}%
  \BibitemOpen
  \bibfield  {author} {\bibinfo {author} {\bibnamefont {Zeldovich},
  \bibfnamefont {Y.}}, \bibinfo {author} {\bibnamefont {Klypin}, \bibfnamefont
  {A.}}, \bibinfo {author} {\bibnamefont {Khlopov}, \bibfnamefont {M.~Y.}}, \
  and\ \bibinfo {author} {\bibnamefont {Chechetkin}, \bibfnamefont {V.}},\
  }\href@noop {} {\bibfield  {journal} {\bibinfo  {journal} {Sov.J.Nucl.Phys.}\
  }\textbf {\bibinfo {volume} {31}},\ \bibinfo {pages} {664} (\bibinfo {year}
  {1980})}\BibitemShut {NoStop}%
%%CITATION = SJNCA,31,664;%%
\bibitem [{\citenamefont {Zimmer}, \citenamefont {Conrad},\ and\ \citenamefont
  {Pinzke}(2011)}]{Zimmer2011}%
  \BibitemOpen
  \bibfield  {author} {\bibinfo {author} {\bibnamefont {Zimmer}, \bibfnamefont
  {S.}}, \bibinfo {author} {\bibnamefont {Conrad}, \bibfnamefont {J.}}, \ and\
  \bibinfo {author} {\bibnamefont {Pinzke}, \bibfnamefont {A.}} (\bibinfo
  {collaboration} {Fermi LAT Collaboration}),\ }in\ \href
  {http://arxiv.org/abs/1110.6863} {\emph {\bibinfo {booktitle} {Proceedings of
  2011 Fermi Symposium}}}\ (\bibinfo {address} {Roma, Italy},\ \bibinfo {year}
  {2011})\ p.~\bibinfo {pages} {5},\ \bibinfo {note} {{arXiv}:
  1110.6863}\BibitemShut {NoStop}%
\bibitem [{\citenamefont {Zitzer}\ \emph {et~al.}(2013)\citenamefont {Zitzer}
  \emph {et~al.}}]{1307.8367v1}%
  \BibitemOpen
  \bibfield  {author} {\bibinfo {author} {\bibnamefont {Zitzer}, \bibfnamefont
  {B.}} \emph {et~al.} (\bibinfo {collaboration} {Veritas Collaboration}),\
  }in\ \href {http://arxiv.org/abs/1307.8367v1} {\emph {\bibinfo {booktitle}
  {Proceedings of the 33rd International Cosmic Ray Conference (ICRC2013), Rio
  de Janeiro (Brazil)}}}\ (\bibinfo {year} {2013})\ \Eprint
  {http://arxiv.org/abs/1307.8367v1} {arXiv:1307.8367v1 [astro-ph.HE]}
  \BibitemShut {NoStop}%
\end{thebibliography}%

\bibliographystyle{aipauth4-1}

\end{document}